\documentclass[11pt]{report}

\usepackage[utf8]{inputenc}   	
\usepackage[english]{babel}
\usepackage{graphicx}
\usepackage{float}
\usepackage{color}      
\usepackage{amssymb}
\usepackage{amsmath}
\usepackage{slashed}
\usepackage{enumerate}
\usepackage{subcaption}
\usepackage[font=small]{caption}    
\usepackage{booktabs} 
\usepackage{setspace}
\usepackage{fancyhdr}
\usepackage{tikz}
\usetikzlibrary{arrows,matrix}
\usepackage{cite}
\usepackage{color}
\usepackage[colorlinks=true
,urlcolor=blue
,anchorcolor=blue
,citecolor=blue
,filecolor=blue
,linkcolor=blue
,menucolor=blue
,pagecolor=blue
,linktocpage=true
,pdfproducer=medialab
,pdfa=true
]{hyperref}

\usepackage{geometry}
\usepackage{multirow}
\usepackage{booktabs}

\newcommand\POWHEGBOX{{\tt POWHEG BOX}}
\newcommand\POWHEG{{POWHEG}}

\newcommand\Pythia{{\tt Pythia}}

\newcommand\PythiaEight{{\tt Pythia8}}

\newcommand\HerwigSevenPone{{\tt Herwig7.1}}

\newcommand\PythiaEightPtwo{{\tt Pythia8.2}}

\newcommand\RES{{\tt POWHEG BOX RES}}
\newcommand\VTWO{{\tt POWHEG BOX V2}}

\newcommand{\mathd}{\mathrm{d}}

\newcommand{\tmop}[1]{\ensuremath{\operatorname{#1}}}

\newcommand\sss{\mathchoice%
{\displaystyle}%
{\scriptstyle}%
{\scriptscriptstyle}%
{\scriptscriptstyle}%
}
\newcommand{\pt}{\ensuremath{p_{\perp}}}

\newcommand{\muF}{\ensuremath{\mu_{\sss \mathrm{F}}}}
\newcommand{\muR}{\ensuremath{\mu_{\sss \mathrm{R}}}}
\newcommand{\KF}{\ensuremath{K_{\sss \mathrm{F}}}}
\newcommand{\KR}{\ensuremath{K_{\sss \mathrm{R}}}}

\def\beq{\begin{equation}}
\def\beqn{\begin{eqnarray}}
\def\eeq{\end{equation}}
\def\eeqn{\end{eqnarray}}

\def\lq{\left[} 
\def\rq{\right]} 
\def\rg{\right\}} 
\def\lg{\left\{} 
\def\({\left(} 
\def\){\right)}

\newcommand\as{\alpha_{\sss\rm S}}

\newcommand\CA{C_{\sss\rm A}}
\newcommand\CF{C_{\sss\rm F}}
\newcommand\TF{T_{\sss\rm F}}
\newcommand\TR{T_{\sss\rm R}}

\newcommand\mb{m_{b}}
\newcommand\mZ{m_{\sss Z}}
\newcommand\mW{m_{\sss W}}
\newcommand\mt{m_{t}}
\newcommand\mtc{m_{t,\, c}}

\newcommand\fourl{e^+\nu_{\sss e}\, \mu^-\bar{\nu}_{\sss \mu}}

\newcommand\pwgopt[1]{{\tt #1}}
\newcommand\bjet{\ensuremath{b}-jet}

\newcommand\mwbj{\ensuremath{m_{Wb_j}}}
\newcommand\mwbjmax{\ensuremath{m_{Wb_j}^{\max}}}
\newcommand\Ebj{\ensuremath{E_{b_j}}}
\newcommand\Ebjmax{\ensuremath{E_{b_j}^{\max}}}
\newcommand\pT{\ensuremath{p_{\perp}}}

\newcount\minutes 
\newcount\scratch 
\def\timestamp{%
\scratch=\time 
\divide\scratch by 60 
\edef\hours{\the\scratch} 
\multiply\scratch by 60 
\minutes=\time 
\advance\minutes by -\scratch 
---$\,$\hours:\null 
\ifnum\minutes< 10 0\fi 
\the\minutes}

\definecolor{mygray}{gray}{0.5}

\newcommand\aprod{\ensuremath{\alpha_{\rm \scriptscriptstyle ISR}}}

\newcommand\obs{{ O}}
\newcommand\obsb{{ \langle O \rangle_{\rm b}}}

\newcommand\ep{\epsilon}
\newcommand{\mpole}{\ensuremath{m}}
\newcommand{\mpolec}{\ensuremath{m^c}}

\newcommand{\mMSB}{\ensuremath{\overline{m}}}
\newcommand{\MSB}{\ensuremath{\overline{\rm MS}}}

\newcommand{\gameul}{{\gamma_{\sss E}}}

\newcommand\PythiaEightPlot{{\tt Py8.2}}
\newcommand\HerwigSevenPlot{{\tt Hw7.1}}

\newcommand\ttbnlodecPlot{$t\bar{t}dec$}
\newcommand\bbllllPlot{$b\bar{b}4\ell$}
\newcommand\hvqPlot{$hvq$}

\newcommand\hvq{\hvqPlot}
\newcommand\bbfourl{\bbllllPlot}
\newcommand\ttNLOdec{\ttbnlodecPlot}
\newcommand\ttbnlodec{\ttNLOdec}

\newcommand\writeApp{Appendix~}

\newlength{\wfigsing}
\newlength{\wfigsingmulti}
\newlength{\wfigdoub}
\newlength{\wtablarge}

\newcommand\fignewline{}

\begin{document}    
\pagenumbering{Roman}
\thispagestyle{empty}
        \vspace{0.5cm}
        \begin{center}
		\begin{large}
			{
			
				\huge 
				Università degli Studi 
				di Milano\,--\,Bicocca \\
				\vskip 0.8 cm
				{\Large Dipartimento di Fisica G. Occhialini \\}	
			}
				\vskip 0.6 cm
				{\Large PhD program in Physics and Astronomy, \; XXXI cycle \\
				Curriculum in Theoretical Physics\\}
			\vspace{1.5cm}
			\includegraphics[width=0.35\textwidth]{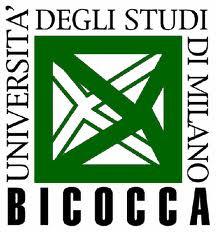} \\
		\end{large}
		\vspace{1.3cm}
		{
			\huge
			\textbf{Top-mass observables: all-orders behaviour,
				renormalons and NLO + Parton Shower effects} \\
			\vspace{0.6cm}
			Silvia Ferrario Ravasio \\
			\vspace{0.4cm}
			{\large Matricola: 735192}
			\vspace{0.6cm}
		}
		{
			\Large 
		}
	\end{center}
	\vskip 0.8 cm 
	\begin{tabular}{ll}
		\Large \bf Advisor:  & \Large Prof. Carlo Oleari \\
		\Large \bf Co-advisor: &\Large Prof. Paolo Nason \\				
		\\
		\large \bf Coordinator: & \large Prof. Marta Calvi \\
		\\
\hspace{3cm} &\hspace{7cm}	\large		Academic year: 2017/2018\\
	\end{tabular}    
\clearpage

\newpage
\thispagestyle{empty}
$ $
\clearpage



\newpage
\chapter*{\centering Declaration}
\thispagestyle{empty}
	This dissertation is a result of my own efforts.
	The work to which it refers is based on my PhD research projects:
	\begin{enumerate}
 		\item ``A Theoretical Study of Top-Mass Measurements at the
                  LHC Using NLO+PS Generators of Increasing Accuracy,'' with T.~Je\v{z}o,
                  P.~Nason and C.~Oleari,\\ {\bf Eur.Phys.J. C78 (2018) no.6, 458}
                  \href{https://arxiv.org/abs/1801.03944}{[arXiv:1801.03944v2 [hep-ph]]}
			
 		\item ``All-orders behaviour and renormalons in top-mass
                  observables'' with P.~Nason and C.~Oleari, \\   JHEP {\bf 1901} (2019) 203
                  \href{https://arxiv.org/abs/1810.10931}{[arXiv:1801.10931 [hep-ph]]}
	\end{enumerate}
	I hereby declare that except where specific reference is made to
	the work of others, the contents of this dissertation are original
	and have not been submitted in whole or in part for consideration
	for any other degree or qualification in this, or any other
	university.
	\begin{flushright}
		Silvia Ferrario Ravasio, \\
		31${}^{\rm st}$ October 2018
	\end{flushright}
\clearpage

\pagenumbering{roman}
\thispagestyle{empty}
\newgeometry{top=-10mm}
\chapter*{\centering Abstract}
{
In this thesis we focus on the theoretical subtleties of the top-quark mass~($m_t$) determination, issue which persists in being highly controversial.  

Typically, in order to infer the top mass, theoretical predictions dependent
on $m_t$ are employed. The parameter $m_t$ is the physical mass, that is
connected with the bare mass though a renormalization procedure. Several
renormalization schemes are possible and the most natural seems to be the
pole-mass one. However, the pole mass is not very well defined for a coloured
object like the top quark. The pole mass is indeed affected by the presence
of infrared renormalons. They manifest as factorially growing coefficients
that spoil the convergence of the perturbative series, leading to ambiguities
of order of $\Lambda_{\sss \rm QCD}$. On the other hand, short-distance mass
schemes, like the \MSB{}, are known to be free from such renormalons.
Luckily, the renormalon ambiguity seems to be safely below the quoted
systematic errors on the pole-mass determinations, so these measurements are
still valuable. In the first part of the thesis, we investigate the presence
of linear renormalons in observables that can be employed to determine the
top mass. We considered a simplified toy model to describe $W^{*}
\rightarrow t \bar{b} \to Wb \bar{b}$. The computation is carried out in the
limit of a large number of flavours~($n_f$), using a new method that allows
to easily evaluate any infrared safe observable at order $\as(\as n_f)^n$ for any $n$. The observables we consider are, in general,
affected by two sources of renormalons: the pole-mass definition and the jet
requirements. We compare and discuss the predictions obtained in the usual
pole scheme with those computed in the \MSB{} one.
We find that the total cross section without cuts, when expressed in terms of
the \MSB{} mass, does not exhibit linear renormalons, but, as soon as
selection cuts are introduced, jet-related linear renormalons arise in any
mass scheme. In addition, we show that the reconstructed mass is affected by
linear renormalons in any scheme. The average energy of the $W$ boson (that
we consider as a simplified example of leptonic observable) has a renormalon
in the narrow-width limit in any mass scheme, that is however screened at
large orders for finite top widths, provided the top mass is in the \MSB{}
scheme.

The most precise determinations of the top mass are the direct ones, i.e.
those that rely upon the reconstruction of the kinematics of the top-decay
products. Direct determinations are heavily based on the use of Monte Carlo
event generators. The generators employed must be as much accurate as
possible, in order not to introduce biases in the measurements. To this
purpose, the second part of the thesis is devoted to the comparison of
several NLO generators, implemented in the {\tt POWHEG BOX} framework, that
differ by the level of accuracy employed to describe the top decay. The
impact of the shower Monte Carlo programs, used to complete the NLO events
generated by {\tt POWHEG BOX}, is also studied. In particular, we discuss the
two most widely used shower Monte Carlo programs, i.e. \PythiaEightPtwo{} and
\HerwigSevenPone{}, and we present a method to interface them with processes
that contain decayed emitting resonances. The comparison of several Monte
Carlo programs that have formally the same level of accuracy is, indeed, a
mandatory step towards a sound estimate of the uncertainty associated with
$m_t$.}
\clearpage
\restoregeometry

\clearpage

\pagestyle{fancy}
\renewcommand{\chaptermark}[1]{%
\markboth{\MakeUppercase{}Chap\ \thechapter.%
\ #1}{}}
\renewcommand{\sectionmark}[1]{%
\markboth{\MakeUppercase{}Sec\ \thesection.%
\ #1}{}}

\lhead{}
\chead{}
\rhead{\leftmark}
\lfoot{}
\cfoot{\thepage}
\rfoot{}
\renewcommand{\headrulewidth}{0.4pt}
\clearpage

\tableofcontents
\newpage
 \newpage
 \clearpage
 
\pagenumbering{arabic}

\chapter*{\centering Introduction}
\addcontentsline{toc}{chapter}{Introduction}

The top quark is the heaviest elementary particle in the Standard Model~(SM)
that has been observed so far. It thus appears clear that its phenomenology
is driven by the large value of its mass $\mt$. Indeed, the top is the only
quark that decays before hadronizing. This provides us the unique occasion to
study the properties of a ``bare'' quark. For these reasons an accurate
determination of $\mt$ is part of the Large Hadron Collider~(LHC) physics
program.

Through radiative corrections, the top-quark mass has a non-negligible impact
on many parameters of the Standard Model, like the masses of the electroweak
bosons and the Higgs self-coupling.  Thus, the value of the $Z$~(or the $W$)
mass is sensitive to the value of the top-quark one. For this reason, the
electroweak data enable us to have a simultaneous determination of the top
and of the $Z$-boson masses and the strong coupling $\as(M_Z)$. The extracted
value of $\mt$ is $176.7\pm2.1$~GeV~\cite{Patrignani:2016xqp-EWreview}, which
is in slight tension with the value of $173.34\pm 0.76$~GeV, i.e.~the latest
Tevatron and the LHC combined results~\cite{ATLAS:2014wva}.  In addition the
top-quark mass is one key ingredient to address the issue of vacuum stability
\cite{Degrassi:2012ry, Buttazzo:2013uya, Andreassen:2017rzq,
  Chigusa:2017dux}.  Under the assumption that there is no new physics up to
the Planck scale, the Higgs self coupling $\lambda(\mu)$ is always positive
during its renormalization-group flow for each scale $\mu$ adopted, if
$\mt<171$~GeV.  If instead $171<\mt<176$~GeV, we are in the metastability
region, since $\lambda(\mu)$ becomes negative only at scales of the order of
the Planck scale. Thus, there is no indication of new physics below the
Planck scale coming from the requirement of vacuum stability.

The most precise determinations of $\mt$ are the so called ``direct
measurements'', which rely upon the full or partial reconstruction of
the top momentum out of its decay products.  Kinematic distributions
sensitive to the top-quark mass are compared to Monte Carlo
predictions and the $\mt$ value that fits the data the best is the
extracted top-quark mass.  The ATLAS and CMS measurements of
Refs.~\cite{Aaboud:2016igd} and~\cite{Khachatryan:2015hba}, yielding
the values $\mt = 172.84 \pm 0.34~{\rm(stat)} \pm 0.61$~(syst)~GeV and
$\mt=172.44 \pm 0.13~{\rm (stat)} \pm 0.47$~(syst)~GeV respectively,
fall into this broad category.  Of course this kind of determinations
is affected by theoretical errors that must be carefully assessed. If
the Monte Carlo used to simulate the distributions is not accurate
enough, it introduces a bias in the determination of $\mt$.  For this
reason, many efforts have been done in order to implement
next-to-leading-order~(NLO) generators capable to handle processes
containing a decayed emitting resonance, like the top quark is. We
will discuss this issue in the second part of the thesis.

However, in contrast with the increasing experimental precision of the
top-mass measurements at the LHC, the theoretical interpretation is still
matter of debate. In Ref.~\cite{Hoang:2014oea} it was argued that the Monte
Carlo mass parameter does not coincide with the top-pole mass and their
difference is unavoidable due to the intrinsic ambiguity of the pole-mass
definition. Indeed, since the top is quark is always colour-connected with
another particle, an isolated top-quark cannot exist. This leads to a
renormalon in the relation of the pole to the \MSB~mass~\cite{Bigi:1994em,
  Beneke:1994sw}.  Nevertheless, the renormalon ambiguity does not seem
severe for the specific case of the top quark, since recent
studies~\cite{Beneke:2016cbu, Hoang:2017btd} have shown that it is in fact
well below the current experimental error.  In any case non-perturbative
corrections to top-mass observables (not necessarily related to the mass
renormalon) are present and must be carefully assessed. The top-quark mass
renormalon and its interplay with the renormalon arising from the use of
jets~\cite{Dasgupta:2007wa} is discussed in the first part of the thesis.


\part{Renormalons and all-orders behaviour in top-mass sensitive observables}

\chapter{Introduction}

The top mass is measured quite precisely at the LHC by both the
ATLAS~\cite{Pearson:2017jck} and the CMS~\cite{Castro:2017yxe}
Collaborations. Up to now, the methods that yield the most accurate results
are the so called ``direct'' methods, where kinematic distribution obtained
reconstructing fully or partially the top decay products are compared to
templates produced with Monte Carlo event generators.

Current uncertainties are now near 500~MeV~\cite{Aaboud:2016igd,
  Khachatryan:2015hba}, so that one can worry whether QCD non-perturbative
effects may substantially affect the result. In fact, the experimental
collaborations estimate these and other effects by varying parameters in the
generators, and eventually comparing different generators.  This method has
been traditionally used in collider physics to estimate theoretical
uncertainties due to the modelling of hadronization and underlying events,
and also to estimate uncertainties related to higher perturbative orders, as
produced by the shower algorithms. We perform a similar study in the second
part of the thesis. This is a valuable strategy, as long as all the
generators under comparison can reproduce faithfully the data.\footnote{If
  such a statement fails to be true, the bad-behaved generator must be
  discarded from the comparison, as discussed in the second part of the
  thesis.}  However, it should not be forgotten that it may only provide a
lower bound on the associated errors.  However, it should not be forgotten
that this statement is true only if all the generators under comparison.  It
is thus important, at the same time, to investigate the associated
uncertainties from a purely theoretical point of view. In consideration of
our poor knowledge of non-perturbative QCD, these investigations can at most
have a qualitative value, but may help us to understand sources of
uncertainties that we might have missed. One such work is presented in
Ref.~\cite{Hoang:2018zrp}, where the authors attempt to relate a
theoretically well-defined mass parameter with a corresponding shower Monte
Carlo one, using as observable the jet mass of a highly boosted top.

We consider the interplay of non-perturbative effects with the behaviour of
perturbative QCD at large orders in the coupling constant, focusing in
particular upon observables that, although quite simple, may be considered of
the kind used in ``direct measurements''.

It is known that, in renormalizable field theories, the renormalization group
flow of the couplings leads to the so called renormalons, i.e. to the
factorial growth of the coefficients of the perturbative expansion as a
function of the order~\cite{Gross:1974jv, Lautrup:1977hs,
 tHooft:1979gj,Parisi:1978az, Mueller:1984vh, Mueller:1992xz,
 Altarelli:1995kz, Beneke:1998ui}. Renormalons lead to a divergence of the
perturbative expansion, that thus becomes asymptotic. In particular, in the
case of \emph{infrared} renormalons in asymptotically-free field theories,
the ambiguity in the summation of the series corresponds to a power
suppressed effect.

Renormalons were originally found in two-point function
diagrams~\cite{Gross:1974jv, Lautrup:1977hs,tHooft:1977xjm}. These
contributions are sometimes identified with renormalons in the so-called
large (and negative) number of flavour $n_f$ limit. We consider a fictitious
process $W^* \rightarrow t\bar{b} \to W b \bar{b}$, where the $W$ boson has
only a vector coupling to quarks, and examine the behaviour of the cross
section, of the reconstructed-top mass and of the energy of the $W$ boson,
order by order in the strong coupling expansion, taking the large-$n_f$
limit. We consider up to one gluon exchange, or emission, and dress this
gluon with an arbitrary number of fermion vacuum-polarization
insertions. Furthermore, we also consider final states where the gluon has
undergone a splitting into a fermion-antifermion pair, corresponding to a cut
vacuum polarization diagram. We assume a finite width for the top quark.

We have devised a method that allowed us to compute in principle any
observable in our process, without further approximations, making use of
simple numerical techniques. We can thus compute the
perturbative expansion at any finite order and infer its asymptotic nature
for any observable, with the only limitation of the numerical
precision.

We focus for simplicity upon simple top-mass observables, such as the
production cross section with or without cuts, the reconstructed-top mass,
defined as the mass of a system comprising the $W$ and a $b$ (not $\bar{b}$)
jet, and, as a simplified example of leptonic observable, the average value
of the energy of the final-state $W$ boson. As discussed earlier, we consider
our reconstructed-top mass as an oversimplified representation of
observables of the kind used in the so called``direct''
measurements. We also stress that we consider the kinematic region where
the top energy is not much larger than its mass, that is the region
typically used in direct measurements.

We know that there are renormalons arising in the computation of the position
of the pole in the top propagators, and we also know that there must be
renormalons associated to jets requirements. Since in our framework we can
compute the perturbative expansion order by order in perturbation theory, we
are in the position to determine explicitly the effects of renormalons in the
perturbative expansion.

Our results can be given in terms of the top mass expressed either in the
pole or in the \MSB{} mass scheme. We know that the expression of the pole
mass in terms of the \MSB{} mass has a linear renormalon. If the \MSB{} mass
is considered a fundamental parameter of the theory, this is to be interpreted
as an uncertainty of the order of a typical hadronic scale associated to the
position of the pole in the top propagator. One may wonder whether the pole
mass could instead be used as a fundamental parameter of the theory, which
would imply that the \MSB{} mass has an uncertainty of the order of a
hadronic scale. In fact, it is well known and clear (but nevertheless we wish
to stress it again) that this last point of view is incorrect. QCD is
characterized by a short distance Lagrangian, and its defining parameters are
short distance parameters. Thus, if we compute an observable in terms of the
\MSB{} mass, and we find that it has no linear renormalons, we can conclude
that the observable has no \emph{physical} linear renormalons, since its
perturbative expansion in terms of the parameters of the short distance
Lagrangian has no linear renormalons. On the other end, in the opposite case
of an observable that has no linear renormalons if expressed in terms of the
pole mass, we must conclude that this observable has a physical renormalon,
that is precisely the one that is contained in the pole mass. We also stress
that it is the \MSB{} mass that should enter more naturally in the
electroweak fits~\cite{Patrignani:2016xqp-EWreview, deBlas:2016ojx,
 Baak:2014ora} and in the calculations relative to the stability of the
vacuum~\cite{Degrassi:2012ry, Buttazzo:2013uya, Andreassen:2017rzq,
 Chigusa:2017dux}, although in practice the pole mass is often used also in
these contexts.

The outline of the first part of the thesis is the following. In
Chap.~\ref{sec:renormalons} we describe some notions strictly connected with
the renormalon issues. In particular we present the physical argument given
by Dyson to show that perturbation expansions are not convergent in quantum
field theory. We then give a formal definition of asymptotic series and of
the Borel transform. In Chap.~\ref{sec:large_nf} we discuss the large-$n_f$
limit, where higher order corrections are accessible up to all orders in the
coupling. As first application, we illustrate the computation of the relation
between the pole and the \MSB{} mass scheme. We also present a possible
solution to move from the large-$n_f$ limit, that portrays QCD as an Abelian
theory, to a more realistic large-$b_0$ limit, where $b_0$ is the first
coefficient of the full QCD $\beta$ function, in order to recover the
asymptotic freedom behaviour of the theory. In
Chap.~\ref{sec:description_calc} we explicitly illustrate the steps for the
computation of the fictitious process $W^* \rightarrow t\bar{b} \to W b
\bar{b}$ in the large number of flavours limit, using the complex pole mass
scheme~\cite{Denner:2005fg, Denner:2006ic} for the normalization of the top
mass. We also show how to rearrange the expression in terms of the \MSB{}
mass, that can be considered as a proxy for all short-distances mass schemes.
In Chap.~\ref{sec:linear-mg} we discuss the presence of infrared linear
renormalons in the inclusive cross section, the reconstructed top-mass and the
energy of the final-state $W$ boson. We also compare the small-momentum
behaviour of such observables computed in the pole scheme with the behaviour
achieved by expressing them in terms of a short-distance mass. In
Chap.~\ref{sec:AllOrderExp} we present the coefficients of the perturbative
expansion in $\as$ of the above-mentioned observables. Finally, we draw
our conclusions in Chap.~\ref{sec:concl_renormalons}. Some technical details
are discussed in Appendices.

The results we present in this first part of the thesis can be found also in
Ref.~\cite{FerrarioRavasio:2018ubr}.

\chapter{Generalities on divergent series and on the renormalon concept}
\label{sec:renormalons}

We now illustrate some basic concepts relative to the physics of infrared renormalons.

\section{Dyson's argument}
\label{sec:dyson}
Dyson in 1952 showed, with a simple and intuitive physical argument, that
perturbative expansions cannot converge in quantum field
theory~\cite{Dyson:1952tj}.
 
We can consider, for example,a generic observable in QED given by a
perturbation expansion in $e^2$:
\begin{equation}
 O(e^2) = \sum_{n=n_{\rm min}}^\infty r_n \,(e^2)^{n}.
\end{equation}
The expansion is performed around the value $e^2=0$. If the series converges,
then there would be a radius of convergence around $e^2=0$. This implies a
convergent result also for small and negative values of $e^2$. Negative
values of $e^2$ would correspond to a force that is repulsive for opposite
charges and attractive for equal charges. ``By creating a large number $N$
of electron-positron pairs, bringing the electrons together in one region of
space and the positrons in another separate region, it is easy to construct a
``pathological'' state in which the negative potential energy of the Coulomb
forces is much greater than the total rest energy and kinetic energy of the
particles''~\cite{Dyson:1952tj}. This corresponds to a state with unbounded
negative energy, that implies the absence of a stable vacuum.
 
We thus conclude that, since a convergence for negative $e^2$ is impossible
because the corresponding theory is meaningless, the radius of convergences
of the series is zero.

\section{Divergent series}
\label{sec:series}
As we have seen in Sec.~\ref{sec:dyson}, perturbative series are divergent in
quantum field theory. In particular, one may ask whether is possible to
assign a ``sum'' to the series. We consider an observable $O$ written in
powers of $\alpha$
\begin{equation}
\label{eq:series}
O \sim \sum_{n=0}^\infty c_n \alpha^{n+1}.
\end{equation}
We interpret the series as an \emph{asymptotic} series in a region
$\mathcal{C}$ of the complex $\alpha$-plane if
for each order $N$ there are numbers $K_N$ that satisfy
\begin{equation}
\left| O - \sum_{n=0}^{N-1} c_n \alpha^{n+1} \right| < K_{N} \alpha^{N+1}
\end{equation}
for all $\alpha$ in $\mathcal{C}$.
Let us consider a factorially divergent series
\begin{equation}
\label{eq_cn_largen}
c_n \xrightarrow{n \to \infty} \mathcal{N} a^n \Gamma(1+b+n)\,,
\end{equation}
with $\mathcal{N}$, $a$ and $b$ constant. When small values of $n$ are
concerned, the terms $c_n\,\alpha^{n+1}$ decrease for increasing $n$.
However, for large values of $n$ the coefficients $c_n$ behaves as
in~(\ref{eq_cn_largen}). When a large value $N^\prime$ is reached such that
\begin{equation}
\left| c_{N^\prime-1} \alpha^{N^\prime} \right| \approx \left| c_{N^\prime} \alpha^{N^\prime+1} \right|,
\end{equation}
i.e. for
\begin{equation}
N^\prime \approx \frac{1}{|a|\alpha},
\end{equation}
the series of $c_n \alpha^{n+1}$ reaches its minimum and then starts growing.
The best approximation of the sum of the series is provided when the truncation error is minimum, i.e. 
\begin{eqnarray}
\label{eq:truncation}
c_{N^\prime} \alpha^{N^\prime+1} &=& \mathcal{N} \alpha (\alpha\, a)^{\frac{1}{|a|\alpha}}
 \Gamma\left(1+b+\frac{1}{|a|\alpha}\right) \nonumber \\
 &\approx& \mathcal{N} \alpha \,\left(|a|\alpha\right)^{-b-\frac{1}{2}} \exp\left(-\frac{1}{|a|\alpha}\right)\,.
\end{eqnarray}
To improve this approximation, we can employ the \emph{Borel summation} technique. 
Given the series in eq.~(\ref{eq:series}) with $\alpha>0$, its Borel transform is given by
\begin{equation}
B\left[O\right](t) = \sum_{n=0} \frac{c_n}{n!} t^n
\end{equation}
and the Borel integral is defined as
\begin{equation}
\widetilde{O} = \int_0^{+\infty} \mathd t\, e^{-t/\alpha} B\left[O\right](t)
= \alpha \int_0^{+\infty} \mathd t\, e^{-t} B\left[O\right](\alpha \,t).
\label{eq:Borel_integral}
\end{equation}
It can be shown that the Borel integral has the same $\alpha$-expansion of~(\ref{eq:series}) and thus,
if $\widetilde{O}$ exists, it can be interpreted as the sum of the divergent series.
This is particularly useful for alternate sign factorially growing series.
We consider the following series and its Borel integral:
\begin{equation}
O = \mathcal{N}\sum_{i=0}^\infty a^n \,\Gamma(1+b+n)\, \alpha^{n+1}
\rightarrow(a\,\alpha)^{-b} \,\frac{\mathcal{N}}{a} \int_0^{+\infty} \mathd t\, e^{-t}\,
\Gamma(1+b)\left[\frac{1}{a\,\alpha}-t\right]^{-1-b}.
\end{equation}
where we have assumed non negative $b$ values. The integral is well defined
if $a<0$, i.e. for an alternated-sign series. For positive $a$ values there
is a pole on the integration path at $t=-1/a\alpha$. We stress that the
location of the pole is independent from the value of $b$. We can give a
meaning to the integral by deforming the integration path above or below the
pole, that thus acquires an imaginary part equal to
\begin{equation}
 \mp \frac{\pi\mathcal{N}}{a} (a\alpha)^{-b}\exp\left(-\frac{1}{a\alpha}\right) ,
\end{equation}
where the sign depends on whether the integration is taken in the upper or
lower complex plane.
The ambiguity can be estimated as $1/2\pi$ times the difference between the
two imaginary parts, i.e.
\begin{equation}
 \frac{\mathcal{N}}{a}(a\alpha)^{-b} \exp\left(-\frac{1}{a\alpha}\right) .
 \label{eq:imagBorel}
\end{equation}
By comparing eqs.~(\ref{eq:imagBorel}) and~(\ref{eq:truncation}) we notice
that, in case of same sign factorially divergent series, a small ambiguity
proportional to $ \exp\left[-1/a\alpha\right]$ is unavoidable.

\section{QCD infrared renormalons}
\label{sec:renorm_intro}
Infrared renormalons~\cite{tHooft:1979gj, Parisi:1978az} provide a connection
between the behaviour of the perturbative expansion at large orders in the
coupling constant and non-perturbative effects. They arise when the last
loop integration in the $(n+1)$-loop order of the the perturbative expansion
acquires the form (see e.g.\cite{Altarelli:1995kz, Beneke:1998ui})
\begin{equation}
\label{eq:factgrowth}
 \as^{n+1}(Q) \frac{1}{Q^p} \int^Q \!\mathd k\, k^{p-1}\, b_0^n\,
\(\log \frac{Q^2}{k^2}\)^n
= \frac{n!}{p}\, \left(\frac{2b_0}{p} \right)^n \as^{n+1}(Q)\equiv c_n\,\as^{n+1}(Q)\,,
\end{equation}
where $Q$ is the typical scale involved in the process and $b_0$ is the first
coefficient of the QCD beta function
\begin{equation}
 \label{eq:b_0}
b_0 = \frac{11\, \CA}{12\pi} - \frac{n_f \,\TR }{3\pi}\,.
\end{equation}
The coefficient $b_0$ arises because the running coupling is the source of
the logarithms in eq.~(\ref{eq:factgrowth}). A naive justification of the
behaviour illustrated in eq.~(\ref{eq:factgrowth}) can be given by
considering the calculation of an arbitrary dimensionless observable,
characterized by a scale $Q$, including the effect of the exchange or
emission of a single gluon with momentum $k$, leading to a correction that,
for small $k$, takes the form
\begin{equation}
 \label{eq:irsensitivity}
 \frac{1}{Q^p} \int^Q \!\mathd k\, k^{p-1} \as,
\end{equation}
where $p$ is an integer greater than zero for the result to be
infrared-finite. Assuming that higher order corrections will lead to the
replacement of $\as$ with the running coupling evaluated at the scale
$l$, given by the geometric expansion
\begin{equation}
 \label{eq:alphsrunning}
 \as(k)=\frac{1}{b_0 \log \frac{k^2}{\Lambda_{\sss \rm QCD}^2}}=\frac{\as(Q)}{1-\as(Q)\, b_0
 \log \frac{Q^2}{k^2}} 
 = \sum_0^\infty \as^{n+1}(Q)\, b_0^n \,\log^n \frac{Q^2}{k^2}\,,
\end{equation}
substituting eq.~(\ref{eq:alphsrunning}) into eq.~(\ref{eq:irsensitivity}),
we obtain the behaviour of eq.~(\ref{eq:factgrowth}).

The coefficients of the perturbative expansion displays a factorial
growth. The series then is not convergent and can at most be
interpreted as an asymptotic series. As anticipated in
Sec.~\ref{sec:series}, the terms of the series are smaller and smaller
for low values of $n$, until they reach a minimum and then they start
to diverge with the order. The minimum is reached when
\begin{equation}
 c_{n-1}\, \as^n(Q) \approx c_{n}\, \as^{n+1}(Q)\,,
\end{equation}
that correspond to $n \approx p/(2 b_0 \as(Q))$, and the size of the minimal
term is
\begin{eqnarray}
 \frac{n!}{p} \left(\frac{2 b_0}{p}\right)^{n} \as^{n+1}(Q) &\approx&
 \frac{Q^p}{p}\,\as(Q)\,n^{-n} \(n^{n+1/2} e^{-n}\)
 \nonumber \\
 &\approx& \as(Q) \, \frac{n^{\frac{1}{2}}}{p}
\exp\left(-\frac{p}{2\, b_0\, \as(Q)}\right) \approx \sqrt{\frac{
 \as(Q)}{2b_0\,p}}\left(\frac{\Lambda_{\sss \rm QCD}}{Q}\right)^p\!\!. \qquad
\end{eqnarray}
If we resum the series whose terms are given in~(\ref{eq:alphsrunning}) using
the Borel summation, we will get an ambiguity 
\begin{equation}
 \frac{1}{2\,b_0} \left(\frac{\Lambda_{\sss \rm QCD}}{Q}\right)^p,
\end{equation} 
where we have used eq.~(\ref{eq:imagBorel}) with $\mathcal{N}=1/p$, $b=0$ and
$a=\frac{2 b_0}{p}$. The value of $p$ depends upon the process under
consideration. In this paper, we are interested in \emph{linear} IR
renormalons, corresponding to $p=1$, that can lead to ambiguities in the
measured mass of the top quark of relative order
$\Lambda_{\sss \rm QCD}/m_t$, i.e.~ambiguities of order
$\Lambda_{\sss \rm QCD}$ in the top mass. Larger values of $p$ lead to
corrections of relative order $\Lambda_{\sss \rm QCD}^{p}/m_t^p$, that are totally
negligible.

We will see in Sec.~\ref{sec:large_nf} that the behaviour of the
perturbative series in eq.~(\ref{eq:factgrowth}) arises when
considering the large number of flavours limit. However, if we
include more refinements, the expected
behaviour~\cite{Altarelli:1995kz, Beneke:1998ui} becomes more complicate:
\begin{equation}
c_{n+1}\, \as^{n+1}(Q) \propto \Gamma(1+b+n) \left(\frac{2b_0}{p}
\right)^n \as^{n+1}(Q)
\end{equation}
being $b$ a positive number. As we
have seen in Sec.~\ref{sec:series}, this does not change the location of the
singularity in the Borel plane and still leads to an ambiguity of the
resummed series proportional to $\exp\left(-\frac{p}{2\, b_0\, \as(Q)}\right)
\approx \left(\frac{\Lambda_{\sss \rm QCD}}{Q}\right)^p$.  Thus our reasoning
is not modified.

\chapter{The large-$\boldsymbol{n_f}$ limit}
\label{sec:large_nf}
The full renormalon structure of QCD is not known. There is however a fully
consistent simplified model where higher order corrections are accessible up
to all orders in the coupling, namely the large number of flavours, $n_f$,
limit of QCD.
In this limit, the only higer-order contributions that must be considered
are the insertion of fermion loops in a gluon propagator, since they involve
powers of $n_f \as$. Examples of computations performed in this limit can be
found in Refs.~\cite{Beneke:1994qe, Nason:1995np}.

Unfortunately, the large-$n_f$ limit of QCD does not yield to an
asymptotically free theory, since the first coefficient of the $\beta$
function would be positive if we neglect self-gauge interactions.

However, it is believed that tracing the fermionic contribution to the
$\beta$ function, and, at the end of the computation, making the replacement
\begin{equation} n_f \to -\frac{11 \CA}{4\TR} + n_l\,,
\end{equation} where $\CA=3$, $\TR=1/2$ and $n_l$ is the number of light
flavours, one recovers the correct results. In this way, the first
coefficient of the $\beta$ function computed in the large-$n_f$,
$b_0$, is matched with its full expression:
\begin{equation}
 b_0= \frac{11}{3\pi}\CA - \frac{4}{3\pi}\TF\,,
 \end{equation}
 with $\TF = \TR n_l$. Since there is no formal proof of this
statement, this is just a working hypothesis. Our strategy to retain
the full QCD $\beta$ function is slightly different from the one above
mentioned and it is described in Sec.~\ref{sec:realistic_b0}.

\section{The dressed gluon propagator}
\label{sec:dressed_gluon}
In this section we address more technical details about the dressed gluon
propagator to all orders in the large-$n_f$ limit.

The insertion of an infinite number of self-energy corrections
\begin{equation}
 \Pi^{\mu\nu}(k,\mu) = (-g^{\mu\nu}k^2 + k^\mu k^\nu) \, i\,
 \Pi(k^2,\mu^2)\,,
 \label{eq:Pimunu}
\end{equation}
where $i\eta$ is a small imaginary part coming from the Feynman prescription to
integrate around the poles, along a gluon propagator of momentum $k$,
gives rise to
\begin{equation}
 \label{eq:dressed_propagator}
 \frac{-i}{k^2+i\eta} \, g^{\mu\nu} + \frac{-i}{k^2+i\eta} \, \Pi^{\mu\nu}(k,\mu)\,
 \frac{-i}{k^2+i\eta} + \ldots = -\frac{i}{k^2+i\eta} \, g^{\mu\nu} \frac{1}{1+\Pi(k^2,\mu^2)}\,,
\end{equation}
where we have dropped all the longitudinal terms. The derivation of the exact
$d$-dimensional expression of $\Pi(k^2, \mu^2)$ can be found e.g. in
Ref.~\cite{Grozin:2005yg}. In the limit of large number of flavours,
i.e.~considering only light-quark loops, $\Pi(k^2,\mu^2)$ is given by
\begin{align}
 \label{eq:Pi_unren_MSB}
 \Pi\left(k^2,\mu^2\right)& = \frac{\as \TF}{ \pi} e^{\epsilon\gameul}
 \frac{\Gamma(1+\ep)\, \Gamma^2(1-\ep)}{\Gamma(1-2\ep)}
 \frac{1-\ep}{(3-2\ep)(1-2\ep)} \, \frac{1}{\ep} \left(\frac{-k^2}{\mu^2}\right)^{-\ep}, \\
 & = \frac{\as \TF}{ \pi} \left[\frac{1}{\ep} - \log\(\frac{-k^2}{\mu^2}\)
 +\frac{5}{3} \right]+\mathcal{O}(\ep),
\end{align}
where $\TF=n_f\TR$, $\gameul$ is the Euler-Mascheroni constant and we
implicitly assume $\as=\as(\mu)$.
Eq.~(\ref{eq:Pi_unren_MSB}) can be obtained replacing
\begin{equation}
 \mu^2 \to \mu^2\frac{e^{\epsilon\gameul}}{4\pi}
 \label{eq:mu_replacement}
\end{equation}
in eq.~(4.21) of Ref.~\cite{Grozin:2005yg}, according to the \MSB{} prescription. 

  If $k^2>0$, we must replace $k^2 \to k^2 + i\eta$, where $i\eta$ is a small
imaginary part coming from the Feynman prescription. As a consequence,
$\Pi(k^2, \mu^2)$ develops an imaginary part equal to
\begin{equation}
 \label{eq:imag_Pi}
 {\rm Im} \,\Pi (k^2, \mu^2) = \frac{\as \TF}{3 }.
\end{equation}
The replacement in eq.~(\ref{eq:mu_replacement}) is particularly convenient
since it enables us to absorb in the counterterm only the (UV) divergent
part of $\Pi$
\begin{equation}
 \label{eq:Pi_ct}
 \Pi_{\tmop{ct}} = \frac{\as \TF}{3 \pi} \frac{1}{\epsilon}\,.
\end{equation}
The renormalized gluon propagator dressed with the sum of all quark-loop
insertion is then given by
\begin{equation}
 \label{eq:dress_ren_prop}
 -\frac{i}{k^2+i\eta} \, g^{\mu\nu} \frac{1}{1+\Pi(k^2,\mu^2)- \Pi_{\tmop{ct}}}\,,
\end{equation}
where
\begin{eqnarray}
 \label{eq:ren_Pi_expansion}
 \Pi (k^2, \mu^2) - \Pi_{\tmop{ct}} &=& \frac{\as \TF}{3 \pi} \lq
 \frac{5}{3} - \log \left(\frac{|k^2|}{\mu^2}\right) + i\pi\,\theta(k^2) \rq+
 {\cal O}(\ep)
 \\
 \label{eq:Pi_C}
 &=& \as \,b_{0,f}\left[
 \log\left(\frac{|k^2|}{\mu^2\,e^C}\right) -i\pi\theta\left(k^2\right) \right]+
 {\cal O}(\ep),
\end{eqnarray}
and we have defined
\begin{equation}
 b_{0,f} \equiv- \frac{\TF}{3\pi }\,,
\end{equation}
and $C=5/3$.

Sometimes it is useful to introduce a fictitious light quark mass $m_q$ to
regulate the behaviour of $\Pi(0,\mu^2)$. The \MSB{}-renormalized vacuum
polarization with a massive quark reads
\begin{align}
  \label{eq:Pi_mq}
  &  \Pi(k^2, m_q^2, \mu^2) -\Pi_{\rm ct}=     \\
  \nonumber
& \as \,b_{0,f} \times 
\begin{cases}
 \displaystyle{ \log\(\frac{\mu^2}{m_q^2} \) 
 } 
 & \hspace{-0.5cm} k^2 = 0\,,
 \\[0.5cm] 
\displaystyle{\lq \log\(\frac{\mu^2}{m_q^2} \)+\frac{5}{3}
 + \rho - \(1+\frac{\rho}{2}\)\sqrt{\rho-1}\(\pi -2\arctan\sqrt{\rho -1}\)\rq
}
& \hspace{-1.7cm} 0<k^2\le4 m_q^2,
\\[0.5cm]
\displaystyle{\lg \log\(\frac{\mu^2}{m_q^2} \)+\frac{5}{3}
 + \rho + \(1+\frac{\rho}{2}\)\sqrt{1-\rho}
 \lq \log \bigg| \frac{1-\sqrt{1-\rho}}{1+\sqrt{1-\rho}}\bigg|+i\pi\theta\(k^2-4m_q^2\) \rq \rg
 }
\\[0.5mm]
&  \hspace{-3cm} k^2 < 0\,\mbox{ and } k^2 > 4m_q^2,
 \end{cases}
\end{align}
where we have defined
\begin{equation}
 \rho = \frac{4\,m_q^2}{k^2}.
\end{equation}
It develops an imaginary part for $k^2>4 m_q^2$ equal to
\begin{equation}
  {\rm Im}\, \Pi(k^2, m_q^2, \mu^2) = \as \, b_{0,f} \, \pi
  \(1+\frac{\rho}{2}\)\sqrt{1-\rho} .
\end{equation}

\section[Pole-\MSB{} mass conversion]{Pole-$\overline{\boldsymbol{\rm MS}}$ conversion}
\label{sec:msbar2pole}

As first example, we compute the difference between the pole
 mass $m$ and the \MSB{} mass $\overline{m}(\mu)$ at all orders in $\as(\as n_f)^n$. The coupling $\as$ is always meant to be evaluated at the
 scale $\mu$.

At $\mathcal{O}(\as)$, the self-energy correction evaluated for the eigenvalue
of $\slashed{p}$ equal to $m$ takes the form
\begin{eqnarray}
 \label{eq:SigmaNLO} 
 {i \,\Sigma^{(1)}(\mpole, \mpole)} =-i\,g^2 \,\CF \left(\frac{\mu^2}{4\pi}
 {\rm e}^{\gameul} \right)^\ep \int \frac{\mathd^d k}{(2\pi)^d}
 \frac{\gamma^\alpha (\slashed{p}+\slashed{ k}+m)\gamma_\alpha
 }{\lq k^2 + i\eta\rq \lq ( k+p)^2-m^2+i\eta \rq} \Bigg|_{\slashed{p}=m}\,,
\end{eqnarray}
being $k$ the gluon momentum. The details of computation of
$\Sigma^{(1)}(\mpole, \mpole)$ can be found in \writeApp{}~\ref{sec:sigma1_0}
and the result is given by eq.~(\ref{eq:Sigma1_0_expanded}). The mass
counterterms defined in the pole and in the \MSB{} schemes (see Sec.~\ref{sec:self-energy}) are given by
\begin{eqnarray}
\mpolec & = & - i \, \Sigma^{(1)}\!\left(\mpole, \mpole \right) \nonumber\\ 
 & = & - \frac{\as}{4\pi}\,\CF\, e^{\ep \gameul}\,\Gamma(\ep)\, \mpole\, \left(
 \frac{\mu^2}{m^2}\right)^\ep \,\frac{3-2\ep}{1-2\ep} \nonumber \\
 & = & - \frac{\as}{4\pi} \,\CF\, \mpole\, \lq \frac{3}{\ep} + 3 \log \left(
 \frac{\mu^2}{m^2}\right) +4 \rq\,+\mathcal{O}(\ep)\,, \\
\mMSB^c(\mu) & = & \left[- i \, \Sigma^{(1)}\!\left(\mMSB(\mu), \mMSB(\mu)
 \right) \right]^{\rm (d)}
\nonumber\\ 
 & = & - \frac{\as}{4\pi} \,\CF\,\mMSB^c(\mu) \,\frac{3}{\ep},
\end{eqnarray}
respectively, where (d) denotes the divergent part according to the \MSB{} definition.
Neglecting terms of the order $\as^2$, the mass difference is given by the finite part of $i \, \Sigma^{(1)}\!\left(\mpole, \mpole \right)$:
\begin{equation}
 \mpole - \mMSB(\mu) = \left[ i \, \Sigma^{(1)}\!\left(\mpole, \mpole
 \right) \right]^{\rm (f)} = \frac{\as}{4\pi} \,\CF\, \mpole\, \lq 3 \log \left( \frac{\mu^2}{m^2}\right)
+4 \rq\,.
\end{equation}
According to Sec.~\ref{sec:self-energy}, to evaluate $\mpole - \mMSB(\mu)$ beyond NLO, we need to compute
\begin{equation}
 i\Sigma(\slashed{p}, \mb) \Big|_{\slashed{p} = \mpole}\,, \quad i\Sigma(\slashed{p}, \mb) \Big|_{\slashed{p} = \mMSB(\mu)},
\end{equation}
being $\mb$ the bare mass. 
Since up to $\as$ corrections
\begin{equation}
 \mb \approx \mpole \approx \mMSB(\mu)
\end{equation}
and $\Sigma(\slashed{p}, \mb)$ already contains a factor $\as$, in the
large-$n_f$ limit we can
just calculate
\begin{equation}
 i\Sigma(\slashed{p}, \mpole) \Big|_{\slashed{p} = \mpole} = i\Sigma(\mpole, \mpole).
\end{equation}
Indeed, if we replace $\mpole$ with $\mb$ or $ \mMSB(\mu)$ in contributions
that are $\mathcal{O}(\as)$, we produce variations of the order $\as^2(\as
n_f)^n$, that are totally negligible in our context.

The all-orders expression of $i\, \Sigma(\mpole, \mpole)$ is obtained by
replacing the free gluon propagator of eq.~(\ref{eq:SigmaNLO}) with the
dressed one, as shown in eq.~(\ref{eq:dressed_propagator}). We thus obtain
\begin{eqnarray}
{i\, \Sigma(\mpole, \mpole)} = -i\,g^2 \,\CF \left(\frac{\mu^2}{4\pi}
 {\rm e}^{\gameul} \right)^\ep \int \frac{\mathd^d k}{(2\pi)^d}
 \frac{\gamma^\alpha (\slashed{p}+\slashed{ k}+m)\gamma_\alpha
 }{k^2\lq ( k+p)^2-m^2 \rq}\frac{1}{1+\Pi(k^2,\mu^2)-\Pi_{\rm
 ct}}\Bigg|_{\slashed{p}=m}.
\end{eqnarray}
By using eq.~(\ref{eq:virt_trick2}), we can write
\begin{align}
i \,\Sigma(\mpole, \mpole) = -\frac{i}{\pi}\int_{0^-}^{+\infty} \mathd \lambda^2
& \left\{ -\,g^2 \,\CF \left(\frac{\mu^2}{4\pi}
 {\rm e}^{\gameul} \right)^\ep \int \frac{\mathd^d k}{(2\pi)^d}
 \frac{\gamma^\alpha (\slashed{p}+\slashed{ k}+m)\gamma_\alpha
 }{\lq k^2 -\lambda^2 \rq \lq ( k+p)^2-m^2 \rq}
\right\} \nonumber \\ & \times {\rm Im}
\lq \frac{1}{\lambda^2}\,\frac{1}{1 + \Pi
 (\lambda^2,\mu^2) - \Pi_{\tmop{ct}}} \rq
\label{eq:Sigma_dressed} 
\end{align}
where we set $m_q=0$ since IR divergences are absent. The expression in the
curly brackets of eq.~(\ref{eq:Sigma_dressed}) is the one-loop self-energy of
a quark of mass $m$, computed with a gluon of mass $\lambda$, that we
denote by $\Sigma^{(1)}_\lambda(\mpole, \mpole)$, whose $\mathcal{O}(\ep)$
expression is given in eq.~(\ref{eq:Sigma1_lambda}). For ease of notation we
introduce
\begin{eqnarray}
\label{eq:rep}
\widetilde{r}(\epsilon, \lambda) & = &
\frac{i\,\Sigma^{(1)}_\lambda(\mpole, \mpole)}{m} \; = \;
 \frac{\as}{4\pi} \CF \left[
 \frac{3}{\ep}+3\log\left(\frac{\mu^2}{m^2}\right)+4 \right. \nonumber \\
 && \left. \hspace{0.5cm} +
 \frac{\lambda^2}{m^2}\left( 1 + \log \left( \frac{\lambda^2}{m^2}\right)
 \right) + \left(2+\frac{\lambda^2}{m^2}\right)H\left(
 \frac{\lambda^2}{m^2}\right) \right] +\mathcal{O}(\ep)\,,\phantom{aaaa}\\
\label{eq:r0}
\widetilde{r}(\epsilon, 0) &= & \frac{i\,\Sigma^{(1)}_0(\mpole, \mpole)}{m} = \frac{\as}{4\pi}
\,\CF\, e^{\ep \gameul}\,\Gamma(\ep)\, \left(
\frac{\mu^2}{m^2}\right)^\ep \frac{3-2\ep}{1-2\ep}\,, \\
\label{eq:rinf}
\widetilde{r}^\infty(\epsilon, \lambda) &= & \frac{i\,\Sigma^{(1)}_{\lambda_\infty}(\mpole,
 \mpole)}{m} = \frac{\as}{4\pi} \,\CF\, e^{\ep
 \gameul}\,\Gamma(\ep)\, \left( \frac{\mu^2}{\lambda^2}\right)^\ep
\frac{2(3-2\ep)}{(1-\ep)(2-\ep)}\,,
\end{eqnarray}
where the function $H$ is defined in eq.~(\ref{eq:H}) and we have used the
expressions of $\Sigma^{(1)}_\lambda(\mpole, \mpole)$,
$\Sigma^{(1)}_0(\mpole, \mpole)$ and $\Sigma_{\lambda_\infty}^{(1)}(\mpole,
\mpole)$ given by eqs~(\ref{eq:Sigma1_lambda}), (\ref{eq:Sigma1_0}) and
(\ref{eq:Sigma1_large}). The parametric dependence on $\mpole$, $\mu$ and
$\as$ of the integrand functions $\widetilde{r}(\ep, \lambda)$,
$\widetilde{r}(\ep, 0)$ and $\widetilde{r}^\infty(\epsilon, \lambda)$ is kept
implicit for ease of notation.
We thus have
\begin{equation}
\label{eq:Sigma_allorders}
\frac{i \,\Sigma(\mpole, \mpole)}{m} = -\frac{1}{\pi}\int_{0^-}^{+\infty} \mathd
\lambda^2 \,\widetilde{r}(\epsilon, \lambda) \times {\rm Im} \lq
\frac{1}{\lambda^2+i\eta}\,\frac{1}{1 + \Pi (\lambda^2,\mu^2) -
 \Pi_{\tmop{ct}}} \rq.
\end{equation}
Since $\widetilde{r}(\ep, \lambda)$ contains a single pole in $\epsilon$ and
does not vanish for large $\lambda$, we need to evaluate the integrand
in $d=4-2\ep$ dimensions, in order to extract its finite part. We can
express $\widetilde{r}(\ep, \lambda)$ as the sum of the following two terms
\begin{eqnarray}
 \label{eq:r_d}
 \widetilde{r}_d\!\left(\ep, \lambda\right) &=& \frac{\mu^2}{\mu^2+\lambda^2} \, \widetilde{r}(\ep,
 0) + \frac{\lambda^2}{\mu^2+\lambda^2} \, \widetilde{r}^\infty\!\left(\ep,
 \lambda\right)\,,\\ 
 \widetilde{r}_f\!\left(\lambda\right) &=& \widetilde{r}\!\left(\ep, \lambda\right)
 -\widetilde{r}_d\!\left(\ep, \lambda\right).
\end{eqnarray}
We dropped the $\ep$ dependence in $\widetilde{r}_f$ since we can safely
perform the $\ep\to 0$ limit, indeed it does not contain any UV $\ep$-pole and
\begin{eqnarray}
\lim_{\lambda \to \infty} \widetilde{r}_f\!\left(\lambda\right) & = & \mathcal{O}\left(
\frac{m^2}{\lambda^2}\right)\,,\\
\label{eq:rf_small_lambda}
\lim_{\lambda \to 0} \widetilde{r}_f\!\left(\lambda\right) & = & - \as
\frac{\CF}{2}\frac{\lambda}{m} +
\mathcal{O}\left(\frac{\lambda^2}{m^2}\right)\,,
\end{eqnarray}
so that we can write
\begin{eqnarray}
 \widetilde{r}_f\!\left(\lambda\right) &=& \frac{\as\CF}{4\pi} \lg -3\log\!
 \left(\frac{\mpole^2}{\mu^2}\right) + \frac{\lambda^2}{m^2}\left(1+\log
 \frac{\lambda^2}{m^2}\right) + 4 + \left(2+\frac{\lambda^2}{m^2}\right)
 H\left(\frac{\lambda^2}{m^2}\right)\right. \nonumber \\ && \left
 . -\frac{\mu^2}{\mu^2+\lambda^2} \, \lq - 3 \log
 \!\left(\frac{m^2}{\mu^2}\right) + 4 \rq - \frac{\lambda^2}{\mu^2+\lambda^2}
 \, \lq - 3 \log \! \left(\frac{\lambda^2}{\mu^2}\right) + \frac{5}{2} \rq
 \rg \,.\phantom{aaaaa}
\end{eqnarray}
We thus rewrite eq.~(\ref{eq:Sigma_allorders}) as
\begin{align}
 \label{eq:rtilde_split}
 \frac{i\, \Sigma(\mpole, \mpole)}{m} &= r_f(\mpole, \mu, \as) + r_d(\mpole,\mu,\as)\,,
 \\
 r_f(\mpole, \mu, \as) &\equiv - \frac{1}{\pi} \int_{0 -}^{\infty}
 \mathd \lambda^2 \, \widetilde{r}_f\!\left(\lambda\right) {\rm Im} \lq
 \frac{1}{\lambda^2+i\eta}\,\frac{1}{1 + \Pi(\lambda^2,\mu^2) -
 \Pi_{\tmop{ct}}} \rq , \\
\label{eq:rtilde_d} 
 r_d(\mpole,\mu,\as) &\equiv - \frac{1}{\pi} \int_{0 -}^{\infty} \mathd
 \lambda^2 \, \widetilde{r}_d\!\left(\ep, \lambda\right) {\rm Im} \lq
 \frac{1}{\lambda^2+i\eta}\,\frac{1}{1 + \Pi(\lambda^2,\mu^2) -
 \Pi_{\tmop{ct}}} \rq,
\end{align}
where we have made explicit the dependence on $\mpole$, $\mu$
 and $\as$ of the terms $r_f(\mpole, \mu, \as)$ and $r_d(\mpole, \mu, \as)$.
 
We manipulate $r_f(\mpole, \mu, \as)$ as follows. The lower
boundary can be moved from $0^-$ to $0$, since $\widetilde{r}_f(0)=0$. We
also have
\begin{equation}
 \label{eq:tointpart}
 \frac{1}{\lambda^2+i\eta} \, \frac{1}{1 + \Pi (\lambda^2,
 \mu^2) - \Pi_{\tmop{ct}}} = \frac{1}{\as\, b_{0,f}}
 \,\frac{\mathd\ }{\mathd \lambda^2} \log \lq 1 + \Pi
 \left(\lambda^2, \mu^2\right) - \Pi_{\tmop{ct}} \rq,
\end{equation} 
so that
\begin{align}
r_f(\mpole, \mu, \as) =& -\frac{1}{\as \, b_{0,f}} \int_{0}^{\infty}
\frac{\mathd \lambda}{\pi} \, \widetilde{r}_f\!\left(\lambda^2\right) {\rm Im} \left\{
\,\frac{\mathd\ }{\mathd \lambda^2} \log \lq 1 + \Pi
\left(\lambda^2, \mu^2\right) - \Pi_{\tmop{ct}} \rq \right\} \nonumber \\
 \label{eq:final_tilderf}
 = & \frac{1}{\as\,b_{0,f}} \int_{0 }^{\infty} \frac{\mathd
 \lambda}{\pi} \, \frac{\mathd\ }{\mathd \lambda} \lq
 \widetilde{r}_f\!\left(\lambda\right) \rq {\rm Im} \lg \log \lq 1 +
 \Pi\left(\lambda^2, \mu^2\right) - \Pi_{\tmop{ct}} \rq \rg \qquad \nonumber \\
 = & - \frac{1}{\as\,b_{0,f}} \int_{0 }^{\infty} \frac{\mathd
 \lambda}{\pi} \, \frac{\mathd\ }{\mathd \lambda} \lq
 \widetilde{r}_f\!\left(\lambda\right) \rq \arctan \lq \frac{\as\,\pi\, b_{0,f}}{1+\as \,b_{0,f}\,
 \log\left( \frac{\lambda^2}{\mu^2 e^C}\right) }\rq,
\end{align}
that can be evaluated numerically. We notice that $r_f(\mpole, \mu, \as)$
contains a linear infrared renormalon since the behaviour of
$\widetilde{r}_f(\lambda)$ for small $\lambda$ is given by
eq.~\ref{eq:rf_small_lambda}.

As far as the integral in eq.~(\ref{eq:rtilde_d}) is concerned, we can split
it into two terms, according to eq.~(\ref{eq:r_d}),
\begin{align}
 r_d(\mpole,\mu,\as) =& r_d^0(\mpole,\mu,\as) + r_d^\infty(\mpole,\mu,\as)\,,
 \\
 r_d^0(\mpole,\mu,\as) \equiv& - \frac{1}{\pi} \int_{0 -}^{\infty}
 \mathd \lambda^2 \, \frac{\mu^2}{\mu^2+\lambda^2} \, \widetilde{r}(\ep, 0)
 \,{\rm Im} \lq \frac{1}{\lambda^2+i\eta}\,\frac{1}{1 + \Pi(\lambda^2,\mu^2) - \Pi_{\tmop{ct}}} \rq,
 \\
\!\!\! r_d^\infty(\mpole,\mu,\as) \equiv& - \frac{1}{\pi} \int_{0^-}^{\infty} \mathd \lambda^2 \, \frac{\lambda^2}{\mu^2+\lambda^2}
 \,\widetilde{r}^\infty\!\left(\ep, \lambda\right)\, {\rm Im} \lq
 \frac{1}{\lambda^2+i\eta}\,\frac{1}{1 + \Pi(\lambda^2,\mu^2)
 - \Pi_{\tmop{ct}}} \rq. \!\!\!
\end{align}
Since the integrand function in $ r_d^0(\mpole,\mu,\as)$ vanishes for large
$\lambda$, the integral of the imaginary part can be replaced with the closed
path integral depicted in Fig.~\ref{fig:integr_path}. Applying the residue
theorem, we have
\begin{equation}
r_d^0(\mpole,\mu,\as) = \widetilde{r}(\ep, 0) \, \frac{1}{1 + \Pi (-\mu^2,\mu^2) -
 \Pi_{\tmop{ct}}}
 = \widetilde{r}(\ep, 0) \sum_{n = 0}^{\infty} \Big[\Pi_{\tmop{ct}} - \Pi\left(-\mu^2,\mu^2\right) \Big]^n. 
\label{eq:final_tilderd_zero}
\end{equation}
In order to deal with the integral in $ r_d^\infty(\mpole,\mu,\as) $, we
need to expose the $\lambda$ dependence of the integrand. From
eq.~(\ref{eq:rinf}), we can write
\begin{equation}
 \widetilde{r}^\infty\!\left(\ep, \lambda\right) =
 \left(\frac{\lambda^2}{\mu^2}\right)^\ep R^\infty(\ep)\,,
\end{equation}
where $R^\infty(\ep) $ depends only on $\epsilon$ and no longer on
$\lambda$. Similarly, using eq.~(\ref{eq:Pi_unren_MSB}), we have
\begin{eqnarray}
 \Pi\left(\lambda^2,\mu^2\right) &=& \frac{\as \TF}{ \pi} e^{\epsilon\gameul}
 \frac{\Gamma(1+\ep)\, \Gamma^2(1-\ep)}{\Gamma(1-2\ep)} 
 \frac{1-\ep}{(3-2\ep)(1-2\ep)} \, \frac{1}{\ep} \left(
 \frac{\lambda^2 }{\mu^2}\right)^{-\ep} e^{i\epsilon\pi}
 \nonumber\\
 &=& \Pi\left(-\mu^2,\mu^2\right) \left(\frac{\lambda^2 }{\mu^2}\right)^{-\ep} e^{i\epsilon\pi},
\end{eqnarray}
and performing a Taylor expansion we can write
\begin{align}
 & \!\!\!\!\!\!\! r_d^\infty(\mpole,\mu,\as) = - \frac{R^\infty(\ep) }{\pi} \int_{0 }^{\infty}
\mathd \lambda^2 \, \frac{1}{\mu^2+\lambda^2} \left(\frac{\lambda^2
}{\mu^2}\right)^{-\ep} \, {\rm Im} \lq \frac{1}{1 + \Pi (\lambda^2,\mu^2) -
 \Pi_{\tmop{ct}}} \rq \nonumber \\& = - \frac{R^\infty(\ep) }{\pi} \,
\sum_{n = 0}^{\infty}\, \int_{0 }^{\infty} \mathd \lambda^2 \,
\frac{1}{\mu^2+\lambda^2}\left(\frac{\lambda^2 }{\mu^2}\right)^{-\ep} \, {\rm
 Im} \lq \Pi_{\tmop{ct}}-
\Pi\left(-\mu^2,\mu^2\right) \left(\frac{\lambda^2 }{\mu^2}\right)^{-\ep}
e^{i\epsilon\pi} \rq^n \nonumber \\ & = - \frac{R^\infty(\ep) }{\pi} \,
\sum_{n = 1}^{\infty}\, \int_{0 }^{\infty} \mathd z \, \frac{z^{-\ep}}{1+z}
\, {\rm Im} \Big[\Pi_{\tmop{ct}}- \Pi\left(-\mu^2,\mu^2\right) z^{-\ep} e^{i\epsilon\pi}\Big]^n \,.
\label{eq:final_tilderd_inf}
\end{align}
By computing the imaginary part of the $n$-th power of the term in the square
brackets, we are let to evaluate integrals of the form
\begin{equation}
\int_0^{\infty} \mathd z \frac{z^{-h}}{1+z} = \Gamma (1 - h) \, \Gamma (h),
\end{equation}
where $h$ is a real number, so that $r_d^\infty(\mpole,\mu,\as)$ can be
straightforwardly evaluated by computer algebraic means at any fixed order in
$\as$.
We emphasize that $r_d(\mpole,\mu,\as)$ has no linear renormalon. Indeed if we
perform an $\epsilon$ expansion and we consider the small-$\lambda$
contribution, by writing $\mathd \lambda^2= 2 \lambda \mathd \lambda$, we
notice that the integrand behaves as $\lambda \log^n(\lambda)$. This signals
the absence of linear renormalons, that come from terms of the type
$\log^n(\lambda)$, without any power of $\lambda$ in front.

As a check, we observe that, at $\mathcal{O}(\as)$, $r_f(\mpole, \mu, \as)$
and $r_d^\infty(\mpole,\mu,\as)$ do not contribute and we recover the correct
NLO result
\begin{equation}
\lq \frac{i \, \Sigma(m,m)}{m} \rq_{\mathcal{O}(\as)} = \widetilde{r}(\ep, 0) = \frac{i \, \Sigma_0^{(1)}(m,0)}{m}= \frac{i \, \Sigma^{(1)}(m,m)}{m} \,.
\end{equation}
From eqs.~(\ref{eq:mpolec}) and~(\ref{eq:MSbar_mass_counterterm}), and
neglecting $\mathcal{O}(\as^2(\as n_f)^n)$ contributions, we have
\begin{align}
& \mpolec = -i \, \Sigma(m,m) = - \, m\, \left[ r_f(\mpole, \mu, \as) + r_d(\mpole,\mu,\as)
 \right]\,, \\
& \mMSB^c(\mu) = -i \, \Sigma^{\rm (d)}\!\(\mpole, \mpole \) = - \mpole\, 
 r^{\rm (d)}_d(\mpole, \mu, \as) \,,
\end{align}
where the superscript (d) denotes the divergent part according to the \MSB{}
scheme. Thus
\begin{eqnarray}
 \mpole - \mMSB(\mu) = -\left[\mpolec - \mMSB^c(\mu)\right]= m\left[
 r_f(\mpole, \mu, \as) + \tilde{r}^{\rm (f)}_d(\mpole, \mu, \as) \right],
 \label{eq:m_diff}
\end{eqnarray}
with (f) denoting the finite part.
We can expand the result of eq.~(\ref{eq:m_diff}) in series of $\as(\mu)$
\begin{eqnarray}
&&r_f(\mpole, \mu, \as) + r^{\rm (f)}_d (\mpole, \mu, \as) =
 \sum_{i=1} c_i(\mpole, \mu)\, \as^i(\mu) \\
&& \mpole - \mMSB(\mu) = \mpole \, \sum_{i=1} c_i(\mpole, \mu)\, \as^i(\mu).
 \label{eq:deltam_as}
\end{eqnarray}
This expression can be employed to evaluate the difference $\mpole -\mMSB(\mu)$
for an arbitrary real value of $\mu$. Furthermore, it can be used both for a complex or a real pole mass.

The authors of Ref.~\cite{Ball:1995ni} performed the same computation, with a
slightly different strategy, for the case of $\mpole$ real and $\mu=\mpole$.
They define\footnote{The definition of $\beta_0^{N_f}$ in~\cite{Ball:1995ni}
 corresponds to $-b_{0,f}$.} 
\begin{equation}
 \label{eq:deltam_ball}
\frac{\mpole -\mMSB(m)}{m} = \frac{4}{3}\frac{\as(\mpole)}{\pi}\left[ 1 +
 \sum_{i=1}^\infty d_i \left(b_{0,f}\, \as(\mpole)\right)^i \right].
\end{equation}
We rearrange eq.~(\ref{eq:deltam_as}) to put it into a form similar to
eq.~(\ref{eq:deltam_ball})
\begin{eqnarray}
 \frac{\mpole - \mMSB(m)}{\mpole} &=& \as(\mpole)\, c_1(m,m) \left[ 1+
 \sum_{i=1}^\infty \frac{c_{i+1}(m,m)}{c_1(m,m)} \as^i(\mpole) \right] \nonumber \\
 & = & \frac{4}{3}\frac{\as(m)}{\pi}\left[ 1+
 \sum_{i=1}^\infty \frac{c_{i+1}(m,m)}{c_1(m,m)}\frac{1}{b_{0,f}^i}\left(b_{0,f}\, \as(\mpole)\right)^i \, \right].
 \end{eqnarray}
Thus the coefficients $d_i$ in eq.~(\ref{eq:deltam_ball}) are given by
\begin{equation}
 d_i = \frac{c_{i+1}(m,m)}{c_1(m,m)}\frac{1}{b_{0,f}^i}\,.
\end{equation}
Given our choice $\mu=m$, the coefficients $c_i$ and $d_i$ are independent
from the value of $m$.
We checked numerically that our results reproduce exactly the coefficients
 $d_i$ reported in the first column of Tab.~2 of Ref.~\cite{Ball:1995ni}.

\section[Realistic large-$b_0$ approximation]{Realistic large-$\boldsymbol{b_0}$ approximation}
\label{sec:realistic_b0}
In order to recover the full QCD one loop $\beta$ function, we will
add to eq.~(\ref{eq:Pi_unren_MSB})
\begin{equation}
 \Pi_g\left(k^2,\mu^2\right) = - \as \frac{11\, \CA}{12\pi} e^{\epsilon\gameul}
 \frac{\Gamma(1+\ep)\, \Gamma^2(1-\ep)}{\Gamma(1-2\ep)} 
 \left( 1+ \ep \, C_g \right) \frac{1}{\ep} \left(
 \frac{-k^2}{\mu^2}\right)^{-\ep}\,,
\end{equation}
where $C_g$ is an arbitrary constant.
Thus we have
\begin{eqnarray}
\Pi\left(k^2,\mu^2\right) & = & \frac{\as}{\pi} e^{\epsilon\gameul} \frac{\Gamma(1+\ep)\,
 \Gamma^2(1-\ep)}{\Gamma(1-2\ep)} \, \frac{1}{\ep} \left( \frac{-k^2}{\mu^2}\right)^{-\ep} \nonumber \\
&& \times \lq n_l \TR\frac{1-\ep}{(3-2\ep)(1-2\ep)} -\frac{11 \CA}{12}\left(1+\ep\, C_g\right) \rq,
\label{eq:Pi_full}
\end{eqnarray}
where we have restored the correct number of light flavour $n_l$. 
In order to cancel the $1/\ep$ pole of $\Pi$, the counterterm must be given by
\begin{equation}
 \label{eq:Pict_full}
 \Pi_{\rm ct} =-\as\frac{b_0}{\ep}\,,
\end{equation}
that allows us to write, adding an infinitesimal positive imaginary part to
$k^2$, 
\begin{eqnarray}
 \label{eq:Pi_ren_full}
 \Pi\left(k^2,\mu^2\right) - \Pi_{\rm ct} &=& \as b_0
 \lq \log \left(\frac{|k^2|}{\mu^2}\right) - i\pi\,\theta(k^2)\rq +
 \frac{\as}{\pi} \lq\frac{n_l \TR}{3} \frac{5}{3} - \frac{11\,
 \CA}{12}\,C_g\rq \qquad \qquad \\
 &=& \as b_0 \lq\log \left(\frac{|k^2|}{\mu^2e^C}\right) - i\pi\,\theta(k^2) \rq
\end{eqnarray}
with
\begin{equation}
 C = \frac{1}{b_0}\lq -\frac{n_l\,\TR}{3\pi}\frac{5}{3} + \frac{11\,\CA}{12\pi}C_g\rq\,.
\end{equation}
In this way, we also get that, for positive $k^2$,
\begin{equation}
 \label{eq:imag_Pi_new}
 {\rm Im} \,\Pi (k^2, \mu^2) = - \as\, b_0\,\pi\,.
\end{equation}
If we choose
\begin{equation}
 \label{eq:C_g}
C_g = \frac{67-3\pi^2}{33}\approx1.133,
\end{equation}
the constant $C$ becomes 
\begin{equation}
 \label{eq:C}
 C = \frac{1}{2\pi\,b_0} K_g\,, \qquad \mbox{where } K_g = \left( \frac{67}{18}-\frac{\pi^2}{6}\right) \CA -\frac{10}{9} n_l \TR.
\end{equation}
Our choice is rather arbitrary and motivated by the fact that the
final expression for the total cross section (or for any infrared safe
obsarvable) computed in the large-$b_0$ limit, that we will derive in
Chap.~\ref{sec:description_calc}, contains a factor
\begin{equation}
 \as(k \, e^{-C/2}) = \frac{\as(k)}{1-\frac{K_g}{2\pi} \as(k)} \approx  \as(k)\left[ 1+ \frac{K_g}{2\pi}\as(k)\right] \equiv \as^{\rm MC}(k),
\label{eq:lambda_eff}
\end{equation}
where MC denotes the Monte Carlo scheme, also known as the CMW scheme,
introduced in Ref.~\cite{Catani:1990rr}. Thus, with our choice of
$C_g$, our formula becomes appropriate to describe a QCD effective
coupling.

Furthermore, one can in principle replace the term $C_g$ with
\begin{equation}
 C_g \to C_g+ \ep \, C_g^\prime + \ep^2 \, C_{g}^{\prime\prime} \ldots
\end{equation}
As we will discuss in Sec.~\ref{sec:description_calc}, the additional terms would not
contribute to the all-orders amplitude computed in the pole scheme, since
there are no UV divergences once the counterterm $\Pi_{\rm ct}$ is
introduced, and thus we are in position to perform the $\ep \to 0$ limit.
On the other hand, as we have seen in Sec.~\ref{sec:msbar2pole}, to
evaluate the $\mpole-\mMSB(\mu)$ difference we need the exact $\ep$
dependence of $\Pi$. However, the leading contribution to this difference,
namely $r_f(\mpole, \mu, \as)$ of eq.~(\ref{eq:final_tilderf}), is computed
in $d=4$ dimensions, so that terms $\ep C_{g}^\prime + \ep^2 \, C_{g}^{\prime\prime} \ldots$ are
dropped. These terms are instead contained in $r_d(\mpole, \mu, \as)$ of
eq.~(\ref{eq:rtilde_d}), but this contribution is subleading, since it does
not involve any infrared renormalon.

\chapter{Description of the calculation}
\label{sec:description_calc}
\begin{figure}[tb]
 \centering
 \begin{subfigure}{0.3\textwidth}
 \includegraphics[width=\textwidth]{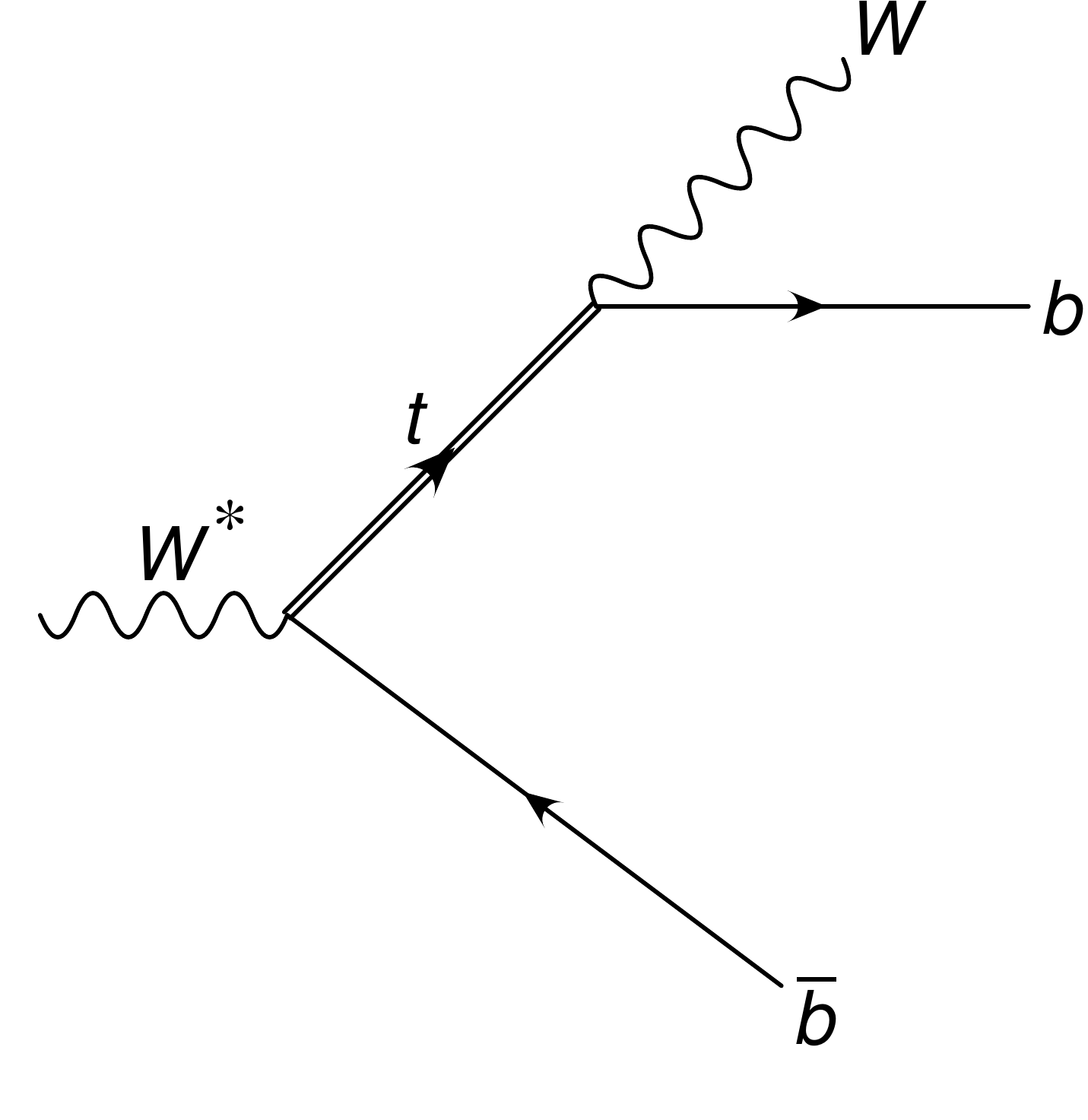}
 \caption{}
 \end{subfigure}
 \begin{subfigure}{0.3\textwidth}
 \includegraphics[width=\textwidth]{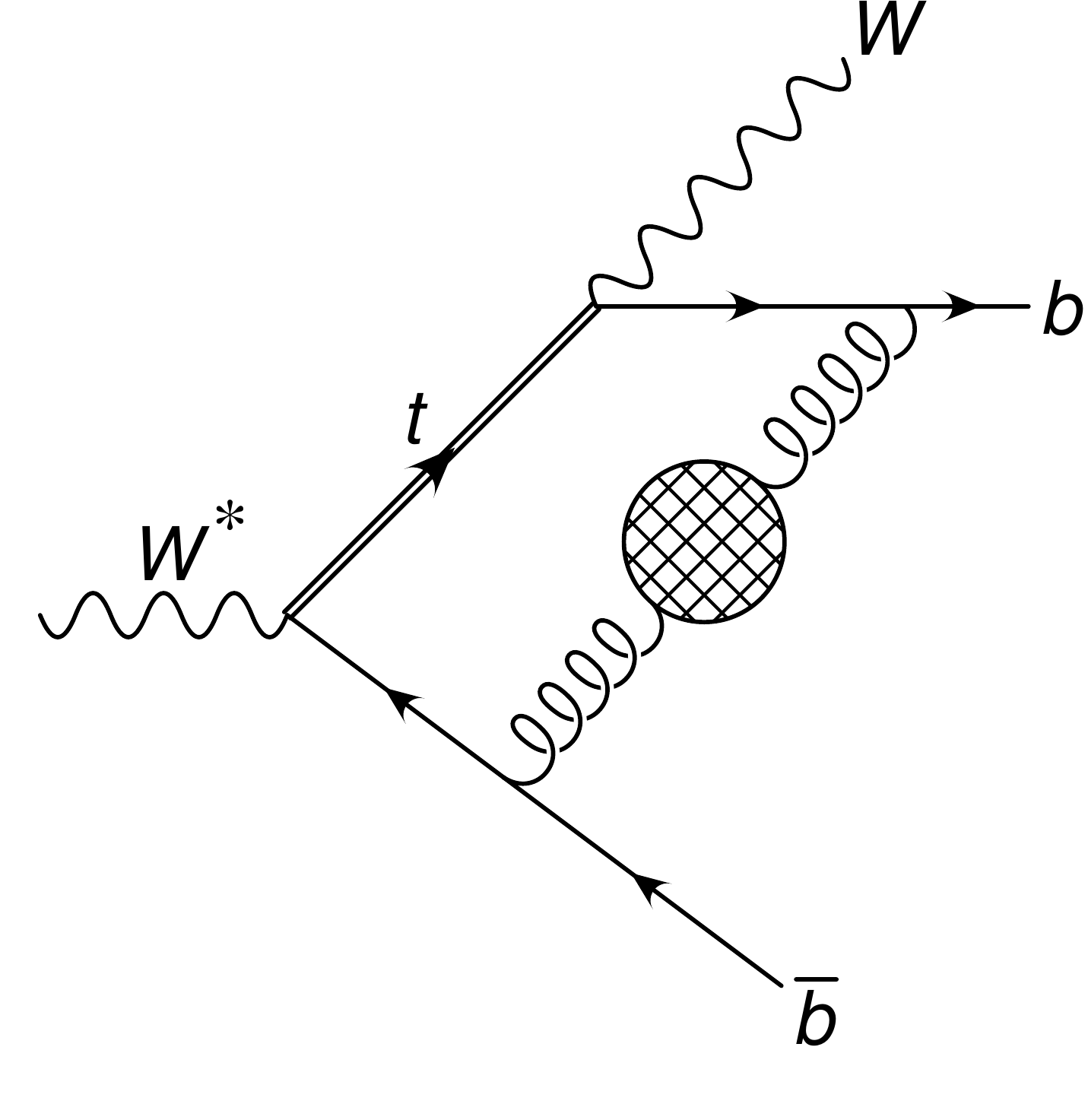}
 \caption{}
 \end{subfigure}

 \begin{subfigure}{0.3\textwidth}
 \includegraphics[width=\textwidth]{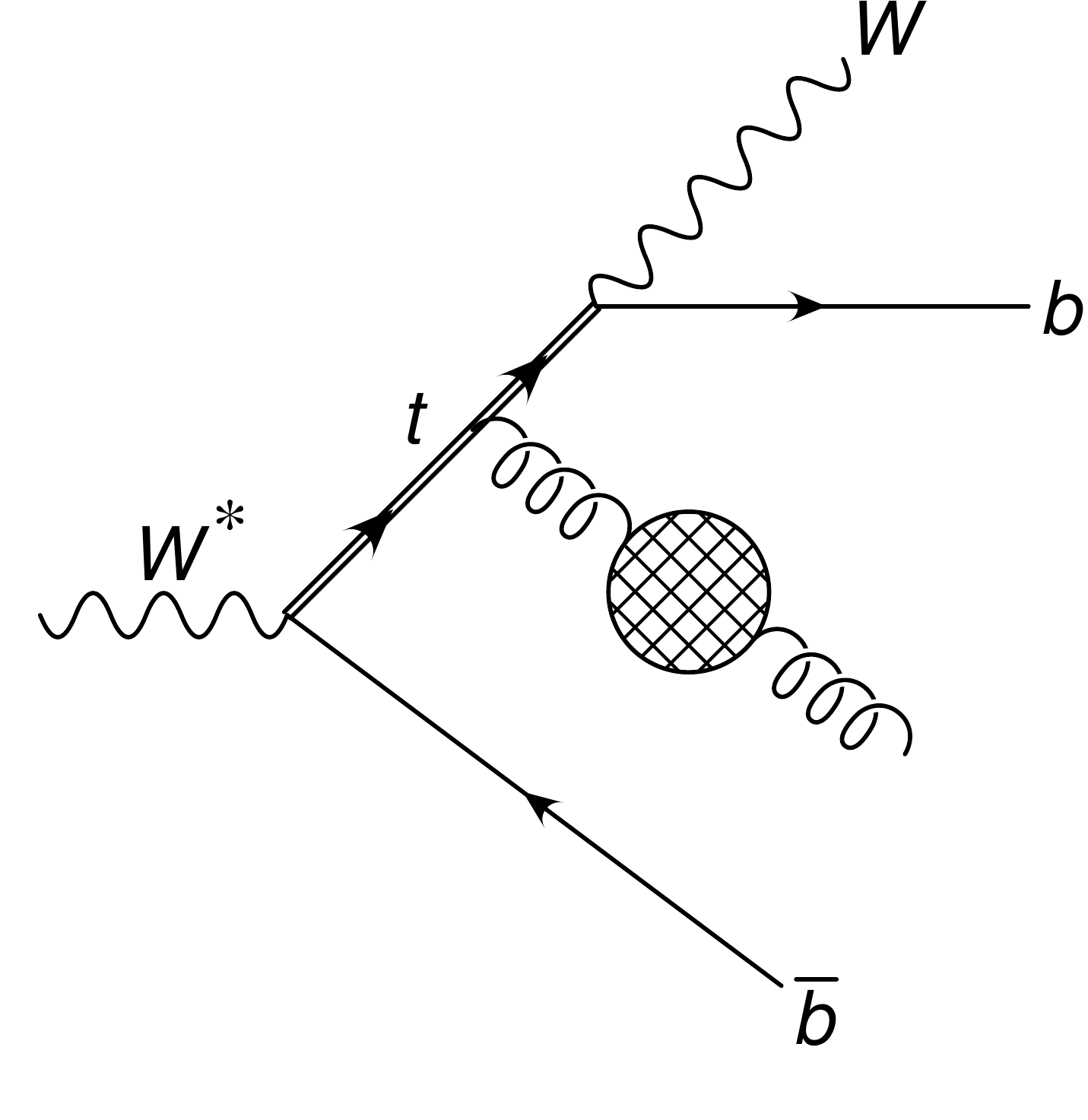}
 \caption{}
 \end{subfigure}
 \begin{subfigure}{0.3\textwidth}
 \includegraphics[width=\textwidth]{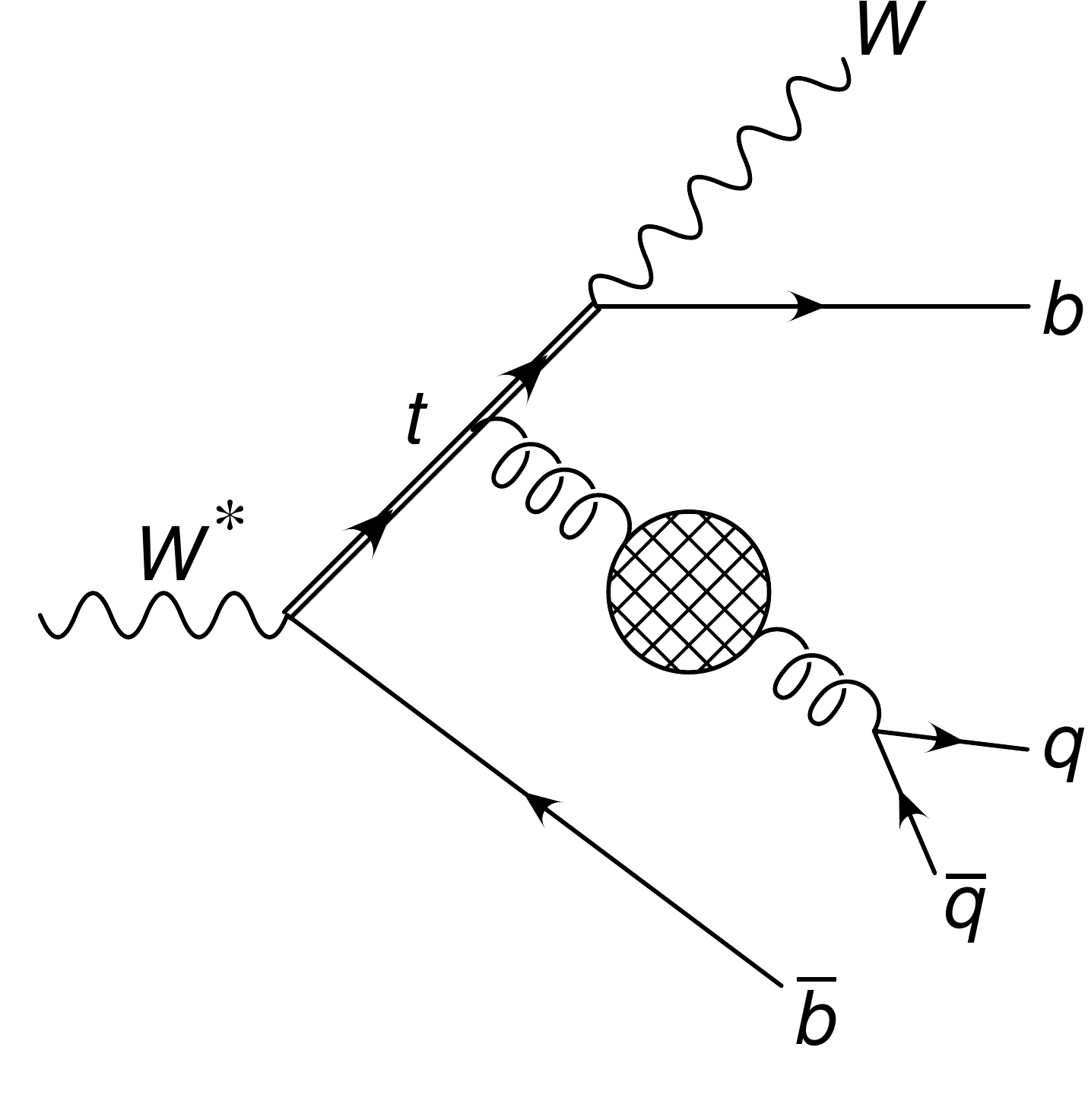}
 \caption{}
 \end{subfigure}
 
 \vspace{1cm}
 \begin{subfigure}{\textwidth}
 \centering
 \includegraphics[width=0.8\textwidth]{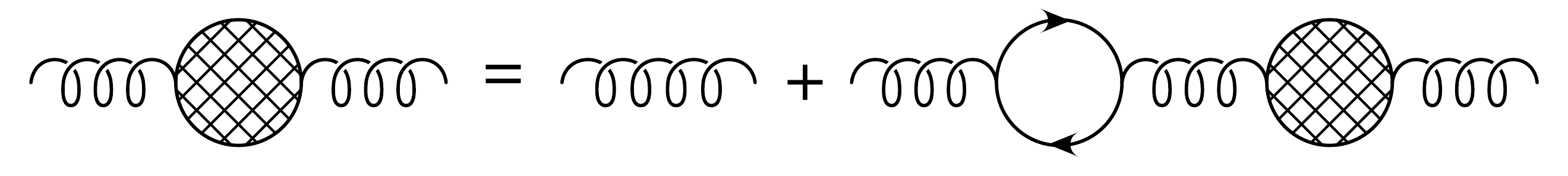}
 \caption{}
 \end{subfigure}
 \caption{Feynman diagram for the Born $W^*\to W b\bar{b}$ process~(a),
 and samples of Feynman diagrams for the virtual contribution~(b), for the
 real-emission contribution~(c) and for $W^*\to W b\bar{b}q\bar{q}$
 production~(d).
 \label{fig:wbbbar}}
\end{figure}

We want to compute the process $W^* \to t\,\bar{b}\to W b\,\bar{b}$, where the
$W$ boson has only a vector coupling to the quarks, at all orders in the
large-number-of-flavour limit. The parameters we choose are
\begin{eqnarray}
 \label{eq:m0}
 m_{\sss 0} & = & 172.5\,{\rm GeV}, \\
 \Gamma_t & = & 1.3279 \,{\rm GeV}, \\
 m & = & \sqrt{m_{\sss 0}^2 - i m_{\sss 0} \Gamma_t}, \\
 m_{\sss W} & = & 80.419\,{\rm GeV}, \\
 E_{\sss\rm CM} & = & 300 \,{\rm GeV}, \\
 \mu & = & m_{\sss 0}\,.
\end{eqnarray}
A sample of Feynman diagrams contributing to this process is depicted in
Fig.~\ref{fig:wbbbar}.
The dashed blob represents the summation of all self-energy insertion in the
large-$n_f$ limit.

We now describe how we compute the total cross section.
We use the complex pole scheme definition for the top mass.
We assume the eventual presence of a set of cuts $\Theta(\Phi)$, 
function of the final state kinematics $\Phi$.
The integrated cross section reads
\begin{eqnarray}
 \label{eq:sigmatot}
 \sigma & = & 
 \int \mathd \Phi_{\tmop{b}} \,\frac{\mathd \sigma_{\tmop{b}} (\Phi_{\tmop{b}})}{\mathd \Phi_{\tmop{b}}} \, \Theta (\Phi_{\tmop{b}})
 + \int \mathd \Phi_{\tmop{b}} \,\frac{\mathd \sigma_{\tmop{v}} (\Phi_{\tmop{b}}) }{\mathd \Phi_{\tmop{b}}}\, \Theta (\Phi_{\tmop{b}})
 + \int \mathd \Phi_g \, \frac{\mathd \sigma_g (\Phi_g)}{\mathd \Phi_g} \, \Theta (\Phi_g) \nonumber \\
 && \hspace{0.7cm} +\int \mathd \Phi_{ q \bar{q}} \,\frac{\mathd \sigma_{q
 \bar{q}} (\Phi_{q \bar{q}})}{\mathd \Phi_{q \bar{q}}} \, \Theta (\Phi_{q \bar{q}}) , 
\end{eqnarray}
where first term represents the Born contribution, the second the virtual
one, the third term represents the contribution due to the emission of a real
gluon and the fourth term represents the contribution with the real
production of $n_f$ $q \bar{q}$ pairs. The last three contributions are
potentially divergent. Equation~(\ref{eq:sigmatot}) implicitly defines our
notation for the different phase space integration volumes.

We always imply that the gluon propagators, in the last three contributions,
include the sum of all vacuum polarization insertions of light quark loops.

We rewrite the total cross section as sum of four contributions:
\begin{eqnarray}
 \label{eq:def_sigma}
 \sigma & = & \sigma_{\tmop{b}} + \sigma_{\tmop{v}} + \sigma_{g^*} + \Delta\sigma_{q\bar{q}}\,, \\
 \sigma_{\tmop{b}} & \equiv &\int \mathd \Phi_{\tmop{b}} \,\frac{\mathd \sigma_{\tmop{b}} (\Phi_{\tmop{b}})}{\mathd \Phi_{\tmop{b}}} \, \Theta (\Phi_{\tmop{b}})\,, \\
 \sigma_{\tmop{v}} & \equiv & \int \mathd \Phi_{\tmop{b}} \,\frac{\mathd \sigma_{\tmop{v}} (\Phi_{\tmop{b}}) }{\mathd \Phi_{\tmop{b}}}\, \Theta (\Phi_{\tmop{v}})\,, 
 \label{eq:sigmav}\\
 \sigma_{g^*} & \equiv & \int \mathd \Phi_g \, \frac{\mathd \sigma_g
 (\Phi_g)}{\mathd \Phi_g} \, \Theta (\Phi_g)
 + \int \mathd \Phi_{q \bar{q}}
 \frac{\mathd
 \sigma_{q\bar{q}} (\Phi_{q \bar{q}})}{\mathd \Phi_{q \bar{q}}} \, \Theta (\Phi_{g^*})\,,
 \label{eq:sigmag}\\
\Delta\sigma_{q\bar{q}} &\equiv& \int \mathd \Phi_{q\bar{q}} \,\frac{\mathd
 \sigma_{q\bar{q}} (\Phi_{q \bar{q}})}{\mathd
 \Phi_{q \bar{q}}}
 \left[\Theta (\Phi_{q \bar{q}})- \Theta (\Phi_{g^*})\right]\,,
 \label{eq:sigmaqq}
\end{eqnarray}
where the selection cuts $\Theta(\Phi_{g^{*}})$ are evaluated with the same
kinematics of the $Wb\bar{b}q\bar{q}$ events but with the $q \bar{q}$ pair
clustered in a single jet~($g^*$).

\section[The ${\Delta\sigma_{q\bar{q}}}$ contribution]{The $\boldsymbol{\Delta\sigma_{q\bar{q}}}$ contribution} 
 
In this section we illustrate how to calculate the term $\Delta
\sigma_{q \bar{q}}$ of eq.~(\ref{eq:def_sigma}). $\Delta
\sigma_{q \bar{q}}$ receives contributions only from the real
graphs with a final state $W b \bar{b} q \bar{q}$, where $q \bar{q}$ is a
pair of light quarks, as depicted in Fig.~\ref{fig:wbbbar}~(d).

Starting from the ${\cal O}(\as^2)$ tree-level cross section for the
process $W^*\to W b \bar{b} q \bar{q}$, that we indicate as
$\mathd \sigma^{(2)}_{q \bar{q}}$, with no vacuum polarization insertions in
the gluon propagator, we obtain the differential cross section
$\mathd \sigma_{q \bar{q}}$ with the insertion of all the light-quark bubbles
by simply replacing the bare gluon propagator with the dressed one of
eq.~(\ref{eq:dress_ren_prop})
\begin{equation}
 \frac{\mathd \sigma_{q \bar{q}}(\Phi_{q\bar{q}})}{\mathd \Phi_{q\bar{q}} } =
\frac{\mathd \sigma^{(2)}_{q \bar{q}}(\Phi_{q\bar{q}})}{\mathd \Phi_{q\bar{q}} }
 \left| \frac{1}{1 + \Pi
 (k^2, \mu^2) - \Pi_{\tmop{ct}} } \right|^2,
\end{equation}
where $k^2$ is the virtuality of the $q\bar{q}$ pair arising after the gluon
splitting.
In order to compute $\Delta\sigma_{q \bar{q}}$, we insert this equation
into~(\ref{eq:sigmaqq}), and we get
\begin{equation}
 \label{eq:sigmaqq_dressed}
 \Delta \sigma_{q \bar{q}} = \int \mathd \lambda^2 \, \delta(k^2-\lambda^2)\,\int \mathd \Phi_{q\bar{q}} \,
\frac{\mathd \sigma^{(2)}_{q \bar{q}}(\Phi_{q\bar{q}})}{\mathd \Phi_{q\bar{q}} } \, \Big[ \Theta (\Phi_{q
 \bar{q}}) - \Theta (\Phi_{g^{*}})\Big] \, \left| \frac{1}{1 + \Pi
 (\lambda^2, \mu^2) - \Pi_{\tmop{ct}} } \right|^2,
\end{equation}
where we have added the dummy integration $\int \mathd \lambda^2 \, \delta(k^2-\lambda^2)=1$.
We remark that, thanks to the subtraction in the square parenthesis in
eq.~(\ref{eq:sigmaqq_dressed}), $ \Delta\sigma_{q \bar{q}}$ is finite as
$\lambda^2 \rightarrow 0$. In fact, if $\lambda^2 \rightarrow 0$ in the
collinear sense, the first clustering of the jet algorithm is the one of the
$q \bar{q}$ pair into a $g^*$ (since $\Phi_{q \bar{q}} \rightarrow
\Phi_{g^{*}}$), unless the gluon three-momentum just happens to lay on the jet
cone. But, in this case, the direction of the pair must be closer to the cone
than the pair aperture, and this leads to a suppression of the cross section
of the order of the pair separation. The finiteness in the case of soft
$\lambda$ is obvious, since the reconstructed mass is insensitive to soft
particles. Notice that the integral is finite in the sense that the integrand
goes like $1/\lambda$, and it is integrated in $\mathd
\lambda^2$. For large $\lambda$, the integrand is zero for kinematic constraints, thus
the integral is finite.

In order to make contact to other contributions we are going to compute, we
write the absolute square of the dressed propagator in terms of the derivative
of an imaginary part. We can perform the following manipulation
\begin{eqnarray}
 \left| \frac{1}{1 + \Pi (\lambda^2, \mu^2) - \Pi_{\tmop{ct}} } \right|^2 & = & -
 \frac{1}{{\rm Im} \, \Pi (\lambda^2, \mu^2) }\, {\rm Im} \left[ \frac{1}{1 + \Pi
 (\lambda^2, \mu^2) - \Pi_{\tmop{ct}}} \right]
 \nonumber\\
 & = & \frac{1}{\as \,\pi\, b_0} \,\tmop{Im} \left[ \frac{1}{1 + \Pi (\lambda^2,
 \mu^2) - \Pi_{\tmop{ct}}} \right]
 \nonumber\\
 & = & \frac{\lambda^2}{\as \,\pi\, b_0 } \,{\rm Im} \left[ \frac{1}{\lambda^2 + i\eta} \frac{1}{1
 + \Pi (\lambda^2, \mu^2) - \Pi_{\tmop{ct}}} \right].\qquad
\end{eqnarray}
since this factor multiplies an expression that is regular for $\lambda=0$, and
thus the $\delta(\lambda)$ arising from the imaginary part (see
eq.~(\ref{eq:imag_den})) does not contribute.
Using eq.~(\ref{eq:tointpart}) we are lead to
\begin{eqnarray}
 \label{eq:asb_square_dress_prop}
 \left| \frac{1}{1 + \Pi (\lambda^2, \mu^2) - \Pi_{\tmop{ct}} } \right|^2
 & = & \frac{\lambda^2}{\as^2 \, \pi\,b_{0,f}^2} \,
 \frac{\mathd\ }{\mathd \lambda^2}\,{\rm Im} \lg \log \lq 1 + \Pi \left(\lambda^2, \mu^2\right) -
 \Pi_{\tmop{ct}}\rq \rg
\end{eqnarray} 
Equation~(\ref{eq:sigmaqq_dressed}) becomes
\begin{eqnarray}
 \Delta\sigma_{q \bar{q}} &=& \int \frac{\mathd \lambda^2}{ \pi}\,
 \delta(k^2-\lambda^2) \int \mathd \Phi_{q\bar{q}} \, \frac{\mathd \sigma^{(2)}_{q \bar{q}}(\Phi_{q\bar{q}})}{\mathd \Phi_{q\bar{q}} }\, \Big[ \Theta (\Phi_{q \bar{q}}) - \Theta (\Phi_{g^{*}})\Big]
 \nonumber \\
 && \phantom{ \int} \times \frac{\lambda^2}{\as^2 \, b_{0,f}^2} \,
 \frac{\mathd }{\mathd \lambda^2} \,{\rm Im} \lg \log \lq 1 + \Pi (\lambda^2+i\eta, \mu^2) -
 \Pi_{\tmop{ct}}\rq \rg.
\label{eq:to_be_integrated}
\end{eqnarray}
Defining
\begin{equation}
 \label{eq:deltak2}
 \Delta\!\left(\lambda\right) \equiv - \frac{\lambda^2}{\as\, b_{0,f}} \int \delta(k^2-\lambda^2)\, \mathd
 \Phi_{q\bar{q}} \,\frac{\mathd \sigma^{(2)}_{q \bar{q}}(\Phi_{q\bar{q}})}{\mathd \Phi_{q\bar{q}} } \, \Big[ \Theta(\Phi_{q \bar{q}}) - \Theta (\Phi_{g^{*}})\Big] ,
\end{equation}
and integrating by parts eq.~(\ref{eq:to_be_integrated}), we are lead to
\begin{equation}
 \label{eq:sigmaqq_final}
 \Delta\sigma_{q \bar{q}} = \frac{1}{\as\,b_{0,f}} \int_0 \frac{\mathd \lambda}{ \pi}\,
 \frac{\mathd\ }{\mathd \lambda}\!\lq \Delta\!\left(\lambda\right) \rq 
 \,{\rm Im} \lg \log \lq 1 + \Pi (\lambda^2+i\eta, \mu^2) - \Pi_{\tmop{ct}}\rq \rg,
\end{equation}
where the integrand function is identically 0 for $\lambda \ge \sqrt{s}-\mW$.
We remind the reader that the cross section $\sigma^{(2)}_{q\bar{q}}$ and $b_{0,f}$
are both proportional to $n_f$, that thus cancels in the definition of
$\Delta(\lambda)$, so when we will move from the large-$n_f$ to the large-$b_0$ limit, $\Delta(\lambda)$ will not change.

\section[The ${ \sigma_{g^*}}$ contribution]
 {The $\boldsymbol{ \sigma_{g^*}}$ contribution}
 
The ${ \sigma_{g^*}}$ term of eq.~(\ref{eq:sigmag}) receives contributions
from final states with both a single real gluon or a $q \bar{q}$ pair. Both
these contributions have collinear divergences related to the $q \bar{q}$
splitting that cancel when integrating the latter over $\lambda^2$ and summing them
up.

\subsection{The gluon contribution}
The first contribution of eq.~(\ref{eq:sigmag}) can be computed starting from
$\sigma_g^{(1)}$, the tree-level cross section for the emission of a single
gluon. The sum over all the polarization insertions gives rise to the the
following identity
\begin{equation}
 \int \mathd \Phi_g\, \frac{\mathd \sigma_g (\Phi_g)}{\mathd \Phi_g} \, \obs (\Phi_g) = \int \mathd \Phi_g
 \, \frac{\mathd \sigma_g^{(1)} (\Phi_g)}{\mathd \Phi_g} \, \obs (\Phi_g) \, \frac{1}{1 + \Pi (0, \mu^2) -
 \Pi_{\tmop{ct}} }.
\end{equation}
Since $\Pi (0, \mu^2)$ is not well-defined, it is convenient to assign to the
quarks in the polarization bubbles a small mass $m_q$. We indicate the
self-energy correction with a massive quark with $\Pi (0, m_q^2, \mu^2)$.
Indeed it can be easily shown that
\begin{equation}
 \label{eq:Pi_mass}
 \Pi(0,m_q^2,\mu^2) = \frac{\as \TF}{3 \pi} 
\lq \frac{1}{\ep} - \log \left(\frac{m_q^2}{\mu^2} \right)\rq +
 {\cal O}(\ep)\,,
\end{equation}
that is well defined and real. The analytic expression of $\Pi(k^2, m_q^2,
\mu^2) - \Pi_{\rm ct}$ for an arbitrary $k^2$ value is given in
eq.~(\ref{eq:Pi_mq}). We can then write
\begin{align}
 \label{eq:gluoncontr}
 \int \mathd \Phi_g \, \frac{\mathd \sigma_g (\Phi_g)}{\mathd \Phi_g} \,
 \Theta (\Phi_g) = \int \mathd \Phi_g \,\frac{\mathd \sigma_g^{(1)}
 (\Phi_g)}{\mathd \Phi_g} \, \obs (\Phi_g) \, \frac{1}{1 + \Pi (0, m_q^2,
 \mu^2) - \Pi_{\tmop{ct}}}.
\end{align}

\subsection[The ${q \bar{q}}$ contribution]
 {The $\boldsymbol{q \bar{q}}$ contribution}
 
In order to treat the second term of eq.~(\ref{eq:sigmag}), we first discuss
the relation between $\sigma^{(2)}_{q \bar{q}}$, the cross section for the
production of $n_f$ $q\bar{q}$ pairs with invariant mass $\lambda$, and
$\sigma^{(1)}_{g^*}$, the cross section for the production of a massive gluon
whose four-momentum $k^\mu$ is equal to the sum of the $q$ and $\bar{q}$
momenta. We have
 \begin{eqnarray}
 \frac{\mathd \sigma^{(1)}_{g^{*}} (\lambda,\Phi_{g^{*}})}{\mathd \Phi_{g^*}}& =&
 \frac{1}{2s} \mathcal{M}_{g^*}^\mu(\Phi_{g^{*}})
 \mathcal{M}_{g^*}^{\nu *}(\Phi_{g^{*}}) \sum_{\rm
 pol}\epsilon_{\mu}\(\lambda\)\epsilon^*_{\nu}\(\lambda\) \nonumber \\
 & = & \frac{1}{2s} \mathcal{M}_{g^*}^\mu(\Phi_{g^{*}})
 \mathcal{M}_{g^*}^{\nu *}(\Phi_{g^{*}}) \left[ -g_{\mu \nu} + \frac{k_\mu
 k_\nu}{\lambda^2}\right]
 \end{eqnarray}
where $1/2s$ is the flux factor and $\mathcal{M}_{g^*}^{\mu}(\Phi_{g^{*}})$ is
the amplitude for the production of a massive gluon $g^*$ of momentum
$k^\mu$, not contracted with its polarization vector $\ep^\mu(k)$.
The real phase space $\Phi_{q\bar{q}}$ can be written as the product of the
phase space for the production of a gluon with virtuality $k^2$, that we call
$\mathd \Phi_{g^{*}}$, and its decay into a $q \bar{q}$ pair, $\mathd
\Phi_{\tmop{dec}}$
\begin{equation}
 \mathd \Phi_{q \bar{q}} = \frac{\mathd k^2}{2 \pi} \,\mathd \Phi_{g^{*}}\,
 \mathd \Phi_{\tmop{dec}}\,.
\end{equation}
Applying the optical theorem we have
\begin{align}
 \int_{4m_q^2} \mathd k^2 \int \mathd \Phi_{\tmop{dec}}& \,\frac{\mathd \sigma^{(2)}_{q \bar{q}}
 (\Phi_{q \bar{q}})}{\mathd \Phi_{q \bar{q}}} \, \delta(k^2 - \lambda^2) \nonumber \\
&= 2 \left[\frac{ \mathcal{M}_{g^*}^\mu(\Phi_{g^{*}}) \mathcal{M}_{g^*}^{\nu,*}(\Phi_{g^{*}})}{2s}
 \frac{1}{\lambda^4}\, {\rm Im}\, \Pi(\lambda^2,m_q^2,\mu^2) \left(-\lambda^2\,g_{\mu\nu}
 + k_\mu k_\nu \right) \right]
 \nonumber \\
&= \frac{\mathd \sigma^{(1)}_{g^{*}} (\lambda,\Phi_{g^{*}})}{\mathd \Phi_{g^*}} 
 \frac{2 \,{\rm Im} \left[\Pi(\lambda^2, m_q^2,\mu^2) \right]}{\lambda^2},
\label{eq:sigmaqq_g*}
\end{align}
where the imaginary part vanishes for $k^2 \le 4m_q^2$.
Using eqs.~(\ref{eq:sigmaqq_dressed}) and~(\ref{eq:sigmaqq_g*}) enables
us to rewrite the $q \bar{q}$ splitting term as

\begin{align}
 \label{eq:qqcontr}
 &\int \mathd \Phi_{q\bar{q}} \,\frac{\mathd \sigma_{q \bar{q}} (\Phi_{q
 \bar{q}})}{\mathd \Phi_{q\bar{q}}}\, \Theta (\Phi_{g^{*}})
 = \int_{4m_q^2} \mathd \lambda^2\, \delta(k^2-\lambda^2)
 \int \mathd \Phi_{q\bar{q}}\,\frac{\mathd \sigma_{q \bar{q}} (\Phi_{q
 \bar{q}})}{\mathd \Phi_{q\bar{q}}}\, \Theta (\Phi_{g^{*}})
 \nonumber \\
 &\hspace{0.5cm}= \int_{4 m_q^2} \frac{\mathd \lambda^2}{2\pi} \int \mathd \Phi_{g^*} \int\mathd
 \Phi_{\rm dec} 
 \, \frac{\mathd \sigma^{(2)}_{q \bar{q}}
 (\Phi_{q \bar{q}})}{\mathd \Phi_{q \bar{q}}} \frac{1}{\left| 1
 + \Pi (\lambda^2+i\eta, m_q^2, \mu^2)
 - \Pi_{\tmop{ct}} \right|^2}\, \Theta (\Phi_{g^{*}})
 \nonumber\\
 &\hspace{0.5cm} =
 \frac{1}{\pi}\int_{4m_q^2} \frac{\mathd \lambda^2}{\lambda^2} \int \mathd \Phi_{g^{*}} \,
 \frac{\mathd \sigma_{g^{*}}^{(1)} (\lambda,\Phi_{g^{*}})}{\mathd \Phi_{g^*}}\, \frac{
 {\rm Im}\, \Pi (\lambda^2, m_q^2, \mu^2)}{\left| 1
 + \Pi (\lambda^2+i\eta, m_q^2, \mu^2) 
 - \Pi_{\tmop{ct}} \right|^2}\, \Theta (\Phi_{g^{*}})\,.\phantom{aaaaaaa} 
\end{align}

\subsection[Combination of the gluon and the $q\bar{q}$ contributions]{Combination of the gluon and $\boldsymbol{q\bar{q}}$ contributions}
Defining
\begin{equation}
 \label{eq:Rlambda}
 R(\lambda)=\int \mathd \Phi_{g^*}\,\frac{
   \sigma_{g^*}^{(1)}(\lambda,\Phi_{g^*} )}{\mathd \Phi_{g^*}}\Theta(\Phi_{g^*}),
\end{equation}
we can combine eq.~(\ref{eq:gluoncontr}) and~(\ref{eq:qqcontr}) and get
\begin{equation}
 \label{eq:sigmag_dressed}
 \sigma_{g^*}=R^{(\epsilon)}(0)\frac{1}{1+\Pi(0,m_q^2,\mu^2)-\Pi_{\rm ct}}-\frac{1}{\pi}\int_{4m_q^2}
 \frac{\mathd \lambda^2}{\lambda^2} R(\lambda)\, {\rm Im}\frac{1}{1+\Pi(\lambda^2,m_q^2,\mu^2)-\Pi_{\rm ct}}\,.
\end{equation}
With the notation $R^{(\epsilon)}(0)$ we remind the reader that for
$\lambda=0$ there are infrared divergences in $R$ that are regulated in
dimensional regularization.

\section[The ${ \sigma_{\rm v}}$ contribution]{The
 $\boldsymbol{ \sigma_{\rm v}}$ contribution}
The NLO differential virtual cross-section can be represented as
\begin{eqnarray}
\frac{ \mathd \sigma^{(1)}_{\rm v}(\Phi_{\rm b})}{\mathd \Phi_{\rm b}} = \int
\frac{\mathd^d k}{(2\pi)^d} \frac{F_{\rm virt}(k,\Phi_{\rm b})}{k^2+i\eta}\,,
\end{eqnarray}
where $k^2$ is the loop momentum and $d=4-2\ep$.
By replacing the free gluon propagator with the dressed one, we obtain the all-orders expression $\sigma_{\rm v}(\Phi_{\rm b})$
\begin{eqnarray}
\frac{ \mathd \sigma_{\rm v}(\Phi_{\rm b})}{\mathd \Phi_{\rm b}} = \int \frac{\mathd^d k}{(2\pi)^d} \frac{F_{\rm virt}(k,\Phi_{\rm b})}{k^2+i\eta}
\frac{1}{1+ \Pi(k^2, \mu^2) - \Pi_{\rm ct}}\,.
\end{eqnarray}
If we use eq.~(\ref{eq:virt_trick}), we obtain
\begin{eqnarray}
\frac{ \mathd \sigma_{\rm v}(\Phi_{\rm b})}{\mathd \Phi_{\rm b}}& =& -\frac{1}{\pi}
\int_{4m_q^2}^{+\infty} \frac{\mathd \lambda^2}{\lambda^2} \int \frac{\mathd^d
 k}{(2\pi)^d} \frac{F_{\rm virt}(k,\Phi_{\rm b})}{k^2-\lambda^2+i\eta} \,{\rm Im} \left\{
\frac{1}{ \Pi(\lambda^2,m_q^2, \mu^2) - \Pi_{\rm ct}} \right\}\!\! \nonumber
\\
&& + \int \frac{\mathd^d k}{(2\pi)^d} \frac{F_{\rm virt}(k,\Phi_{\rm b})}{k^2+i\eta}\frac{1}{ \Pi(0,m_q^2, \mu^2) - \Pi_{\rm ct}},
\end{eqnarray}
where $m_q$ is the light quark mass that has been introduced to regulate the bad behaviour of $\Pi(0,\mu^2)$.
We define
\begin{equation}
\label{eq:Vlambda}
V(\lambda) \equiv \int \mathd \Phi_{\rm b} \Theta (\Phi_{\rm b}) \int
\frac{\mathd^d k}{(2\pi)^d} \frac{F_{\rm virt}(k,\Phi_{\rm b})}{k^2-\lambda^2+i\eta} =
\int \mathd \Phi_{\rm b} \Theta (\Phi_{\rm b}) \,\frac{\mathd
 \sigma^{(1)}_{\rm v}\left(\lambda, \Phi_{\rm b}\right)}{\mathd \Phi_{\rm b}} \,,
\end{equation}
where $\mathd \sigma^{(1)}_{\rm v}(\lambda, \Phi_{\rm b})$ is the
differential virtual cross section computed with a gluon of mass $\lambda$.
The ${ \sigma_{\rm v}}$ of eq.~(\ref{eq:sigmav}) can be finally rewritten
as
\begin{equation}
 \label{eq:sigmav_dressed}
\sigma_{\rm v} = V^{(\epsilon)}(0)\frac{1}{1+\Pi(0,m_q^2,\mu^2)-\Pi_{\rm ct}}\,-\,\frac{1}{\pi}\int
 \frac{\mathd \lambda^2}{\lambda^2} V(\lambda)\, {\rm Im}\frac{1}{1+\Pi(\lambda^2,m_q^2,\mu^2)-\Pi_{\rm ct}}\,,
\end{equation}
where the notation $V^{(\epsilon)}(0)$ signals the presence of the leftover
IR divergences for $\lambda=0$, that are handled in dimensional
regularization. If a finite gluon mass $\lambda$ is employed, IR
singularities are regulated by single and double logarithms of $\lambda$. If
we choose the pole mass scheme, $V(\lambda) \to \frac{1}{\lambda^2}$ for
large $\lambda$. Furthermore, $V(\lambda)/\as$ does not depend on $\mu$.
This signals that there are no UV divergences. Thus $V(\lambda)$, with
$\lambda>0$, can be evaluated performing the $\ep \to 0$ limit since the
integral in eq.~(\ref{eq:sigmav_dressed}) is finite.

\section{Combining the virtual and the real contributions}
We define
\begin{equation}
S(\lambda) \equiv  R(\lambda) + V(\lambda)\,,
\end{equation}
with $R$ and $V$ given by eqs.~(\ref{eq:Rlambda}) and (\ref{eq:Vlambda})
respectively.
As long as $\lambda$ is not zero, $R$ and $V$ are separately well defined in
$d=4$ dimensions, but they are IR divergent for $\lambda \to 0$.
However, as required by the KLN theorem, their sum $S$ is finite in that
limit, indeed
\begin{equation}
S(0) = \lim_{\lambda \to 0 }\Big[ R(\lambda) +V(\lambda) \Big]= \lim_{\epsilon
  \to 0 } \lq R^{(\ep)}(0)+V^{(\ep)}(0) \rq = \sigma^{(1)},
\end{equation}
being  $\sigma^{(1)}$ the
integrated NLO cross section. 
Thus, summing eqs.~(\ref{eq:sigmav_dressed}) and
(\ref{eq:sigmag_dressed}), we have
\begin{eqnarray}
 \label{eq:RpV}
 \sigma_{\rm v} + \sigma_{g^*} &=& S(0) \frac{1}{1 + \Pi (0,m_q^2,\mu^2) -
 \Pi_{\tmop{ct}}} \nonumber \\
 && - \frac{1}{\pi}\int_{4 m_q^2}^{+\infty}
\frac{\mathd \lambda^2 }{\lambda^2}\, S (\lambda) \, {\rm Im} \lq \frac{1}{1 + \Pi (\lambda^2,m_q^2,\mu^2) -
 \Pi_{\tmop{ct}}}\rq
 \nonumber \\
 &=& - \int_{0^-}^{+\infty}
\frac{\mathd \lambda^2 }{\pi}\, S (\lambda) \, {\rm Im} \lq \frac{1}{\lambda^2+
 i\eta}\, \frac{1}{1 + \Pi (\lambda^2,m_q^2,\mu^2) -
 \Pi_{\tmop{ct}}}\rq, \phantom{aaaa}
\end{eqnarray}
where in the last line we have used eq.~(\ref{eq:imag_den}) and the fact that
the limit $\lambda \to 0$ is well defined for $S$, while this was not the case
for $V$ and $R$ separately.
We can check that at $\mathcal{O}(\as)$ we recover the NLO result
\begin{equation}
\lq \sigma_{\rm v} + \sigma_{g^*} \rq_{\mathcal{O}(\as)} = - \int_{0^-}^{+\infty} \frac{\mathd \lambda^2 }{\pi}\, S (\lambda) \,
 {\rm Im} \lq \frac{1}{\lambda^2+ i\eta}\, \rq 
 = - \int \frac{\mathd \lambda^2 }{\pi}\, S (\lambda) \lq -\pi \, \delta(\lambda^2) \rq 
 = \sigma^{(1)}\,. 
\end{equation}
We now illustrate a procedure to safety take the limit $m_q \to
0$ in eq.~(\ref{eq:RpV}).
We can write eq.~(\ref{eq:RpV}) by adding and subtracting the same term
\begin{align}
 \label{eq:sigma_g_plus_obs_v}
 \sigma_g + \sigma_{\tmop{v}} = & - \int_{0^-}^{\infty} \frac{\mathd
 \lambda^2}{\pi} \left[S(\lambda)- S (0) \,\frac{\mu^2}{\lambda^2 + \mu^2}
 \right] {\rm Im} \left[ \frac{1}{\lambda^2+i\eta} \, \frac{1}{1 + \Pi
 (\lambda^2, m_q^2, \mu^2) - \Pi_{\tmop{ct}}} \right]
 \nonumber\\ & - \int_{0^-}^{\infty} \frac{\mathd \lambda^2}{\pi} S (0)
 \,\frac{\mu^2}{\lambda^2 + \mu^2} \, {\rm Im} \left[ \frac{1}{\lambda^2+i\eta}
 \frac{1}{1 + \Pi (\lambda^2, m_q^2, \mu^2) - \Pi_{\tmop{ct}}}
 \right],
\end{align}
where $\mu^2$ is a real positive number. The first integral is regular for
$\lambda^2 \rightarrow 0$ also if $m_q = 0$, since the term in the square
brackets is ${\cal O}(\lambda)$ in this limit. This also enables us to move
the lower bound from $0^-$ to $0$. Using the identity already employed in
eq.~(\ref{eq:tointpart}), we are left to
\begin{align}
 \label{eq:S_subtracted}
 & \int_0^{\infty} \frac{\mathd \lambda^2}{\pi} \left[S(\lambda)- S (0)\,
 \frac{\mu^2}{\lambda^2 + \mu^2} \right]{\rm Im} \left[ \frac{1}{\lambda^2+i\eta} \, \frac{1}{1 +
 \Pi (\lambda^2, \mu^2) - \Pi_{\tmop{ct}}} \right]
 \nonumber \\ 
 &\hspace{0.2cm} = \int_0^{\infty} \frac{\mathd \lambda^2}{\pi} \left[S(\lambda)- S (0)
 \,\frac{\mu^2}{\lambda^2 + \mu^2} \right]{\rm Im} \lg \frac{1}{\as \, b_{0,f}}\,
 \frac{\mathd\ }{\mathd \lambda^2} \log \lq 1 + \Pi\left(\lambda^2, \mu^2\right) -
 \Pi_{\tmop{ct}} \rq \rg
 \nonumber \\
 &\hspace{0.2cm} = -\int_0^{\infty} \frac{\mathd \lambda^2}{\pi} \frac{\mathd\ }{\mathd \lambda^2}
 \left[S(\lambda)- S (0)\, \frac{\mu^2}{\lambda^2 + \mu^2} \right] \frac{1}{\as \, b_{0,f}
 }\, {\rm Im} \lg \log \lq 1 + \Pi\left(\lambda^2, \mu^2\right) -
 \Pi_{\tmop{ct}} \rq \rg,
 \nonumber\\
\end{align}
where the boundary terms vanish, since $S(\lambda)$ vanishes for large
$\lambda$ (if the pole scheme is adopted),
and the difference in the square brackets is zero for $\lambda=0$.

\begin{figure}[tb]
 \centering
 \includegraphics[width=0.55\textwidth]{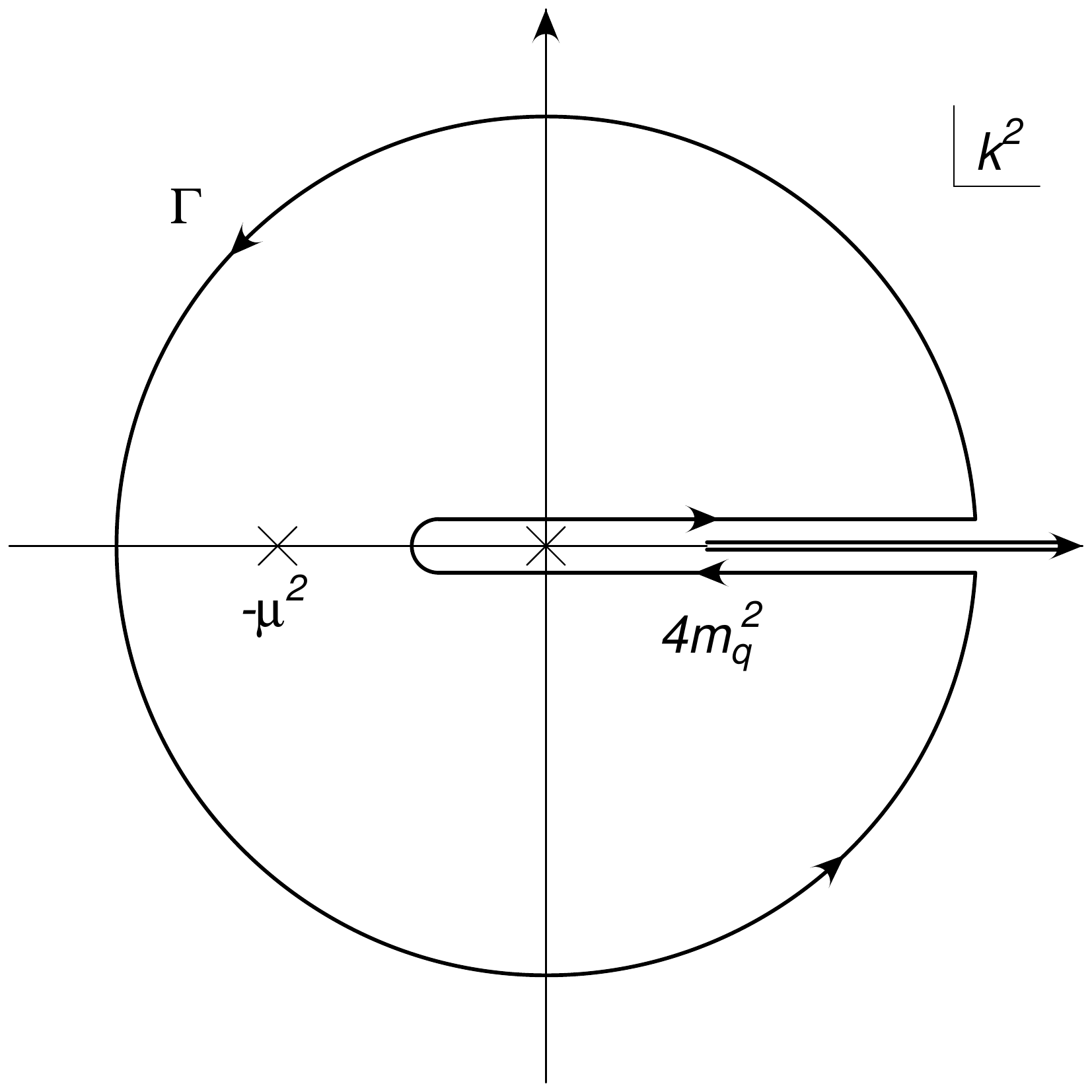}
 \caption{Integration path.}
 \label{fig:integr_path1}
\end{figure}

The second integral can be extended below from $0^-$ to $- \mu^2 / 2$, since
the imaginary part is zero for $\lambda^2 <0$. We can rewrite
\begin{eqnarray}
 && \int_{0^-}^{\infty} \frac{\mathd \lambda^2}{\pi} S (0)\, \frac{\mu^2}{\lambda^2 + \mu^2}\,
 {\rm Im} \left[ \frac{1}{\lambda^2+i\eta} \, \frac{1}{1 + \Pi (\lambda^2, m_q^2, \mu^2) -
 \Pi_{\tmop{ct}}} \right]
 \nonumber\\
 &&{}\hspace{0.5cm} =
\int_{- \mu^2 / 2}^{\infty} \frac{\mathd \lambda^2}{\pi} S (0)\, \frac{\mu^2}{\lambda^2 + \mu^2}\,
 {\rm Im} \left[ \frac{1}{\lambda^2+i\eta} \, \frac{1}{1 + \Pi (\lambda^2, m_q^2, \mu^2) -
 \Pi_{\tmop{ct}}} \right]
\nonumber \\
 &&{}\hspace{0.5cm} = \frac{1}{2 i} \oint \frac{\mathd \lambda^2}{\pi} S(0)\,
\frac{\mu^2}{\lambda^2 + \mu^2} \,
 \frac{1}{\lambda^2} \, \frac{1}{1 + \Pi (\lambda^2, m_q^2, \mu^2) - \Pi_{\tmop{ct}}}\,,
\end{eqnarray}
where the contour is depicted in Fig.~\ref{fig:integr_path1}.
The integral in the last line is equal to the residue at $\lambda^2=-\mu^2$, that is
well-defined also for $m_q=0$. This allows us to safely take the limit $m_q
\rightarrow 0$.
By using eq.~(\ref{eq:tointpart}), and integrating by parts, we are lead to

\begin{eqnarray}
 & & \frac{1}{2 i} \oint \frac{\mathd \lambda^2}{\pi} S (0) \, \frac{m^2}{\lambda^2 +
 m^2} \, \frac{1}{\lambda^2} \, \frac{1}{1 + \Pi (\lambda^2, \mu^2) -
 \Pi_{\tmop{ct}}}
 \nonumber \\
 &&\hspace{0.5cm} = \frac{1}{2 i} \oint \frac{\mathd \lambda^2}{\pi} S(0)\, \frac{m^2}{\lambda^2 +
 m^2} \, \frac{1}{\as \, b_{0,f}}\, \frac{\mathd\ }{\mathd \lambda^2} \log \lq 1 + \Pi
 \left(\lambda^2, \mu^2\right) - \Pi_{\tmop{ct}}\rq
 \nonumber \\
 &&\hspace{0.5cm} = -\frac{1}{2 i} \oint \frac{\mathd \lambda^2}{\pi} S(0)\,
 \frac{\mathd\ }{\mathd \lambda^2} \left[ \frac{\mu^2}{\lambda^2 + \mu^2} \right]
 \frac{1}{\as \, b_{0,f}}\, \log \lq 1 + \Pi\left(\lambda^2, \mu^2\right) - \Pi_{\tmop{ct}})\rq. \qquad
\end{eqnarray}
If we set to infinity the larger radius of the boundary in
Fig.~\ref{fig:integr_path1}, its contribution vanishes. The same holds
if the radius around $\lambda^2=-m^2/2$ becomes infinitesimal.
We are thus left with
\begin{align}
 & - \frac{1}{2i}\int_{-\mu^2/2}^{\infty} \frac{\mathd \lambda^2}{\pi} S(0)\,
 \frac{\mathd \ }{\mathd \lambda^2} \left[ \frac{\mu^2}{\lambda^2 + \mu^2} \right]
 \frac{1}{\as \, b_{0,f}} \nonumber \\
 &\hspace{1.5cm} \times \Big\{ \log \lq 1 + \Pi\left(\lambda^2+i\eta, \mu^2\right) - \Pi_{\tmop{ct}} \rq -\log \lq 1 + \Pi\left(\lambda^2-i\eta, \mu^2\right) - \Pi_{\tmop{ct}} \rq \Big\} \nonumber \\
 & \hspace{0.5cm} = -\int_{-\mu^2/2}^{\infty} \frac{\mathd \lambda^2}{\pi} S(0)\,
 \frac{\mathd \ }{\mathd \lambda^2} \left[ \frac{\mu^2}{\lambda^2 + \mu^2} \right]
 \frac{1}{\as \, b_{0,f}}
\, {\rm Im} \lg \log \lq 1 + \Pi\left(\lambda^2, \mu^2\right) - \Pi_{\tmop{ct}} \rq \rg \nonumber \\ 
 &\hspace{0.5cm} = -\int_0^{\infty} \frac{\mathd \lambda^2}{\pi} S(0)\,
 \frac{\mathd \ }{\mathd \lambda^2} \left[ \frac{\mu^2}{\lambda^2 + \mu^2} \right]
 \frac{1}{\as \, b_{0,f}}
 \, {\rm Im} \lg \log \lq 1 + \Pi\left(\lambda^2, \mu^2\right) - \Pi_{\tmop{ct}} \rq \rg , \phantom{aaaaaa}
 \label{eq:Szero}
\end{align}
where in the last step we used the fact that the the imaginary part vanishes for negative values of $\lambda$. Using the results of
eqs.~(\ref{eq:S_subtracted}) and~(\ref{eq:Szero}), we can write
eq.~(\ref{eq:sigma_g_plus_obs_v}) as
\begin{eqnarray}
 \label{eq:sigma_g_plus_obs_v_final}
 \sigma_g + \sigma_{\tmop{v}} = \frac{1}{\as \, b_{0,f}}
 \int_0^{\infty} \frac{\mathd \lambda}{\pi} \, \frac{\mathd\ }{\mathd \lambda}\!
 \left[S(\lambda)\right] {\rm Im} \lg \log \lq 1 + \Pi\left(\lambda^2+i\eta, \mu^2\right) -
 \Pi_{\tmop{ct}} \rq \rg.
\end{eqnarray}

\section{Calculation summary}
Here we summarize our findings.
If we combine eqs.~(\ref{eq:sigma_g_plus_obs_v_final}) and (\ref{eq:sigmaqq_final}), we have
\begin{eqnarray}
 \!\!\!\!\!\!\!\!\!
 \sigma & = & \int \mathd \Phi_{\tmop{b}} \,\frac{\mathd \sigma_{\tmop{b}} (\Phi_{\tmop{b}})}{\mathd \Phi_{\tmop{b}}} \, \Theta (\Phi_{\tmop{b}})
 + \int \mathd \Phi_{\tmop{b}} \,\frac{\mathd \sigma_{\tmop{v}}
 (\Phi_{\tmop{b}})}{\mathd \Phi_{\tmop{b}}}\, \Theta (\Phi_{\tmop{b}})
 + \int \mathd \Phi_g \, \frac{\mathd \sigma_g (\Phi_g)}{\mathd \Phi_g} \, \Theta (\Phi_g) \nonumber \\
 && +\int \mathd \Phi_{ q \bar{q}} \,\frac{\mathd \sigma_{q
 \bar{q}} (\Phi_{q \bar{q}})}{\mathd \Phi_{q \bar{q}}} \, \Theta (\Phi_{q \bar{q}}) \label{eq:sigma_initial} \\ 
& = & \sigma_{\rm b} +\frac{1}{\as \, b_{0,f}}\int_0^{\infty} \frac{\mathd k}{\pi} \, \frac{\mathd\ }{\mathd k}\!
 \left[ T(\lambda) \right] \, {\rm Im} \lg \log \lq 1 + \Pi\left(\lambda^2, \mu^2\right) -
 \Pi_{\tmop{ct}} \rq \rg\!\nonumber \\
& = & \sigma_{\rm b} -\frac{1}{\as \, b_{0,f}}\int_0^{\infty} \frac{\mathd \lambda}{\pi} \, \frac{\mathd\ }{\mathd \lambda}\!
 \left[ T(\lambda) \right] \,
 \arctan \lq \frac{\as \,\pi\,b_{0,f}}{1+\as \,b_{0,f}
 \log\left( \frac{\lambda^2}{\mu^2 e^C}\right) }\rq,
 \label{eq:sigma_final}
\end{eqnarray}
with
\begin{eqnarray}
 \sigma_{\tmop{b}} & = &\int \mathd \Phi_{\tmop{b}} \,\frac{\mathd
 \sigma_{\tmop{b}} (\Phi_{\tmop{b}})}{\mathd
 \Phi_{\tmop{b}}} \, \Theta (\Phi_{\tmop{b}})\,, \\ 
\label{eq:Tk2_final} 
T(\lambda) & = & V(\lambda) + R(\lambda) + \Delta(\lambda)\,,\\
V(\lambda) & = & \int \mathd \Phi_{\rm b} \,\frac{\mathd
 \sigma^{(1)}_{\rm v}\left(\lambda, \Phi_{\rm
 b}\right)}{\mathd\Phi_{\rm b} }\Theta (\Phi_{\rm b})\,, \\
R(\lambda) & = & \int \mathd \Phi_{g^{*}}\,
 \frac{\mathd
 \sigma^{(1)}_{g^*}\left(\lambda,\Phi_{g^{*}}\right)}{\mathd
 \Phi_{g^{*}}} \,
 \Theta (\Phi_{g^{*}})\,, \\ 
\Delta\!\left(\lambda\right) &=& - \frac{\lambda^2}{\as\, b_{0,f}} \int \delta(k^2-\lambda^2)\, \mathd
 \Phi_{q\bar{q}} \,\frac{\mathd \sigma^{(2)}_{q \bar{q}}(\Phi_{q\bar{q}})}{\mathd \Phi_{q\bar{q}} } \, \Big[ \Theta(\Phi_{q \bar{q}}) - \Theta (\Phi_{g^{*}})\Big] .
\label{eq:Deltak2_final}
\end{eqnarray}
To obtain the final expression in eq.~(\ref{eq:sigma_final}) we have employed
the following identity
\begin{equation}
 {\rm Im} \lg \log \lq 1 + \Pi\left(\lambda^2, \mu^2\right) -
 \Pi_{\tmop{ct}} \rq \rg =  -\arctan \lq \frac{\as \,\pi\,b_{0,f}}{1+\as \,b_{0,f}
   \log\left( \frac{\lambda^2}{\mu^2 e^C}\right) }\rq,
 \label{eq:arctan}
\end{equation}
where in the right-hand side we have neglected the contribution
\begin{equation}
  \label{eq:thetaLandau}
-\pi \Theta \left(\mu^2\exp\left(-\frac{1}{b_0\as}+C\right) -\lambda^2 \right).
\end{equation}
We stress that the perturbative expansion in $\as$ of
formula~(\ref{eq:sigma_final}) is an asymptotic one, and only its
coefficients are unambiguously defined, and are the subject of the present
work. Thus, for our purposes, eq.~(\ref{eq:sigma_final}) is defined up to
corrections that have a vanishing perturbative expansion in $\as$, as are,
for instance, the exponentials of the negative inverse of $\as$. For this
reason, the contribution in eq.~(\ref{eq:thetaLandau}) can be neglected and
eq.~(\ref{eq:arctan}) becomes exact, since the left and the right-hand side
have the same perturbative expansion in $\as$.  In ref.~\cite{Ball:1995ni},
eqs.~(2.24) and (2.25), the form of the resummed expression for typical
euclidean quantities is given by taking the inverse Borel transform of the
Borel transform of the perturbatve expansion, with the prescription that the
singularities in the Borel integration should be bypassed above the positive
real axis. The form of their result is simlar to ours, except for corrections
that yield powers of $\exp(-1/(b_0\as))$.

Eq.~(\ref{eq:sigma_final}) can be rewritten as
\begin{equation}
\sigma = \sigma_{\rm b} -\frac{1}{ b_{0,f}}\int_0^{\infty} \frac{\mathd
 \lambda}{\pi} \, \frac{\mathd\ }{\mathd \lambda}\! \left[
 \frac{T(\lambda)}{\as(\mu)} \right] \, \arctan \lq b_{0,f}
\,\pi\,\as\(\lambda \,e^{-C/2}\)\rq.
\end{equation} 
From this expression it is evident that the resummed result does
not depend on $\mu$, since the factor $\as(\mu)$ cancels in the expression in
the squared brackets.

In order to evaluate numerically $T(\lambda)$ we performed the following
steps. We computed separately $V(\lambda)$, $R(\lambda)$ and
$\Delta(\lambda)$ for several values of $\lambda$, using the \RES{}
framework~\cite{Jezo:2015aia}, to integrate over the phase space. Indeed, as
long as $\lambda>0$, these contributions are finite.  The $\lambda=0$ point
is also computed in the \RES{}, that automatically implements the
subtraction of infrared singularities in the real cross section. The
scalar integrals appearing in the virtual amplitude are evaluated using the
{\tt COLLIER}~\cite{Denner:2016kdg} library. The calculation of the top
pole-mass counterterm and of the bottom field normalization constant in
presence of a finite gluon mass is detailed in
\writeApp~\ref{sec:self-energy}.

In order to obtain the analytic expression of $T$, we performed a
polynomial fit for small-$k$ values, specifically for $\lambda<5$~GeV,
\begin{equation}
 T(\lambda) = p_0 \,+\,p_1\, \lambda \,+\, p_2\, \lambda^2\, +\,
 \ldots\,,
 \label{eq:T_polynom}
\end{equation}
while we adopted a cubic spline interpolation for larger values of $\lambda$,
imposing that both $T$ and its derivative are continuous for
$\lambda=5$~GeV. The fitting functions that we find are seen to represent
sufficiently well the numerical results for $T$, with the only caveat that,
for small $\lambda$, these have non-negligible errors. These errors strongly
affect the coefficient $p_1$, and have negligible effects on the other
coefficients. In fact, $p_0$ is computed directly for massless gluons, and
has a totally negligible error. The $p_2$ and higher order coefficients are
controlled by the larger values of $\lambda$, where our computation has a
smaller error. Furthermore, only $p_1$ is responsible for the presence of
linear renormalons, thus, at higher-orders, it dominates the value of the
integral in~(\ref{eq:sigma_final}). We thus propagated only the error on the
$p_1$ coefficient to the calculation of the coefficients of the
perturbative expansion.

\section{Infrared-safe observables}
We are also interested in evaluating the average value of a generic infrared-safe observable, function of the phase space kinematics, $O(\Phi)$:
\begin{align}
 \label{eq:aveM1}
 \langle O \rangle \, = \, N_\Theta &\lg \int \mathd \Phi_{\tmop{b}}
 \,\frac{\mathd \sigma_{\tmop{b}}(\Phi_{\tmop{b}}) }{\mathd \Phi_{\tmop{b}}}\, O_\Theta (\Phi_{\tmop{b}})
 + \int \mathd \Phi_{\tmop{b}} \,\frac{\mathd \sigma_{\tmop{v}}
 (\Phi_{\tmop{b}})}{\mathd \Phi_{\tmop{b}}} \, O_\Theta (\Phi_{\tmop{b}})
 \right.\nonumber \\
 &\; \left. + \int \mathd \Phi_g \, \frac{\mathd \sigma_g (\Phi_g)}{\mathd \Phi_g} \, O_\Theta (\Phi_g)
 + \int \mathd
 \Phi_{q\bar{q}} \,\frac{\sigma_{q \bar{q}} (\Phi_{q \bar{q}})}{\mathd \Phi_{q \bar{q}}} \, O_\Theta (\Phi_{q \bar{q}}) \rg, \qquad
\end{align}
where 
\begin{equation}
O_\Theta \left(\Phi\right) = \Theta\left(\Phi\right) \times O\left(\Phi\right)
\end{equation}
and $N_\Theta$ is a normalization factor whose expression is given by
\begin{align}
 \label{eq:def_N}
 N_\Theta^{- 1}  = &\int \mathd \Phi_{\tmop{b}} \,\frac{\mathd \sigma_{\tmop{b}} (\Phi_{\tmop{b}})}{\mathd \Phi_{\tmop{b}}} \, \Theta (\Phi_{\tmop{b}})
 + \int \mathd \Phi_{\tmop{b}} \,\frac{\mathd \sigma_{\tmop{v}} (\Phi_{\tmop{b}}) }{\mathd \Phi_{\tmop{b}}}\, \Theta (\Phi_{\tmop{b}})
 + \int \mathd \Phi_g \, \frac{\mathd \sigma_g (\Phi_g)}{\mathd \Phi_g} \, \Theta (\Phi_g) \nonumber \\
 &
+\int \mathd
 \Phi_{q\bar{q}} \,\frac{\sigma_{q \bar{q}} (\Phi_{q \bar{q}})}{\mathd
   \Phi_{q \bar{q}}} \, \Theta (\Phi_{q \bar{q}}) \, = \, \sigma^{-1}.
\end{align}
The factor $N_\Theta$ that multiplies the virtual, the real and the $q\bar{q}$ contributions is in
fact simply the inverse of the Born cross section, since the quantities it
multiplies are already at NLO level. Thus, in these cases,
\begin{equation}
 \label{eq:N0}
 N_\Theta \rightarrow N^{(0)}_\Theta = \lg \int\mathd \Phi_{\tmop{b}}\,
 \frac{\mathd \sigma_{\tmop{b}} (\Phi_{\tmop{b}})}{\mathd \Phi_{\tmop{b}}}\,\Theta\(\Phi_{\tmop{b}}\)\rg^{-1}. 
\end{equation}
The factor of $N_\Theta$ in front of the Born term, on the other hand, must be expanded in series
\begin{eqnarray}
 N_\Theta 
 & = & N^{(0)}_\Theta \lg 1 - N^{(0)}_\Theta
 \lq
 \int \mathd \Phi_{\tmop{b}} \,\frac{\mathd \sigma_{\tmop{v}} (\Phi_{\tmop{b}}) }{\mathd \Phi_{\tmop{b}}}\, \Theta (\Phi_{\tmop{b}})
 + \int \mathd \Phi_g \, \frac{\mathd \sigma_g (\Phi_g)}{\mathd \Phi_g} \, \Theta (\Phi_g) 
 \right.\right.
\nonumber \\
 && \hspace{2.5cm}\left.\left. + \int \mathd \Phi_{q\bar{q}}\,\sigma_{q \bar{q}} (\Phi_{q
 \bar{q}})\Theta\left(\Phi_{q\bar{q}}\right) 
\rq \rg + \mathcal{O}\left(\as^2 \left(\as n_f \right)^n\right). \qquad \qquad
\end{eqnarray}
This gives rise to a constant Born term of the form
\begin{equation}
 \obsb \equiv N^{(0)}_\Theta \int \mathd
 \Phi_{\tmop{b}}\, \frac{\mathd \sigma_{\tmop{b}} 
 (\Phi_{\tmop{b}})}{\mathd \Phi_{\tmop{b}}}\, O_\Theta (\Phi_{\tmop{b}})\, , 
\end{equation}
plus an NLO correction equal to
\begin{equation}
 - N^{(0)}_\Theta \, \obsb \lq \int \mathd \Phi_{\tmop{b}} \, \frac{\mathd
 \sigma_{\tmop{v}}(\Phi_{\tmop{b}})}{\mathd \Phi_{\tmop{b}}}
 + \int \mathd \Phi_g \, \frac{\mathd \sigma_g (\Phi_g)}{\mathd \Phi_g} \, 
 + \int\mathd \Phi_{q\bar{q}}\, \frac{\mathd \sigma_{q \bar{q}} (\Phi_{q
 \bar{q}})}{\mathd \Phi_{q\bar{q}}} \rq . 
\end{equation}
In summary, eq.~(\ref{eq:aveM1}) becomes
\begin{align}
 \label{eq:Mave_initial}
 \langle O \rangle = \obsb &{} + N^{(0)}_\Theta \int \mathd
 \Phi_{\tmop{b}} \, \frac{ \mathd \sigma_{\tmop{v}}
 (\Phi_{\tmop{b}})}{\mathd \Phi_{\tmop{b}}} \Big[ O_\Theta (\Phi_{\tmop{b}}) - \obsb\Theta(\Phi_{\tmop{b}}) \Big]
 \nonumber\\
 & {}+ N^{(0)}_\Theta 
 \int \mathd \Phi_g \, \frac{\mathd \sigma_g(\Phi_g)}{\mathd \Phi_g} \Big[ O_\Theta (\Phi_g) - \obsb \Theta(\Phi_{g})\Big]
 \nonumber \\
 & {}+ N^{(0)}_\Theta \int \mathd \Phi_{q\bar{q}} \, \frac{\mathd \sigma_{q \bar{q}}
 (\Phi_{q \bar{q}})}{\mathd \Phi_{q\bar{q}}} \,
 \Big[ O_\Theta(\Phi_{q \bar{q}}) - \obsb
 \Theta(\Phi_{q \bar{q}})\Big]\,.
\end{align}
We notice that eqs.~(\ref{eq:Mave_initial}) and (\ref{eq:sigma_initial}) are
similar: the expression of the higher order corrections of $\langle O \rangle$
can obtained from the the expression of the higher order corrections of the
total cross section by replacing
\begin{equation}
 \label{eq:aveObs_prescription}
\Theta(\Phi) \to N^{(0)}_\Theta \big[ O_\Theta(\Phi) - \obsb \Theta(\Phi) \big].
\end{equation}
Thus, starting from eqs.~(\ref{eq:sigma_final})--(\ref{eq:Deltak2_final}), we
can write

\begin{equation}
 \label{eq:final_M_wcuts}
 \langle O \rangle = \obsb{} - \frac{1}{\as \, b_{0,f}}\! \int_0^{\infty}
 \!\frac{\mathd \lambda}{\pi} \, \frac{\mathd\ }{\mathd k} \!\lq
 \widetilde{T}(\lambda) \rq \arctan \lq \frac{\as\,\pi\, b_{0,f}}{1+\as \,b_{0,f}\,\log\left( \frac{\lambda^2}{\mu^2 e^C}\right) }\rq,
\end{equation}
where
{ \allowdisplaybreaks
\begin{eqnarray}
 \label{eq:Mbdef}
 \obsb{} &=& N^{(0)}_\Theta \int \mathd \Phi_{\tmop{b}}\, \sigma_{\tmop{b}}
 (\Phi_{\tmop{b}}) \, O (\Phi_{\tmop{b}}) \, \Theta(\Phi_{\tmop{b}}) \,,
 \\
 \label{eq:Ttdef}
 \widetilde{T}(\lambda) & = &
 \widetilde{V}(\lambda) + \widetilde{R}(\lambda) + 
 \widetilde{\Delta}(\lambda)
 \\
 \label{eq:final_Vtilde}
 \widetilde{V}(\lambda) &=& N^{(0)}_\Theta \int \mathd \Phi_{\rm b} \, \sigma_{\rm v}^{(1)}
 (\lambda,\Phi_{\rm b})\, \Big[ O (\Phi_{\tmop{b}}) - \obsb{}\Big] \Theta(\Phi_{\tmop{b}})\, ,
 \\
 \label{eq:final_Rtilde}
 \widetilde{R}(\lambda) &=& N^{(0)}_\Theta \int \mathd \Phi_{g^{*}}\, \sigma^{(1)}_{g^{*} } (\lambda,
 \Phi_{g^{*}}) \Big[
 O(\Phi_{g^{*}}) - \obsb{}\Big] \Theta(\Phi_{g^{*}})\,,
 \\
\widetilde{\Delta}(\lambda) &=& - N^{(0)}_\Theta \frac{\lambda^2}{\as\, b_{0,f}} \int \delta(k^2-\lambda^2)\, \mathd
 \Phi_{q\bar{q}} \,\frac{\mathd \sigma^{(2)}_{q \bar{q}}(\Phi_{q\bar{q}})}{\mathd \Phi_{q\bar{q}} } \nonumber\\ \label{eq:final_Deltaqqtilde}
&& \hspace{1cm} \times \Big\{
 \Big[ O(\Phi_{q \bar{q}}) - \obsb{}\Big] \Theta(\Phi_{q \bar{q}})
 - \Big[ O(\Phi_{g^{*}}) - \obsb{}\Big] \Theta(\Phi_{g^{*}})\Big\}\,. \qquad
\end{eqnarray}
}
We notice that when computing inclusive quantities or quantities that do not
depend upon the jet kinematics, the $\widetilde{\Delta}(\lambda)$ and
$\Delta(\lambda)$ terms of eqs.~(\ref{eq:final_Deltaqqtilde})
and~(\ref{eq:Deltak2_final}) are zero. In these cases, our results can just
be expressed as functions of the NLO differential cross sections, computed
with a non-zero gluon mass. In general, however, the
$\widetilde{\Delta}(\lambda)$ and $\Delta(\lambda)$ contributions cannot be
neglected, since observables built with the full kinematics may differ from
those obtained by clustering the $q\bar{q}$ pair into a massive gluon. This
was first discussed in Ref.~\cite{Nason:1995np}, in the context of $e^+ e^-$
annihilation into jets.\footnote{In Refs.~\cite{Dokshitzer:1997iz,
 Dokshitzer:1998pt} it was shown that, for a large set of jet-shape
 observables, in order to account for the effect of the $\Delta$ terms, the
 naive predictions computed considering only the $V+R$ contributions must be
 rescaled by a factor, dubbed the ``Milan factor'', to get the correct
 coefficient for the $1/Q$ non-perturbative effects.}

The strategy we adopted to extract the analytic expression of
$\widetilde{T}(\lambda)$ is the same we employed for $T(\lambda)$.

\section{Changing the mass scheme}
\label{sec:changescheme}
The relation between the pole mass $m$ and the \MSB{} mass $\mMSB$ is given
by the formula (see Sec.~\ref{sec:msbar2pole})
\begin{equation}
 \label{eq:msbpole}
 m - \mMSB(\mu) = \mpole \Big[r_f(\mpole, \mu, \as) +
 r^{\rm (f)}_d(\mpole,\mu,\as) \Big]\,,
\end{equation}
where $r_f(\mpole, \mu, \as)$ and $r_d(\mpole,\mu,\as)$ are defined in
eqs.~(\ref{eq:rtilde_split}) and~(\ref{eq:rtilde_d}) respectively and
(f) denotes the finite part according to the \MSB{} prescription.
The $r_f(\mpole, \mu, \as)$ term can be manipulated as in
eq.~(\ref{eq:final_tilderf}), that we report here for ease of reading,
\begin{equation}
r_f(\mpole, \mu, \as) = -\frac{1}{\as b_{0,f}}
\int_{0 }^{\infty}
\frac{\mathd \lambda}{\pi} \, \frac{\mathd }{\mathd \lambda}
\lq \widetilde{r}_f(\lambda) \rq
 \arctan \lq \frac{\as \, \pi\,b_{0,f}}{1+\as \, b_{0,f}
 \log\left( \frac{\lambda^2}{\mu^2 e^C}\right) }\rq,
\end{equation}
where (see eq.~(\ref{eq:rf_small_lambda}))
\begin{equation}
 \widetilde{r}_f(\lambda) = -\as \frac{\CF}{2} \frac{\lambda}{m} + {\cal O}(\lambda)\,.
\end{equation}
As stressed in App.~\ref{sec:msbar2pole}, the linear dependence of
$\widetilde{r}_f(\lambda^2)$ from $\lambda$ is responsible for the presence of a linear
renormalon in the expression of the pole mass in terms of the \MSB{}
mass, while $r_d(\mpole,\mu,\as)$ is free from linear renormalons.\footnote{The relation between the pole and the \MSB{} mass
in the large-$n_f$ limit is well-known (see e.g.~\cite{Beneke:1994rs,
 Beneke:1994qe, Beneke:1994sw}). Here we have re-derived it so as to put it
in a form similar to eqs.~(\ref{eq:final_M_wcuts})
and~(\ref{eq:sigma_final}).}

In the present work we deal with the finite width of the top by using the
complex mass scheme~\cite{Denner:2005fg, Denner:2006ic}. Thus, in our mass
relation, both $m$ and $\mMSB$ are complex, and also $r_f(\mpole, \mu, \as)$ and
$r_d(\mpole,\mu,\as)$. 

Given a result for a quantity $Q$ expressed in terms of the pole mass,
in order to find its expression in terms of the \MSB{} mass we need to
Taylor-expand its mass dependence in its leading order expression, and
multiply it by the appropriate mass correction. In order to do so, we
express $Q$ in terms of the pole mass and its complex conjugate, as if
they were independent variables (one can think of $m$ appearing in the
amplitude, and $m^*$ appearing in its complex conjugate). Denoting
with $Q_{\rm b}$ the LO prediction, we can write
\begin{align}
 \label{eq:changescheme}
 \!\!\!\! Q_{\rm b}(m,m^*)=&\,Q_{\rm b}(\mMSB,\mMSB^*) +\lg \frac{\partial
 Q_{\rm b}(\mpole,\mpole^*) }{\partial \mpole}\left(m-\mMSB\right)+{\rm
 cc}\rg + \mathcal{O}\(\as^2 \(\as n_f\)^n\) \nonumber \\
 =&\,Q_{\rm b}(\mMSB,\mMSB^*) + 2\,{\rm Re}\lg\frac{\partial
 Q_{\rm b}(\mpole,\mpole^*) }{\partial \mpole}\left(m-\mMSB\right) \rg
 \nonumber \\
 =&\, Q_{\rm b}(\mMSB,\mMSB^*)
 +2\,{\rm Re}\lg \frac{\partial Q_{\rm b}(\mpole,\mpole^*) }{\partial
 \mpole}\, \mpole \, \Big[ r_f(\mpole, \mu, \as) +
 r_d\left(\as,\mu,\mpole\right) \Big]\rg, \!\!\!
\end{align}
where cc means complex conjugate.
If $Q_{\rm b} = \sigma_{\rm b}$, we have
\begin{equation}
 \label{eq:sigma_derivative}
 \frac{\partial}{\partial m} \sigma_{\rm b}\left(m,m^*\right)= \int \mathd
 \Phi_{\tmop{b}}\, \frac{\partial \sigma_{\tmop{b}} \left(m,m^*;
 \Phi_{\tmop{b}}\right)}{\partial m} \Theta(\Phi_{\tmop{b}})\,,
\end{equation}
that corresponds to the coefficient of the pole-mass counterterm of
the interference between the virtual and Born amplitude (before
taking two times the real part).

If $Q_{\rm b} = \obsb$ and we explicit the $m$ dependence of the
normalization factor $N^{(0)}_\Theta$, we obtain
\begin{align}
 \frac{\partial \langle \obs \rangle_{\rm b}\left(m,m^*\right)}{\partial m} =
 &\frac{\partial}{\partial m} \lq N^{(0)}_\Theta\left(m,m^*\right) \int \mathd
 \Phi_{\tmop{b}}\, \sigma_{\tmop{b}} \left(m,m^*; \Phi_{\tmop{b}}\right) \, O
 (\Phi_{\tmop{b}}) \, \Theta(\Phi_{\tmop{b}}) \rq \nonumber \\ = &
 N^{(0)}_\Theta\left(m,m^*\right) \int \mathd \Phi_{\tmop{b}}\, \frac{\partial
 \sigma_{\tmop{b}} \left(m,m^*; \Phi_{\tmop{b}}\right)}{\partial m} \, \Big[
 O (\Phi_{\tmop{b}}) - \langle \obs \rangle_{\rm b}\left(m,m^*\right)\Big] \,
 \Theta(\Phi_{\tmop{b}})\,, \qquad \qquad
 \label{eq:aveObs_derivative}
\end{align}
that, again, corresponds to the pole mass counterterm coefficient.
Notice that we could have obtained eq.~(\ref{eq:aveObs_derivative})
from eq.~(\ref{eq:sigma_derivative}) by applying the replacement in
eq.~(\ref{eq:aveObs_prescription}). Thus the term
\begin{equation}
2\,{\rm
 Re}\lg\frac{\partial Q_{\rm b}(\mpole,\mpole^*) }{\partial
 \mpole}\left(\mpole-\mMSB\right) \rg =-2\,{\rm Re}\lg\frac{\partial Q_{\rm
 b}(\mpole,\mpole^*) }{\partial \mpole}\left(\mpolec -\mMSB^c(\mu)\right)
\rg
\end{equation}
in eq.~(\ref{eq:changescheme}) tells us to subtract the contribution arising
from the insertion of the mass counterterm defined in the pole scheme and add
the one computed using the \MSB{} definition of the counterterm.

Notice that, for the term linear in $\lambda$, we get the simplified form
\begin{eqnarray}
Q_{\rm b}(m,m^*)\!&=& \! Q_{\rm b}(\mMSB,\mMSB^*)+\left[\frac{\partial Q_{\rm b}(\mpole,\mpole^*) }{\partial
 \mpole}+{\rm cc}\right] 
\nonumber \\
&\times&\!\!\left\{-\frac{1}{\as \, b_{0,f}}\! \int_0^{\infty} \!\frac{\mathd
 \lambda}{\pi} \, \frac{\mathd\ }{\mathd \lambda} \!\lq - \as\frac{\CF}{2} \lambda \rq
 \arctan \lq \frac{\as\, \pi\,b_{0,f}}{1+\as \,b_0\,
 \log\left( \frac{\lambda^2}{\mu^2 e^C}\right) }\rq
\!\!\right\}\!. \phantom{aaaaaaa}
\end{eqnarray}
 Furthermore, we have
\begin{equation}
 \left[\frac{\partial Q_{\rm b}(\mpole,\mpole^*) }{\partial
 \mpole}+{\rm cc}\right] =\frac{\partial Q_{\rm b}(\mpole,\mpole^*) }{\partial
 {\rm Re}(\mpole)}.
\end{equation}
Thus, when going from the pole to the \MSB{} mass scheme, the
definitions for $T$ and $\widetilde{T}$ are modified for small $\lambda$
into
\begin{eqnarray} 
 T(\lambda) & \to & T(\lambda) - \frac{\partial \sigma_{\rm b}(\mpole,\mpole^*)}{\partial
 {\rm Re}(\mpole)} \,\frac{\CF \as}{2} \lambda + {\cal O}(\lambda^2)\,,
 \label{eq:Tchangescheme} \\
 \widetilde{T}(\lambda) & \to &\widetilde{T}(\lambda) - \frac{\partial \obsb(\mpole,\mpole^*)}{\partial
 {\rm Re}(\mpole)} \,\frac{\CF \as}{2} \lambda + {\cal O}(\lambda^2)\,.
 \label{eq:Ttchangescheme}
\end{eqnarray}
We stress that eqs.~(\ref{eq:Ttchangescheme})
and~(\ref{eq:Tchangescheme}) also apply to any so called ``short distance''
mass
schemes~\cite{Czarnecki:1997sz,Beneke:1998rk,Hoang:1998ng,Pineda:2001zq,Fleming:2007qr,Jain:2008gb,Hoang:2008yj}.
These schemes are such that no mass renormalon affects their definition, and
of course in order for this to be the case, their small-$\lambda$ behaviour should
be the same one of the \MSB{} scheme.

\chapter{Evaluation of the linear sensitivity in top-mass dependent observables}
\label{sec:linear-mg}
As we have seen from eq.~(\ref{eq:sigma_final}), in order to compute the
all-orders total cross section we need to evaluate
\begin{equation}
\sigma - \sigma_b = -\frac{1}{\as \, b_{0}}\int_0^{\infty} \frac{\mathd \lambda}{\pi} \, \frac{\mathd\ }{\mathd \lambda}\!
 \left[ T(\lambda) \right] \,
 \arctan \lq \frac{\as \,\pi\,b_{0}}{1+\as \,b_0\,
 \log\left( \frac{\lambda^2}{\mu^2 e^C}\right) }\rq,
\end{equation}
where we have performed the adjustments described in
Sec.~\ref{sec:realistic_b0} to obtain a semi-realistic large-$b_0$ expansion. 
The small-$\lambda$ the contribution to the integral is given by
\begin{align}
& -\frac{1}{\as \, b_{0}}\int_0^{\mu} \frac{\mathd \lambda}{\pi} \, \left[
    T^\prime(0)+ T^{\prime\prime}(0)\lambda + \cdots \right] \nonumber \\
  & \quad \times \sum_{m=0}^{\infty}\frac{(-1)^{m}}{2m+1} \lg \as\,\pi\, b_0 \left[ \sum_{n=0}^{\infty} \left( -\as \,b_0\,
 \log\left( \frac{\lambda^2}{\mu^2 e^C}\right)\right)^n\right] \rg^{2m+1} \nonumber \\
&\approx  -\int_0^{\mu} \mathd \lambda \, T^\prime(0) \sum_{n=0}^{\infty} \left[ -2\,\as\,b_0 \log\left( \frac{\lambda}{\mu }\right) \right]^n \nonumber \\
& =  - T^\prime(0) \sum_{n=0}^{\infty} \left(2\,\as\,b_0 \right)^n n!,
\label{eq:sigma_series}
\end{align}
where we have neglected subleading powers of $\log(\lambda/\mu)$.
The Borel transform of the series in eq.~(\ref{eq:sigma_series}) is given by
\begin{equation}
\mathcal{B}\left[O\right](t) = - \frac{T^\prime(0)}{\as} \sum_{n=0}^{\infty} \left( 2\,t\,b_0\right)^n = - \frac{T^\prime(0)}{\as} \frac{1}{1-2\,t \,b_0} ,
\end{equation}
and eq.~(\ref{eq:sigma_series}) can be rewritten as
\begin{equation}
\as \int_0^{+\infty} \mathd t \,e^{-t}\,\mathcal{B}[O](\as\,t) = -
 T^\prime(0)\int_0^{+\infty} \mathd t \frac{e^{-t}}{1-2\as \,t\, b_0}.
\end{equation}
The integrand function has a pole located at $t= 1/\(2\as b_0\)$ whose
residue is proportional to
\begin{equation}
\exp \left[ \frac{-1}{2\,\as \,b_0} \right] =\frac{\Lambda_{\rm \sss
 QCD}}{\mu}=\frac{\Lambda_{\rm \sss QCD}}{m_0}
.
\end{equation}
Thus, if $T^\prime(0)$ is non zero, we have infrared linear renormalons.
A very similar situation appears if we investigate an infrared safe observable, where $T$ is replaced by $\widetilde{T}$.

If the quantity $Q$ is computed in the pole-mass scheme, to obtain the linear sensitivity in the \MSB{}-mass scheme we need to add to $T^\prime(0)$ (or to $\widetilde{T}^\prime(0)$) the term
\begin{equation}
-\frac{\CF \as}{2} \frac{\partial Q_b(m,m^*)}{\partial {\rm Re}(m)}\,,
\end{equation}
being $Q_b$ the leading order prediction, as it is discussed in Sec.~\ref{sec:changescheme}.

We now investigate the presence of linear terms in the expression of $T/\widetilde{T}$ for the total cross section, for the reconstructed-mass and for the energy of the final-state $W$ boson, expressed in terms of the pole mass and in terms of the \MSB{} one.

\section{Inclusive cross section}
\label{sec:total-xsec}
The formula for the total cross section is given in
eq.~(\ref{eq:sigma_final}). We now study the presence of linear $\lambda$
terms in the expression of $T(\lambda)$ in eq.~(\ref{eq:Tk2_final}), both for
the inclusive process or in presence of selection cuts.

\subsection{Selection cuts}
\label{sec:cuts}
In order to mimic the experimental selections adopted at hadron colliders, at
times we introduce selection cuts for our cross sections, requiring the
presence of a $b$ jet and a (separated) $\bar{b}$ jet, both having energy
greater than 30~GeV. Jets are reconstructed using the {\tt
 Fastjet}~\cite{Cacciari:2011ma} implementation of the anti-$k_t$
algorithm~\cite{Cacciari:2008gp} for $e^+ e^-$ collisions, with a variable
$R$ parameter.

\subsection{Total cross section without cuts}
\label{sec:Xtot_wocuts}
\begin{figure}[tb!]
 \centering
 \includegraphics[width=0.6\textwidth]{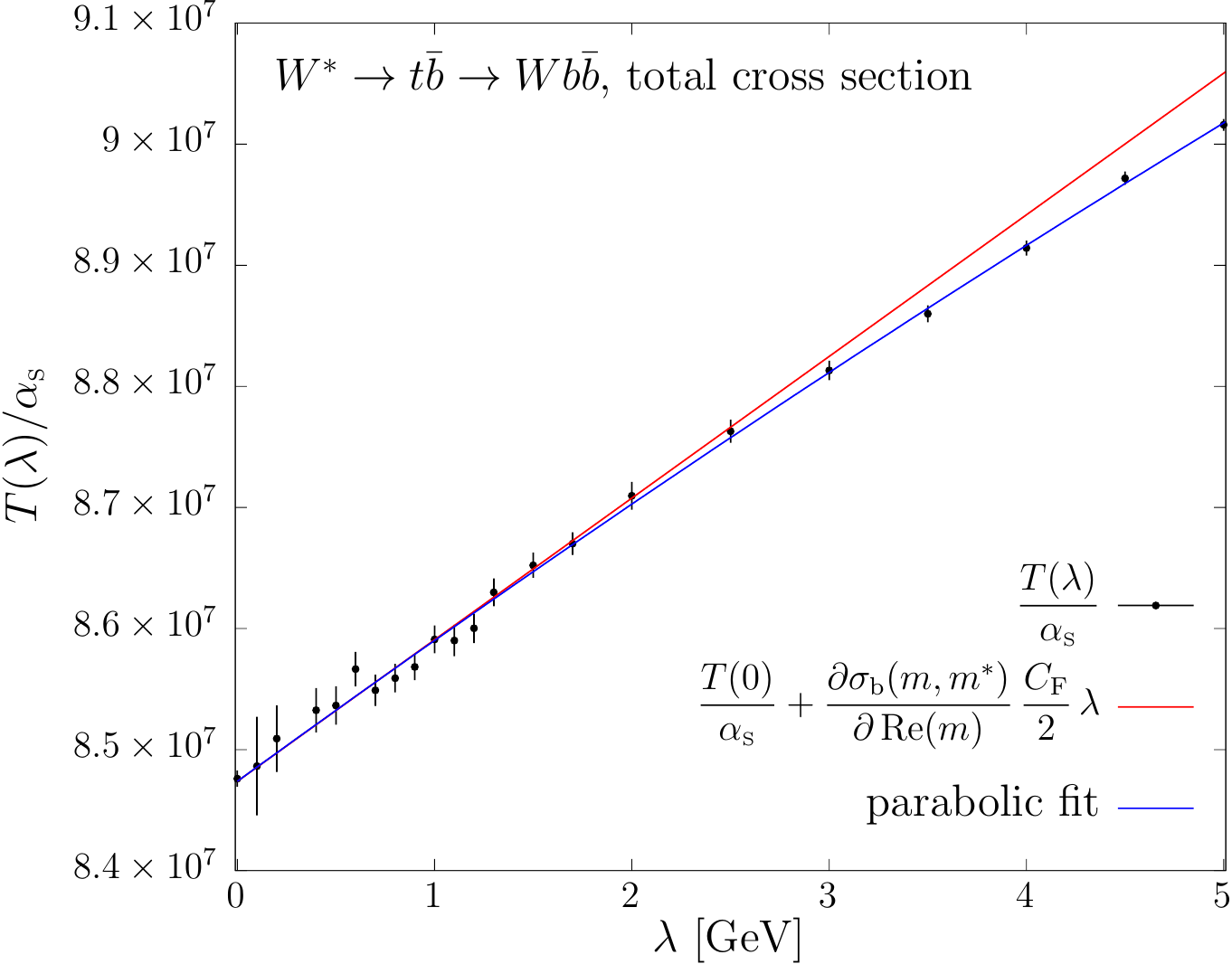}
 \caption{Small-$\lambda$ behaviour of $T(\lambda)$ for the total cross
   section, where $\lambda$ is the gluon mass. In black the data points
   extracted with our numerical simulations, in red the linear $\lambda$
   dependence and in blue the parabolic fit of the points.}
 \label{fig:sigtot_smallk}
\end{figure}

In the absence of cuts, the expression for $T(\lambda)$ in
eq.~(\ref{eq:Tk2_final}) simplifies, since $\Delta(\lambda)$, given by
eq.~(\ref{eq:Deltak2_final}) is identically zero. Its small-$\lambda$ behaviour is
shown in Fig.~\ref{fig:sigtot_smallk}.
As discussed in Sec.~\ref{sec:changescheme}, the same calculation performed
in the \MSB{} mass scheme would yield, for the total cross section, to the
replacement given in eq.~(\ref{eq:Tchangescheme})
\begin{equation}
T(\lambda)\to T(\lambda)-\frac{\partial \sigma_b}{\partial \,{\rm Re}(m)} \frac{\CF \as}{2}
\lambda+{\cal O}(\lambda^2).
\end{equation}
So, in the same figure, we also plot (in red) the expression
\begin{equation}
T(\lambda)=T(0)+\frac{\partial \sigma_b}{\partial \,{\rm Re}(m)} \frac{\CF \as}{2}
\lambda\,.
\end{equation}
It is then clear that the \MSB{} result would have no linear term in $\lambda$ for
small $\lambda$, and thus that there is no linear renormalon in this scheme. From
the figure it is also clear that this holds for both $\lambda\lesssim \Gamma_t$ and
for $\lambda\gg \Gamma_t$, where $\Gamma_t$ is the top width. The $\lambda\lesssim
\Gamma_t$ behaviour is justified by the fact that, because of the finite
width, phase-space points where the top is on shell are never reached (see
Appendix~\ref{sec:Ew_linear_sensitivity}). Thus,
no linear renormalon is present \emph{unless} one uses the pole-mass scheme,
that has a linear renormalon in the counterterm.

\begin{figure}[tb!]
 \centering
 \includegraphics[width=0.62\textwidth]{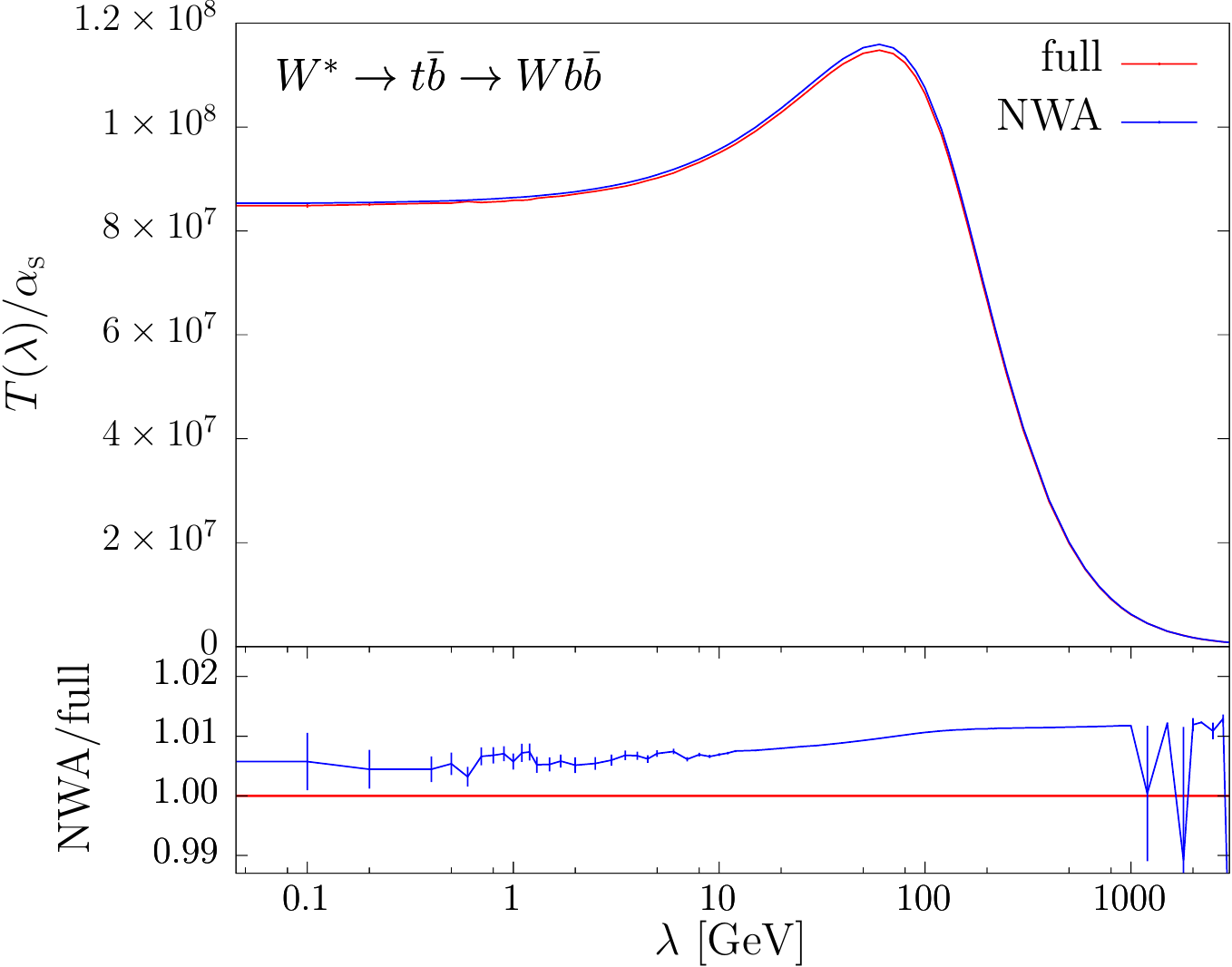}
 \caption{$T(\lambda)$ for the NLO total cross section, where $\lambda$ is the
 gluon mass, computed in the pole-mass scheme using the exact full matrix
 elements~(red) and the narrow-width approximation, NWA~(blue).}
 \label{fig:sigNWA}
\end{figure}

As far as the $\lambda\gg \Gamma_t$ limit, we notice that the $\lambda$ behaviour should
be the same as that of the narrow-width approximation~(NWA), where the cross
section factorizes in terms of the on-shell top-production cross section, and
its decay partial width:
\begin{equation}
 \sigma\!\left(W^* \rightarrow W b\bar{b}\right) = \sigma\!\left(W^* \rightarrow
 t\bar{b}\right) \frac{\Gamma(t\rightarrow W b)}{\Gamma_t} +
 \mathcal{O}\left(\frac{\Gamma_t}{\mpole}\right).
\end{equation}
The behaviour of $T(\lambda)$, computed either exactly or in the NWA, is shown in
Fig.~\ref{fig:sigNWA}. 

The factor $\sigma(W^* \rightarrow t\bar{b})$ is clearly free of linear
renormalons, since it a totally inclusive decay of a colour-neutral
system. Although less obvious, this is also the case for the factor
$\Gamma(t\rightarrow W b)$ (see ~\cite{Bigi:1994em, Beneke:1994bc,
 Beneke:1994qe}).

The computation of $T(\lambda)$ using the NWA is discussed in
Appendix~(\ref{sec:NWA}), where we show in Figs.~\ref{fig:W_tb_xsec}
and~\ref{fig:Gammat} that the linear $\lambda$ behaviour is due to the
choice of the pole-mass scheme.

\subsection{Total cross section with cuts}
\label{sec:xsec_cuts}
When the selection cuts discussed earlier are imposed, the cross section
depends explicitly upon the jet radius~$R$. We expect jets requirement to
induce the presence of linear renormalons, and thus linear small-$\lambda$
behaviour of $T$, with a slope that goes like $1/R$ for small
$R$~\cite{Korchemsky:1994is,Dasgupta:2007wa}.
\begin{figure}[tb!]
 \centering
 \includegraphics[width=0.6\textwidth]{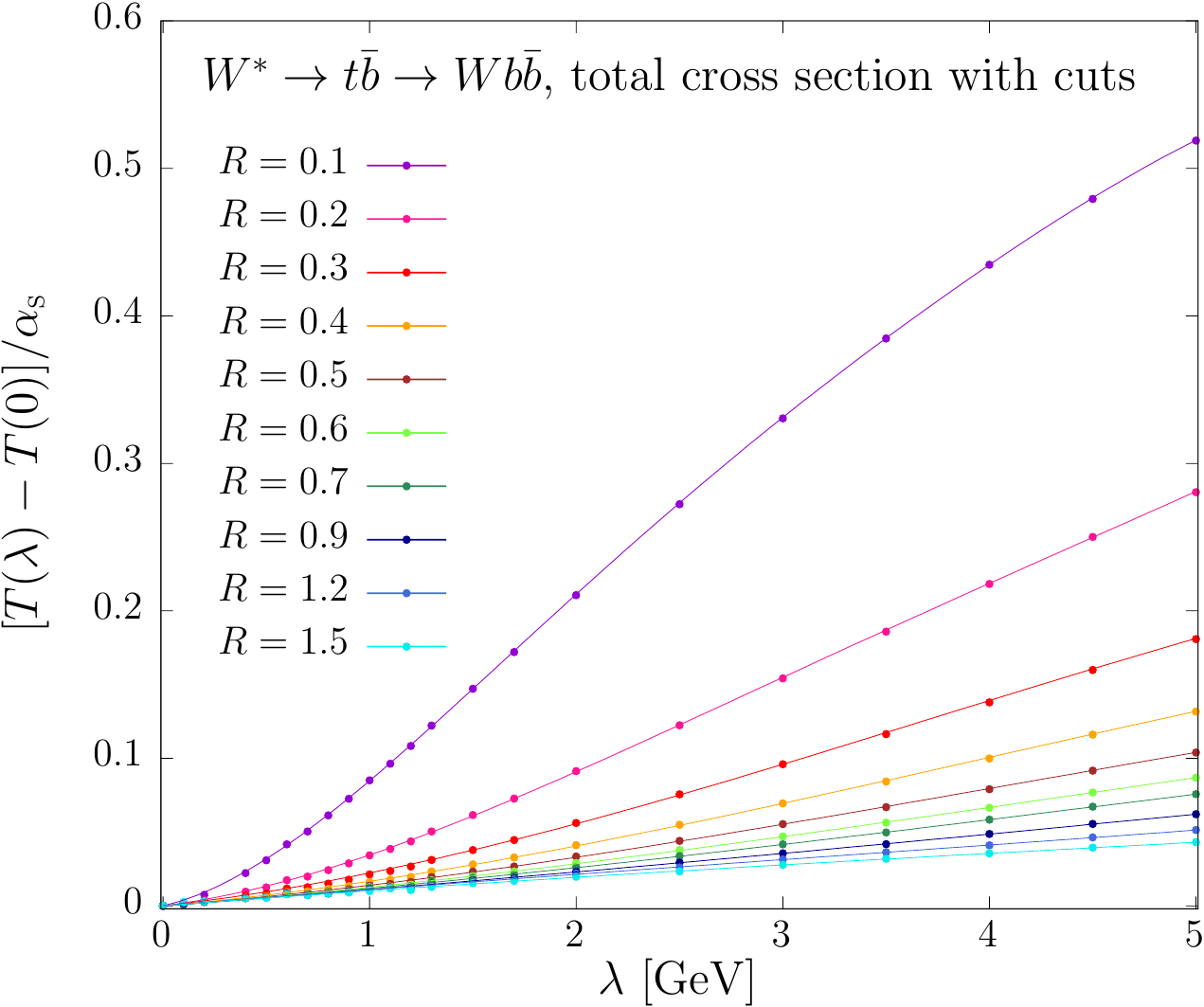}
 \caption{Small-$\lambda$ behaviour for $T\!\(\lambda\)$ for the total cross section
 with cuts, for several jet radii. The points represent the data obtained
 with our numerical calculations, while in solid lines their polynomial
 fit. The fitting functions are order 5, 4 and 3 polynomials for
 $R=0.1$, $R=0.2$ and $R \ge 0.3$ radii respectively.}
 \label{fig:sigvrqq_cut2}
\end{figure}
In Fig.~\ref{fig:sigvrqq_cut2} we display the small-$\lambda$ behaviour for
$T(\lambda)$ for the total cross section with cuts, for several jet radii.
Together with the results of our simulation, we plot also, for each value of
$R$, a polynomial fit to the data.

When changing from the pole to the \MSB{}-mass scheme, we only expect a mild
$R$ dependent correction to the slope of $T(\lambda)$ at $\lambda=0$\footnote{The
 change of scheme is governed by formula~(\ref{eq:Tchangescheme}), where the
 only radius dependence comes from the derivative of the LO value of the
 observable, and this is mild for small $R$.}, and thus we cannot expect the
same benefit that we observed for the cross section without cuts.
\begin{figure}[tb!]
 \centering
 \includegraphics[width=0.6\textwidth]{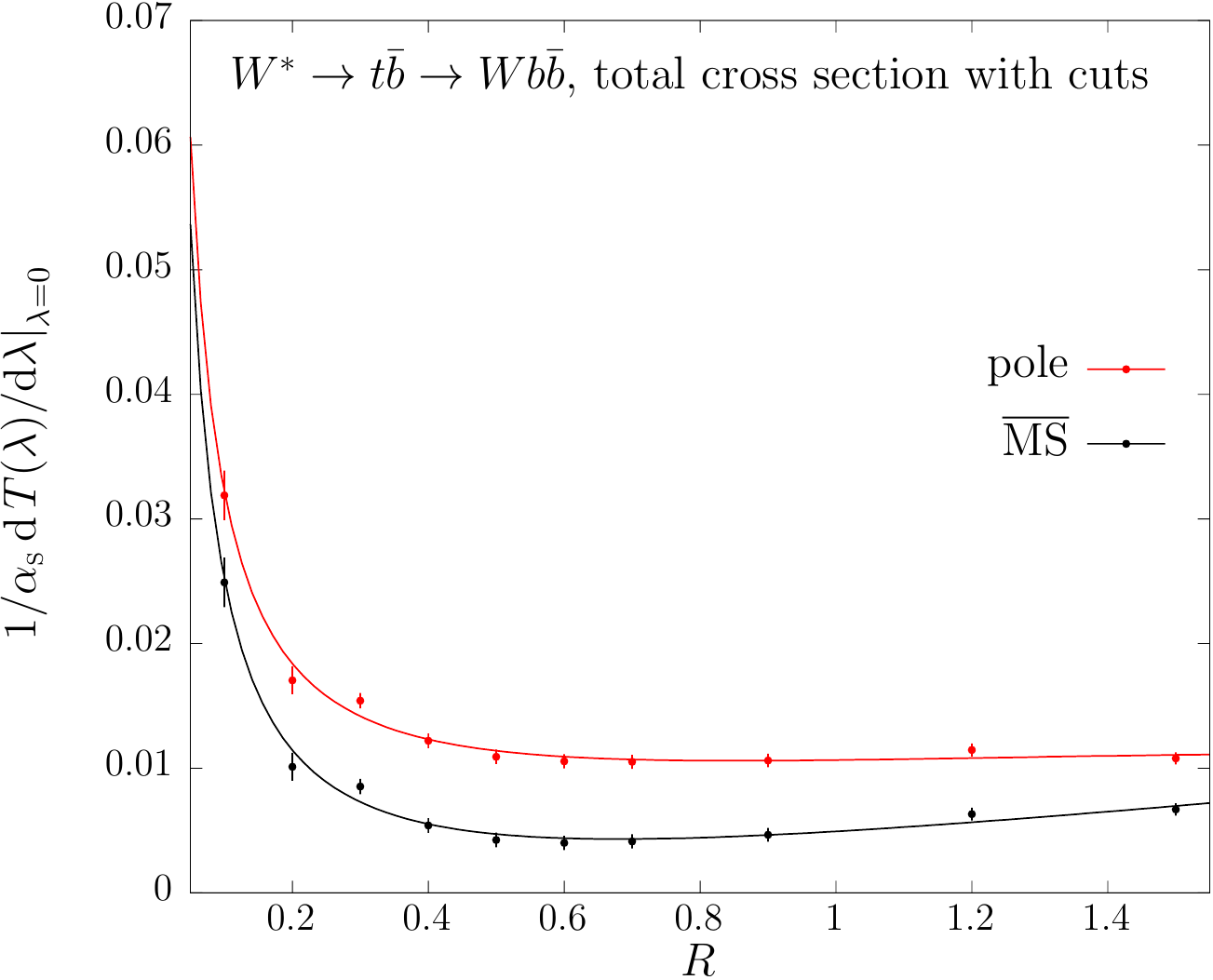}
 \caption{$R$ dependence of the slope of $T(\lambda)$ for the total cross
 section, at $\lambda=0$, using the pole~(red) or the \MSB{} mass
 scheme~(black). The solid lines represent fits of parametric form
 $a/R+b+c\,R+d\,R^2$.}
 \label{fig:sigtot-linear-allR}
\end{figure}
This is illustrated in Fig.~\ref{fig:sigtot-linear-allR} for several jet
radii. The $1/R$ behaviour is clearly visible. In addition, for relatively
large-$R$ values, the use of the \MSB{} scheme brings about some reduction to
the slope of the linear term. This may be due to the fact that the cross
section with cuts captures a good part of the cross section without cuts, and
thus it partially inherits its benefits when changing scheme. However, it is
also clear that linear non-perturbative ambiguities remain important also in
the \MSB{} scheme when cuts are involved.

\section{Reconstructed-top mass}
\label{sec:rec-top-mass}
In this section we consider the average value $\langle M \rangle$,
where $M$ is the mass of the system comprising the $W$ boson and the $b$
jet. Such an observable is closely related to the top
mass, and, on the other hand, is simple enough to be easily computed in our
framework. We use the same selection cuts described previously.

We computed $\langle M \rangle$ also in the narrow width limit, by simply
setting the top width to 0.001~GeV. In this limit, top production and decay
factorize, so that we have an unambiguous assignment of the final state
partons to the top decay products. We first compute $\langle M \rangle$ in
the narrow width limit, using only the top decay products, and without
applying any cuts. We then compute it again, still using only the top decay
products, but introducing our standard cuts. Finally we compute it again
using all decay products and our standard cuts.
\begin{figure}[tb]
 \centering
 \includegraphics[width=0.48\textwidth]{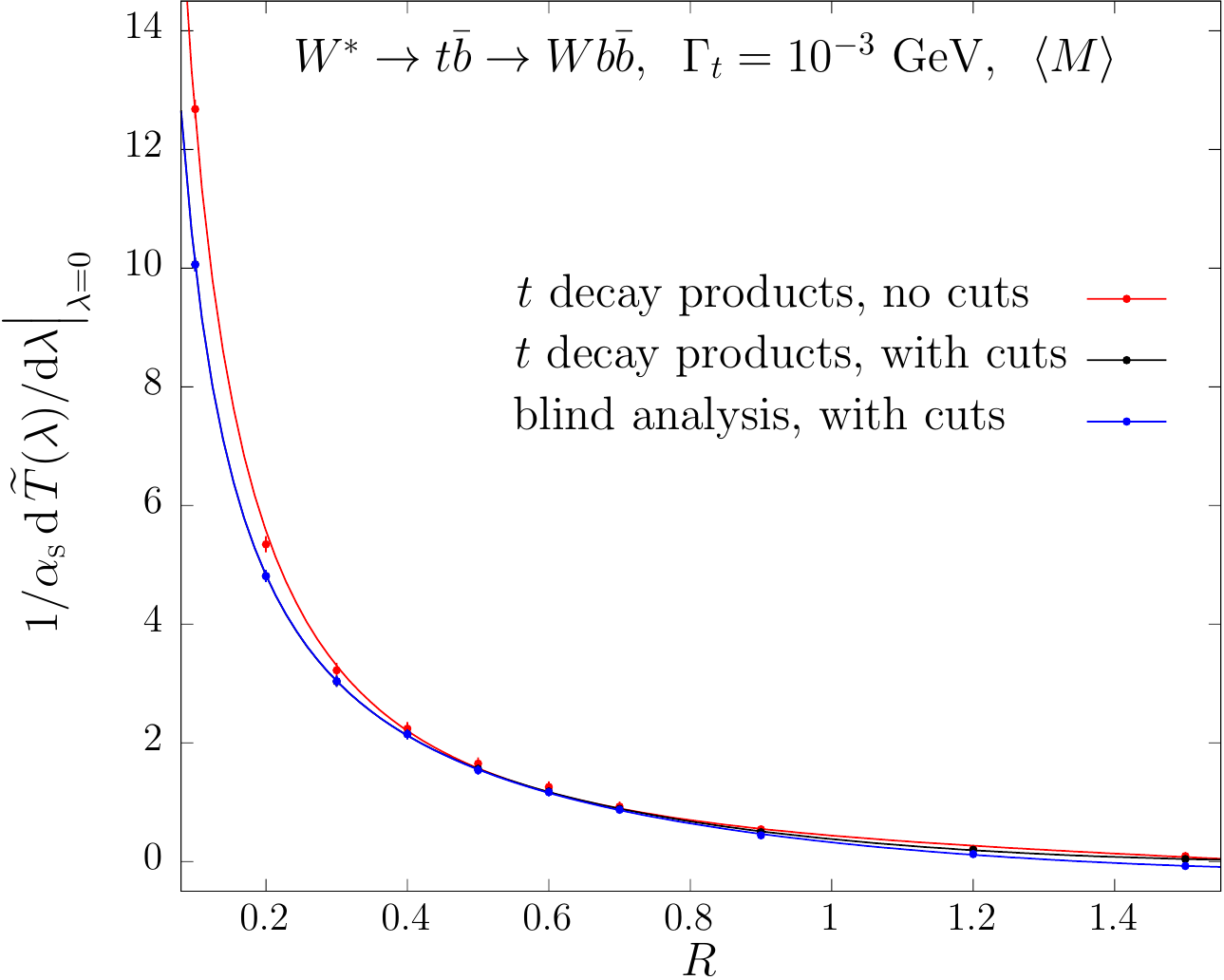}
 \includegraphics[width=0.48\textwidth]{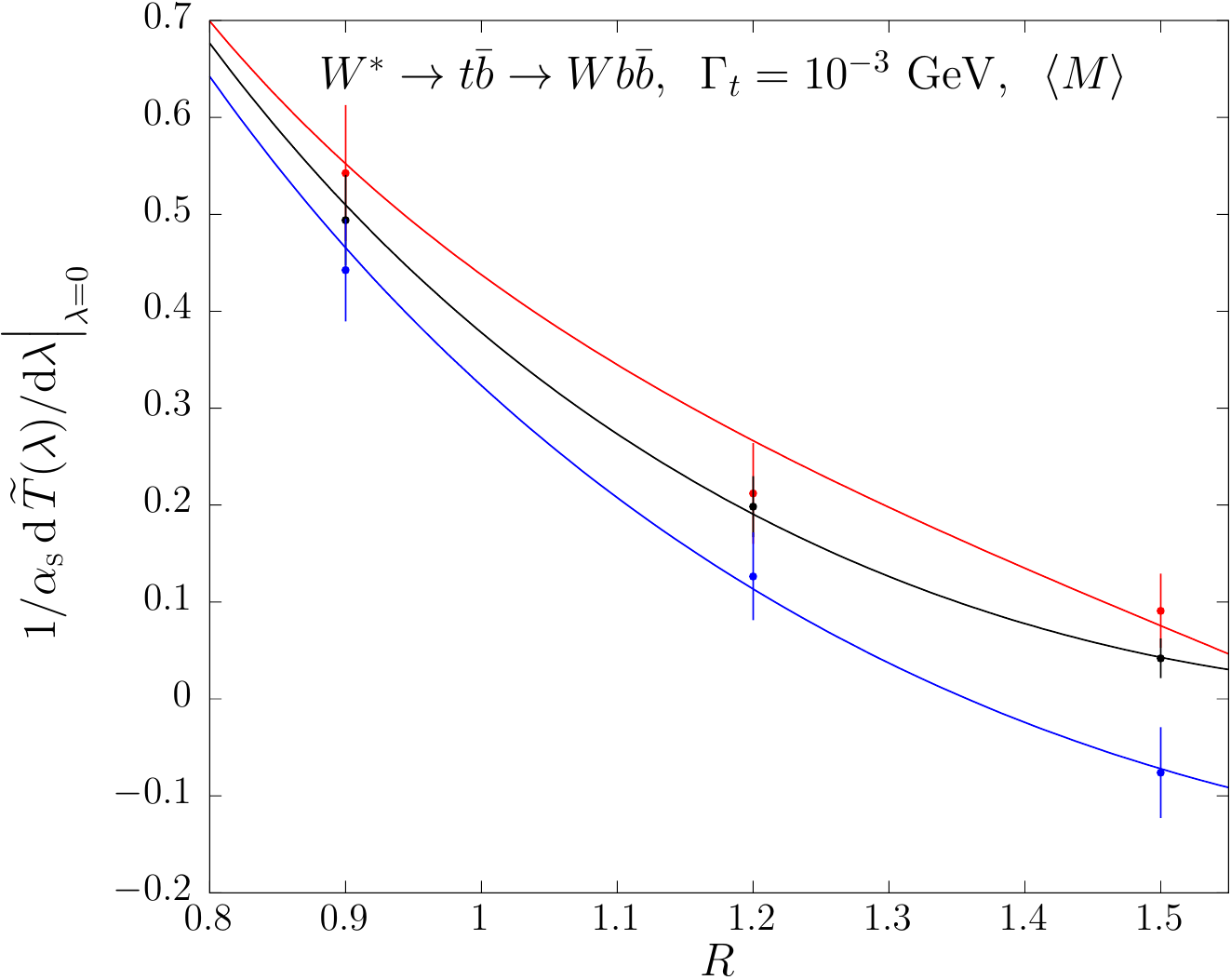} 
 \caption{ $R$ dependence of the slope of $\widetilde{T}(\lambda)$, computed with
 $\Gamma_t=10^{-3}$~GeV, for the averaged reconstructed mass $M$, when
 using only the top decay products and no cuts~(red), when using only the
 top decay products and applying our cuts~(black) and when using all
 final-state particles and applying our selection cuts~(blue). The solid
 lines are the result of a fit of the form $a/R+b+c\,R+d\,R^2$. The black
 and the blue curves are almost completely overlapping and are
 indistinguishable in the plot on the left. A blowup of the high-$R$
 region is illustrated in the plot on the right.}
 \label{fig:NW-mctruth}
\end{figure}
The results of these calculations are reported in Fig.~\ref{fig:NW-mctruth},
where the slope at $\lambda=0$ of $\widetilde{T}$ for our observable
is plotted as a function of the jet radius $R$. As expected we see the
shape proportional to $1/R$ for small $R$~\cite{Korchemsky:1994is,Dasgupta:2007wa}.

In the case of the calculation of $\langle M \rangle$ performed using only the top decay
products, and without any cuts, we expect that, for large values of $R$, the
average value of $M$ should get closer and closer to the input top pole mass,
irrespective of the value of $\lambda$. Thus, the slope of $\widetilde{T}\!\(\lambda\)$ for $\lambda=0$ should become
smaller and smaller. We find in this case that, for the largest value of $R$
we are using ($R=1.5$), the slope has a value around 0.09. When cuts are
introduced this value becomes even smaller, around 0.04. This curve is
fairly close to the one obtained using all final-state particles and
including cuts. The large-$R$ value in this case is $-0.08$.

\begin{figure}[h!]
 \centering
 \includegraphics[width=0.6\textwidth]{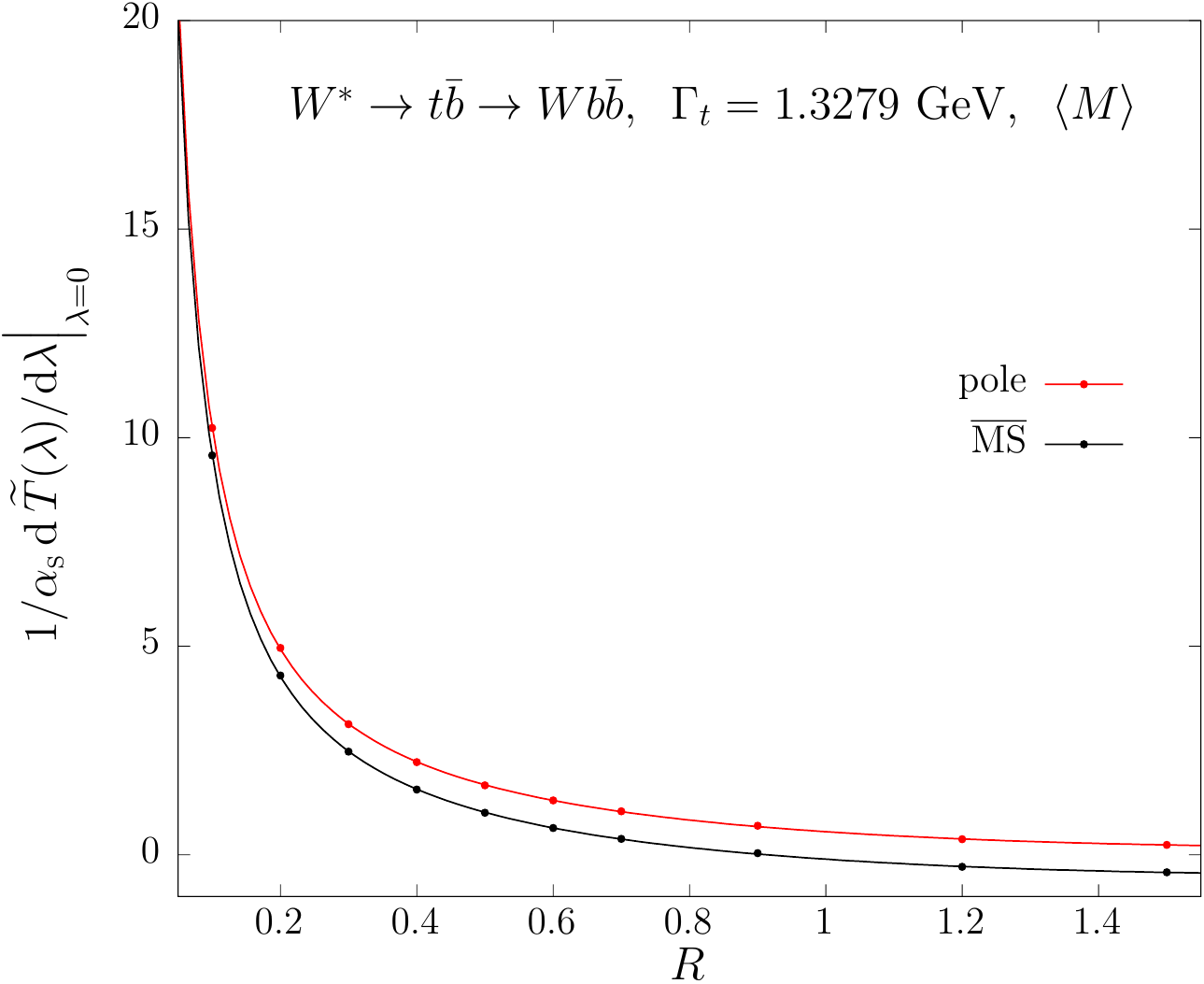}
 \caption{ $R$ dependence of the slope of $\widetilde{T}(\lambda)$ for the averaged
 reconstructed mass $M$. The solid lines are the result of a fit of the
 form $a/R+b+c\,R+d\,R^2$.}
 \label{fig:Mrec_slope}
\end{figure}

If we change scheme from the pole mass to the \MSB{} one, the corresponding
change of $\widetilde{T}$ is given by eq.~(\ref{eq:Ttchangescheme}), and for
the observable at hand the derivative term it is very near~1. The change in
slope when going to the \MSB{} scheme is roughly $-\CF/2\approx-0.67$. Thus,
if we insisted in using the \MSB{} mass for the present observable, for large
jet-radius parameters, we would get an ambiguity larger than if we used the
pole mass scheme.
The same holds even if we employ a finite top width. The $R$ dependence of
the $\widetilde{T}(\lambda)$ slope for $\Gamma_t=1.3279$~GeV is shown in
Fig.~\ref{fig:Mrec_slope}.
\begin{figure}[h!]
 \centering
 \includegraphics[width=0.48\textwidth]{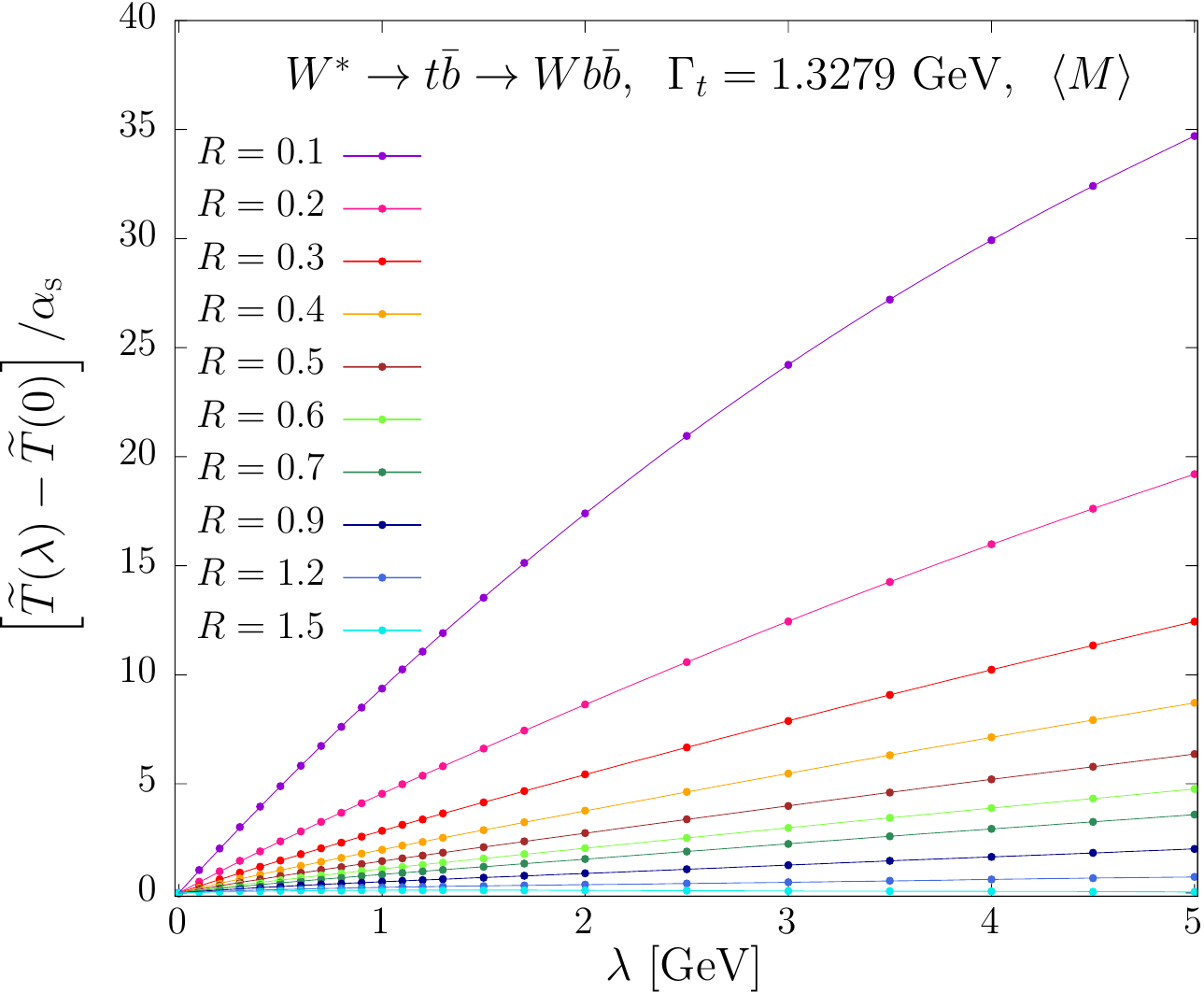}
 \includegraphics[width=0.485\textwidth]{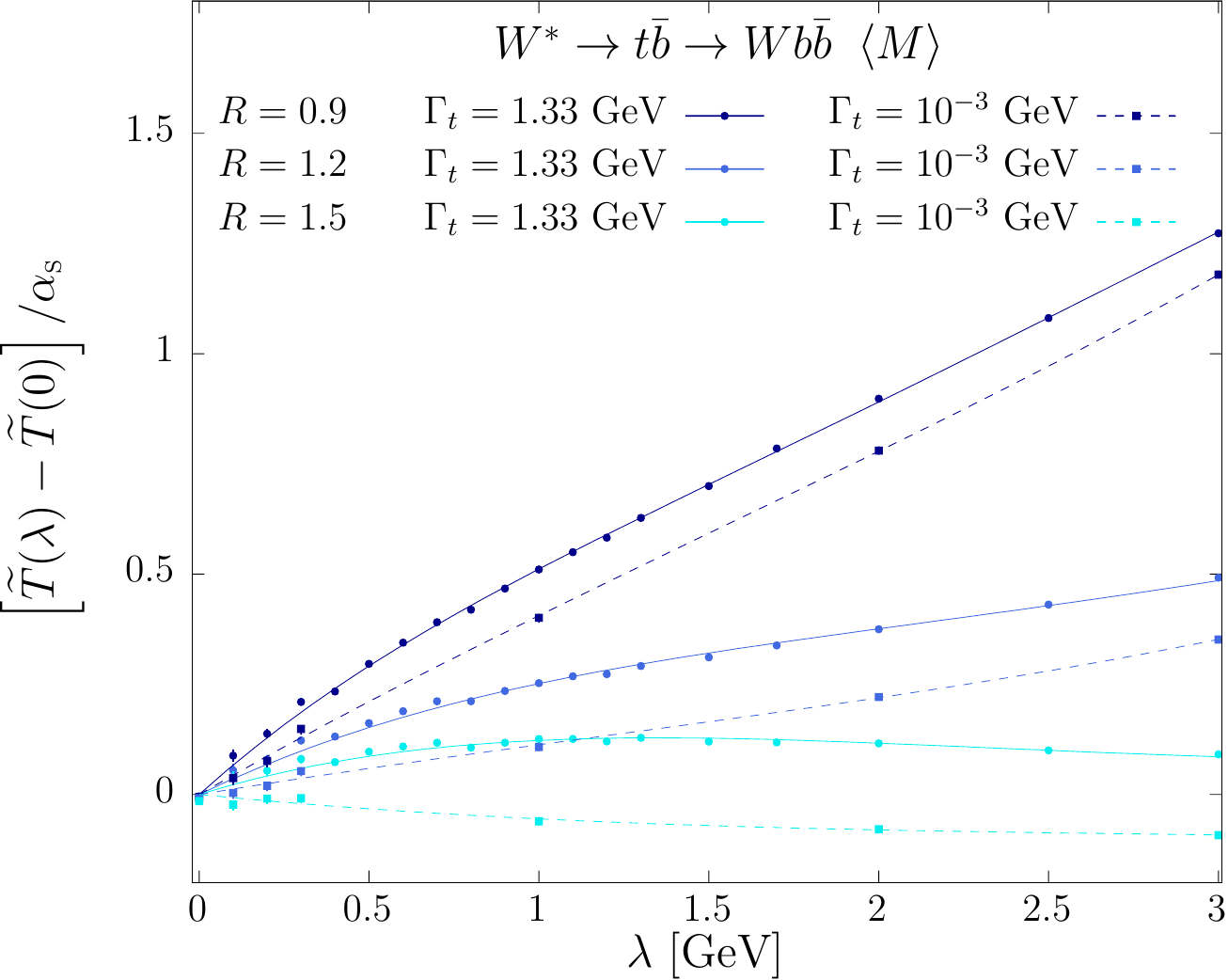}
 \caption{Small-$\lambda$ behaviour of $\widetilde{T}(\lambda)$ for the
 averaged reconstructed-top mass, for several values of the jet radius
 $R$. In the left pane, the results obtained with $\Gamma_t=1.3279$~GeV
 are shown. The solid lines represent the polynomial fit of the points.
 For $R \ge 1.2$ a $4^{\rm th}$ order polynomial is adopted, while for the
 other $R$ values a $5^{\rm th}$ order polynomial is employed. In the
 right pane, only the results corresponding to large jet radii are
 displayed, together with the ones obtained with $\Gamma_t=10^{-3}$, that
 are interpolated with a cubic fit~(dashed lines).}
 \label{fig:Mrec_smallk}
\end{figure}
We notice that, in the present case, for values of $R$ below 1, the \MSB{}
scheme seems to be better, because of a cancellation of the $R$ dependent
renormalon and the mass one. From our study, however, it clearly emerges that
such cancellation is accidental, and one should not rely upon it to claim an
increase in accuracy.

In the left pane of Fig.~\ref{fig:Mrec_smallk} we plot the small-$\lambda$ behaviour of
$\widetilde{T}(\lambda)$ for the reconstructed-top mass, computed with the
finite top width, for several values of the jet radius $R$. It is clear that
our observable is strongly affected by the jet renormalon. The same plot for
only the three largest values of $R$ is shown in
the right pane.
The figure shows clearly that the $\lambda$ slope computed with
$\Gamma_t=1.3279$~GeV changes when $\lambda$ goes below $1$~GeV, that is to say,
when it goes below the top width. This behaviour is expected, since the top
width act as a cutoff on soft radiation. In the figure we also report the
$\lambda$ behaviour in the narrow-width approximation. It is evident that the
slopes computed in this limit are similar to the slopes with
$\Gamma_t=1.3279$~GeV, for values of $\lambda$ larger than the top width. It is
also clear that the slopes that we find here for the largest $R$ value are
considerably smaller than the slope change induced by a change to a short
distance mass scheme, that amounts to $-0.67$. In other words, the pole mass
scheme is more appropriate for this observable, irrespective of finite width
effects.

\section[$W$ boson energy]{$\boldsymbol{W}$ boson energy}
\label{sec:Ew}
In this section we study the behaviour of the average value of the $W$ energy,
$E_{\sss W}$, since this is another top-mass sensitive observable. This
observable is chosen since it is a case of an observable that does not depend
upon the jet definition. It can thus be considered to be a representative of
pure ``leptonic'' observables in top-mass measurements. In this study, we do
not apply any cut, in order to avoid all possible jet or hadronic biases.
Our goal is to see if this observable is free of renormalons in some mass
scheme.

\begin{figure}[tb]
 \centering
 \includegraphics[width=0.55\textwidth]{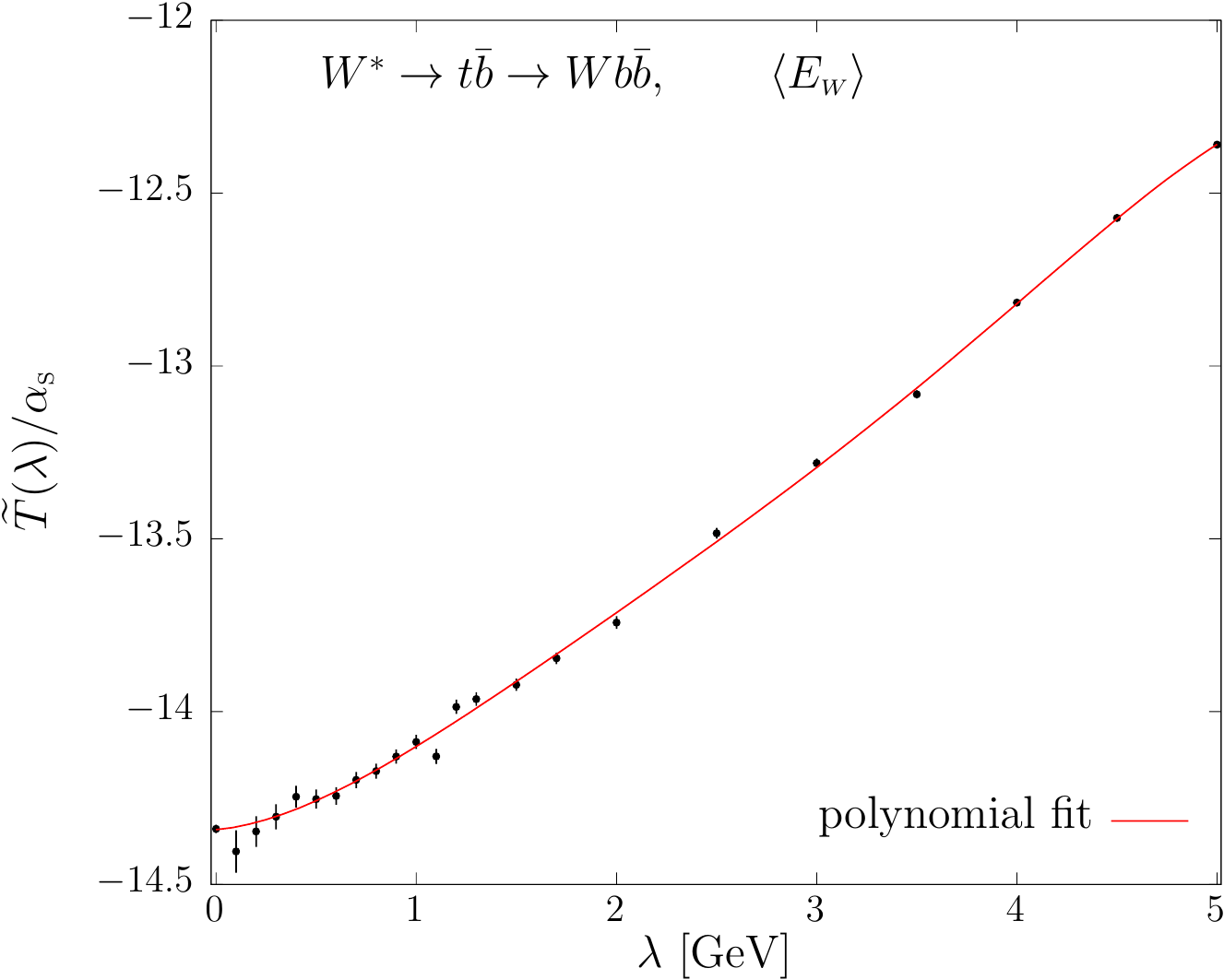}
 \caption{Small-$\lambda$ behaviour of $\widetilde{T}(\lambda)$ for $\langle
 E_{\sss W} \rangle$. The solid line represents a $5^{\rm th}$ order
 polynomial fit.}
 \label{fig:Ew_smallk}
\end{figure}
In order to change scheme, according to eq.~(\ref{eq:Ttchangescheme}), we
need the derivative of the Born value of the observable with respect to the
real part of the top mass. We have computed numerically this term and its
value is given by
\begin{equation}
 \label{eq:Ewmass-slope}
 \frac{\partial \langle E_{\sss W} \rangle_b}{\partial\, {\rm Re}(m)}=0.0980
 \,(8) \,.
\end{equation}
The small-$\lambda$ dependence of the corresponding $\widetilde{T}$ function is
shown in Fig.~\ref{fig:Ew_smallk}: for values of $\lambda$ much larger than the
width, the slope of the curve is roughly 0.45. Thus, under these conditions,
a renormalon is clearly present whether we use the pole or the \MSB{} scheme,
since the correction in slope due to the use of the latter would be
$-0.098\times\CF/2=-0.065$.

\begin{figure}[h!]
 \centering
 \includegraphics[width=0.6\textwidth]{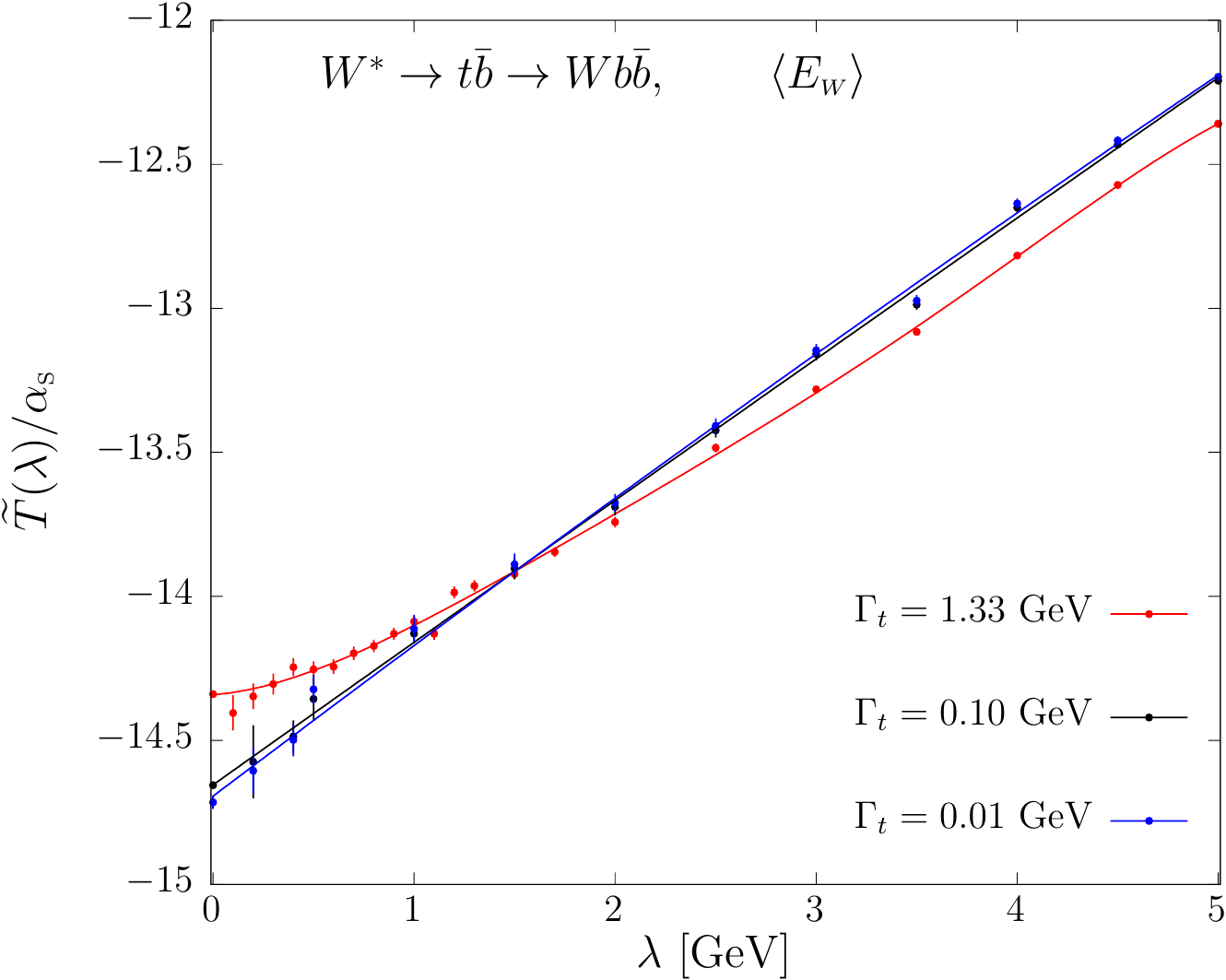}
 \caption{Small-$\lambda$ behaviour of $\widetilde{T}(\lambda)$ for
 $\langle E_{\sss W} \rangle$, for increasingly smaller values of
 $\Gamma_t$. The blue and the black solid lines are given by a parabolic
 fit, the red line is the same one displayed in
 Fig.~\ref{fig:Ew_smallk}. }
 \label{fig:Ew-nocuts-smallwidth}
\end{figure}

For $\lambda$ below the top width we see a reduction in slope, that is too
difficult to estimate because of the lack of statistics. Since the change in
slope is clearly related to the top finite width, we carried out the
following tests: we run the program with a reduced $\Gamma_t$, expecting to
see a constant slope extending down to smaller values of $\lambda$. This is
illustrated in Fig.~\ref{fig:Ew-nocuts-smallwidth}.
We clearly see that, as $\Gamma_t$ becomes smaller, the slope of the $\lambda$
dependence remains constant, near the value $0.45$ found before, down to
smaller values of $\lambda$.
Since we have that 
\begin{eqnarray}
 \frac{\partial \langle E_{\sss W} \rangle_b}{\partial\, {\rm
 Re}(m)} &=& +0.098\,(4)\,, \quad {\rm for }\quad \Gamma_t=0.1~{\rm
 GeV}, 
 \\
 \frac{\partial \langle E_{\sss W} \rangle_b}{\partial\, {\rm
 Re}(m)} &=& +0.10\,(3) \,, \phantom{0} \quad {\rm for }\quad
 \Gamma_t=0.01~{\rm GeV}, 
\end{eqnarray}
it is clear that, for a vanishing top width, the \MSB{} scheme, as well as
the pole scheme, is still affected by the presence of a linear renormalon.

\begin{figure}[h!]
 \centering
 \includegraphics[width=0.6\textwidth]{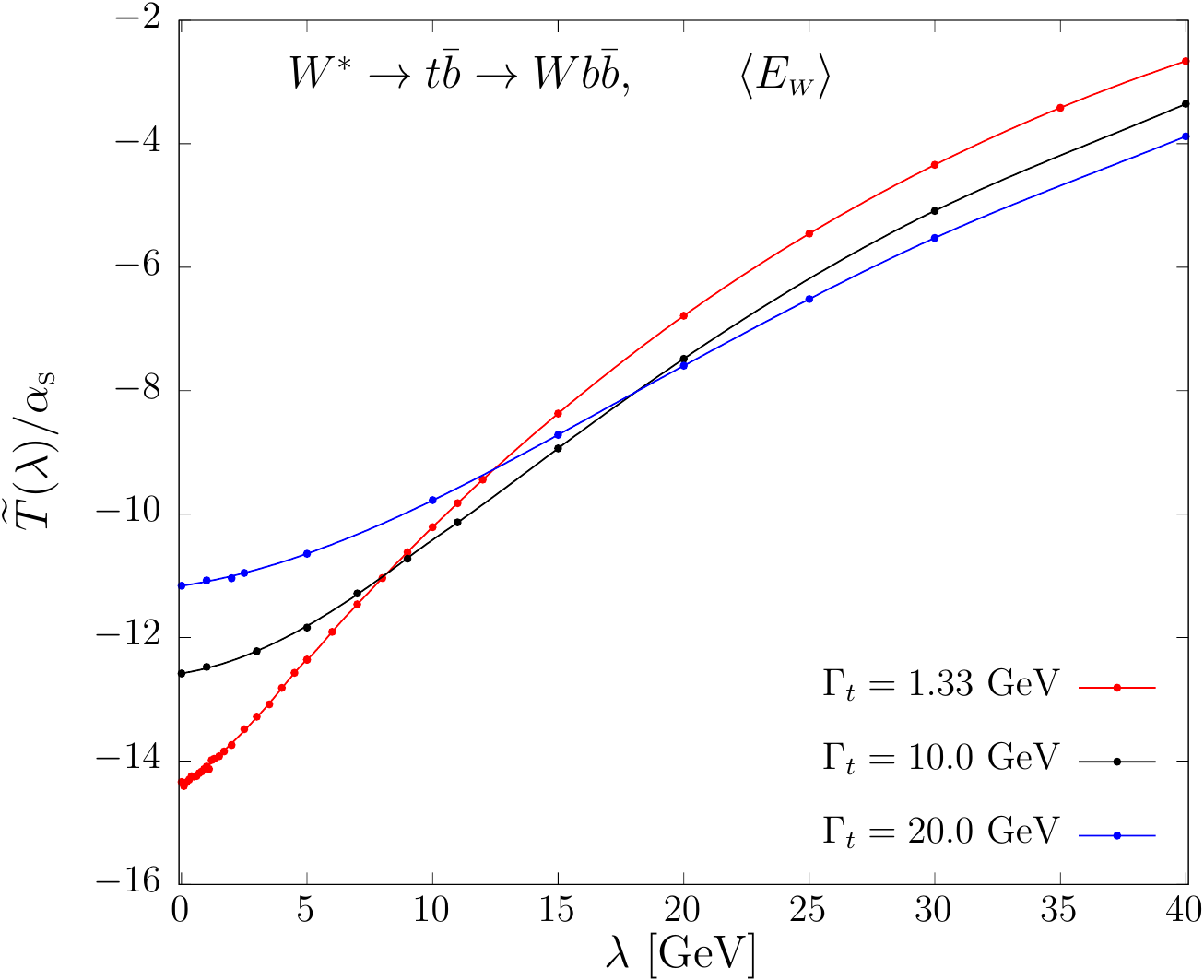}
 \caption{Results for the small-$\lambda$ behaviour of $\widetilde{T}$ for
 $\langle E_{\sss W}\rangle$, at different values of $\Gamma_t$. The
 error bar associated to each point computed at a given value of $\lambda$ is
 also plotted, but is too small to be visible on the scale of the
 figure. 
 The red line~($\Gamma_t=1.33$~GeV) is a $5^{\rm th}$ order
 polynomial fit for $\lambda \le 5$~GeV and a spline for larger $\lambda$ values. The
 blue and the black solid lines, that interpolates the results obtained
 with $\Gamma_t=10$~GeV and $\Gamma_t=20$~GeV respectively, are a cubic
 fit for $\lambda<\Gamma_t$ and a spline for $\lambda>\Gamma_t$.}
 \label{fig:Ew-nocuts-largewidth}
\end{figure}

We also performed a run with $\Gamma_t=10$~GeV and $\Gamma_t=20$~GeV, in
order to estimate more accurately the value of the slope for $\lambda\ll
\Gamma_t$.  The result is shown in Fig.~\ref{fig:Ew-nocuts-largewidth}. In
Tab.~\ref{tab:slopes_Ew} we illustrate the slopes of $\widetilde{T}(\lambda)$
for small $\lambda$, obtained from the polynomial interpolation displayed in
Fig.~\ref{fig:Ew-nocuts-largewidth}, and the corresponding value in the
\MSB{} scheme, obtained by adding $ -\frac{\CF}{2}\frac{\partial \langle
  E_{\sss W} \rangle_b}{\partial\, {\rm Re}(m)} $ to the fitted slope. This
shows that the linear sensitivity largely cancels in the \MSB{} scheme.

One may now wonder if the cancellation of the linear sensitivity in the
\MSB{} scheme is exact, or just accidental. In fact, we show in
App.~\ref{sec:Ew_linear_sensitivity} that the cancellation is exact.

\begin{table}[h!]
 \centering
\begin{tabular}{|c|c|c|c|c|}
 \hline
 $\Gamma_t$ & slope (pole) & $ \displaystyle \frac{\partial \langle E_{\sss
 W}^{\phantom{\big(}}
 \rangle_b}{\partial\, {\rm Re}(m)}$ & $\displaystyle -\frac{\CF}{2}\frac{\partial
 \langle E_{\sss W} \rangle_b}{\partial\, {\rm Re}(m)}_{\phantom{\big(}}$ & slope (\MSB)\\
 \hline
 10~GeV & $0.058\,(8)$ & $0.0936\,(4)$ & $-0.0624\,(3)$ & $0.004\,(8)$\\
 \hline
 20~GeV & $0.061\,(2)$ & $0.0901\,(2)$ & $-0.0601\,(1)$ &
 $0.001\,(2)$\\
 \hline
\end{tabular}
\caption{Slopes of $\widetilde{T}(\lambda)/\as$, evaluated ad $\lambda=0$,
  computed for $\langle E_{\sss W} \rangle$ in the pole-mass scheme and the
  derivative terms needed to change to the \MSB{} one, for large top widths.}
\label{tab:slopes_Ew}
\end{table}

\chapter{All-order expansions in $\boldsymbol{\as}$}
\label{sec:AllOrderExp}
We will now consider the all-order expansion of various quantities, in order
to see how the infrared renormalon affects the large-order behaviour, both in
the pole mass scheme and in the \MSB{} scheme.

One may think that in our framework we may even compare quantities computed
in different mass schemes, and thus assess the reliability of the methods
used to estimate the resummation of divergent series, and the corresponding
ambiguity. In fact, within our large-$n_f$ approximation, if the method
adopted to resum the perturbative expansion is linear, as is the case of the
Borel transform method, we should find identical results (always in the large
$n_f$ sense) in the \MSB{} and the pole-mass schemes. This is shown as
follows. The relation between the pole and \MSB{} scheme for a generic
observable is given by, following eq.~(\ref{eq:changescheme}),
\begin{align}
 \obsb(m,m^*)+\langle O \rangle ^{(1)}(m,m^*)\as
 =\,&\obsb(\mMSB,\mMSB^*)+\bigg\{\left[\frac{\partial \obsb(\mpole,\mpole^*)}{\partial\mpole} 
 (m-\mMSB) + {\rm cc}\right] \nonumber \\
 &+\langle O \rangle ^{(1)}(m,m^*)\as\bigg\} + {\cal O}(\as^2 (\as
 n_f)^n).
\end{align}
Neglecting subleading terms, this is an identity, since the expansion of
$\obsb$ in the mass difference stops at the first order in the large-$n_f$
limit. When performing the calculation in the pole mass scheme, we need to
resum the expansion of $\langle O \rangle^{(1)}$, while if we perform the
calculation in the \MSB{} scheme, we are resumming the expansion of the sum of
terms in the curly bracket. If the resummation method is linear this last
resummation can be performed on the individual terms inside the curly
bracket. This is exactly what we would do on the left-hand side if, after the
resummation, we wanted to express the same result in the \MSB{} scheme. In
other words, if one uses the Borel method to perform the resummation, and
defines the pole mass to be the sum of the mass relation formula
eq.~(\ref{eq:msbpole}), all results obtained in the \MSB{} scheme would be
identical to those obtained in the pole mass scheme up to terms of relative
order $\as \CF$ or $\as\CA$, provided the same Borel sum method is used
also for the observables.

In the following we will try to estimate the terms of the perturbative
expansion using our large-$n_f$ results. In order to do this, we will perform
the replacement $-\TF/(3\pi) \to b_0$ accompanied by some other minor
adjustments, as described in Sec.~\ref{sec:realistic_b0}. Needless to say,
with these realistic values, the large-$n_f$ approximation breaks down, and
terms of relative order $\as \CF$ or $\as\CA$ may be sizeable. We thus expect
that by changing scheme we will generate difference of relative orders
$\as\CF$ or $\as\CA$, that are not negligible. These differences should not
therefore be interpreted as due to large ambiguities related to the choice of
mass scheme, but rather to the large-$n_f$ approximation.

The procedure we adopt in order to compute the terms of the perturbative
expansion follows from eq.~(\ref{eq:final_M_wcuts}). We fit numerically the
$\lambda$ dependence of the appropriate $T$ or $\widetilde{T}$ function, and we
take the derivative of the fit. The arctangent factor is instead expanded
analytically, and the integration is performed numerically for each
perturbative order. In order to have a semi-realistic result for the
perturbative coefficients we perform the following replacement
\begin{align}
 \Pi(\lambda^2+i\eta, \mu^2) - \Pi_{\rm ct} = \frac{\as
 \TF}{3\pi}\left(\frac{5}{3}-\log\frac{\lambda^2}{\mu^2} + i \pi\right) 
 \equiv \frac{\as n_f\TR}{3\pi}\left(\frac{5}{3}-\log\frac{\lambda^2}{\mu^2} + i \pi\right)
 \nonumber \\
 \to - \frac{\as 11 \CA }{12\pi}\left(C_g-\log\frac{\lambda^2}{\mu^2} + i \pi\right)
 + \frac{\as n_l\TR}{3\pi}\left(\frac{5}{3}-\log\frac{\lambda^2}{\mu^2} + i
 \pi\right), \qquad
 \label{eq:Pi_replacement}
\end{align}
where $C_g$ is given in eq.~(\ref{eq:C_g}). As a consequence, in
eqs.~(\ref{eq:final_M_wcuts}), (\ref{eq:sigma_final}),
and~(\ref{eq:final_tilderf}) the constant $C$, introduced in
eq.~(\ref{eq:Pi_C}), is replaced with eq.~(\ref{eq:C}) and the overall
$1/b_{0,f}$ factor with $1/b_0=(\frac{\as n_l\TR}{3}-\frac{\as 11 \CA
}{12})^{-1}$. For the computation of our observables no further modification
is required, since the factors $\Delta(\lambda)$ of
eq.~(\ref{eq:final_Deltaqqtilde}) and $\widetilde{\Delta}(\lambda)$ of
eq.~(\ref{eq:final_Deltaqqtilde}) do not depend on $n_f$, that cancels in the
ratio $\sigma_{q\bar{q}}^{(2)}/b_{0,f}$.

However, given the fact that the pole-\MSB{} mass relation involves
ultraviolet divergent quantities, it must be carried out in $d=4-2\ep$
dimension. For this reason, in this case, we cannot simply use the
$\mathcal{O}(\ep^0)$ expansion of $\lq \Pi(\lambda^2+i\eta, \mu^2)-\Pi_{\rm
  ct}\rq$ given in eq.~(\ref{eq:Pi_replacement}) to evaluate
eqs.~(\ref{eq:final_tilderd_zero}) and~(\ref{eq:final_tilderd_inf}), but we
need to use eqs.~(\ref{eq:Pi_full}) and~(\ref{eq:Pict_full}). We remark that
these contributions do not contain any infrared renormalon, conversely to
$r_f(\mpole, \mu, \as)$, that can be computed in $d=4$ dimensions. For this
reason, as already discussed in Sec.~\ref{sec:realistic_b0}, the presence of
a second term $C_g^\prime \ep^2$, that accompanies $C_g \ep$ in
eq.~(\ref{eq:Pi_full}), is totally negligible for the estimate of the leading
linear renormalons.

\section{Mass-conversion formula}
The procedure for the calculation of the mass-conversion formula is described
in Sec.~\ref{sec:msbar2pole}. Here we switch to the realistic $b_0$ and $C$
values as discussed in the previous section. The expansion of the mass
conversion formula reads
\begin{equation}
\label{eq:mpoleMSBnumer} 
\mMSB(\mu)= \mpole \(1-\sum_{i=1}^\infty c_i \, \as^i\)\,,
\end{equation}
and the $c_i$ coefficients are tabulated in Tab.~\ref{tab:ck}, with 
 $\mu^2={\rm Re}(\mpole^2)=m_0^2$, where $m_0$ is given in eq.~(\ref{eq:m0}).
\begin{table}[tb]
 \centering
  \begin{tabular}{|c|c|c|c|c|}
  \cline{1-5}
 \multicolumn{5}{|c|}{ $\displaystyle \phantom{\Big|} \mpole - \mMSB(\mu)  \phantom{\Big|} $}
 \\ \cline{1-5}
  $ \phantom{\Big|} i \phantom{\Big|}$ & ${\rm Re}\(c_i\)$   & ${\rm Im}\(c_i\)$   & ${\rm Re}\(\mpole\,c_i\,\as^i\)$ & ${\rm Im}\(\mpole\,c_i\,\as^i\)$
 \\ \cline{1-5}
 $ \phantom{\Big|} 1 \phantom{\Big|}$ & $4.244\times 10^{-1}$ & $2.450\times 10^{-3}$ & $7.919\times 10^{+0} $ & $+1.524\times 10^{-2} $
 \\ \cline{1-5}
 $ \phantom{\Big|} 2 \phantom{\Big|}$& $6.437\times 10^{-1}$ & $2.094\times 10^{-3}$ & $1.299\times 10^{+0} $ & $-7.729\times 10^{-4} $
 \\ \cline{1-5}
 $ \phantom{\Big|} 3 \phantom{\Big|}$& $1.968\times 10^{+0} $ & $8.019\times 10^{-3}$ & $4.297\times 10^{-1}$ & $+9.665\times 10^{-5} $
 \\ \cline{1-5}
 $ \phantom{\Big|} 4 \phantom{\Big|}$& $7.231\times 10^{+0} $ & $2.567\times 10^{-2}$ & $1.707\times 10^{-1}$ & $-5.110\times 10^{-5} $
 \\ \cline{1-5}
 $ \phantom{\Big|} 5 \phantom{\Big|}$& $3.497\times 10^{+1} $ & $1.394\times 10^{-1}$ & $8.930\times 10^{-2}$ & $+1.240\times 10^{-5} $
 \\ \cline{1-5}
 $ \phantom{\Big|} 6 \phantom{\Big|}$& $2.174\times 10^{+2} $ & $8.164\times 10^{-1}$ & $6.005\times 10^{-2}$ & $-5.616\times 10^{-6} $
 \\ \cline{1-5}
 $ \phantom{\Big|} 7 \phantom{\Big|}$& $1.576\times 10^{+3} $ & $6.133\times 10^{+0} $ & $4.709\times 10^{-2}$ & $+2.009\times 10^{-6} $
 \\ \cline{1-5}
 $ \phantom{\Big|} 8 \phantom{\Big|}$& $1.354\times 10^{+4} $ & $5.180\times 10^{+1} $ & $4.376\times 10^{-2}$ & $-1.031\times 10^{-6} $
 \\ \cline{1-5}
 $ \phantom{\Big|} 9 \phantom{\Big|}$& $1.318\times 10^{+5} $ & $5.087\times 10^{+2} $ & $4.608\times 10^{-2}$ & $+4.961\times 10^{-7} $
 \\ \cline{1-5}
$ \phantom{\Big|} 10 \phantom{\Big|}$& $1.450\times 10^{+6} $ & $5.572\times 10^{+3} $ & $5.481\times 10^{-2}$ & $-2.909\times 10^{-7} $
 \\ \cline{1-5}
 \end{tabular}

\caption{Real and imaginary parts of the coefficients $c_i$ of the mass
 relation~(\ref{eq:mpoleMSBnumer}), up to the tenth order in the strong
 coupling constant $\as(\mu)$, with $\mu^2={\rm Re}(\mpole^2)$ and
 $\mpole=172.5$~GeV. The numbers shown in the last two columns are expressed in GeV.}
\label{tab:ck}
\end{table}
Since we are using the complex mass scheme, they are complex, with a small
imaginary part, and they have a slight dependence upon the ratio
$\Gamma_t/{\rm Re}(m)$. For small $\Gamma_t$ they become independent on $m$
and $\Gamma_t$, and their imaginary part vanishes.

The value of the \MSB{} mass we adopt in the following is found by
truncating the series in eq.~\ref{eq:mpoleMSBnumer} at the smallest term
before the series starts diverging, that corresponds to $i=8$, as shown in Tab.~\ref{tab:ck}. We thus find
that for a complex pole mass
\begin{equation}
\mpole = (172.50-0.66\,i)~{\rm GeV}
\end{equation}
the value of the corresponding \MSB{} complex mass is 
\begin{equation}
\mMSB(\mu) = \mpole \(1-\sum_{i=1}^8 c_i \, \as^i\)\ =(162.44 -0.68\,i)~{\rm GeV},
\end{equation}
with $\mu=\sqrt{{\rm Re}(m^2)}$.

\section{The total cross section}
In this section we deal with the perturbative expansion of the total cross
section, first without cuts, and then with cuts.

\subsection{Total cross section without cuts}
As discussed in Sec.~\ref{sec:Xtot_wocuts},
$T(\lambda)$~(\ref{eq:Tk2_final}) for the total cross section does
not have any term linear in $\lambda$, if expressed in terms of the \MSB{} mass. It
follows that the total cross section computed in the \MSB{} scheme should not
have any $\Lambda_{\sss \rm QCD}/\mMSB$ renormalon and should display a
better behavior at large orders.

\begin{table}[tb]
 \centering
  \begin{tabular}{c|c|c|c|c|}
  \cline{2-5}
  & \multicolumn{4}{|c|}{$\phantom{\Big|} \sigma/\sigma^{\rm nocuts}_{\rm b}(\mpole{}) \phantom{\Big|}$}
 \\ \cline{2-5}
  & \multicolumn{2}{|c|}{ \phantom{\Big|}pole scheme \phantom{\Big|}}& \multicolumn{2}{|c|}{\MSB{} scheme}
 \\ \cline{1-5}
 \multicolumn{1}{|c|}{$\phantom{\Big|} i \phantom{\Big|}$}  & $c_i $ & $ c_i\, \as^i$  & $c_i $ & $ c_i\, \as^i$ 
 \\ \cline{1-5}
 \multicolumn{1}{|c|}{$\phantom{\Big|}$ 0 $\phantom{\Big|}$}& 1.00000000 &1.0000000& 0.86841331 &0.8684133
 \\ \cline{1-5}
 \multicolumn{1}{|c|}{ $\phantom{\Big|}$           1 $\phantom{\Big|}$}& $5.003 \, (0) \times 10^{   -1}$ & $5.411 \, (0) \times 10^{   -2}$& $1.480 \, (0) \times 10^{    0}$ & $1.601 \, (0) \times 10^{   -1}$
 \\ \cline{1-5}
 \multicolumn{1}{|c|}{ $\phantom{\Big|}$           2 $\phantom{\Big|}$}& $-6.20 \, (2) \times 10^{   -1}$ & $-7.25 \, (2) \times 10^{   -3}$& $4.42 \, (2) \times 10^{   -1}$ & $5.17 \, (2) \times 10^{   -3}$
 \\ \cline{1-5}
 \multicolumn{1}{|c|}{ $\phantom{\Big|}$           3 $\phantom{\Big|}$}& $-3.03 \, (2) \times 10^{    0}$ & $-3.83 \, (3) \times 10^{   -3}$& $6.4 \, (2) \times 10^{   -1}$ & $8.1 \, (3) \times 10^{   -4}$
 \\ \cline{1-5}
 \multicolumn{1}{|c|}{ $\phantom{\Big|}$           4 $\phantom{\Big|}$}& $-1.25 \, (2) \times 10^{    1}$ & $-1.70 \, (3) \times 10^{   -3}$& $0 \, (2) \times 10^{   -2}$ & $0 \, (3) \times 10^{   -6}$
 \\ \cline{1-5}
 \multicolumn{1}{|c|}{ $\phantom{\Big|}$           5 $\phantom{\Big|}$}& $-6.4 \, (2) \times 10^{    1}$ & $-9.4 \, (3) \times 10^{   -4}$& $1 \, (2) \times 10^{   -1}$ & $1 \, (3) \times 10^{   -5}$
 \\ \cline{1-5}
 \multicolumn{1}{|c|}{ $\phantom{\Big|}$           6 $\phantom{\Big|}$}& $-3.9 \, (1) \times 10^{    2}$ & $-6.2 \, (2) \times 10^{   -4}$& $0 \, (1) \times 10^{    0}$ & $0 \, (2) \times 10^{   -6}$
 \\ \cline{1-5}
 \multicolumn{1}{|c|}{ $\phantom{\Big|}$           7 $\phantom{\Big|}$}& $-2.9 \, (1) \times 10^{    3}$ & $-5.0 \, (2) \times 10^{   -4}$& $0 \, (1) \times 10^{    1}$ & $0 \, (2) \times 10^{   -6}$
 \\ \cline{1-5}
 \multicolumn{1}{|c|}{ $\phantom{\Big|}$           8 $\phantom{\Big|}$}& $-2.5 \, (1) \times 10^{    4}$ & $-4.6 \, (2) \times 10^{   -4}$& $0 \, (1) \times 10^{    2}$ & $0 \, (2) \times 10^{   -6}$
 \\ \cline{1-5}
 \multicolumn{1}{|c|}{ $\phantom{\Big|}$           9 $\phantom{\Big|}$}& $-2.4 \, (1) \times 10^{    5}$ & $-4.9 \, (2) \times 10^{   -4}$& $0 \, (1) \times 10^{    3}$ & $0 \, (2) \times 10^{   -6}$
 \\ \cline{1-5}
 \multicolumn{1}{|c|}{ $\phantom{\Big|}$          10 $\phantom{\Big|}$}& $-2.6 \, (1) \times 10^{    6}$ & $-5.8 \, (2) \times 10^{   -4}$& $0 \, (1) \times 10^{    4}$ & $-1 \, (2) \times 10^{   -6}$
 \\ \cline{1-5}
 \end{tabular}

 \caption{Coefficients of the $\as$ expansion of the inclusive cross section
 to all orders, computed in the large-$b_0$ limit, normalized to the Born
 cross section computed in the pole-mass scheme.
 The errors reported in parenthesis are due to the uncertainty
 on the linear coefficient of the fit (i.e. $p_1$ in eq.~(\ref{eq:T_polynom})).
 }
 \label{tab:sigtot_expansion}
\end{table}
The coefficients $c_i$ of the expansion of eq.~(\ref{eq:sigma_final}) in
terms of $\as$
\begin{equation}
 \label{eq:sigma_expansion}
 \sigma = \sigma_{\rm b}^{\rm nocuts}(\mpole)\left( c_0 + \sum_{i=1}^{\infty}
 c_i\, \as^i \right)
\end{equation}
are collected in Tab.~\ref{tab:sigtot_expansion}, in the pole~(left) and in
the \MSB{}~(right) schemes. At large orders, the \MSB{} total cross section
receives much smaller contributions. On the other hand we see that the
$\rm N^3LO$ contribution to $\sigma(\mpole)$ is already affected by a
factorial growth. The minimum of the series is reached for $i=8$ (that
corresponds to an $\mathcal{O}(\as^8)$ correction), and it is two orders of
magnitude larger than the corresponding contribution computed in the \MSB{}
scheme. We also notice that the \MSB{} result has an NLO correction larger
than the pole mass result, an NNLO correction that is similar, and smaller
N$^3$LO and higher order corrections. We also expect that the apparent
convergence of the expansion for the first few orders should depend upon the
available phase space for radiation.

\subsection{Total cross section with cuts}
As we have seen in Sec.~\ref{sec:xsec_cuts}, the presence of selection cuts
introduces a renormalon in the total cross section whose magnitude goes like
$1/R$.

\begin{table}[tb]
 \centering
\resizebox{\textwidth}{!}
{  \begin{tabular}{c|c|c|}
  \cline{2-3}
  & \multicolumn{2}{|c|}{$\phantom{\Big|} \sigma/\sigma^{\rm nocuts}_{\rm
    b}(\mpole{}) \;\; R=0.1 \phantom{\Big|}$}
 \\ \cline{2-3}
  & \multicolumn{1}{|c|}{ \phantom{\Big|}pole scheme \phantom{\Big|}}& \multicolumn{1}{|c|}{\MSB{} scheme}
 \\ \cline{1-3}
 \multicolumn{1}{|c|}{$\phantom{\Big|} i \phantom{\Big|}$} & $ c_i\, \as^i$ & $ c_i\, \as^i$ 
 \\ \cline{1-3}
 \multicolumn{1}{|c|}{$\phantom{\Big|}$ 0 $\phantom{\Big|}$} &0.9985836&0.8666708
 \\ \cline{1-3}
 \multicolumn{1}{|c|}{ $\phantom{\Big|}$           1 $\phantom{\Big|}$}& $-7.953 \, (0) \times 10^{   -2}$& $2.650 \, (0) \times 10^{   -2}$
 \\ \cline{1-3}
 \multicolumn{1}{|c|}{ $\phantom{\Big|}$           2 $\phantom{\Big|}$}& $-7.22 \, (2) \times 10^{   -2}$& $-5.98 \, (2) \times 10^{   -2}$
 \\ \cline{1-3}
 \multicolumn{1}{|c|}{ $\phantom{\Big|}$           3 $\phantom{\Big|}$}& $-3.71 \, (2) \times 10^{   -2}$& $-3.24 \, (2) \times 10^{   -2}$
 \\ \cline{1-3}
 \multicolumn{1}{|c|}{ $\phantom{\Big|}$           4 $\phantom{\Big|}$}& $-1.97 \, (2) \times 10^{   -2}$& $-1.80 \, (2) \times 10^{   -2}$
 \\ \cline{1-3}
 \multicolumn{1}{|c|}{ $\phantom{\Big|}$           5 $\phantom{\Big|}$}& $-1.13 \, (2) \times 10^{   -2}$& $-1.04 \, (2) \times 10^{   -2}$
 \\ \cline{1-3}
 \multicolumn{1}{|c|}{ $\phantom{\Big|}$           6 $\phantom{\Big|}$}& $-7.0 \, (2) \times 10^{   -3}$& $-6.4 \, (2) \times 10^{   -3}$
 \\ \cline{1-3}
 \multicolumn{1}{|c|}{ $\phantom{\Big|}$           7 $\phantom{\Big|}$}& $-4.8 \, (1) \times 10^{   -3}$& $-4.3 \, (1) \times 10^{   -3}$
 \\ \cline{1-3}
 \multicolumn{1}{|c|}{ $\phantom{\Big|}$           8 $\phantom{\Big|}$}& $-3.6 \, (1) \times 10^{   -3}$& $-3.1 \, (1) \times 10^{   -3}$
 \\ \cline{1-3}
 \multicolumn{1}{|c|}{ $\phantom{\Big|}$           9 $\phantom{\Big|}$}& $-3.1 \, (1) \times 10^{   -3}$& $-2.7 \, (1) \times 10^{   -3}$
 \\ \cline{1-3}
 \multicolumn{1}{|c|}{ $\phantom{\Big|}$          10 $\phantom{\Big|}$}& $-3.2 \, (2) \times 10^{   -3}$& $-2.6 \, (2) \times 10^{   -3}$
 \\ \cline{1-3}
 \end{tabular}

  \begin{tabular}{c|c|c|}
  \cline{2-3}
  & \multicolumn{2}{|c|}{$\phantom{\Big|} \sigma/\sigma^{\rm nocuts}_{\rm
    b}(\mpole{}) \;\; R=0.5 \phantom{\Big|}$}
 \\ \cline{2-3}
  & \multicolumn{1}{|c|}{ \phantom{\Big|}pole scheme \phantom{\Big|}}& \multicolumn{1}{|c|}{\MSB{} scheme}
 \\ \cline{1-3}
 \multicolumn{1}{|c|}{$\phantom{\Big|} i \phantom{\Big|}$} & $ c_i\, \as^i$ & $ c_i\, \as^i$ 
 \\ \cline{1-3}
 \multicolumn{1}{|c|}{$\phantom{\Big|}$ 0 $\phantom{\Big|}$} &0.9783310&0.8511828
 \\ \cline{1-3}
 \multicolumn{1}{|c|}{ $\phantom{\Big|}$           1 $\phantom{\Big|}$}& $-4.992 \, (0) \times 10^{   -3}$& $9.705 \, (0) \times 10^{   -2}$
 \\ \cline{1-3}
 \multicolumn{1}{|c|}{ $\phantom{\Big|}$           2 $\phantom{\Big|}$}& $-2.966 \, (5) \times 10^{   -2}$& $-1.779 \, (5) \times 10^{   -2}$
 \\ \cline{1-3}
 \multicolumn{1}{|c|}{ $\phantom{\Big|}$           3 $\phantom{\Big|}$}& $-1.267 \, (6) \times 10^{   -2}$& $-8.22 \, (6) \times 10^{   -3}$
 \\ \cline{1-3}
 \multicolumn{1}{|c|}{ $\phantom{\Big|}$           4 $\phantom{\Big|}$}& $-5.37 \, (6) \times 10^{   -3}$& $-3.73 \, (6) \times 10^{   -3}$
 \\ \cline{1-3}
 \multicolumn{1}{|c|}{ $\phantom{\Big|}$           5 $\phantom{\Big|}$}& $-2.58 \, (5) \times 10^{   -3}$& $-1.66 \, (5) \times 10^{   -3}$
 \\ \cline{1-3}
 \multicolumn{1}{|c|}{ $\phantom{\Big|}$           6 $\phantom{\Big|}$}& $-1.44 \, (4) \times 10^{   -3}$& $-8.5 \, (4) \times 10^{   -4}$
 \\ \cline{1-3}
 \multicolumn{1}{|c|}{ $\phantom{\Big|}$           7 $\phantom{\Big|}$}& $-9.8 \, (4) \times 10^{   -4}$& $-5.0 \, (4) \times 10^{   -4}$
 \\ \cline{1-3}
 \multicolumn{1}{|c|}{ $\phantom{\Big|}$           8 $\phantom{\Big|}$}& $-8.1 \, (4) \times 10^{   -4}$& $-3.7 \, (4) \times 10^{   -4}$
 \\ \cline{1-3}
 \multicolumn{1}{|c|}{ $\phantom{\Big|}$           9 $\phantom{\Big|}$}& $-8.0 \, (4) \times 10^{   -4}$& $-3.4 \, (4) \times 10^{   -4}$
 \\ \cline{1-3}
 \multicolumn{1}{|c|}{ $\phantom{\Big|}$          10 $\phantom{\Big|}$}& $-9.2 \, (5) \times 10^{   -4}$& $-3.7 \, (5) \times 10^{   -4}$
 \\ \cline{1-3}
 \end{tabular}
}
 \caption{Values of the $c_i\,\as^i$ terms of the perturbative expansion for
   the average value of the cross section with cuts, normalized to the
   inclusive Born cross section computed in the pole-mass scheme (see
   eq.~(\ref{eq:sigma_expansion})), for two different values of the jet
   radius ($R=0.1$ in the left pane and $R=0.5$ in the right one).  The
   errors reported in parenthesis are due to the uncertainty on the linear
   coefficient of the fit (i.e. $p_1$ in eq.~(\ref{eq:T_polynom})).  }
 \label{tab:sigtot_expansion_cut2}
\end{table}
In Tab.~\ref{tab:sigtot_expansion_cut2} we present the results for the total
cross section, in the pole and in the \MSB-mass scheme, for a small jet
radius, $R=0.1$, and a more realistic value, $R=0.5$. For small radii, the
perturbative expansion displays roughly the same bad behaviour, either when
we use the pole or the \MSB{} scheme. For larger values of $R$, the
size of the coefficients are typically smaller than the corresponding ones
with smaller values or $R$. In particular if we compare the coefficients for
$R=0.1$ and $R=0.5$, the second ones are one order of magnitude smaller than
the first ones. Furthermore, for $R=0.5$, the coefficients computed in the
\MSB{}-mass scheme are roughly half of the ones computed in the pole-mass
scheme. As remarked earlier, this reduction is due to an accidental
cancellation of the pole-mass associated renormalon and the $1/R$, jet
related one, and cannot be used to imply that the \MSB{} scheme should be
favoured in this case.

\section{Reconstructed-top mass}
In this section, we discuss the terms of the perturbative expansion for the
average reconstructed mass~$\langle M\rangle$
 \begin{equation}
 \label{eq:M_expansion}
 \langle M \rangle = \sum_{i=0}^\infty c_i\, \as^i\,,
\end{equation}
for three values of the $R$ parameter. We apply the cuts described in
Sec.~\ref{sec:total-xsec} and the results are collected in Tab.~\ref{tab:M-coeffs}.

\begin{table}[h!]
 \resizebox{\textwidth}{!}
 {\begin{tabular}{c|c|c|c|c|c|c|c|}
  \cline{2-7}
  & \multicolumn{6}{|c|}{$\phantom{\Big|}$ $\langle M \rangle$ [GeV]$\phantom{\Big|}$ } 
  \\ \cline{2-7}
  & \multicolumn{2}{|c|}{$\phantom{\Big|} R=0.1 \phantom{\Big|}$ }
  & \multicolumn{2}{|c|}{$\phantom{\Big|} R=0.5 \phantom{\Big|}$ }
  & \multicolumn{2}{|c|}{$\phantom{\Big|} R=1.5 \phantom{\Big|}$ }
 \\ \cline{1-7}
 \multicolumn{1}{|c|}{$i$} & \multicolumn{1}{|c|}{ \phantom{\Big|} pole  \phantom{\Big|}}& \multicolumn{1}{|c|}{\MSB{}} & \multicolumn{1}{|c|}{ \phantom{\Big|} pole\phantom{\Big|}}& \multicolumn{1}{|c|}{\MSB{}} & \multicolumn{1}{|c|}{ \phantom{\Big|} pole \phantom{\Big|}}& \multicolumn{1}{|c|}{\MSB{} }
 \\ \cline{1-7}
 \multicolumn{1}{|c|}{  $\phantom{\Big|}$           0  $\phantom{\Big|}$}
 & $172.8280                                                                                            $& $163.0146                                                                                            $
 & $172.8201                                                                                            $& $163.0040                                                                                            $
 & $172.7533                                                                                            $& $162.9244                                                                                            $
 \\ \cline{1-7}
 \multicolumn{1}{|c|}{  $\phantom{\Big|}$           1  $\phantom{\Big|}$}
 & $-7.597 \, (0) \times 10^{    0}                                                                     $& $2.163 \, (0) \times 10^{   -1}                                                                      $
 & $-2.785 \, (0) \times 10^{    0}                                                                     $& $5.030 \, (0) \times 10^{    0}                                                                      $
 & $4.446 \, (0) \times 10^{   -1}                                                                      $& $8.268 \, (0) \times 10^{    0}                                                                      $
 \\ \cline{1-7}
 \multicolumn{1}{|c|}{  $\phantom{\Big|}$           2  $\phantom{\Big|}$}
 & $-4.136 \, (2) \times 10^{    0}                                                                     $& $-2.852 \, (2) \times 10^{    0}                                                                     $
 & $-1.255 \, (1) \times 10^{    0}                                                                     $& $2.9 \, (1) \times 10^{   -2}                                                                        $
 & $1.029 \, (8) \times 10^{   -1}                                                                      $& $1.387 \, (1) \times 10^{    0}                                                                      $
 \\ \cline{1-7}
 \multicolumn{1}{|c|}{  $\phantom{\Big|}$           3  $\phantom{\Big|}$}
 & $-2.397 \, (2) \times 10^{    0}                                                                     $& $-1.973 \, (2) \times 10^{    0}                                                                     $
 & $-5.96 \, (2) \times 10^{   -1}                                                                      $& $-1.72 \, (2) \times 10^{   -1}                                                                      $
 & $1.4 \, (1) \times 10^{   -2}                                                                        $& $4.38 \, (1) \times 10^{   -1}                                                                       $
 \\ \cline{1-7}
 \multicolumn{1}{|c|}{  $\phantom{\Big|}$           4  $\phantom{\Big|}$}
 & $-1.505 \, (2) \times 10^{    0}                                                                     $& $-1.337 \, (2) \times 10^{    0}                                                                     $
 & $-3.13 \, (2) \times 10^{   -1}                                                                      $& $-1.44 \, (2) \times 10^{   -1}                                                                      $
 & $-6 \, (1) \times 10^{   -3}                                                                         $& $1.63 \, (1) \times 10^{   -1}                                                                       $
 \\ \cline{1-7}
 \multicolumn{1}{|c|}{  $\phantom{\Big|}$           5  $\phantom{\Big|}$}
 & $-1.038 \, (2) \times 10^{    0}                                                                     $& $-9.50 \, (2) \times 10^{   -1}                                                                      $
 & $-1.88 \, (2) \times 10^{   -1}                                                                      $& $-1.00 \, (2) \times 10^{   -2}                                                                      $
 & $-9.7 \, (9) \times 10^{   -3}                                                                       $& $7.86 \, (9) \times 10^{   -2}                                                                       $
 \\ \cline{1-7}
 \multicolumn{1}{|c|}{  $\phantom{\Big|}$           6  $\phantom{\Big|}$}
 & $-7.94 \, (2) \times 10^{   -1}                                                                      $& $-7.35 \, (2) \times 10^{   -1}                                                                      $
 & $-1.33 \, (1) \times 10^{   -1}                                                                      $& $-7.3 \, (1) \times 10^{   -2}                                                                       $
 & $-1.05 \, (8) \times 10^{   -2}                                                                      $& $4.89 \, (8) \times 10^{   -2}                                                                       $
 \\ \cline{1-7}
 \multicolumn{1}{|c|}{  $\phantom{\Big|}$           7  $\phantom{\Big|}$}
 & $-6.79 \, (2) \times 10^{   -1}                                                                      $& $-6.33 \, (2) \times 10^{   -1}                                                                      $
 & $-1.09 \, (1) \times 10^{   -1}                                                                      $& $-6.3 \, (1) \times 10^{   -2}                                                                       $
 & $-1.12 \, (7) \times 10^{   -2}                                                                      $& $3.53 \, (7) \times 10^{   -2}                                                                       $
 \\ \cline{1-7}
 \multicolumn{1}{|c|}{  $\phantom{\Big|}$           8  $\phantom{\Big|}$}
 & $-6.51 \, (2) \times 10^{   -1}                                                                      $& $-6.08 \, (2) \times 10^{   -1}                                                                      $
 & $-1.04 \, (1) \times 10^{   -1}                                                                      $& $-6.1 \, (1) \times 10^{   -2}                                                                       $
 & $-1.25 \, (7) \times 10^{   -2}                                                                      $& $3.08 \, (7) \times 10^{   -2}                                                                       $
 \\ \cline{1-7}
 \multicolumn{1}{|c|}{  $\phantom{\Big|}$           9  $\phantom{\Big|}$}
 & $-6.99 \, (2) \times 10^{   -1}                                                                      $& $-6.54 \, (2) \times 10^{   -1}                                                                      $
 & $-1.12 \, (1) \times 10^{   -1}                                                                      $& $-6.7 \, (1) \times 10^{   -2}                                                                       $
 & $-1.47 \, (7) \times 10^{   -2}                                                                      $& $3.09 \, (7) \times 10^{   -2}                                                                       $
 \\ \cline{1-7}
 \multicolumn{1}{|c|}{  $\phantom{\Big|}$          10  $\phantom{\Big|}$}
 & $-8.37 \, (2) \times 10^{   -1}                                                                      $& $-7.83 \, (2) \times 10^{   -1}                                                                      $
 & $-1.35 \, (1) \times 10^{   -1}                                                                      $& $-8.1 \, (1) \times 10^{   -2}                                                                       $
 & $-1.85 \, (9) \times 10^{   -2}                                                                      $& $3.57 \, (9) \times 10^{   -2}                                                                       $
 \\ \cline{1-7}
 \end{tabular}
}
 \centering
 \caption{Values of the $c_i\, \as^i$ terms of the perturbative expansion for
   the average value of the reconstructed-top mass, defined in
   eq.~(\ref{eq:M_expansion}), for three different jet radii in the pole-mass
   and \MSB{}-mass scheme. The errors
   reported in parenthesis are due to the uncertainty on the linear
   coefficient of the fit (i.e. $p_1$ in eq.~(\ref{eq:T_polynom})).}
 \label{tab:M-coeffs}
\end{table}
From the table we can see that, for very small jet radii, the asymptotic
character of the perturbative expansion is manifest in both the pole and
\MSB{} scheme. For the realistic value $R=0.5$, the \MSB{} scheme seems to
behave slightly better. In fact, this is only a consequence of the fact that
the jet-renormalon and the mass-renormalon corrections have
opposite signs, with the mass correction in the \MSB{} scheme largely
prevailing at small orders, yielding positive effects.

As the radius becomes very large, the jet renormalon becomes less and less
pronounced, in the pole-mass scheme, leading to smaller corrections at all
orders. This is consistent with the discussion given in
Sec.~\ref{sec:rec-top-mass}, where we have seen that, for large radii, the
reconstructed mass becomes strongly related to the top pole mass, since it
approaches what one would reconstruct from the ``true'' top decay
products.\footnote{We recall here that, in the narrow width limit, and in perturbation
theory, the concept of a ``true'' top decay final state is well defined.}

\section[${W}$ boson energy]{$\boldsymbol{W}$ boson energy}
The coefficients of the perturbative expansion of the average energy of the
$W$ boson in the pole and \MSB{} schemes
 \begin{equation}
 \label{eq:Ew_expansion}
 \langle E_{\sss W}\rangle = \sum_{i=0}^\infty c_i\, \as^i.
 \end{equation}
are displayed in Tab.~\ref{tab:Wexp}.
We notice that the perturbative expansions are similarly behaved in both
schemes up to $i\approx 6$, while, for higher orders, the \MSB{} scheme
result is clearly better convergent. This supports the observation, done in
Sec.~\ref{sec:Ew}, that the top width screens the renormalon effect if the
\MSB{} mass is used. In fact, the $6^{\rm th}$ order renormalon contribution
is dominated by scales of order $m_t \, e^{-5}\approx 1.16$,
as illustrated in Sec.~\ref{sec:renorm_intro}, very near the top width.

By looking at the $i=0$ row, we notice that a variation of roughly 10~GeV in
the value of the top mass, corresponding to the pole-\MSB{} mass difference,
leads to a variation of less than 1~GeV in $\langle E_{\sss W}
\rangle_b$. This implies that the sensitivity of the $W$-boson energy
$E_{\sss W}$ to the top mass is much weaker than for the reconstructed-top
mass $M$. Indeed, in Secs.~\ref{sec:rec-top-mass} and~\ref{sec:Ew} we already
noticed that
\begin{equation}
  \frac{\partial \langle E_{\sss W} \rangle_b}{\partial\, {\rm Re}(m)} \approx 0.1\,,\qquad \frac{\partial \langle M \rangle_b}{\partial\, {\rm Re}(m)} \approx  1. 
\end{equation}
The value of the $E_{\sss W}$ derivative is strongly affected by our choice
of the rest frame energy $E=300$~GeV, that corresponds to a boost
$\beta_{\sss top}=|\vec{p}_{\sss top}|/E_{\sss top}=0.5$ for an on-shell
top-quark.  Thus, despite the fact that $E_{\sss W}$ is free from physical
renormalons, if the top quark has substantial kinetic energy, the weak
sensitivity of such observable to the value of the top mass may in practice
reduce the precision of the measurement.  

\begin{table}[h!]
 \centering
  \begin{tabular}{c|c|c|c|c|}
  \cline{2-5}
  & \multicolumn{4}{|c|}{ $\phantom{\Big|}$ $\displaystyle \langle E_W
    \rangle$ $\phantom{\Big|}$ [GeV]}
 \\ \cline{2-5}
  & \multicolumn{2}{|c|}{ \phantom{\Big|}pole scheme \phantom{\Big|}}& \multicolumn{2}{|c|}{\MSB{} scheme}
 \\ \cline{1-5}
 \multicolumn{1}{|c|}{$\phantom{\Big|} i \phantom{\Big|}$}  & $c_i $ & $ c_i\,\as^i$  & $c_i $ & $ c_i \, \as^i$ 
 \\ \cline{1-5}
 \multicolumn{1}{|c|}{ $\phantom{\Big|}$ 0 $\phantom{\Big|}$}& $121.5818$ & $121.5818$& $120.8654$ & $120.8654$
 \\ \cline{1-5}
 \multicolumn{1}{|c|}{$\phantom{\Big|}$           1 $\phantom{\Big|}$}& $-1.435 \, (0) \times 10^{    1}$ & $-1.552 \, (0) \times 10^{    0}$ & $-7.192 \, (0) \times 10^{    0}$ & $-7.779 \, (0) \times 10^{   -1}$ 
 \\ \cline{1-5}
 \multicolumn{1}{|c|}{$\phantom{\Big|}$           2 $\phantom{\Big|}$}& $-4.97 \, (4) \times 10^{    1}$ & $-5.82 \, (4) \times 10^{   -1}$ & $-3.88 \, (4) \times 10^{    1}$ & $-4.54 \, (4) \times 10^{   -1}$ 
 \\ \cline{1-5}
 \multicolumn{1}{|c|}{$\phantom{\Big|}$           3 $\phantom{\Big|}$}& $-1.79 \, (5) \times 10^{    2}$ & $-2.26 \, (6) \times 10^{   -1}$ & $-1.45 \, (5) \times 10^{    2}$ & $-1.84 \, (6) \times 10^{   -1}$ 
 \\ \cline{1-5}
 \multicolumn{1}{|c|}{$\phantom{\Big|}$           4 $\phantom{\Big|}$}& $-6.9 \, (4) \times 10^{    2}$ & $-9.4 \, (6) \times 10^{   -2}$ & $-5.7 \, (4) \times 10^{    2}$ & $-7.8 \, (6) \times 10^{   -2}$ 
 \\ \cline{1-5}
 \multicolumn{1}{|c|}{$\phantom{\Big|}$           5 $\phantom{\Big|}$}& $-2.9 \, (3) \times 10^{    3}$ & $-4.4 \, (5) \times 10^{   -2}$ & $-2.4 \, (3) \times 10^{    3}$ & $-3.5 \, (5) \times 10^{   -2}$ 
 \\ \cline{1-5}
 \multicolumn{1}{|c|}{$\phantom{\Big|}$           6 $\phantom{\Big|}$}& $-1.4 \, (3) \times 10^{    4}$ & $-2.2 \, (4) \times 10^{   -2}$ & $-1.0 \, (3) \times 10^{    4}$ & $-1.7 \, (4) \times 10^{   -2}$ 
 \\ \cline{1-5}
 \multicolumn{1}{|c|}{$\phantom{\Big|}$           7 $\phantom{\Big|}$}& $-8 \, (2) \times 10^{    4}$ & $-1.3 \, (4) \times 10^{   -2}$ & $-5 \, (2) \times 10^{    4}$ & $-8 \, (4) \times 10^{   -3}$ 
 \\ \cline{1-5}
 \multicolumn{1}{|c|}{$\phantom{\Big|}$           8 $\phantom{\Big|}$}& $-5 \, (2) \times 10^{    5}$ & $-9 \, (4) \times 10^{   -3}$ & $-2 \, (2) \times 10^{    5}$ & $-4 \, (4) \times 10^{   -3}$ 
 \\ \cline{1-5}
 \multicolumn{1}{|c|}{$\phantom{\Big|}$           9 $\phantom{\Big|}$}& $-3 \, (2) \times 10^{    6}$ & $-7 \, (4) \times 10^{   -3}$ & $-1 \, (2) \times 10^{    6}$ & $-2 \, (4) \times 10^{   -3}$ 
 \\ \cline{1-5}
 \multicolumn{1}{|c|}{$\phantom{\Big|}$          10 $\phantom{\Big|}$}& $-3 \, (2) \times 10^{    7}$ & $-6 \, (5) \times 10^{   -3}$ & $0 \, (2) \times 10^{    6}$ & $-1 \, (5) \times 10^{   -4}$ 
 \\ \cline{1-5}
 \multicolumn{1}{|c|}{$\phantom{\Big|}$          11 $\phantom{\Big|}$}& $-3 \, (3) \times 10^{    8}$ & $-7 \, (6) \times 10^{   -3}$ & $0 \, (3) \times 10^{    6}$ & $0 \, (6) \times 10^{   -5}$ 
 \\ \cline{1-5}
 \multicolumn{1}{|c|}{$\phantom{\Big|}$          12 $\phantom{\Big|}$}& $-4 \, (3) \times 10^{    9}$ & $-9 \, (9) \times 10^{   -3}$ & $0 \, (3) \times 10^{    8}$ & $1 \, (9) \times 10^{   -3}$ 
 \\ \cline{1-5}
 \end{tabular}

 \caption{Coefficients of the perturbative expansion of the average
 $W$-boson energy in the pole and \MSB{} schemes (see
 eq.~(\ref{eq:Ew_expansion})). The errors reported in parenthesis are due
 to the uncertainty on the linear coefficient of the fit (i.e. $p_1$ in
 eq.~(\ref{eq:T_polynom})).}
 \label{tab:Wexp}
\end{table}

\chapter{Summary and conclusions}
\label{sec:concl_renormalons}
In this first part of the thesis we have examined non-perturbative
corrections related to infrared renormalons relevant to typical top-quark
mass measurements, in the simplified context of a
$W^* \rightarrow t\bar{b} \to W b \bar{b}$ process, with an on-shell
final-state $W$ boson and massless $b$ quarks.  As a further simplification,
we have considered only vector-current couplings.  We have however fully
taken into account top finite width effects.

We have investigated non-perturbative corrections that arise from the
resummation of light-quark loop insertions in the gluon propagator,
corresponding to the so called $\mbox{large-}n_f$ limit of QCD. The
$\mbox{large-}n_f$ limit result can be turned into the so called large-$b_0$
approximation, by replacing the large-$n_f$ beta function coefficient with
the true QCD one. This approximation has been adopted in several contexts for
the study of non-perturbative effects (see e.g.~Refs.~\cite{Beneke:1994qe,
  Beneke:1994sw, Beneke:1994bc, Bigi:1994em, Ball:1995ni, Seymour:1994df}).

In this paper we have developed a method to compute the $\mbox{large-}n_f$
results exactly, using a combination of analytic and numerical methods. The
latter is in essence the combination of four parton level generators, that
allowed us to compute kinematic observables of arbitrary complexity. We
stress that, besides being able to study the effect of the leading
renormalons, we can also compute numerically the coefficients of the
perturbative expansion up and beyond the order at which it starts to diverge.

Although our findings have all been obtained in the simplified context just
described, we can safely say that all effects that we have found are likely
to be present in the full theory, although we are not in a position to
exclude the presence of other effects related to the non-Abelian nature of
QCD, or to non-perturbative effects not related to renormalons.

Our findings can be summarized as follows:
\begin{itemize}
\item The total cross section for the process at hand is free of
  \emph{physical} linear renormalons, i.e. its perturbative expansion in terms
  of a short distance mass is free of
  linear renormalons. This result holds both for
  finite top width and in the narrow-width limit.
  In the former case, the absence of a linear renormalon is due to the
  screening effect of the top finite width, while,
  in the latter case, it is
  a straightforward consequence of the fact that both the top production
  cross section and the decay partial width are free of physical linear
  renormalons.

  By examining the perturbative expansion
  order by order, we find that, already at the NNLO level, the \MSB{} scheme
  result for the cross section is much more accurate than the pole-mass-scheme one.

  We stress that our choice of 300~GeV for the incoming
  energy corresponds to a momentum of 100~GeV for the top quark, that
  in turn roughly corresponds to the peak value of the transverse momentum
  of the top quarks produced at the LHC. Thus, the available phase space
  for soft radiation at the LHC is similar to the case of the process considered
  here, so that it is reasonable to assume that our result gives an indication
  in favour of using the \MSB{} scheme for the total cross section
  without cuts at the LHC.
\item As soon as jet requirements are imposed on the final state,
  corrections of order $\Lambda_{\rm \sss QCD}$ arise. They have a
  leading behaviour proportional to $1/R$, where $R$ is the jet
  radius, for small $R$~\cite{Korchemsky:1994is,
    Dasgupta:2007wa}. These corrections are present irrespective of
  the top-mass scheme being used. They are however reduced if the
  efficiency of the cuts is increased, for example by increasing the
  jet radius, giving an indication in favour of the use of the \MSB{}
  scheme for the total cross section calculation also in the presence
  of cuts.  It should be stressed, however, that with a typical jet
  radius of 0.5 the behaviour of the perturbative expansion in the
  \MSB{} and Pole-mass scheme are very similar, with a rather small
  advantage of the first one over the latter.
\item The reconstructed-top mass, defined as the mass of the system
  comprising the $W$ and the $b$ jet, has the characteristic power
  correction due to jets, with the typical $1/R$ dependence. No benefit,
  i.e.~reduction of the power corrections, seems to be associated with the
  use of a short-distance mass. In particular, at large jet radii, when the
  jet renormalon becomes particularly small, in the pole mass scheme the
  linear renormalon coefficient is smaller. This observation is justified if one considers
  that, in the narrow-width limit, the production and decay processes
  factorize to all orders in the perturbative expansion, yielding a clean
  separation of radiation in production and decay. In this limit, the
  system of the top decay products is well defined, and its mass
  is exactly equal to the pole mass.
  Consistently with this observation, we have shown that, for very
  large jet radii, the linear renormalon coefficient for the
  reconstructed-top mass is quite small (if the observables is expressed in terms
  of the pole mass).
  One may then worry that, when reconstructing
  the top mass from the full final state, renormalons associated
  with soft emissions in production from the top and from the $\bar{b}$ quark
  may affect the reconstructed mass, since these soft emissions may enter the $b$-jet
  cone. By comparing the reconstructed mass obtained using only the top decay products
  to the one obtain using all final state particles,
  we have shown that these effects are in fact small.

  We should also add, however, that the benefit of using very large jet
  radii cannot be exploited at hadron colliders, since we expect
  other renormalon effects, due to soft-gluon radiation in production
  entering the jet cone. This problem can in principle be investigated with our
  approach, by applying it to the process of $t\bar{t}$ production in
  hadronic collisions.
\item We have considered, as a prototype for a leptonic observable relevant
  for top mass measurement, the average energy of the $W$ boson. We have
  found two interesting results:
  \begin{itemize}
  \item In the narrow-width limit, this observable has a linear renormalon,
    irrespective of the mass scheme being used for the top. This finding
    does not support the frequent claim that leptonic observables should be
    better behaved as far as non-perturbative QCD corrections are concerned.
    It also reminds us that, even if we wanted to measure the top-production
    cross section by triggering exclusively upon leptons, we may induce
    linear power corrections in the result that cannot be eliminated by
    going to the \MSB{} scheme.

    The presence of renormalons in leptonic observables seems to be in contrast
    with what is found in inclusive semileptonic decays of heavy flavours~\cite{Bigi:1994em,Beneke:1994bc}.
    We have however verified that there is no contradiction with this case.
    If the average value of the $W$ energy is computed in the top rest frame (which
    makes it fully analogous to a leptonic observable in $B$ decay) then no
    renormalon is present if the result is expressed in terms of the \MSB{} mass.
  \item For finite widths, if a short-distance mass is used, there is no
    linear renormalon. We verified this numerically, and furthermore we were
    also able to give a formal proof of this finding.  What this means in
    practice is that the perturbative expansion for this quantity will have
    factorial growth up to an order $n\approx 1+\log(m/\Gamma_t)$, that will
    stop for higher orders.  In practice, for realistic values of the width,
    this turns out to be a relatively large order. Thus, although in
    principle we cannot exclude a useful direct determination of the top
    short-distance mass from leptonic observables, it seems clear that
    finite-order calculations should be carried out at relatively high orders
    (up to the fourth or fifth order) in order to exploit it. Although it
    seems unlikely that results at these high orders may become available in
    the foreseeable future, perhaps it is not impossible to devise methods to
    estimate their leading renormalon contributions, still allowing a viable
    mass measurement (this assumning that the weaker sensitivity of leptonic
    obervables to the top-mass value does not prove to be too strong a
    limitation).
  \end{itemize}  
\end{itemize}
In this work we have made several simplifying assumptions.  These assumptions
were motivated by the fact that the computational technique is new, and we
wanted to make it as simple as possible.  Some of these restrictions may be
removed in future works. For example, we could consider hadronic collisions,
the full left-handed coupling for the $W$, the $W$ finite width and the
effects of a finite $b$ mass.  Although removing these limitations can lead
to interesting results, we should not forget that our calculation does not
exhaust all sources of non-perturbative effects that can affect the mass
measurement. As an obvious example, we should consider that confinement
effects are not present in our large-$b_0$ approximation, while, on the other
hand, it is not difficult to show that they may give rise to linear power
corrections. It is clear that theoretical problems of this sort should be
investigated by different means.


\part{Top-quark mass extraction with Monte Carlo event generators}
\newcommand\diffdiffRttdec{$   55$}
\newcommand\diffdiffRhvq{$   34$}
\newcommand\pyminushwRfour{$  830$}
\newcommand\pyminushwRsix{$ 1267$}
\newcommand\pyminushwdeltaRfoursix{$  437$}
\newcommand\Bfrommwbjbbfourl{     1.008}
\newcommand\Berrfrommwbjbbfourl{     0.002}
\newcommand\Bfrommwbjsmearbbfourl{     0.958}
\newcommand\Berrfrommwbjsmearbbfourl{     0.001}
\newcommand\Bfrommwbjttdec{     1.000}
\newcommand\Berrfrommwbjttdec{     0.002}
\newcommand\Bfrommwbjsmearttdec{     0.957}
\newcommand\Berrfrommwbjsmearttdec{     0.001}
\newcommand\Bfrommwbjhvq{     1.002}
\newcommand\Berrfrommwbjhvq{     0.002}
\newcommand\Bfrommwbjsmearhvq{     0.949}
\newcommand\Berrfrommwbjsmearhvq{     0.001}
\newcommand\diffttdecbbfourl{$ 140$}
\newcommand\diffhvqbbfourl{$-147$}
\newcommand\diffaverage{$ 140$}
\newcommand\varPDF{$   9$}
\newcommand\varscalemax{  86}
\newcommand\varscalemin{  53}
\newcommand\varscaleothers{$   7$}
\newcommand\deltaalphashvq{$  18$}
\newcommand\deltaalphasttdec{$ 108$}
\newcommand\deltaalphasbbfourl{$ 128$}
\newcommand\deltaalphasunsmear{$   8$}
\newcommand\pdferrorhvqsmear{$   5$}
\newcommand\pdferrorhvqnosmear{$   3$}
\newcommand\pyminushwhvq{$ 240$}
\newcommand\MECmPOWHEG{$  356$}
\newcommand\hwTSnosmear{$  48$}
\newcommand\hwTSnoTSnosmear{$ 152$}
\newcommand\hwTSsmearbbfourl{$ 130$}
\newcommand\hwTSsmearttdec{$ 132$}
\newcommand\hwTSsmearbbfourlttdec{$ 130$}

\newcommand\Bforptlep{   0.19}
\newcommand\pdferrorptlep{ 130}

\newcommand\BfromEbjbbfourl{   0.54}
\newcommand\BerrfromEbjbbfourl{   0.07}
\newcommand\BfromEbjttdec{   0.50}
\newcommand\BerrfromEbjttdec{   0.03}
\newcommand\BfromEbjhvq{   0.50}
\newcommand\BerrfromEbjhvq{   0.03}
\newcommand\diffEbjcml{    3.4}
\newcommand\diffEbjumc{    3.3}
\newcommand\Ebjbbfourlmhvq{$  460$}
\newcommand\Ebjbbfourlmhvqerr{$  100$}

\chapter{Introduction}
\label{sec:intro-topmass}
The main results presented in the following sections can be also found
in Ref.~\cite{Ravasio:2018lzi} and refer to the problem of the
determination of the theoretical uncertainty associated with the use
of Monte Carlo~(MC) event generators to infer the top-quark mass.

The question on how precisely we can measure the top mass at hadron
colliders is related to our understanding of QCD and collider physics.
In view of the large abundance of top-pair production at the LHC, it
is likely that precise measurements will be performed with very
different methods, and that comparing them will give us confidence in
our ability to handle hadron-collider physics problems.

Top-mass measurements are generally performed by fitting $\mt$-dependent
kinematic distributions to MC predictions. The most precise ones, generally
called \emph{direct measurements}, rely upon the full or partial
reconstruction of the system of the top-decay products.  The ATLAS and CMS
measurements of Refs.~\cite{Aaboud:2016igd} and~\cite{Khachatryan:2015hba},
yielding the value $172.84 \pm 0.34~{\rm (stat)} \pm 0.61$~(syst)~GeV and
$\mt=172.44 \pm 0.13~{\rm (stat)} \pm 0.47$~(syst)~GeV respectively, fall
into this broad category.

The MC event generators employed for the \emph{direct
  measurements} make use of the pole mass that is, as we have seen,
affected by non perturbative corrections of order $\Lambda_{\rm QCD}$
due to the presence of infrared renormalons.

The theoretical problems raised upon the top-quark mass measurement
issues have induced several theorists to study and propose alternative
methods.  The total cross section for $t\bar{t}$ production is
sensitive to the top mass, and has been computed including NNLO QCD
corrections~\cite{Czakon:2013goa}, that have been recently combined
with NLO electroweak ones~\cite{Czakon:2017wor}. These computations can
be employed to extract a top mass value~\cite{Khachatryan:2016mqs,
  Aad:2014kva, Langenfeld:2009wd}. If it is computed in the \MSB{}
scheme, it has the advantage of being free from the pole-mass
renormalons.

In Refs.~\cite{Alioli:2013mxa, Bevilacqua:2017ipv} observables related
to the $t{\bar t}+{\rm jet}$ kinematics are considered in NWA and with
full off-shell matrix elements in the dilepton channel, respectively.
The authors of Ref.~\cite{Kawabataa:2014osa} presented a method based
upon the charged-lepton energy spectrum, that is not sensitive to top
production kinematics, but only to top decay, arguing that, since this
has been computed at NNLO accuracy~\cite{Gao:2012ja,
  Brucherseifer:2013iv}, a very accurate measurement may be achieved.
Some authors have advocated the use of boosted top jets (see
Ref.~\cite{Hoang:2017kmk} and references therein).  In
Ref.~\cite{Agashe:2016bok}, the authors make use of the \bjet{} energy
peak position, that is claimed to have a reduced sensitivity to
production dynamics.  In Ref.~\cite{Frixione:2014ala}, the use of
lowest Mellin moments of lepton kinematic distributions is discussed.
In the leptonic channel, it is also possible to use distributions
based on the ``stransverse'' mass variable~\cite{Sirunyan:2017idq},
which generalizes the concept of transverse mass for a system with two
identical decay branches~\cite{Lester:1999tx, Barr:2009jv}.  Some of
these methods have in fact been exploited~\cite{CMS-PAS-TOP-13-006,
  CMS-PAS-TOP-15-002, Aad:2015waa, Sirunyan:2017idq, Aaboud:2017ujq}
to yield alternative determinations of $\mt$.

It turns out, however, that the direct methods yield smaller errors at
the moment, and it is likely that alternative methods, when reaching
the same precision level, will face similar theoretical problems.
Because of this, and given the fact that recent
studies~\cite{Beneke:2016cbu, Hoang:2017btd} have shown that the
renormalon ambiguity in the top-mass definition is not as large as
previously anticipated, being in fact well below the current
experimental error.\footnote{In fact, values in this range were
  obtained much earlier in Refs.~\cite{Pineda:2001zq, Bali:2013pla},
  mostly in a bottom physics context, but since the renormalon
  ambiguity does not depend upon the heavy quark mass, they also apply
  to top.} it is still worthwhile to employ \emph{direct measurements}
and to try to implement more and more accurate MC event
generators, to avoid biased measurements.
  
Recently, many efforts have been done in order to implement NLO+PS
generators capable of handling a decayed coloured resonance, like the
top quark in the contest of the \POWHEGBOX~\cite{Jezo:2015aia} and
{\tt MadGraph5\_aMC@NLO}~\cite{Frederix:2016rdc}. In these references
two alternatives of the standard Frixione-Kunszt-Signer subtraction
method~\cite{Frixione:1995ms} are discussed taking as example the case
of single top production. In Ref.~\cite{Hoche:2018ouj} the same
problem, applied to the process of top-pair production, is discussed
for the Catani-Seymour subtraction method~\cite{Catani:1996vz}.

In this work, we exploit the availability of the new \POWHEGBOX{}
generators for top-pair production, in order to perform a
theoretical study of uncertainties in the top-mass determination. In
particular, we are in a position to assess whether NLO corrections
in top decay and finite width effects, non-resonant contributions
and interference of radiation generated in production and decay can
lead to sizeable corrections to the extracted value of the top mass.
Since the old
\hvq{} generator~\cite{Frixione:2007nw}, that implements NLO
corrections only in production, is widely used by the experimental
collaborations in top-mass analyses, we are particularly interested in
comparing it with the new generators, and in assessing to what extent
it is compatible with them. We will consider variations in the scales,
parton distribution functions~(PDFs) and the jet radius parameter to
better assess the level of compatibility of the different generators.

We are especially interested in effects that can be important in the
top-mass determination performed in direct measurements. Thus, the
main focus of our work is upon the mass of a reconstructed top, that
we define as a system comprising a hard lepton, a hard neutrino and a
hard $b$ jet. We will assume that we have access to the particle truth
level, i.e.~that we can also access the flavour of the $b$ jet, and
the neutrino momentum and flavour. We are first of all interested in
understanding to what extent the mass peak of the reconstructed top
depends upon the chosen NLO+PS generator. This would be evidence that
the new features introduced in the most recent generators are
mandatory for an accurate mass extraction.

We will also consider the inclusion of detector effects in the form of
a smearing function applied to our results. Although this procedure is
quite crude, it gives a rough indication of whether the overall
description of the process, also outside of the reconstructed
resonance peak, affects the measurement.

Besides studying different NLO+PS generators, we have also attempted
to give a first assessment of ambiguities associated with shower and
non-perturbative effects, by interfacing our NLO+PS generators to two
shower MC programs:
\PythiaEightPtwo{}~\cite{Sjostrand:2014zea} and
\HerwigSevenPone{}~\cite{Bahr:2008pv, Bellm:2015jjp}. Our work focuses
upon NLO+PS and shower matching.  We thus did not consider further
variations of parameters and options within the same parton shower~(PS),
nor variations on the observables aimed at reducing the dependence
upon those.\footnote{An interesting example of work along this
  direction can be found in Refs.~\cite{Wicke:2008iz}
  and~\cite{Sjostrand:2013cya}, where the impact of the colour
  reconnection model on top-mass measurement is analysed.  In
  Ref.~\cite{Andreassen:2017ugs}, a study is performed to determine
  whether the use of jet-grooming techniques in top-mass measurement
  can reduce the MC tune dependence.}

We have also considered two alternative proposals for top-mass
measurements: the position of the peak in the $b$-jet
energy~\cite{Agashe:2016bok} and the leptonic observables of
Ref.~\cite{Frixione:2014ala}.  The first proposal is an example of a
hadronic observable that should be relatively insensitive to the
production mechanism, but may be strongly affected by NLO corrections
in decay.  The second proposal is an example of observables that
depend only upon the lepton kinematics, and that also depend upon
production dynamics, thus stronger sensitivity to scale variations and
PDFs may be expected. It is also generally assumed that leptonic
observables should be insensitive to the $b$-jet modelling.  One should
remember, however, that jet dynamics affects lepton momenta via recoil
effects, so it is interesting to study whether there is any ground to
this assumption.

The study presented in this work was triggered by the availability of new
NLO+PS generators describing top decay with increasing accuracy.\footnote{A
  fixed order study concerning the impact of the top-decay implementation on
  top mass determinations can be found in Ref.~\cite{Heinrich:2017bqp}: NLO
  QCD predictions for the $W^+W^-b\bar{b}$ process are compared to those
  obtained using the narrow width approximation, where the process of
  $t\bar{t}$ production is described at NLO QCD and the decay of the top
  quarks is implemented at LO, at NLO or by a PS.} As such, its initial aim
was to determine whether and to what extent these new generators, and the
associated new effects that they implement, may impact present top-mass
measurements.  As we will see, had we limited ourselves to the study of the
NLO+PS generators interfaced to \PythiaEightPtwo{}, we would have found a
fairly consistent picture and a rather simple answer to this question.

Since another modern shower generator that can be interfaced to our
NLO+PS calculation is available, namely \HerwigSevenPone{}, we have
developed an appropriate interface to it, and have also carried out
our study using it as our shower model.  Our results with
\HerwigSevenPone{} turn out to be quite different from the
\PythiaEightPtwo{} ones, to the point of drastically altering the
conclusions of our study. In fact, variations in the extracted top
mass values due to switching between \PythiaEightPtwo{} and
\HerwigSevenPone{} prevail over all variations that can be obtained
within \PythiaEightPtwo{} by switching among different NLO+PS
generators, or by varying scales and matching parameters within
them. Moreover, the comparison of the various NLO+PS generators, when
using \HerwigSevenPone{}, does not display the same degree of
consistency that we find within \PythiaEightPtwo{}. If, as it seems,
the differences found between \PythiaEightPtwo{} and
\HerwigSevenPone{} are due to the different shower models (the former
being a dipole shower, and the latter an angular-ordered one), the
very minimal message that can be drawn from our work is that, in order
to assess a meaningful theoretical error in top-mass measurements, the
use of different shower models, associated with different NLO+PS
generators, is mandatory.

This part is organized as follows.  In Chap.~\ref{sec:generators} we
briefly review the features of the POWHEG generators.  We also briefly
discuss the interfaces to the parton-shower programs
\PythiaEightPtwo{} and \HerwigSevenPone{}. More details about the
NLO+PS matching are discussed in appendix~\ref{sec:interfacePS}.
In Chap.~\ref{sec:pheno}, we detail the setup employed for the
phenomenological studies presented in the subsequent sections and we describe
how we relate the computed value of our observables to the corresponding
value of the top mass that would be extracted in a measurement.
In Chap.~\ref{sec:reconstructedpeak}, we perform a generic study of the
differences of our generators focusing upon the mass distribution of
the $W\, b$-jet system. The aim of this section is to show how this
distribution is affected by the different components of the generators
by examining results at the Born level, after the inclusion of NLO
corrections, after the PS, and at the hadron level.
In Chap.~\ref{sec:mwbj} we consider as our top-mass sensitive observable the
peak position in the mass distribution of the $W\, b$-jet system.  We study
its dependence upon the NLO+PS generator being used, the scale choices, the
PDFs, the value of $\as$ and the jet radius parameter.  Furthermore, we
present and compare results obtained with the two shower MC generators
\PythiaEightPtwo{} and \HerwigSevenPone{}.
We repeat these studies for the peak of the $b$-jet energy
spectrum~\cite{Agashe:2016bok} in Chap.~\ref{sec:Ebjet}, and for the
leptonic observables~\cite{Frixione:2014ala} in Chap.~\ref{sec:LepObs}.
In Chap.~\ref{sec:Summary} we summarize our results, and in
Chap.~\ref{sec:Conc} we present our conclusions. In the appendices we
give some technical details.
 
\chapter{\POWHEGBOX{} generators for top-pair production}
\label{sec:generators}
In this section we describe the features of the NLO generators
implemented in the \POWHEGBOX{} framework that describe the process of
top-pair production and decay, i.e.~the \hvq{}, the \ttNLOdec{} and
the \bbfourl{} generators\footnote{The \hvq{} and \ttbnlodec{}
  generators can be found under the {\tt User-Processes-V2} directory
  of the \VTWO{} repository in the {\tt hvq} and {\tt ttb\_NLO\_dec}
  directories, respectively.  The \bbfourl{} generator can be found
  under the {\tt User-Processes-RES/b\_bbar\_4l} directory of the
  \RES{} code. Detailed instructions are found at
  \url{powhegbox.mib.infn.it}.}.  The POWHEG method is described in
\writeApp\ref{sec:powheg}.

The first top-pair production generator implemented in the \POWHEGBOX{} and the
most widely used, up to now, is the \hvq{} program~\cite{Frixione:2007nw}.
It describes the process of production of $t\bar{t}$ pairs at NLO.  The top
decay is introduced in an approximate way according to the method presented
in Ref.~\cite{Frixione:2007zp} that makes it possible to take into account
approximatively off-shell and spin correlations effects.  Radiation off the
top-quark decay-products is fully handled by the PS.  The ones that we
consider, \PythiaEightPtwo{} and \HerwigSevenPone{}, implement internally
matrix-element corrections~(MEC) for top decay.  Furthermore,
\HerwigSevenPone{} also optionally includes a POWHEG-style hardest-radiation
generation.  Thus, the accuracy in the description of top decay is, for our
purposes, equivalent to the NLO level.

The second in time generator implemented is the \ttNLOdec{}
code~\cite{Campbell:2014kua}.  NLO corrections and spin correlations
are implemented exactly using the narrow-width method, thus
interference of radiation generated in production and decay is not
included.  Off-shell effects are implemented via a reweighting method,
such that the LO cross section includes them exactly.  It also
contains contributions of associated top-quark and $W$-boson
production at LO.

In \ttbnlodec{} the POWHEG method was adapted to deal with radiation in
resonance decays. As it is discussed in \writeApp\ref{app:powhegMethod},
this offers the possibility to modify the standard POWHEG single-radiation
approach.  For the case of $t\bar{t}$ production followed by leptonic top
decay, the ``resonance aware'' POWHEG formalism enables us to generate events
that contains up to three emissions: from the production process and from the
$b$ and $\bar{b}$ quarks arising after the $t$ and $\bar{t}$ decay.  This
represents an improvement with respect to the single-emission formalism,
where only the hardest emission would have been kept. Indeed, for the
$t\bar{t}$ production process, the hardest emission is most likely the one
from the production process. This would leave to the PS the task to generate
the hardest emission off the top quark, limiting the effective accuracy
employed for the description of the top decay.  On the other hand, the
multi-emission formalism allows to overcome this problem since the hardest
emission of each decayed resonance is already included in the LH event.

A general procedure for dealing with decayed emitting resonances has
been implemented in a fully general and automatic way in a new version
of the \POWHEGBOX{} code, the \RES~\cite{Jezo:2015aia}. This framework
allows for the treatment of off-shell effects, non-resonant
subprocesses including full interference, and for the treatment of
interference between radiation generated in production and from the
resonances decay. Further details are given in
\writeApp\ref{sec:powheg}.  The first generator implemented in \RES{}
is the \bbfourl{} code~\cite{Jezo:2016ujg} , that describes the
process $pp\to b\bar{b}\,\fourl$, that is dominated by top-pair
production with leptonic decay, including all QCD NLO corrections in
the 4-flavour scheme, i.e.~accounting for finite $b$-mass effects.
Furthermore, double-top, single-top and non-resonant\footnote{By
  non-resonant we mean processes that do not contain an intermediate
  top quark, e.g.~$pp \to b\,\bar{b} \,Z\to b\,\bar{b} \,W^+\, W^- \to
  b\, \bar{b}\,\fourl $.}  diagrams are all included with full
spin-correlation effects, radiation in production and decays, and
their interference.

The interfaces between the \POWHEGBOX{} generators and standard shower
MC programs that we have employed, i.e.~the
\PythiaEightPtwo{} and \HerwigSevenPone{}, are detailed in
\writeApp{}\ref{sec:interfacePS}.

\chapter{Phenomenological analysis setup}
\label{sec:pheno}
We simulate the process $p\,p \to b\, \bar{b}\,\fourl$, which is
available in all the three NLO generators we are investigating.  It is
dominated by top-pair production, with a smaller contribution by $Wt$
topologies.

The \hvq{} and \ttbnlodec{} generators employ the pole-mass scheme for
the renormalization of the top mass $m_t$, while in \bbfourl{} the
complex mass scheme~\cite{Jezo:2016ujg} is adopted, with the complex
mass defined as $\sqrt{\mt^2-i\,\mt \,\Gamma_t}$.

The center-of-mass energy of our simulated sample is $\sqrt{s}=8$~TeV.
The parton distribution function~(PDF) used is the central member of
the {\tt MSTW2008nlo68cl} set~\cite{Martin:2009iq}.  PDF variations to
assess the theoretical uncertainty are also performed.  Using the
internal reweighting facility of the \POWHEGBOX{}, we produced predictions
for the central member of the following PDF sets:\\ \\
\begin{minipage}{0.495\textwidth}
\begin{itemize}
\item {\tt PDF4LHC15\_nlo\_30\_pdfas}~\cite{Butterworth:2015oua}\,,
\item {\tt CT14nlo}~\cite{Dulat:2015mca}\,,
\end{itemize}
\end{minipage}\begin{minipage}{0.495\textwidth}
\begin{itemize}
\item {\tt MMHT2014nlo68cl}~\cite{Harland-Lang:2014zoa}\,,
\item {\tt NNPDF30\_nlo\_as\_0118}~\cite{Ball:2014uwa}\,.
\end{itemize}
\end{minipage}
\\ \\ We also generated a sample using the central parton-distribution
function of the \\{\tt PDF4LHC15\_nlo\_30\_pdfas} set, and, by
reweighting, all its members, within the \hvq{} generator.  In this
case, our error is given by the sum in quadrature of all deviations.
We find that the variation band obtained in this way contains the
central value results for the different PDF sets that we have
considered. It thus makes sense to use this procedure for the estimate
of PDF uncertainties.  Since reweighting for the 30 members of the set
in the \bbfourl{} or in the \ttbnlodec{} case is quite time consuming
and since the dependence on the PDF is mostly due to the
implementation of the production processes, and all our generators
describe it at NLO accuracy, we thus assume that the PDF uncertainties
computed in the \hvq{} case are also valid for the \bbfourl{} and
\ttbnlodec{} cases. We have indeed checked that by reweighting to
several PDF sets we get very similar variations for all generators.

In the \POWHEGBOX{}, the scale used to generate the real emissions is the
transverse momentum of the radiated parton with respect to the emitter.
Variations of this scheme can lead to different radiation pattern around the
$b$ jet, that can in turn have a non-negligible effect on the reconstructed
mass and thus must be evaluated to assess the theoretical uncertainty.
Since, at the moment, the \POWHEGBOX{} does not offer the possibility to vary
the definition of the scale of the emission, the simplest way at our disposal
for studying the sensitivity to the intensity of radiation from the $b$ quark
is by varying the value of $\as(\mZ)$. To this end we use the {\tt
  NNPDF30\_nlo\_as115} and {\tt NNPDF30\_nlo\_as121} sets, where
$\as(\mZ)$=0.115 and $\as(\mZ)$=0.121, respectively.  In the reweighting
procedure, only the inclusive \POWHEG{} cross section is recomputed.  The
Sudakov form factor is not recomputed, so that the radiated partons retain
the same kinematics.  For this reason, it can not be employed to evaluate the
$\as$ dependence since it changes the Sudakov form factor.  For this reason,
we generated two dedicated samples.

The typical scale of radiation in top decay can be estimated to be
30~GeV, that corresponds to one-half of the typical $b$ energy in the
top rest frame.  The ratio of the $\as(\mu)$ values for the two PDF
sets considered is 1.052 at $\mu=\mZ$ and it becomes 1.06 at
$\mu=30$~GeV.  On the other hand, a scale variation of a factor of two
above and below $30$~GeV yields a variation in $\as$ of about 26\%.
This can be taken as a rough indication that a standard scale
variation would yield to a variation in the peak position that is
$26/6\approx 4$ times larger than the one obtained by varying $\as$.

The central renormalization and factorization scale ($\muR$ and
$\muF$) is given by the quantity $\mu$, defined, following
Ref.~\cite{Jezo:2016ujg}, as the geometric average of the transverse
masses of the top and anti-top
\begin{equation}
\label{eq:centralscale}  
\mu= \sqrt[4]{\left(E^2_t -p_{z,t}^2\right)\left(E^2_{\bar{t}}
  -p_{z,\bar{t}}^2\right)}\,,
\end{equation}
where the top and anti-top energies $E_{t/\bar{t}}$ and longitudinal
momenta $p_{z,t/\bar{t}}$ are evaluated at the underlying-Born level.
In the \bbfourl{} case, there is a tiny component of the cross section
given by the topology
\begin{equation}
pp\to Z g \to (W^+ \to e^+ \nu_e) (W^- \to \mu^- \bar{\nu}_\mu) (g \to
b \bar{b}).
\end{equation}
In this case $\mu$ is taken as
\begin{equation}
  \label{eq:centralscaleZ}
  \mu= \frac{\sqrt{p_{\sss Z}^2}}{2}\,,
\end{equation}
where $p_{\sss Z}=p_{\mu^-}+p_{\bar{\nu}_\mu} + p_{e^+} +p_{\nu_{\sss
    e}}$.

We studied the dependence of our results on $\muR$ and $\muF$, that
gives an indication of the size of higher-orders corrections.  We
varied $\muR$ and $\muF$ around the central scale $\mu$ defined in
eqs.~\eqref{eq:centralscale} and~(\ref{eq:centralscaleZ})
\begin{equation}
  \muR = \KR\, \mu \, , \quad \muF = \KF \, \mu \, ,
\end{equation}
where $(\KR,\KF)$ are varied over the following combinations
\beq \label{eq:scalechoices} \bigg\{ (1,1), \, (2,2),\, \left(
\frac{1}{2}, \frac{1}{2} \right), \, (1,2), \, \left( 1, \frac{1}{2}
\right), \, (2,1), \, \left( \frac{1}{2},1 \right) \bigg\}.  \eeq The
scale variations have been performed using the reweighting technique.

The parameter {\tt hdamp} controls the separation of remnants, see
eq.~(\ref{eq:hdamp}), in the production of $t\bar{t}$ pairs with large
transverse momentum\footnote{We have only ISR remnants, this separation is
  not performed for radiation in decay.}.  We set it to the value of the top
mass.\footnote{See Appendix~\ref{app:hdamp} for a discussion
  concerning the value of {\tt hdamp} employed.}

The central predictions, together with the scale and PDF set
variations, for a total of 12 weights, have been obtained for each
generator under study for three top mass values: 169.5~GeV, 172.5~GeV,
175.5~GeV.  The $\as$ variations and the \hvq{} sample containing the
30 members of the {\tt PDF4LHC15\_nlo\_30\_pdfas} set have been
produced only for $\mt=172.5$~GeV.  The number of events for each
generated sample, together with an indicative computational time, are
reported in Tab.~\ref{tab:samples}.
\begin{table*}[tb]
  \centering
  \resizebox{\textwidth}{!}
{ \begin{tabular}{l|c|c||c|c||c|c||c|c||c|c|} \cline{2-11}
    &\multicolumn{10}{|c|}{Generated samples} \\ \cline{2-11}
    &\multicolumn{6}{|c||}{$m_t$~[GeV]} &
    \multicolumn{4}{|c|}{$\as(\mZ)$} \\ \cline{2-11} &
    \multicolumn{2}{|c||}{$172.5$} & \multicolumn{2}{|c||}{$169.5$} &
    \multicolumn{2}{|c||}{$175.5$} & \multicolumn{2}{|c||}{$0.115$} &
    \multicolumn{2}{|c|}{$0.121$} \\ \cline{2-11} & \phantom{\Big|}\#
    events & time &\# events & time & \# events & time & \# events &
    time & \# events & time \\ \cline{1-11}
    \multicolumn{1}{|c|}{\hvq{}} & \phantom{\Big|}12~M & 10~h & 3M &
    2.5~h & 3~M & 2.5~h & 12~M & 9~h & 12~M & 9~h \\ \cline{1-11}
    \multicolumn{1}{|c|}{\ttbnlodec{}} & \phantom{\Big|}12~M & 46~d &
    3M & 11.5~d & 3~M & 11.5~d & 12~M & 25~d & 12~M & 25~d
    \\ \cline{1-11} \multicolumn{1}{|c|}{\bbfourl{}} &
    \phantom{\Big|}20~M & 4600~d & 1.7M & 390~d & 1.7~M & 390~d & 3~M
    & 64~d & 3~M & 64~d \\ \cline{1-11}
  \end{tabular}
  }
  \caption{Number of events and total CPU time of the generated
    samples.  The samples used for the $\as$ variations were obtained
    in a relatively smaller time, since in this case only the central
    weight was computed. This leads to a difference that can be
    sizeable, depending upon the complexity of the virtual
    corrections.}
\label{tab:samples}
\end{table*}

\section{Physics objects}
\label{sec:physicsObjects}
In our analyses, we set the $B$ hadrons stable, in order to simplify
the definitions of $b$ jets.  Jets are reconstructed using the
Fastjet~\cite{Cacciari:2011ma} implementation of the anti-$k_{\rm\sss
  T}$ algorithm~\cite{Cacciari:2008gp} with $R=0.5$.  We denote as
$B$~(${\bar B}$) the hardest (i.e.~largest \pT{}) $b$~($\bar{b}$)
flavoured hadron. The $b$~($\bar{b}$) jet is the jet that contains the
hardest $B$~($\bar{B}$).\footnote{Note that this notation is the
  opposite of what is commonly adopted for $B$ mesons, where $B$
  refers to the meson containing the $\bar{b}$ quark.}  It will be
indicated as $j_B$~($j_{\bar{B}}$).  The hardest $e^+$~($\mu^-$) and
the hardest $\nu_e$~($\bar{\nu}_{\mu}$) are paired to reconstruct the
$W^+$~($W^-$).  The reconstructed top~(anti-top) quark is identified
with the corresponding $W^+j_B$ ($W^-j_{\bar{B}}$) pair.  In the
following we will refer to the mass of this system as \mwbj{}.

We require the presence of a $b$ jet and a separated $\bar{b}$ jet,
that satisfy
\begin{equation}
 \pT>30\mbox{~GeV}\,, \qquad |\eta|<2.5\,.
\end{equation}
These cuts suppress the $Wt$ topologies, that are not included by the
\hvq{} generator and included only at LO by \ttbnlodec{}.  The hardest
$e^+$ and the hardest $\mu^-$ must satisfy
\begin{equation}
 \pT>20\mbox{~GeV}\,, \qquad |\eta|<2.4\,.
\end{equation}

\section{Methodology}
\label{sec:methodology}
In the following we will focus upon three observables, the peak of the
reconstructed-top mass distribution $\mwbj$, the peak of the \bjet{}
energy spectrum $\Ebj$~\cite{Agashe:2016bok} and the average value of
the leptonic observable of Ref.~\cite{Frixione:2014ala}, and we will
examine several sources of theoretical uncertainty in the top-mass
extraction.

Our observables are sensitive to the value of the top-mass and they
bear a simple relation with it
\begin{equation}
  O = O_c + B \(\mt-\mtc\) + {\cal O}\left(\(\mt-\mtc\)^2\right),
  \label{eq:linearfitfunc}
\end{equation}
where $\mt$ is the input mass parameter in the generator, and
$\mtc=172.5$~GeV is our reference central value for the top mass.  The
parameters $O_c$ and $B$ can be extracted using a MC generator and in general
depend on the generator and the setups employed.  Given an experimental
result for the observable $O$, $O_{\rm exp}$, the extracted top-mass value is
\begin{equation}
  \mt = \mtc+\frac{O_{\rm exp}-O_{c}}{B}\,.
\end{equation}
Changing the generator (or its setup), leads to different parameters $O_{\rm
  c}'$ and $B'$, and thus to a different value for the top-mass $m'_t$.  The
difference between the two extracted masses reads
\begin{equation}
  {m'_t-\mt} = \frac{O_{\rm c}-O_{\rm c}'}{B} + \(O_{\rm exp}-O_{\rm
    c}'\)\frac{B-B'}{BB'}\,,
\end{equation}
where the second term is parametrically smaller if we assume that at
least one of them yields a $\mt$ value sufficiently close to
$\mtc$. If this is the case, we are allowed to write
\begin{equation}
  \label{eq:delta_mt}
  m'_t-\mt \approx \frac{O_{\rm c}-O_{\rm c}'}{B}\,.
\end{equation}
In practice, in the following, we will compute the $B$ parameter using
the \hvq{} generator, that is the one that requires less computation
time.  We also checked that using the other generators for this
purpose yields results that differ by at most 10\%{}, confirming the
validity of our approximation.

\chapter{Anatomy of the reconstructed-top mass distribution}
\label{sec:reconstructedpeak}

Before comparing the predictions of the several generators employed,
we investigate the impact of the several ingredients of a typical
NLO+PS on the kinematic distribution of the reconstructed-top mass
\mwbj{}.  Despite the fact this is a simplified observable, it can be
considered a proxy of all top-mass sensitive observables that rely
upon the full or partial reconstruction of the top quark.  As we will
see in Chap.~\ref{sec:mwbj}, where the extraction of the peak of the
distribution is discussed, our very crude approach allows us to
concentrate more on theoretical issues rather than experimental ones.

On the perturbative side, we can compare the predictions obtained with
our three NLO generators, that describe the top decay with different
level of accuracy and assess the impact of the PS employed.
On the non-perturbative side, we illustrate the effect of including
hadronization and underlying event in the simulation.

\section{Les Houches event level comparison of the generators}
We begin by comparing the three generators at the Les Houches
event~(LHE) level, i.e.~before applying the PS. These
events contain only the POWHEG hardest emission(s).

\begin{figure}[tb!]
\centering
\includegraphics[width=\wfigsing]{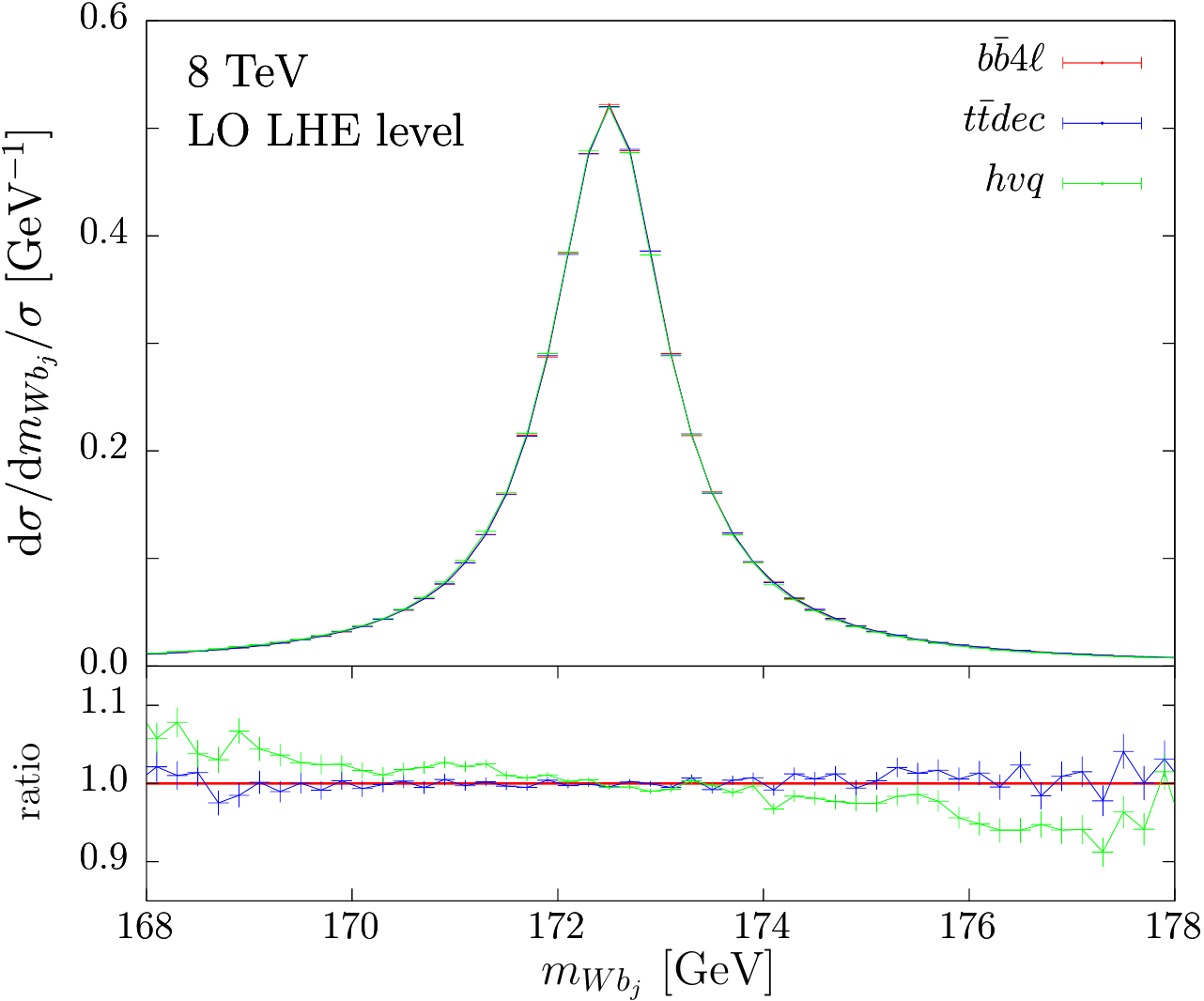}
   \caption{${d\sigma}/{d \mwbj}$ distribution at LO obtained with
     \bbfourl{}~(red), \ttNLOdec{}~(blue) and \hvq{}~(green),
     normalized to 1 in the displayed range. In the bottom panel the
     ratio with the \bbfourl{} prediction is shown.}
   \label{fig:allLO}
\end{figure}

We first compare the LO distributions, i.e.~without the inclusion of
radiative corrections.  This is illustrated in Fig.~\ref{fig:allLO}, where we
see that a non-negligible (although not dramatic) difference in shape is
present also at the LO level between the \hvq{} and the other two generators.
This difference between the \hvq{} and the other two generators is due
to the different description of how off-shell effects.  The \bbfourl{}
and \ttNLOdec{} generators are guaranteed to yield the correct
top-virtuality distribution at the NLO and LO level, respectively.
This is not the case for the \hvq{} generator, where the resonance
structure is recovered by a reweighting procedure that does not
guarantee LO accuracy.

\begin{figure}[tb!]
\centering
\includegraphics[width=\wfigdoub]{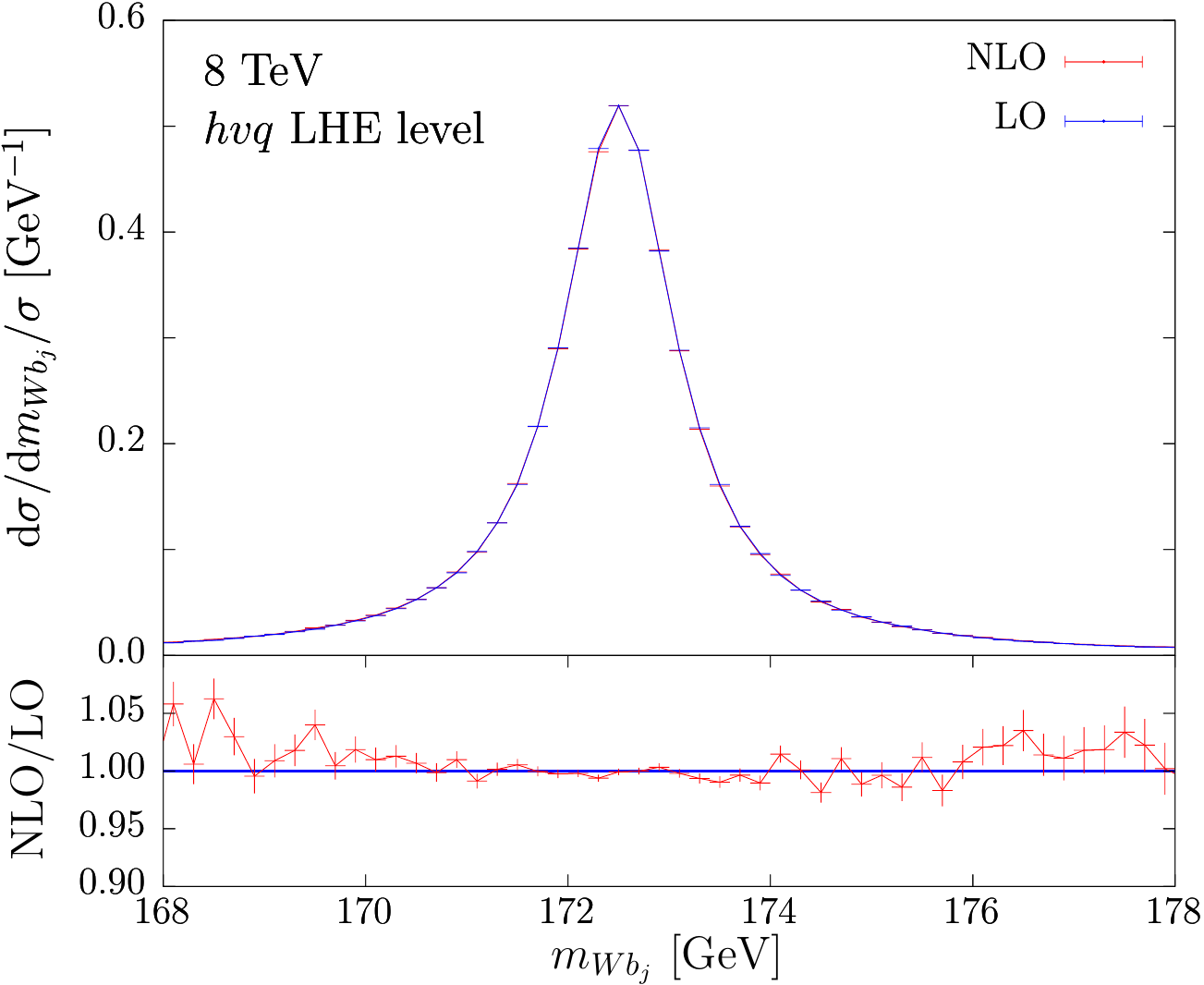}
\includegraphics[width=\wfigdoub]{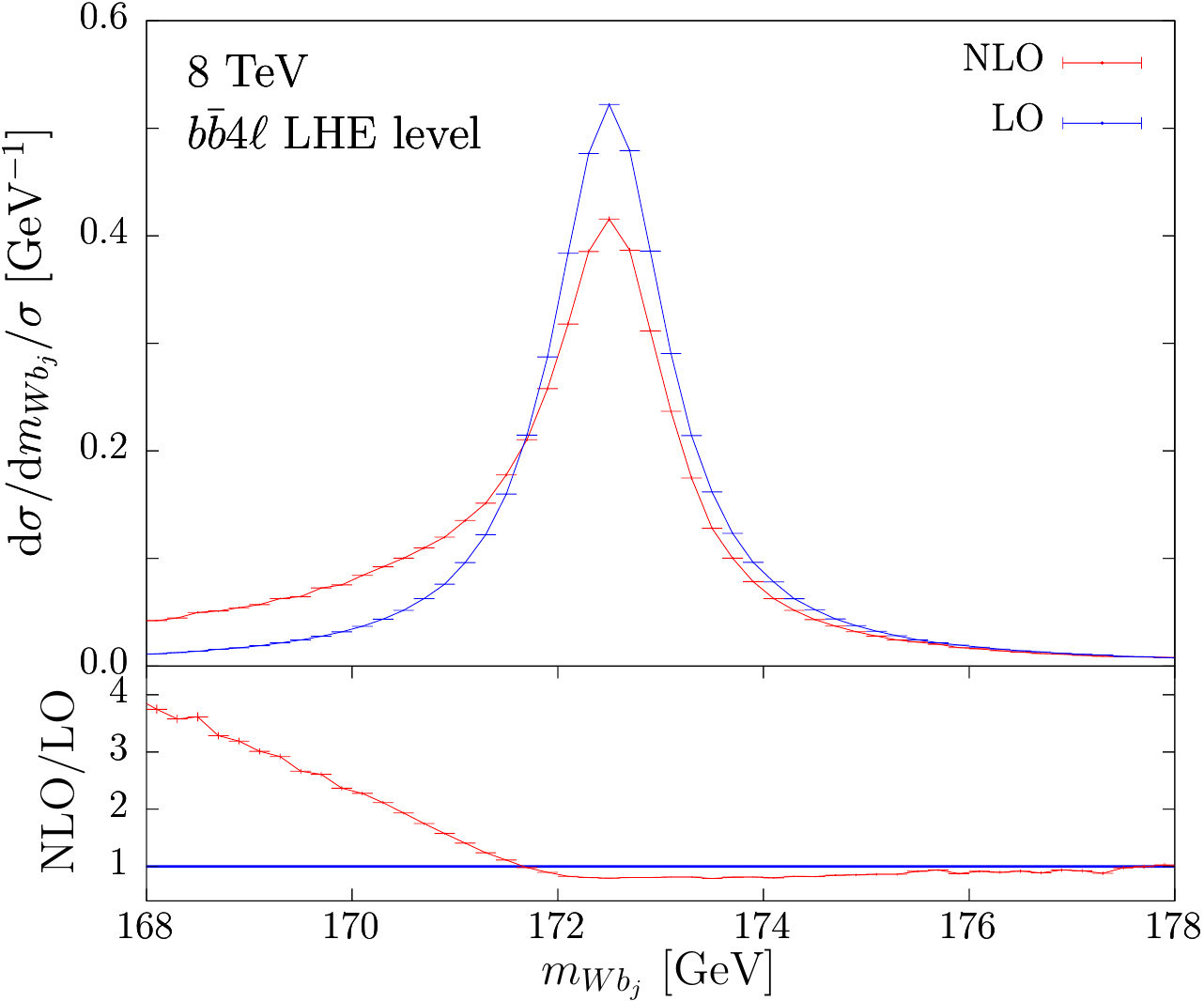}
\caption{${d\sigma}/{d \mwbj}$ distribution at LO~(blue) and at
  NLO~(red) obtained with the \hvq{}~(left) and \bbfourl{}~(right)
  generator, normalized to 1 in the displayed range. In the bottom
  panel the ratio with the LO prediction is shown. }
\label{fig:hvqbb4lLHE}
\end{figure}

\begin{figure}[tb!]
\centering
\includegraphics[width=\wfigsing]{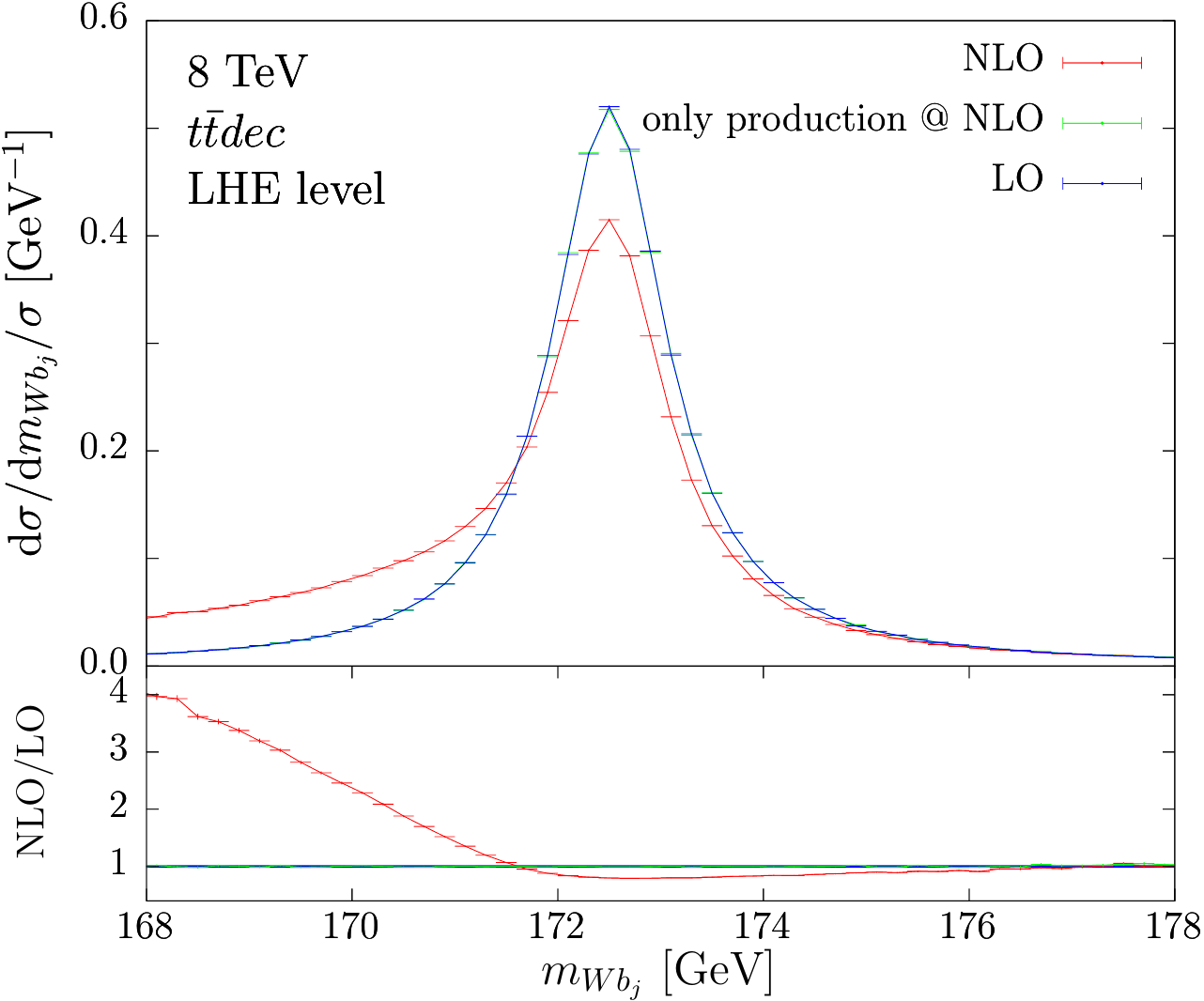}
   \caption{${d\sigma}/{d \mwbj}$ distribution with NLO accuracy in
     production and decay~(red), only in production~(green) and with
     LO accuracy~(blue) obtained with the \ttNLOdec{} generator,
     normalized to 1 in the displayed range. In the bottom panel the
     ratio with the LO prediction is shown.}
   \label{fig:ttdecLHE}
\end{figure}

We now investigate the impact of the POWHEG hardest emission(s) on the
LO distributions.  In Fig.~\ref{fig:hvqbb4lLHE} we compare \mwbj{},
normalized to 1 in the displayed range, at LO and NLO accuracy using
the \hvq{}~(left panel) and the \bbfourl{}~(right panel) generators.
Since the \hvq{} generator implements NLO corrections only in the
production process, the \mwbj{} distribution is not significantly
modified when moving from the LO to the NLO prediction. On the other
hand, the \bbfourl{} generator includes radiative corrections also to
the decay process. Thus, when comparing the NLO and the LO curves,
large differences below the peak region, that can be easily
interpreted as due to radiation outside the \bjet{} cone, arise.
The \ttNLOdec{} generator enables us to specify whether NLO accuracy
is required both in production and decay (default behaviour), or just
in production (by using the \pwgopt{nlowhich 1} option). A graphical
display of both options is given in Fig.~\ref{fig:ttdecLHE}.  As we
have already seen, NLO corrections in production leads to a roughly
constant $K$-factor, while radiation from top decay affects the shape
of the distribution.

\section{Shower effects}
We now investigate how the PS, that provides radiations beyond the
hardest one, affect the shape of the \mwbj{} distributions.

\begin{figure}[bt!]
\centering
\includegraphics[width=\wfigdoub]{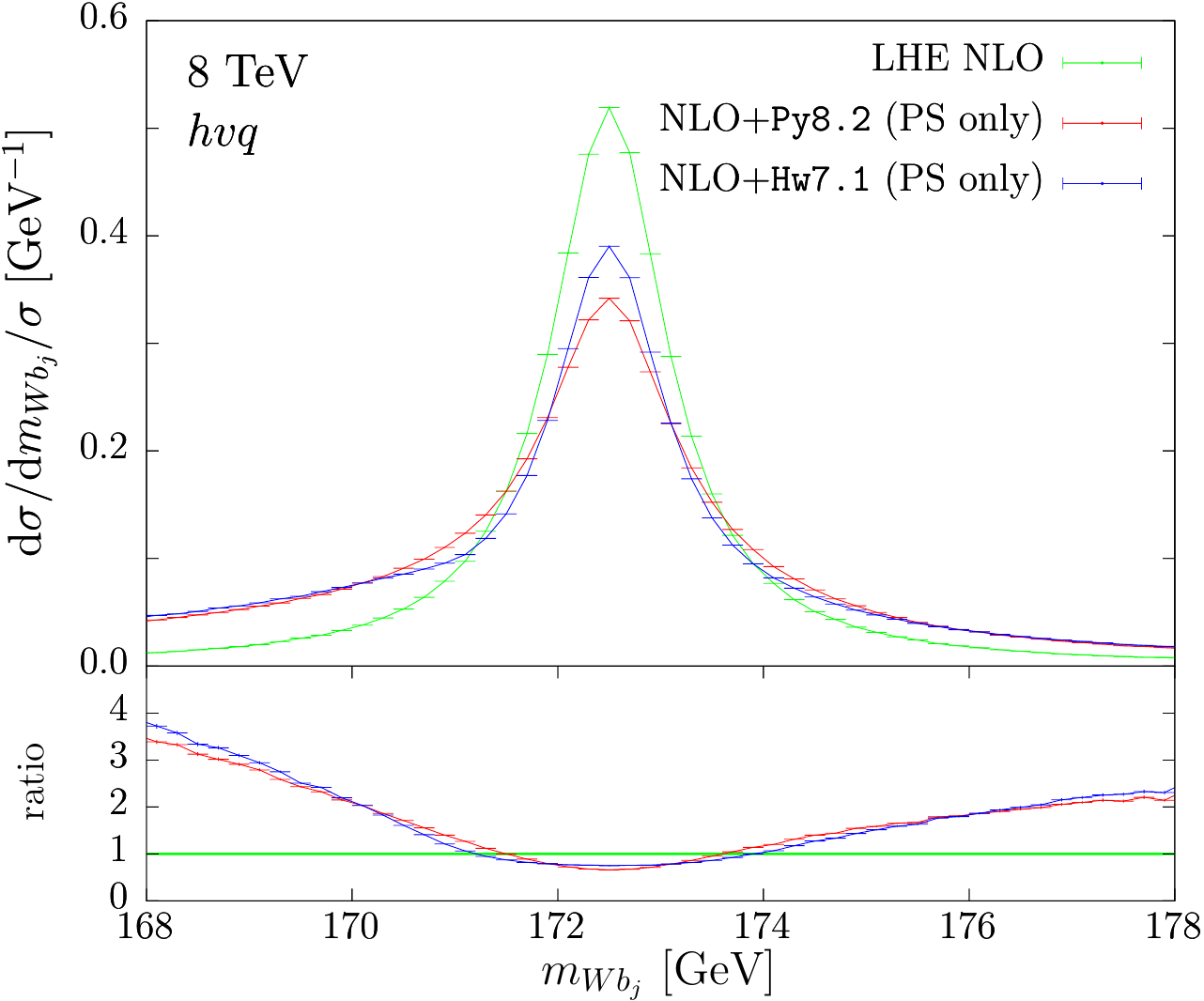}
\,
\includegraphics[width=\wfigdoub]{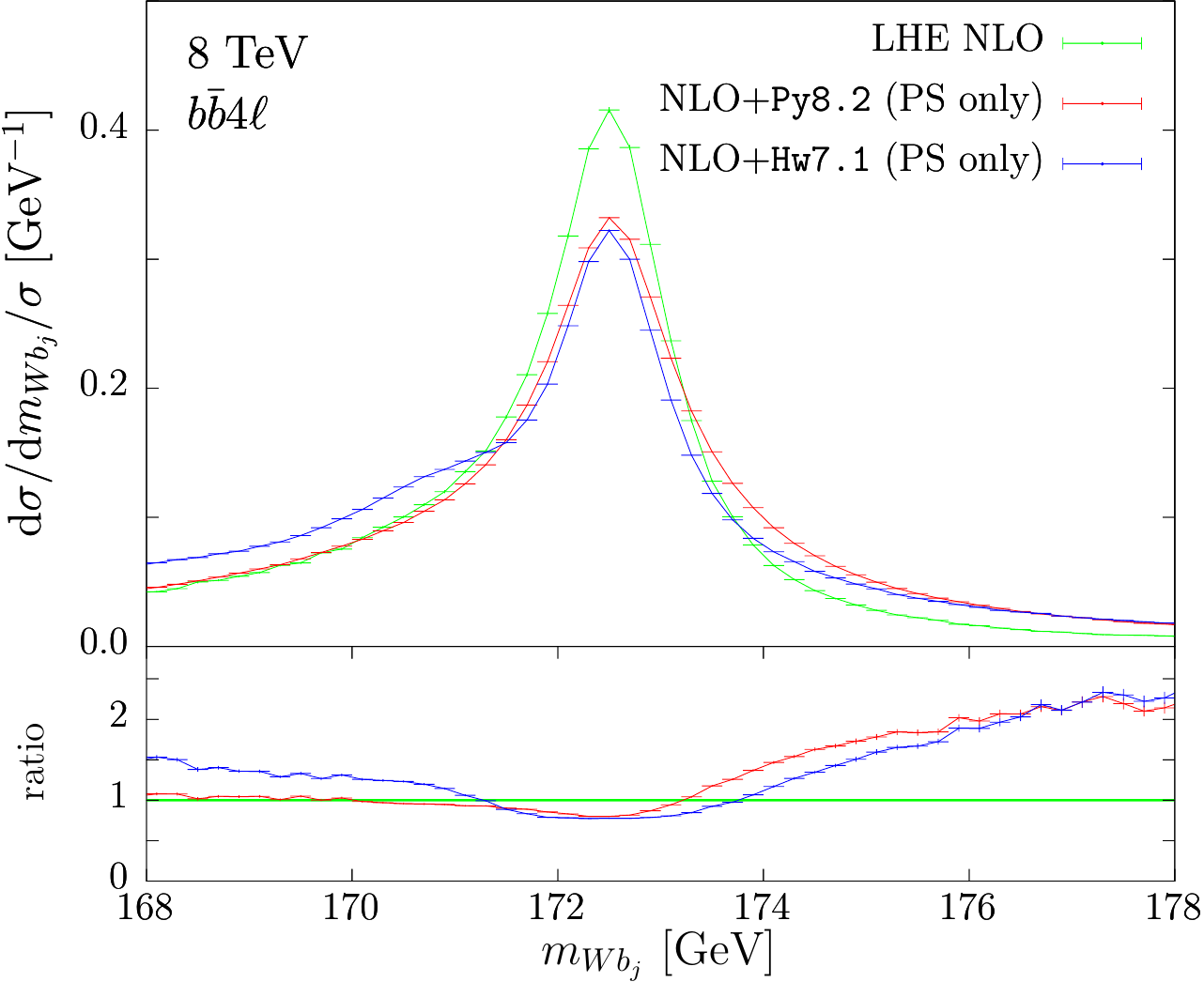}
\caption{${d\sigma}/{d \mwbj}$ distribution obtained with \hvq{}~(left
  pane) and \bbfourl{}~(right pane) at the NLO LHE level~(green), and
  at NLO+shower (in red \PythiaEightPtwo{} and in blue
  \HerwigSevenPone{}), normalized to 1 in the displayed range. In the
  bottom panel the ratio with the NLO LHE is shown.}
\label{fig:NLO-PS}
\end{figure}

In Fig.~\ref{fig:NLO-PS} we can see the impact of the PS provided by
\PythiaEightPtwo{}~(red) and by \HerwigSevenPone{}~(blue) on the
\hvq{}~(left) and \bbfourl{}~(right) generators.  In \hvq{}, we notice
a large effect in the low tail of the distribution, since radiation in
decay is fully generated by the shower.  Conversely for \bbfourl{} we
get smaller shower corrections, since the hardest radiation in decay
is already included at the LHE level.  In both cases, we see an
enhancement in the region above the peak. This is attributed to shower
radiation that is captured by the \bjet{} cone.

\begin{figure}[tb!]
\centering
\includegraphics[width=\wfigsing]{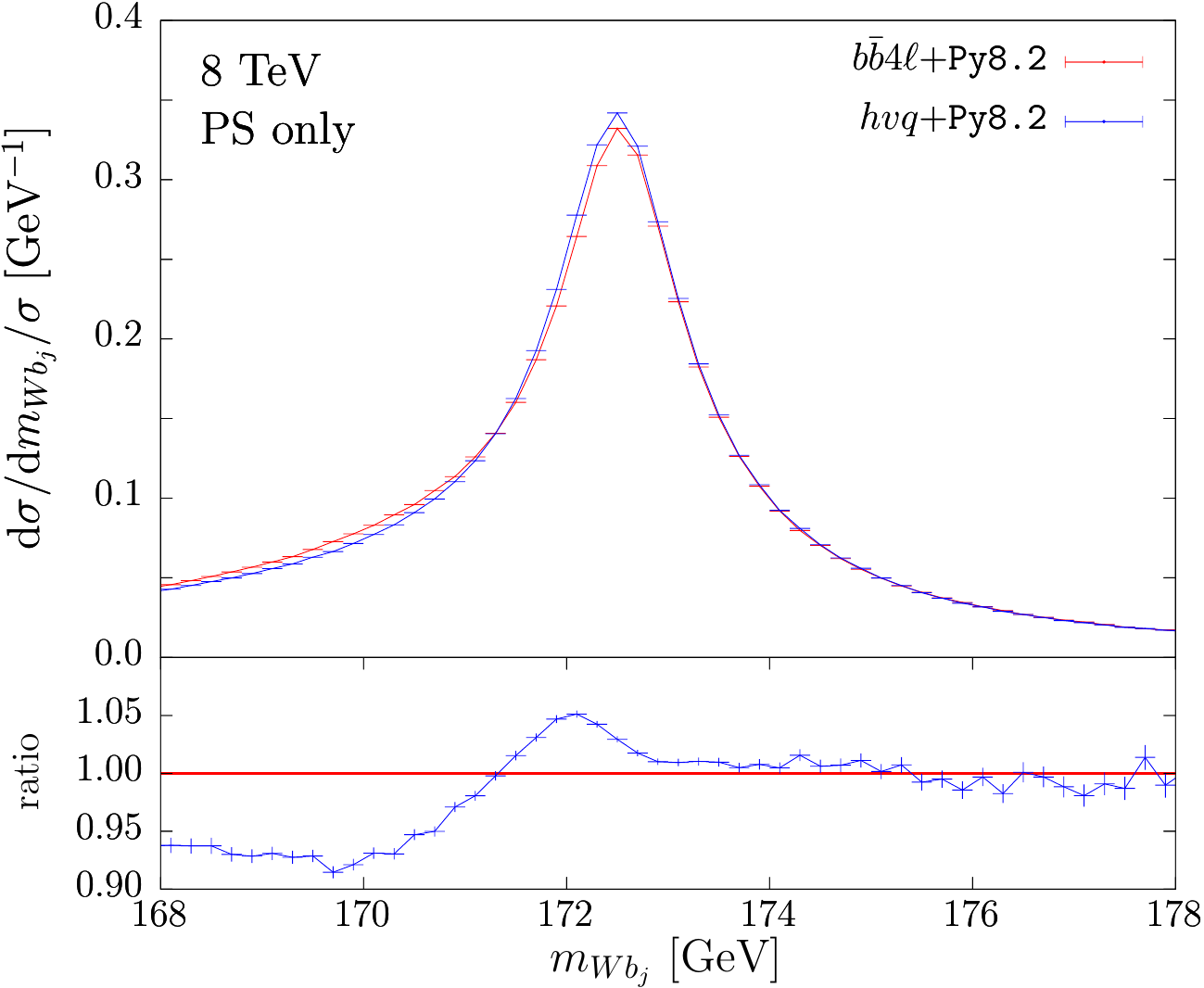}%
\caption{${d\sigma}/{d \mwbj}$ distribution, normalized to 1 in the
  displayed range, obtained with \bbfourl{}~(red) and \hvq{}~(blue) at
  the NLO+PS level using \PythiaEightPtwo{}.}
\label{fig:bb4l+hvq-PS}
\end{figure}

 We observe that, after shower, the \hvq{} result becomes
 qualitatively very similar to the \bbfourl{} one, as shown in
 Fig.~\ref{fig:bb4l+hvq-PS}.

The inclusion of the shower in \ttNLOdec{} leads to effects similar to
those observed in \bbfourl{}.

\section{Hadronization and underlying events}

\begin{figure}[tb]
\centering
\includegraphics[width=0.975\wfigdoub]{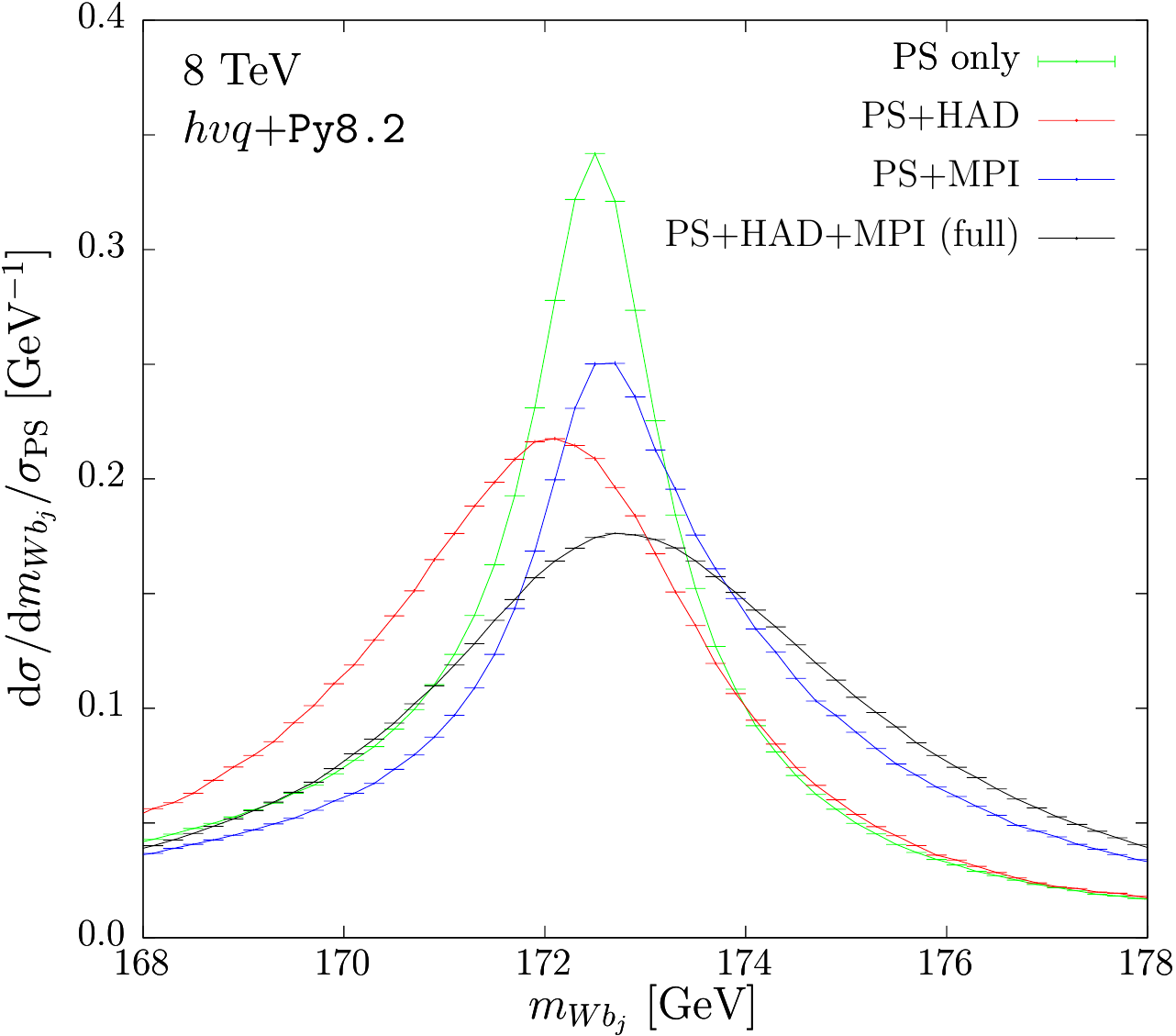}\,
\includegraphics[width=0.975\wfigdoub]{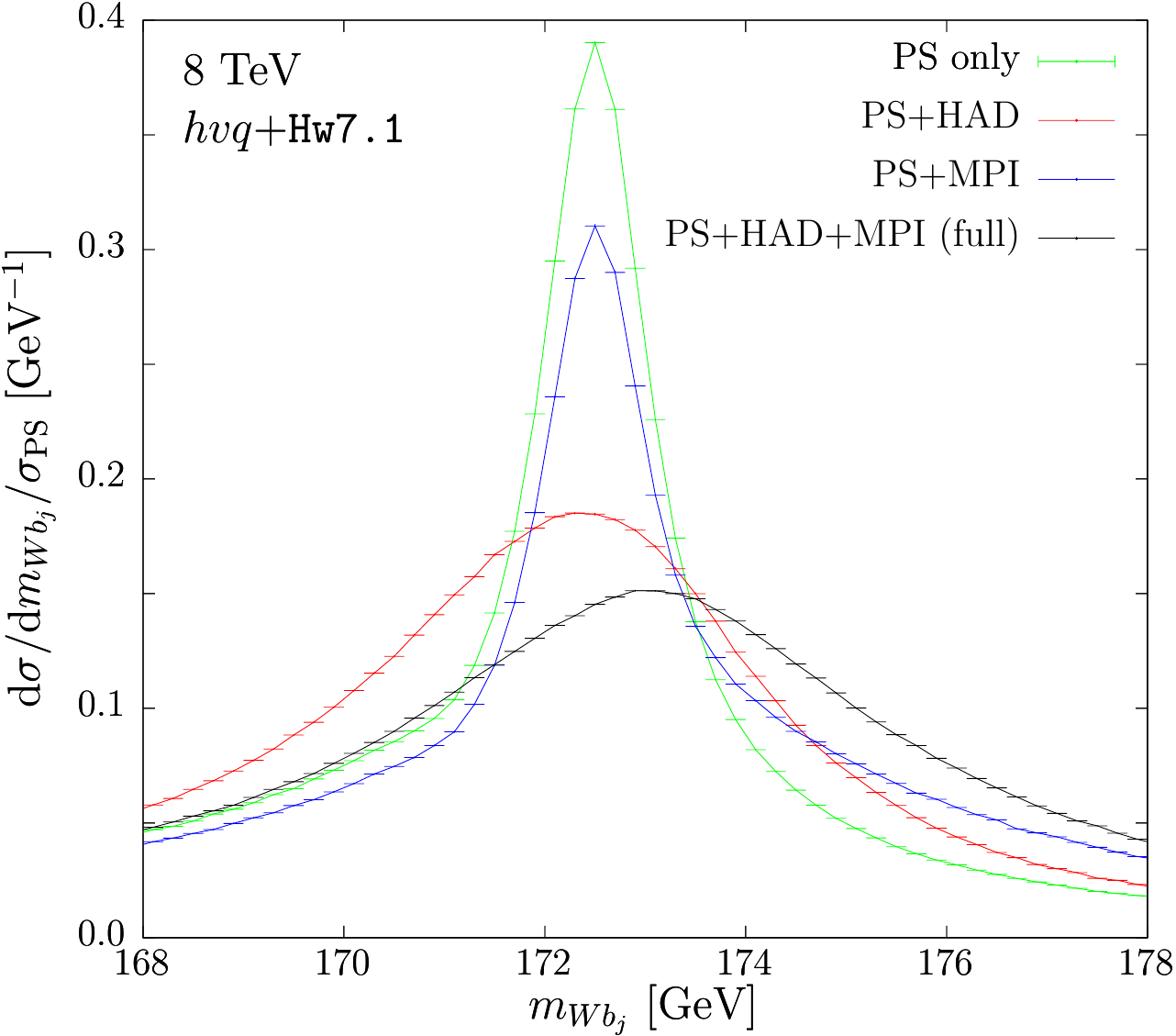}
\caption{${d\sigma}/{d \mwbj}$ distribution obtained with \hvq{}
  interfaced with \PythiaEightPtwo{}~(left panel) and
  \HerwigSevenPone{}~(right panel). In green, the NLO+PS results; in
  red, hadronization effects are included; in blue, NLO+PS with
  multi-parton interactions~(MPI); and in black, with hadronization
  and MPI effects. The curves are normalized using the NLO+PS cross
  section in the displayed range.}
\label{fig:hvq-PS-NP}
\end{figure}

The effect of hadronization and multi-parton interactions~(MPI), as
modelled by \PythiaEightPtwo{} and \HerwigSevenPone{}, when interfaced
to the \hvq{} generator, is showed in Fig.~\ref{fig:hvq-PS-NP}.
The hadronization~(red) has a large impact on the final
distributions. It widens the peak for both generators. However, in the
\HerwigSevenPone{} case, we also observe a clear enhancement of the
high mass region, that is not as evident in the \PythiaEightPtwo{}
case.
The MPI~(blue), that generate particles that can deposit in the
\bjet{} cone, raise the tail of the distributions above the peak.
In the combined effect of hadronization and MPI~(black),
\HerwigSevenPone{} has a wider peak. On the other hand, the high tail
enhancement seems similar in the two generators.

\begin{figure}[tb]
\centering
\includegraphics[width=0.975\wfigdoub]{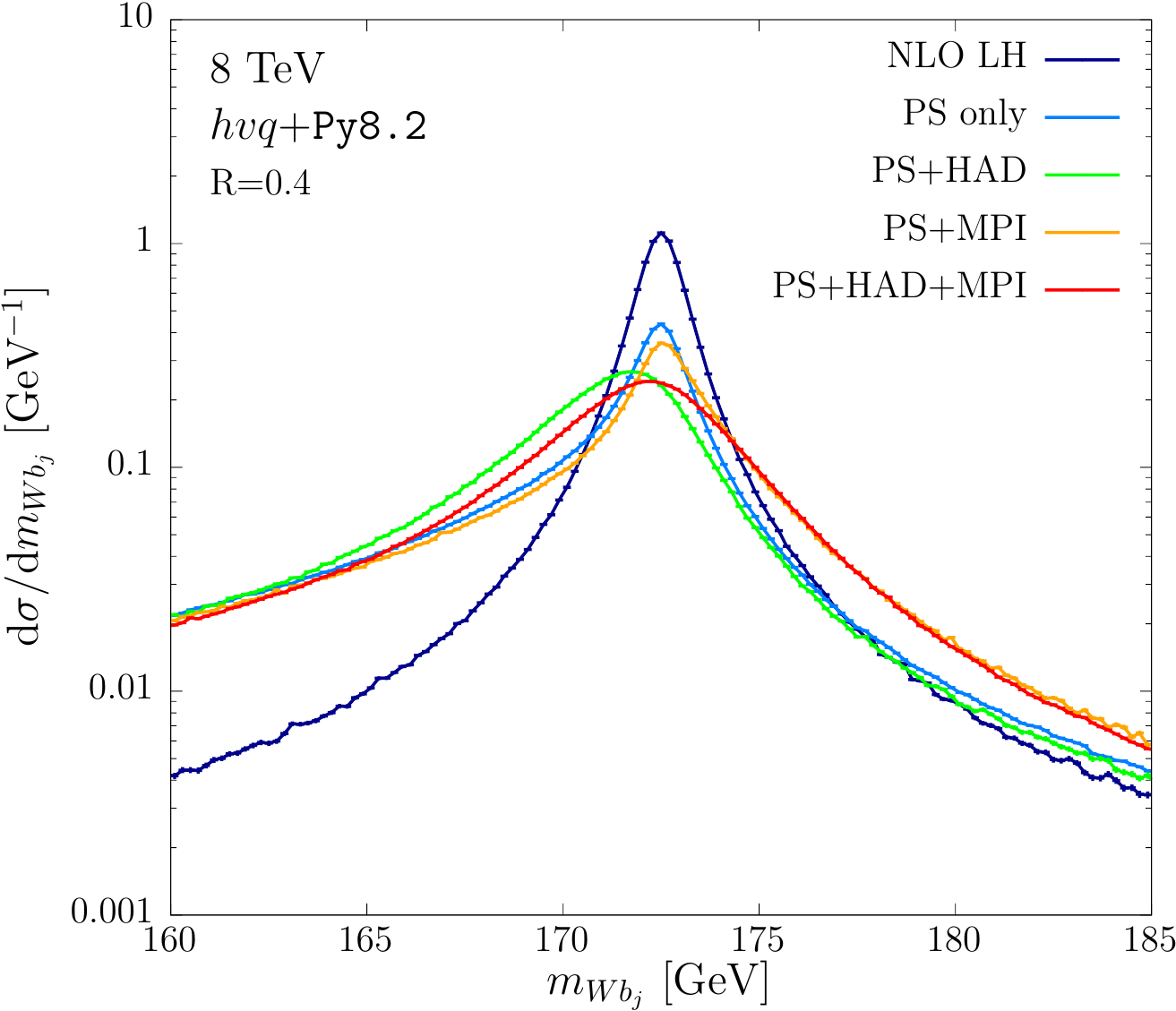}
\,
\includegraphics[width=0.975\wfigdoub]{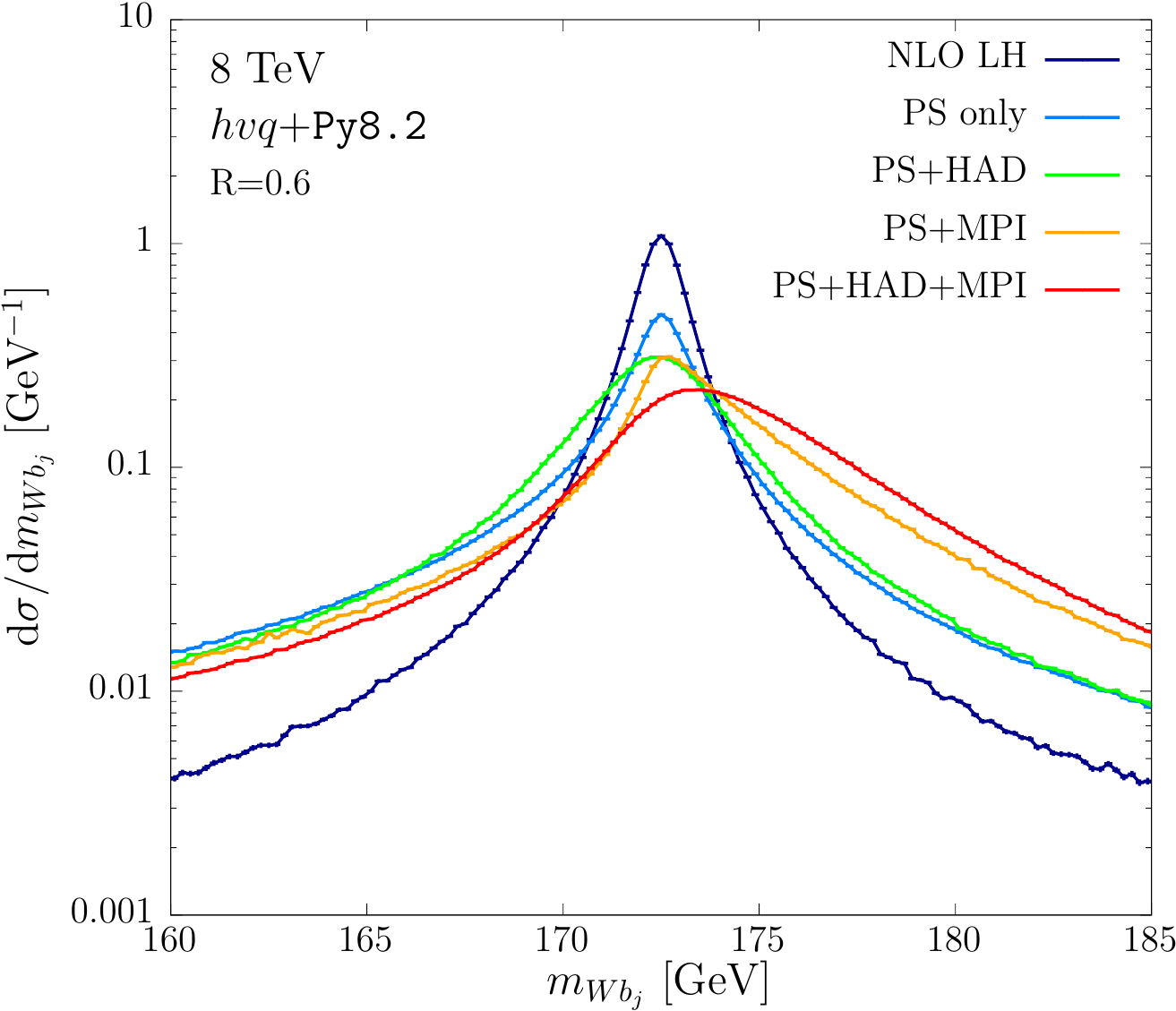}
\\ \includegraphics[width=0.975\wfigdoub]{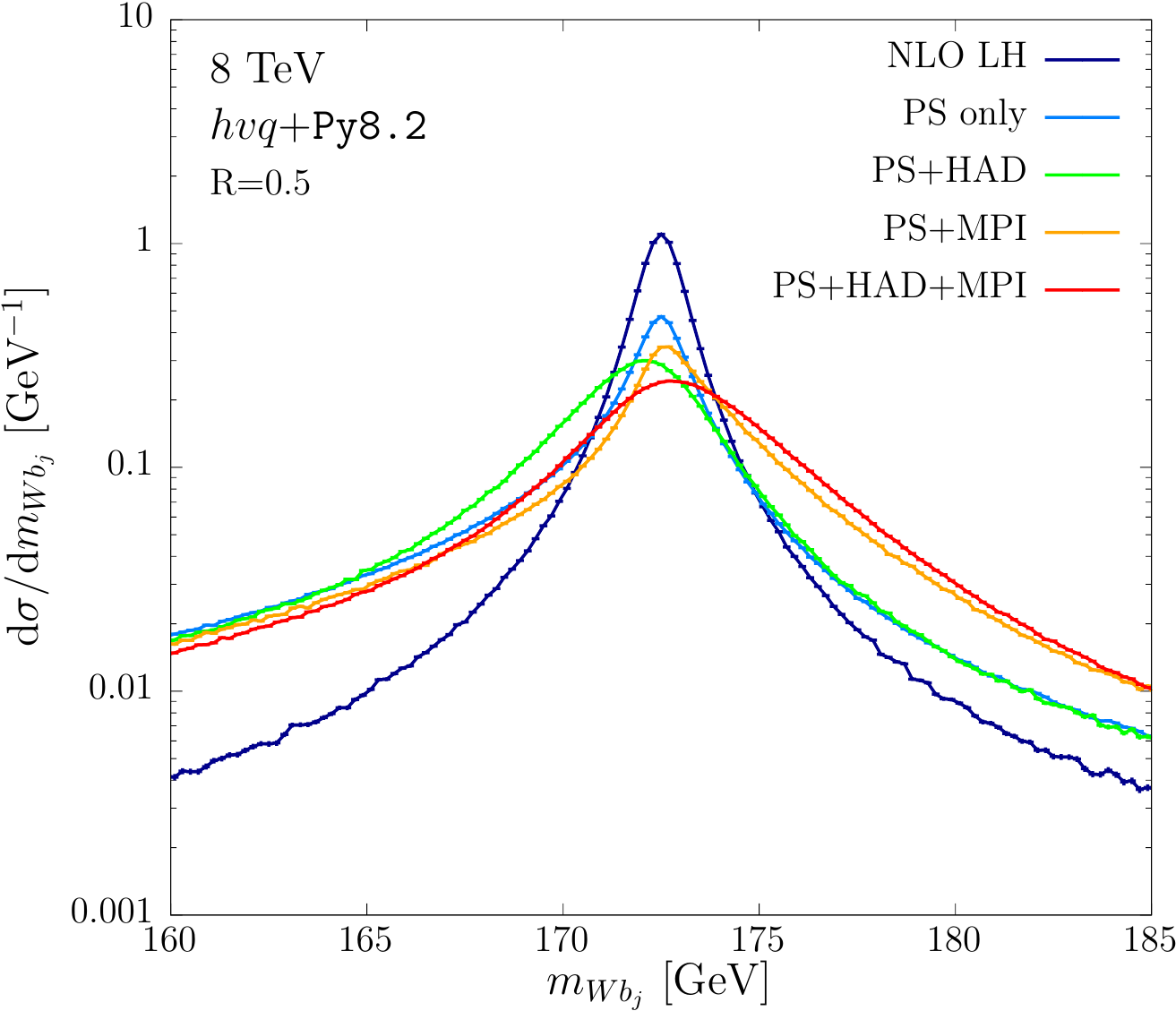}
\caption{${d\sigma}/{d \mwbj}$ distribution obtained with \hvq{}
  interfaced with \PythiaEightPtwo{} for $R=0.4$~(upper-left),
  $R=0.5$~(down) and $R=0.6$~(upper-right). In blue, the NLO result
  (LH level); in azure, the PS is included; in green, the PS and the
  hadronization are included; in orange, the PS and the MPI are
  included; in red, the full results including PS, hadronization and
  the underlying event.}
\label{fig:hvq-PY8-R}
\end{figure}

The impact of the several components we have analysed, i.e.~radiative
corrections, the hadronization and the MPI, strongly depend on the jet-radius
parameter $R$. By increasing (or decreasing) $R$, the peak position is
shifted to the left (or right) as it is shown in Fig.~\ref{fig:hvq-PY8-R}.
Furthermore, as we will see in Chap.~\ref{sec:mwbj}, differences in the
implementation of radiation from the resonances, the hadronization
model and the underlying events can also shift the peak, leading
eventually to a displacement of the extracted top mass, that should be
carefully assessed for a sound estimate of the top-quark mass
theoretical uncertainty.  \\
\vspace{15cm} $ $
 
\chapter{Reconstructed-top mass distribution}
\label{sec:mwbj}
The peak of the reconstructed mass \mwbj{}, defined in
Sec.~\ref{sec:physicsObjects}, is a representative of all the direct
measurement methods.  To simplify our analysis, we assume we can
distinguish between the $b$ and the $\bar{b}$ flavoured jets and that
we can fully reconstruct the neutrinos momenta together with their
sign.  Our approach is very crude but it enables us to focus on
theoretical issues rather then experimental ones.  Indeed, if we find
differences in the extracted mass using our ideal \mwbj{} observable,
we would be forced to conclude that there is an irreducible
theoretical error (i.e.~an error that cannot be reduced by increasing
the experimental accuracy) on the mass measurement.

In order to mimic the experimental systematics, we introduce a Gaussian
smearing to the \mwbj{} distributions. If differences among the generators
are found in this latter case, the problem is less severe and may be reduced
once experimental resolution is improved.

However, we need to remark that also ``irreducible'' errors (according
to the definition given above) may be reduced in practice. This is the
case if one of the generators at hand does not fit satisfactorily
measurable distributions related to top production. If we change the
allowed range for the parameters of the generator in order to
reproduce the data fairly. This procedure can reduce the error
associated with top-quark mass measurements.

In Sec.~\ref{sec:Mwbjmax_extraction} we describe the procedure for the
extraction of the reconstructed-top mass peak, for both the ideal
\mwbj{} distribution and the smeared one.
In Sec.~\ref{sec:mwbj-cmp} we will compare the three \POWHEGBOX{}
generators for $t\bar{t}$ production interfaced to
\PythiaEightPtwo{}. Scale and PDF variations are also taken into
account, together with $\as$ variations to investigate the sensitivity
to the intensity of radiation from the $b$ quark.
In Sec.~\ref{sec:mwbj_hw7} we investigate the differences between the
\PythiaEightPtwo{} and \HerwigSevenPone{} predictions and the various
alternative settings that the \HerwigSevenPone{} program
offers\footnote{Unless specified otherwise, \PythiaEightPtwo{} and
  \HerwigSevenPone{} are setup to run in full hadron mode including
  shower, hadronization and multi-parton interactions.}.

\section[\mwbjmax{} extraction]{$\boldsymbol{\mwbjmax{}}$ extraction} 
\label{sec:Mwbjmax_extraction}
We first fit the \mwbj{} distribution around the peak region.  We call
$Y(\mwbj)$ the histogram of our distribution, and $y(\mwbj,\{a\})$ our
fitting functional form, where $\{a\}$ represent the fitting
parameters
\begin{equation}
  y(\mwbj,\{a\})=\frac{a_2[1+a_4(\mwbj-a_1)]}{(\mwbj-a_1)^2+a_3^2}+a_5\,.
\end{equation}
To extract the parameters $\{a\}$ we proceed as follows:
\begin{itemize}
  \item We find the bin with the highest value, and assign its center
    to the variable $\mwbjmax$.
  \item We find all surrounding bins whose value is not less than
    $Y(\mwbjmax)/2$. We assign to the variable $\Delta$ the range
    covered by these bins divided by two.
  \item
    We minimize the $\chi^2$ computed from the difference of the
    integral of $y(\mwbj,\{a\})$ in each bin, divided by the bin size,
    with respect to $Y(\mwbj)$, choosing as a range all bins that
    overlap with the segment $[\mwbjmax-\Delta, \mwbjmax+\Delta]$.
  \item
    From the fitted function we extract the maximum position and
    assign it to $\mwbjmax$.
  \item If the reduced $\chi^2$ of the fit is less than 2, we keep
    this result.  If not, we replace $\Delta\to 0.95\times \Delta$ and
    repeat the operation until this condition is met.
\end{itemize}
Once the parameters $\{a\}$ are determined, the extracted peak
position $\mwbjmax$ is found by solving
\begin{equation}
  \frac{ d \, y(\mwbj, \{a\})}{ d \mwbj} \Big|_{\mwbj=\, \mwbjmax} = 0\,.
\end{equation}
Its error is derived by propagating the errors of the parameters
$\{a\}$ extracted using our fitting procedures, in the expression of
the peak.  This procedure is applied to both the ideal $\mwbj$
distribution and the smeared one.

The histogram of the smeared distribution $Y_{\rm s}(\mwbj)$ is
obtained convoluting $Y(\mwbj)$ with a Gaussian of width
$\sigma=15$~GeV (that is the typical experimental resolution on the
reconstructed top mass)
\begin{equation}
Y(\mwbj) = \mathcal{N} \sum_{m_i} \Delta m_i\; y(m_i) \exp
\left(-\frac{(\mwbj - m_i)^2}{2\sigma^2}\right)\,,
\end{equation}
where $m_i$ is the central value of the $i^{\rm th}$ bin and $\Delta m_i$
its width. $\mathcal{N}$ is a normalization constant.

The $B$ coefficient of eq.~(\ref{eq:linearfitfunc}) that links $\mwbjmax$ and
$\mt$ is found performing a linear fit for the three values of
$\mt=$169.5~GeV, 172.5~GeV, 175.5~GeV.  Since \mwbj{} is the mass of the
reconstructed top, we expect $B\approx 1$ for the \mwbjmax{} observable.

The values for the $B$ coefficients that we have obtained with the
three generators showered with \PythiaEightPtwo{} are collected in
Tab.~\ref{tab:mwbj_B_values}, and confirm our expectation.
\begin{table}[tb]
  \centering
 { \begin{tabular}{l|c|c|}
    \cline{2-3}
   $\phantom{\Big|}$ & $B$, no smearing & $B$, smearing \\
    \cline{1-3} \multicolumn{1}{ |c|}{\hvq{}} &
    $\phantom{\Big|}\Bfrommwbjhvq \pm \Berrfrommwbjhvq$ &
    $\Bfrommwbjsmearhvq \pm \Berrfrommwbjsmearhvq$ \\ \cline{1-3}
    \multicolumn{1}{ |c|}{\ttbnlodec{}} & $\phantom{\Big|}
    \Bfrommwbjttdec \pm \Berrfrommwbjttdec$ & $\Bfrommwbjsmearttdec
    \pm \Berrfrommwbjsmearttdec$ \\ \cline{1-3} \multicolumn{1}{
      |c|}{\bbfourl{}} & $\phantom{\Big|} \Bfrommwbjbbfourl \pm
    \Berrfrommwbjbbfourl$ & $\Bfrommwbjsmearbbfourl \pm
    \Berrfrommwbjsmearbbfourl$ \\ \cline{1-3}
  \end{tabular}}
  \caption{Values for the $B$ coefficients of eq.~(\ref{eq:linearfitfunc})
    for the \mwbj{} peak position, for the non-smeared and smeared
    distributions~(see Sec.~\ref{sec:mwbj-cmp} for details), obtained with
    the \hvq{}, \ttbnlodec{} and \bbfourl{} generators showered with
    \PythiaEightPtwo{}.}
  \label{tab:mwbj_B_values}
\end{table}
Thus, we can safely assume
\begin{equation}
\Delta \mt = - \Delta \mwbjmax{}.
\end{equation}

\section{Comparison among different NLO+PS generators}
\label{sec:mwbj-cmp}
We begin by showing comparisons of our three generators, interfaced
with \PythiaEightPtwo{}, for our reference top-mass value of
172.5~GeV.

\begin{figure}[tb!]
\centering
\includegraphics[width=\wfigdoub]{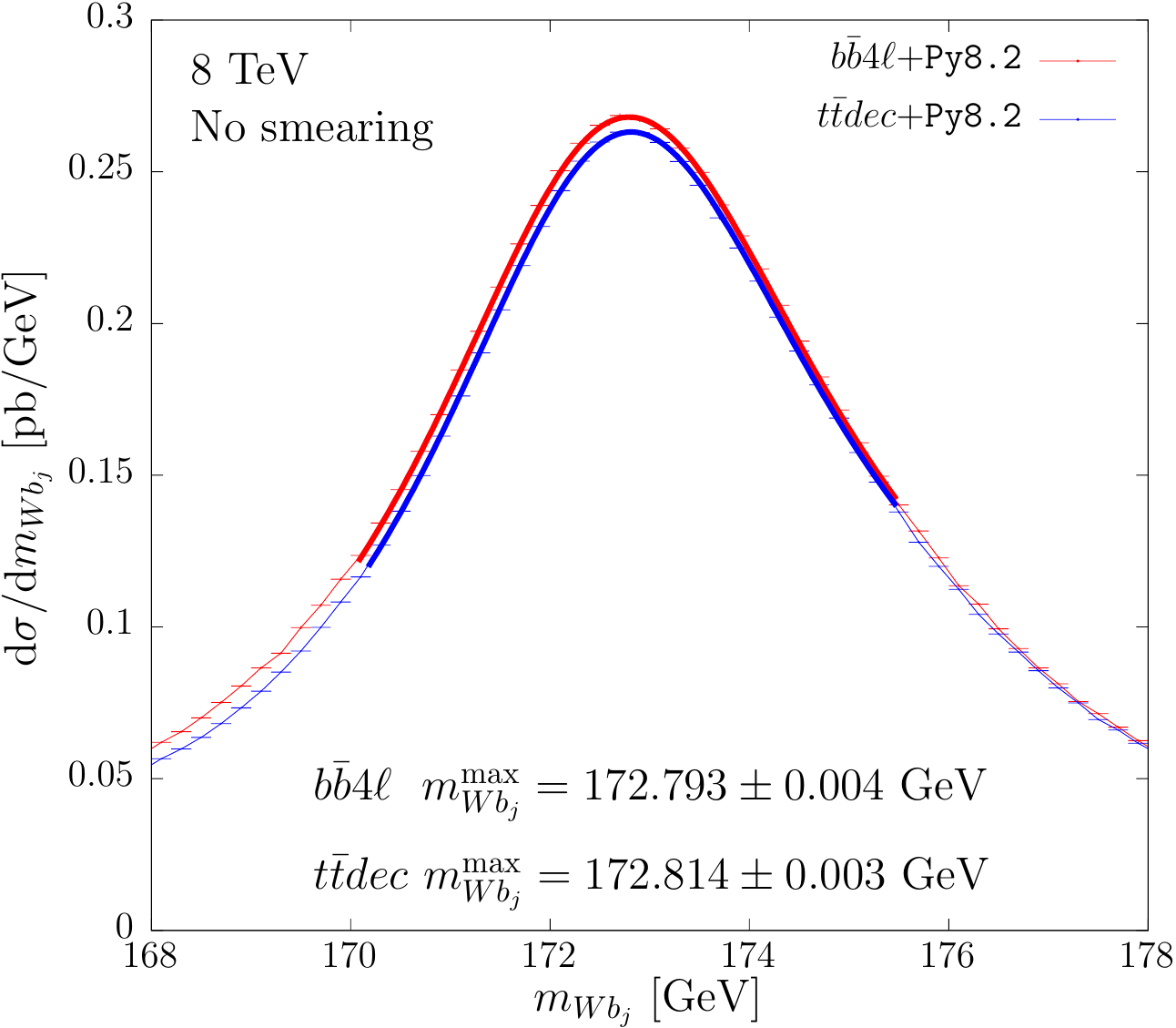}%
\includegraphics[width=\wfigdoub]{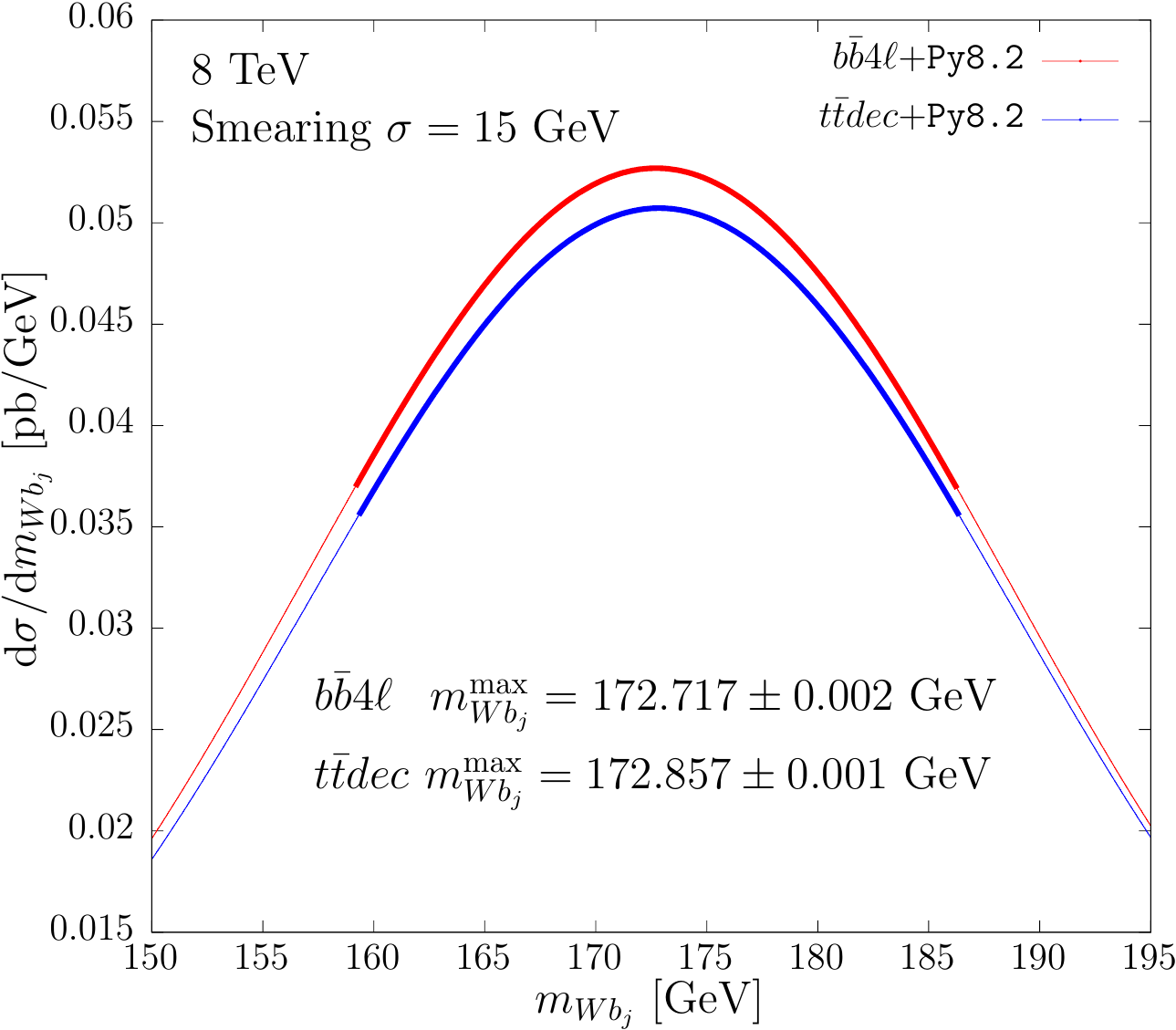}
\caption{${d\sigma}/{d \mwbj}$ distribution obtained with the
  \bbfourl{} and \ttNLOdec{} generators interfaced with
  \PythiaEightPtwo{}, for $\mt=172.5$~GeV, before~(left) and
  after~(right) applying the Gaussian smearing.}
\label{fig:MassPeaks-py8-bb4l-ttb}
\end{figure}

The \mwbj{} distributions of the \bbfourl{} and \ttNLOdec{} generators
are compared in Fig.~\ref{fig:MassPeaks-py8-bb4l-ttb}, before~(left)
and after~(right) applying the Gaussian smearing.
In the left plot we see that the two generators yield a very similar shape
and the results concerning \mwbj{} are very similar.  This is an indication
that interference effects in radiation and other off-shell effects, that are
included in \bbfourl{} but not in \ttNLOdec{}, have a very minor impact on
the peak position, at least if we consider a measurement with an ideal
resolution.  The results obtained applying a 15 GeV smearing on the
distribution are shown in the right plot of
Fig.~\ref{fig:MassPeaks-py8-bb4l-ttb}.  The smearing procedure correlates
points near the peak together with the tails of the \mwbj{} distribution,
increasing the impact on \mwbjmax{} of the region away from the peak, where
there are larger differences between the two generators.  This leads to a
difference in the peak position of \diffttdecbbfourl~MeV.

\begin{figure}[tb]
\centering
\includegraphics[width=\wfigdoub]{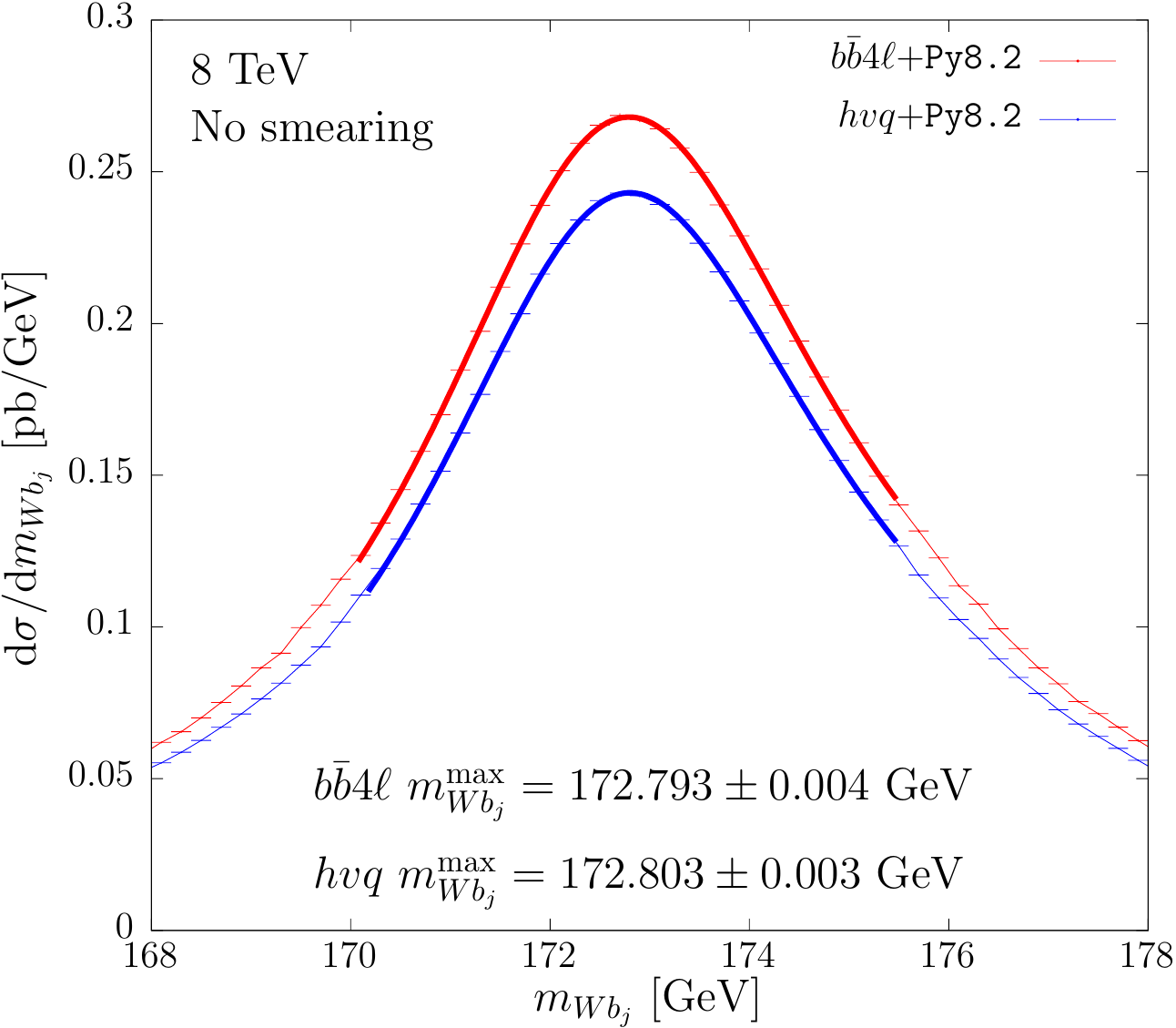}
\includegraphics[width=\wfigdoub]{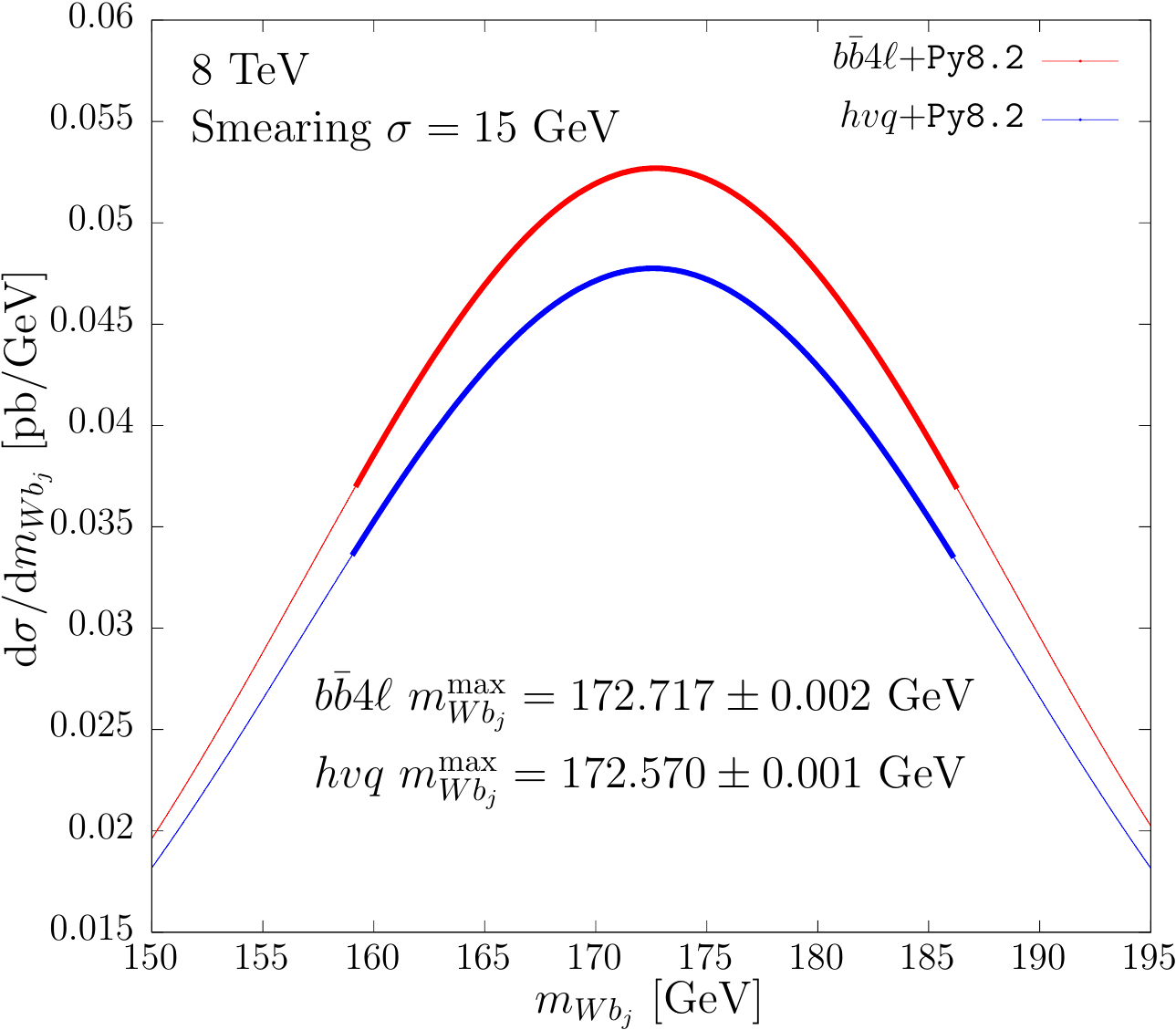}
\caption{${d\sigma}/{d \mwbj}$ distribution obtained with the
  \bbfourl{} and \hvq{} generators interfaced with \PythiaEightPtwo{},
  for $\mt=172.5$~GeV, before~(left) and after~(right) applying the
  Gaussian smearing.}
\label{fig:MassPeaks-py8-bb4l-hvq}
\end{figure}
In Figs.~\ref{fig:MassPeaks-py8-bb4l-hvq} we compare the \bbfourl{}
and the \hvq{} generators.  We see a negligible
difference in the peak position in the non-smeared case, while, in the
smeared case, the \hvq{} generator differs from \bbfourl{} by
\diffhvqbbfourl~MeV, similar in magnitude to the case of \ttbnlodec{},
but with opposite sign.

Our findings are summarized in Tab.~\ref{tab:mwbj_showerOnly}, where
we also include results obtained at the shower level, i.e.~ without
the inclusion of the underlying event and of the hadronization.
\begin{table}[tb]
\centering \resizebox{\textwidth}{!}
           { \begin{tabular}{l|c|c|c|c|}
 \cline{2-5}
 &  \multicolumn{2}{ |c|}{PS only}
 &  \multicolumn{2}{ |c|}{ \phantom{\Big|} full}\\
 \cline{2-5}
 & \phantom{\Big|} No smearing & 15~GeV smearing
 & \phantom{\Big|} No smearing & 15~GeV smearing \\
 \cline{1-5}
 \multicolumn{1}{ |c|  }{ \phantom{\Big|}  \bbfourl{}}
 & $172.522\pm  0.002$~GeV
 & $171.403\pm  0.002$~GeV
 & $172.793\pm  0.004$~GeV
 & $172.717\pm  0.002$~GeV
 \\ \cline{1-5}
 \multicolumn{1}{ |c|  }{ \phantom{\Big|}\ttbnlodec{} ${}-$ \bbfourl{}}
  &  $         -18 \pm            2 $~MeV
  &  $+         191 \pm            2 $~MeV
  &  $+          21 \pm            6 $~MeV
  &  $+         140 \pm            2 $~MeV
 \\ \cline{1-5}
 \multicolumn{1}{ |c|  }{ \phantom{\Big|}\hvq{} ${}-$ \bbfourl{}}
  &  $         -24 \pm            2 $~MeV
  &  $         -89 \pm            2 $~MeV
  &  $+          10 \pm            6 $~MeV
  &  $        -147 \pm            2 $~MeV
 \\ \cline{1-5}
\end{tabular}
}
\caption{Differences in the $\mwbj{}$ peak position for
  $\mt$=172.5~GeV for \ttbnlodec{} and \hvq{} with respect to
  \bbfourl{}, showered with \PythiaEightPtwo{}, at the NLO+PS level
  and at the full hadron level.}
\label{tab:mwbj_showerOnly}
\end{table}

We notice that \hvq{}, even if it does not implement NLO corrections
in top decay, yields to results similar to those of the most accurate
\bbfourl{} generator.  This is due to the inclusion of matrix-element
corrections~(MEC) in top decay by the \PythiaEightPtwo{} PS. MEC are
indeed equivalent, up to an irrelevant normalization factor, to
next-to-leading order corrections in decay.  This observation is
confirmed by examining Tab.~\ref{tab:mwbj_MEC}, where predictions
obtained with and without MEC are compared.
\begin{table}[t!]
\centering \resizebox{\textwidth}{!}
           { \begin{tabular}{l|c|c|c|c|}
 \cline{2-5}
 &  \multicolumn{2}{ |c|}{ \phantom{\Big|} No smearing}
 &  \multicolumn{2}{ |c|}{15~GeV smearing} \\
 \cline{2-5}
 & \phantom{\Big|} MEC  & MEC ${}-$ no  MEC
 &  MEC   & MEC ${}-$ no MEC\\
 \cline{1-5}
 \multicolumn{1}{ |c|  }{ \phantom{\Big|}\bbfourl{}}
 & $172.793\pm   0.004$~GeV
  &  $         -12 \pm            6 $~MeV
 & $172.717\pm   0.002$~GeV
  &  $+          55 \pm            2 $~MeV
 \\ \cline{1-5}
 \multicolumn{1}{ |c|  }{ \phantom{\Big|}\ttbnlodec{}}
 & $172.814\pm   0.003$~GeV
  &  $          -4 \pm            5 $~MeV
 & $172.857\pm   0.001$~GeV
  &  $         -26 \pm            2 $~MeV
 \\ \cline{1-5}
 \multicolumn{1}{ |c|  }{ \phantom{\Big|}\hvq{}}
 & $172.803\pm   0.003$~GeV
  &  $+          61 \pm            5 $~MeV
 & $172.570\pm   0.001$~GeV
  &  $+         916 \pm            2 $~MeV
 \\ \cline{1-5}
\end{tabular}
}
\caption{$\mwbj{}$ peak position for $\mt$=172.5~GeV obtained with the
  three different generators, showered with
  \PythiaEightPtwo{}+MEC~(default). We also show the differences
  between \PythiaEightPtwo{}+MEC and \PythiaEightPtwo{} without MEC.
}
\label{tab:mwbj_MEC}
\end{table}
When MEC are not included, the peak position in the smeared
distribution provided by the \hvq{} generator receives a considerable
shift, near 1~GeV.  On the other hand, the \bbfourl{} and \ttbnlodec{}
generators, that already include the hardest emission off $b$ quarks,
display a reduced sensitivity on the MEC.  This leads to the
conclusion that the MEC in \PythiaEightPtwo{} do a decent job in
simulating top decay as far as the \mwbj{} distribution is
concerned. The remaining uncertainty of roughly \diffaverage~MeV in
the case of both \hvq{} and \ttbnlodec{} generators, pulling in
opposite directions, is likely to be due to the approximate treatment
of off-shell effects.

\subsection{Matching uncertainties}
The {\tt FSREmission} veto procedure described in
\writeApp{}\ref{app:pythia} represents the most accurate way to
perform the vetoed shower on the \POWHEGBOX{} generated events
interfaced with the \PythiaEightPtwo{}, because it uses the \POWHEG{}
definition of transverse momentum rather than the \PythiaEightPtwo{}
one.

There is also an alternative, the {\tt ScaleResonance} procedure,
always described in \writeApp{}\ref{app:pythia}, where the scale of
the POWHEG emissions from the $t$ and the $\bar{t}$ are set as initial
scales for the showers evolutions.  The {\tt ScaleResonance} procedure
can introduce a mismatch that we take as an indication of the size of
the matching uncertainties.  The extracted peak position for the
\bbfourl{} and \ttbnlodec{} with the two matching procedures are
summarized in Tab.~\ref{tab:mass_extraction-matching}.
\begin{table}[tb]
\centering
{ \begin{tabular}{l|c|c|c|c|}
 \cline{2-5}
 &  \multicolumn{2}{ |c|}{ \phantom{\Big|} No smearing}
 &  \multicolumn{2}{ |c|}{15~GeV smearing} \\
 \cline{2-5}
 & \phantom{\Big|} {\tt SR} & {\tt SR} ${}{-}$ {\tt FSR}
 & \phantom{\Big|} {\tt SR} & {\tt SR} ${}{-}$ {\tt FSR}\\
 \cline{1-5}
 \multicolumn{1}{ |c|  }{ \phantom{\Big|}\bbfourl{}}
 & $172.816\pm   0.004$~GeV 
  &  $+          23 \pm            6 $~MeV
 & $172.737\pm   0.002$~GeV
  &  $          20 \pm            2 $~MeV
 \\ \cline{1-5}
 \multicolumn{1}{ |c|  }{ \phantom{\Big|}\ttbnlodec{}}
 & $172.812\pm   0.004$~GeV 
  &  $          -1 \pm            5 $~MeV
 & $172.878\pm   0.001$~GeV
  &  $          21 \pm            2 $~MeV
 \\ \cline{1-5}
\end{tabular}
}
\caption{$\mwbj{}$ peak position for $\mt$=172.5~GeV obtained with the
  \bbfourl{} and \ttbnlodec{} generators, showered with
  \PythiaEightPtwo{}, for the {\tt ScaleResonance}~({\tt SR}) veto
  procedure. The differences with {\tt FSREmission}~({\tt FSR}), that
  is our default, are also shown.}
\label{tab:mass_extraction-matching}
\end{table}
We can see that these differences are roughly 20~MeV in \bbfourl{} for
both the no-smearing and smearing case, and in \ttbnlodec{} they are a
few MeV for the no-smearing case, and 20~MeV with smearing.

This is due to the fact that the first emissions of the decayed top
has already a small transverse momentum and in the collinear limit the
POWHEG and the \PythiaEightPtwo{} definitions of $\pT$ are equivalent.

We can also compare the default behaviour for dealing with radiation
in the production process with the results obtained with the {\tt
  PowhegHooks} veto machinery, that is activated with the setting
\begin{verbatim}
POWHEG:veto = 1.
\end{verbatim}
The results are shown in Tab.~\ref{tab:pwgveto_mwbj}.
\begin{table}[tb]
\centering  \begin{tabular}{|c|c|c|c|}
\cline{1-4}
\multicolumn{4}{|c|}{ \phantom{\Big|} {\tt PowhegHooks} ${}{-}$ no {\tt PowhegHooks} [MeV]} \\
\cline{1-4}
\phantom{\Big|} observable & \bbfourl & \ttbnlodec & \hvq \\
\cline{1-4}
 $\phantom{\Big|} \mwbjmax$ no smearing                                & $          35 \pm           6 $& $          18 \pm           5 $& $          17 \pm           5 $\\ \cline{1-4}
 $\phantom{\Big|} \mwbjmax$ smearing                                   & $          77 \pm           2 $& $          78 \pm           2 $& $          71 \pm           2 $\\ \cline{1-4}
 \end{tabular}

\caption{Differences between the \mwbjmax{} predictions obtained using
  the {\tt POWHEG:veto = 1} and the {\tt POWHEG:veto = 0} settings for
  the three generators interfaced with \PythiaEightPtwo.}
\label{tab:pwgveto_mwbj}
\end{table}
A slightly increased dependence on the veto procedure is found
concerning radiation in the production process, since it is in general
much more harder than the radiation in production and thus the two
$\pT$ definitions may differ.  However, we notice that even for the
smeared mass distribution the differences between the $\mwbjmax{}$
obtained with {\tt PowhegHooks} and the default ones are rather small
and equivalent for all the NLO generators. Thus, our choice of not
using the {\tt PowhegHooks} settings as default does not alter the
comparison among the three codes.

\subsection{Scale, PDF and strong-coupling variations}
\label{sec:mwbjSummary}

\begin{table}[tb]
\centering \resizebox{\textwidth}{!}
           { \begin{tabular}{l|c|c|c|c|c|c|c|c|}
 \cline{2-9}
 &  \multicolumn{4}{ |c|}{\phantom{\Big|} No smearing} &   \multicolumn{4}{ |c|}{15 GeV smearing} \\
 \cline{2-9}
 & \phantom{\Big|} $\%$ ${}-{}$ \bbfourl{} & $(\muR, \muF)$ & PDF & $\as$    
 & \% ${}-{}$ \bbfourl{} & $(\muR, \muF)$ & PDF & $\as$  \\
 \cline{1-9}
 \multicolumn{1}{ |c|  }{ \phantom{\Big|}\bbfourl{}}
 & $+           0 $~MeV
 & $ {}_{-          17 }^{+          26 }$~MeV & -                                                                                                 & $\pm            8 $~MeV
 & $+           0 $~MeV
 & ${}_{-          53 }^{+          86 }$~MeV & -                                                                                                 & $\pm           64 $~MeV\\ \cline{1-9}
 \multicolumn{1}{ |c|  }{ \phantom{\Big|}\ttbnlodec{}}
 & $+          21 $~MeV
 & $ {}_{-          10 }^{+           2 }$~MeV & -                                                                                                 & $\pm            8 $~MeV
 & $+         140 $~MeV
 & ${}_{-           6 }^{+           6 }$~MeV & -                                                                                                 & $\pm           54 $~MeV\\ \cline{1-9}
 \multicolumn{1}{ |c|  }{ \phantom{\Big|}\hvq{}}
 & $+          10 $~MeV
 & $ {}_{-           6 }^{+           2 }$~MeV & $\pm      3$~MeV                                                                                  & $\pm            2 $~MeV
 & $        -147 $~MeV
 & ${}_{-           7 }^{+           7 }$~MeV & $\pm      5$~MeV                                                                                  & $\pm            9 $~MeV\\ \cline{1-9}
\end{tabular}
}
\caption{Theoretical uncertainties associated with the $\mwbj{}$ peak
  position extraction for $\mt$=172.5~GeV for the three different
  generators, showered with \PythiaEightPtwo{}. The PDF uncertainty on
  the \bbfourl{} and \ttbnlodec{} generators is assumed to be equal to
  the \hvq{} one, as explained in Chap.~\ref{sec:pheno}.}
\label{tab:mass_extraction-errors}
\end{table}
In Tab.~\ref{tab:mass_extraction-errors} we summarize the
uncertainties due to scale, PDF and strong-coupling variations,
connected with the extraction of the \mwbj{} peak position, for the
input mass $\mt=172.5$~GeV, for all the generators showered with
\PythiaEightPtwo{}.

The upper (lower) error due to scale variation reported in the table
is obtained by taking the maximum (minimum) position of the \mwbj{}
peak for each of the seven scales choices of
eq.~(\ref{eq:scalechoices}), minus the one obtained for the central
scale.

In the PDF case, as discussed in Chap.~\ref{sec:pheno}, we compute the
PDF uncertainties only for the \hvq{} generator, and assume that they
are the same for \bbfourl{} and \ttbnlodec{}.

We consider a symmetrized strong-coupling dependence uncertainty,
whose expression is given by
\begin{equation}
\Delta { \mwbj\left(\as(\mZ)\right)} = \pm \frac{\left|
  \mwbj\left(\as(\mZ)=0.115\right)-\mwbj\left(\as(\mZ)=0.121\right)
  \right|}{2}\,.
\end{equation}
We stress that these variations have only an indicative meaning. In a
realistic analysis, experimental constraints may reduce these
uncertainties.  We also stress that these are not the only theoretical
uncertainties. Others may be obtained by varying MC
parameters. Here we focus specifically on those uncertainties that are
associated with the NLO+PS generators.

As we have already discussed, the use of the \hvq{} and the
\ttbnlodec{} generators would lead to a negligible bias in the \mwbj{}
distribution if we were able to measure it without any resolution
effects. However, if we introduce a smearing to mimic them, the
description of the region away from the peak plays an important role,
and the \hvq{} and \ttbnlodec{} generators yield predictions for the
mass peak position that are shifted by roughly \diffaverage~MeV in the
downward and upward direction respectively with respect to \bbfourl{}.

We also notice that the \bbfourl{} generator is the most affected by
theoretical uncertainties. In particular, the \ttbnlodec{} and \hvq{}
generators have an unrealistically small scale dependence of the peak
shape, due to the way in which off-shell effects are approximately
described.  The \ttbnlodec{} generator displays a non-negligible
sensitivity only to the strong-coupling constant.  The theoretical
errors that we have studied here lead to very small effects for the
\hvq{} generator, since it does not include radiative corrections in
the top decay. On the other hand, the \hvq{} generator is bound to be
more sensitive to variation of parameters in \PythiaEightPtwo{}, that
in this case fully controls the radiation from the $b$ quark.

\subsection{Radius dependence}
In this section we investigate the stability of the previous results
with respect to the choice of the jet radius. The results are
summarized in Tab.~\ref{tab:mass_extraction-radius}.
\begin{table}[tb]
\centering \resizebox{\textwidth}{!}
           { \begin{tabular}{l|c|c|c|c|c|c|}
 \cline{2-7}
 &  \multicolumn{2}{ |c|}{ \phantom{\Big|} $R=0.4$}
 &  \multicolumn{2}{ |c|}{ \phantom{\Big|} $R=0.5$}
 &  \multicolumn{2}{ |c|}{ \phantom{\Big|} $R=0.6$} \\
 \cline{2-7}
 & \phantom{\Big|} No smearing & 15~GeV smearing
 & \phantom{\Big|} No smearing & 15~GeV smearing
 & \phantom{\Big|} No smearing & 15~GeV smearing \\
 \cline{1-7}
 \multicolumn{1}{ |c|  }{ \phantom{\Big|} \bbfourl{} [GeV]}

 & $ 172.156\pm  0.004$ & $ 171.018\pm  0.002$

 & $ 172.793\pm  0.004$ & $ 172.717\pm  0.002$

 & $ 173.436\pm  0.005$ & $ 174.378\pm  0.002$
\\ \cline{1-7}
\multicolumn{1}{ |c|  }{ \phantom{\Big|}\ttbnlodec{} ${}-$  \bbfourl{}}
 & $ +    35
\pm      5$~MeV
 & $ +   195
\pm      2$~MeV
 & $ +    21
\pm      6$~MeV
 & $ +   140
\pm      2$~MeV
 & $ +     1
\pm      7$~MeV
 & $ +    97
\pm      2$~MeV
\\ \cline{1-7}
\multicolumn{1}{ |c|  }{ \phantom{\Big|}\hvq{} ${}-$  \bbfourl{}}
 & $ +    47
\pm      5$~MeV
 & $    -113
\pm      2$~MeV
 & $ +    10
\pm      6$~MeV
 & $    -147
\pm      2$~MeV
 & $      -7
\pm      6$~MeV
 & $    -174
\pm      2$~MeV
\\ \cline{1-7}
\end{tabular}
 }
\caption{ \mwbj{} peak position obtained with the \bbfourl{} generator for
  three choices of the jet radius. The differences with the \ttbnlodec{} and
  the \hvq{} generators are also shown.}
\label{tab:mass_extraction-radius}
\end{table}
For the distributions without smearing, the differences between the three
generators are small and decrease as $R$ increases.  For the smeared
distributions, the differences between \ttbnlodec{} and \bbfourl{} decrease
as the radius increases, while the difference between the \hvq{} and the
\bbfourl{} generator increases.

The small differences in the $R$ dependence among the three generators in the
non-smeared cases can be understood if we consider that differences in the
$b$ radiation do not affect much the peak position in the non-smeared
distribution, but rather they affect the strength of the tail on the left
side of the peak. On the other hand, the peak position is affected by
radiation in production and by the underlying-event structure, that is very
similar in the three generators.

It should be noticed that the difference between the displacements of the
\ttbnlodec{} and \hvq{} with respect to \bbfourl{} is less than
\diffdiffRttdec~MeV and \diffdiffRhvq~MeV, respectively, below the current
statistical precision of top-mass measurements. Thus, the good agreement
found among the three generators persists also for different $R$ values.

\section{Comparison with \HerwigSevenPone{}}
\label{sec:mwbj_hw7}
In order to assess uncertainties due to the shower Monte Carlo~(SMC)
program, in this section we compare the results obtained using
\HerwigSevenPone{} and \PythiaEightPtwo{}.
\begin{table}[tb]
\centering \resizebox{\textwidth}{!}
           { \begin{tabular}{l|c|c|c|c|}
 \cline{2-5}
 &  \multicolumn{2}{ |c|}{ \phantom{\Big|} No smearing}
 &  \multicolumn{2}{ |c|}{15~GeV smearing} \\
 \cline{2-5}
 & \phantom{\Big|} {\HerwigSevenPlot} & {\PythiaEightPlot} ${}-$ {\HerwigSevenPlot} 
 & \phantom{\Big|} {\HerwigSevenPlot} & {\PythiaEightPlot} ${}-$ {\HerwigSevenPlot}\\
 \cline{1-5}
 \multicolumn{1}{ |c|  }{ \phantom{\Big|}\bbfourl{}}
 & $172.727\pm   0.005$~GeV 
  &  $+          66 \pm            7 $~MeV
 & $171.626\pm   0.002$~GeV 
  &  $+        1091 \pm            2 $~MeV
 \\ \cline{1-5}
 \multicolumn{1}{ |c|  }{ \phantom{\Big|}\ttbnlodec{}}
 & $172.775\pm   0.004$~GeV 
  &  $+          39 \pm            5 $~MeV
 & $171.678\pm   0.001$~GeV 
  &  $+        1179 \pm            2 $~MeV
 \\ \cline{1-5}
 \multicolumn{1}{ |c|  }{ \phantom{\Big|}\hvq{}}
 & $173.038\pm   0.004$~GeV 
  &  $        -235 \pm            5 $~MeV
 & $172.319\pm   0.001$~GeV 
  &  $+         251 \pm            2 $~MeV
 \\ \cline{1-5}
\end{tabular}
}
\caption{$\mwbj{}$ peak position for $\mt$=172.5~GeV obtained with the
  three different generators, showered with
  \HerwigSevenPone{}~({\HerwigSevenPlot}). The differences with
  \PythiaEightPtwo{}~({\PythiaEightPlot}) are also shown.}
\label{tab:mass_extraction-shower}
\end{table}
In Tab.~\ref{tab:mass_extraction-shower} we compare the \mwbj{} peak
position extracted for the input mass $\mt = 172.5$~GeV using the
three generators showered with \PythiaEightPtwo{} and
\HerwigSevenPone{}.  For the \hvq{} generator, the differences are of
the order of \pyminushwhvq~MeV for both the smeared and non-smeared
case, but with opposite signs.  In the smeared case, both the
\ttbnlodec{} and \bbfourl{} generators yield much larger differences,
of more than 1~GeV.

In Tab.~\ref{tab:mass_extraction-shower-showerOnly} we report the
differences between the \HerwigSevenPone{} and \PythiaEightPtwo{}
predictions for all the generators, at the NLO+PS level and at the
full hadron level.
\begin{table}[tb]
\centering
{ \begin{tabular}{l|c|c|c|c|}
 \cline{2-5}
 &  \multicolumn{4}{ |c|}{ $\phantom{\Big|}$ \PythiaEightPtwo{} ${}-$ \HerwigSevenPone{}} \\
 \cline{2-5}
 &  \multicolumn{2}{ |c|}{PS only}
 &  \multicolumn{2}{ |c|}{ \phantom{\Big|} full}\\
 \cline{2-5}
 & \phantom{\Big|} No smearing & 15~GeV smearing
 & \phantom{\Big|} No smearing & 15~GeV smearing \\
 \cline{1-5}
 \multicolumn{1}{ |c|  }{ \phantom{\Big|}\bbfourl{}}
  &  $+          10 \pm            2 $~MeV
  &  $+         984 \pm            2 $~MeV
  &  $+          66 \pm            7 $~MeV
  &  $+        1091 \pm            2 $~MeV
 \\ \cline{1-5}
 \multicolumn{1}{ |c|  }{ \phantom{\Big|}\ttbnlodec{}}
  &  $+           5 \pm            2 $~MeV
  &  $+        1083 \pm            2 $~MeV
  &  $+          39 \pm            5 $~MeV
  &  $+        1179 \pm            2 $~MeV
 \\ \cline{1-5}
 \multicolumn{1}{ |c|  }{ \phantom{\Big|}\hvq{}}
  &  $-           0 \pm            2 $~MeV
  &  $+         113 \pm            2 $~MeV
  &  $        -235 \pm            5 $~MeV
  &  $+         251 \pm            2 $~MeV
 \\ \cline{1-5}
\end{tabular}
}
\caption{Differences between \PythiaEightPtwo{} and \HerwigSevenPone{}
  in the extracted $\mwbj{}$ peak position for $\mt$=172.5~GeV
  obtained with the three different generators, at the NLO+PS level
  (PS only) and including also the underlying events, the multi-parton
  interactions and the hadronization~(full).}
\label{tab:mass_extraction-shower-showerOnly}
\end{table}
We notice that at the NLO+PS level and without smearing, the
differences between the two parton-shower programs are negligible. For
the smeared distributions, at both the NLO+PS and full level, the
differences are roughly 1~GeV for the \bbfourl{} and the \ttbnlodec{}
generator. For \hvq{} the differences are considerably smaller,
although not quite negligible.

\begin{figure}[tb!]
\centering
\includegraphics[width=\wfigdoub]{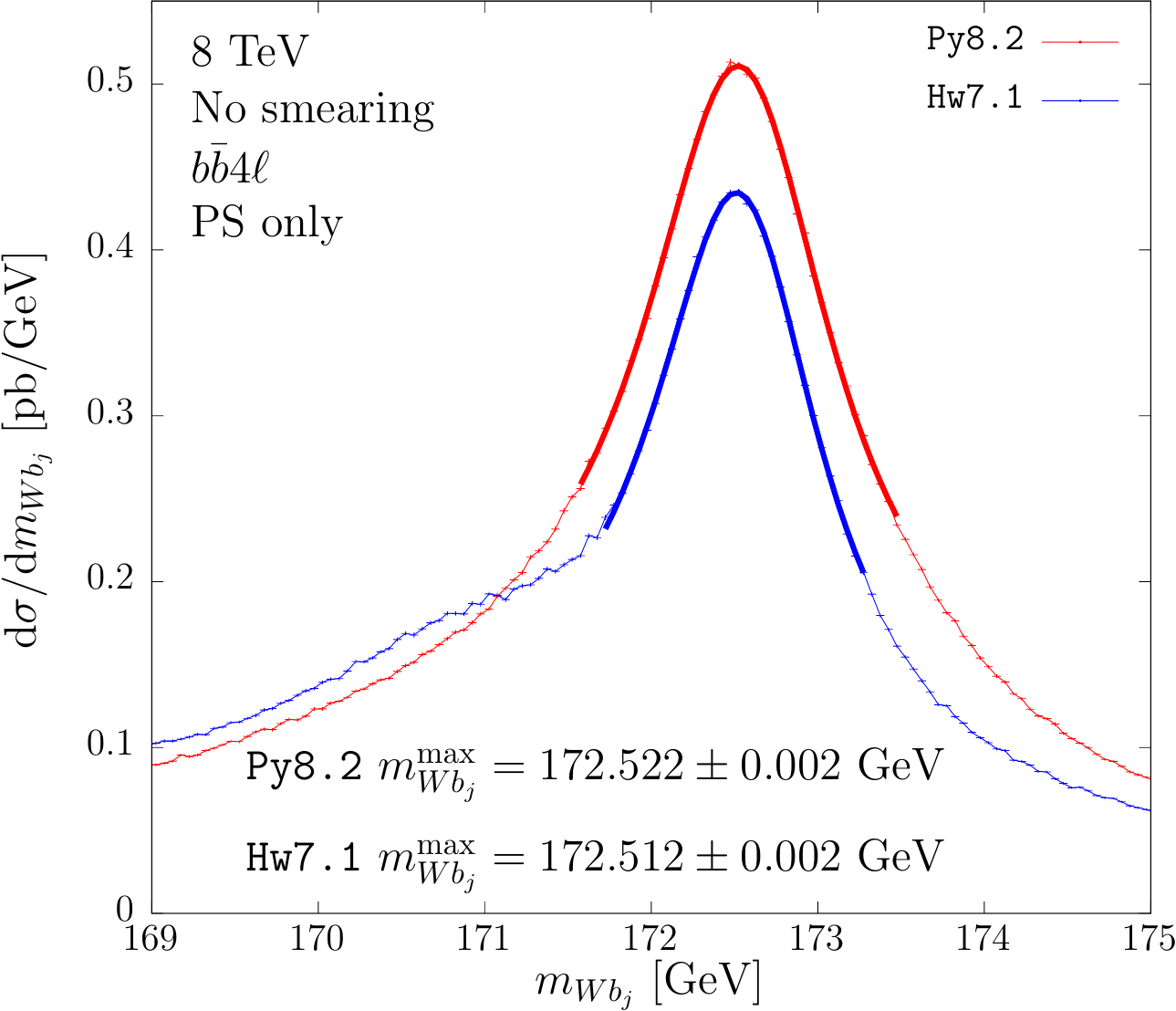}
\includegraphics[width=\wfigdoub]{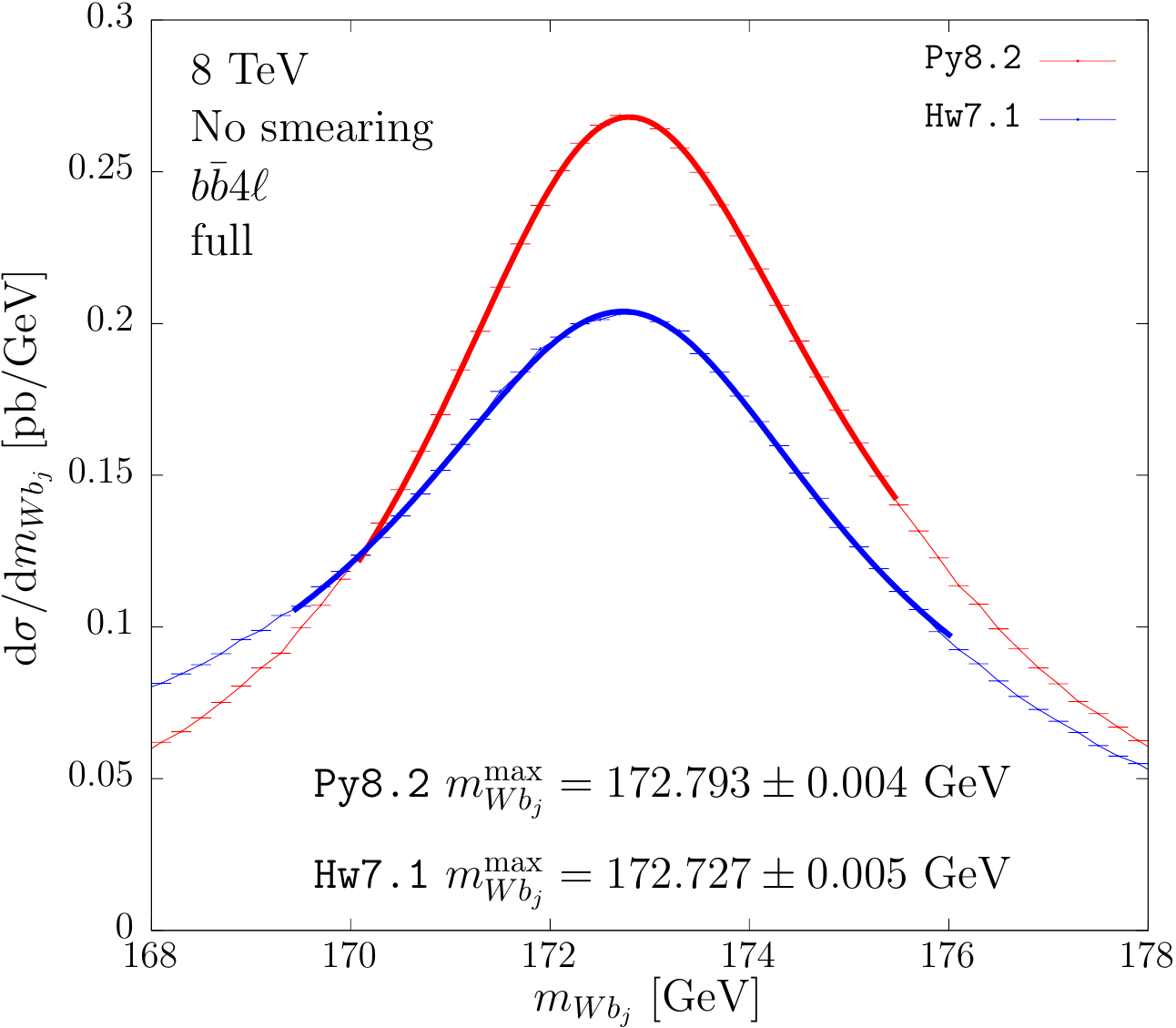}
  \caption{${d\sigma}/{d \mwbj}$ distribution obtained by showering
    the \bbfourl{} results with \PythiaEightPtwo{} and
    \HerwigSevenPone{}, at parton-shower level~(left) and with
    hadronization and underlying events~(right).}
\label{fig:mwbjshapespyh7}
\end{figure}

\begin{figure}[tb!]
\centering
\includegraphics[width=\wfigsing]{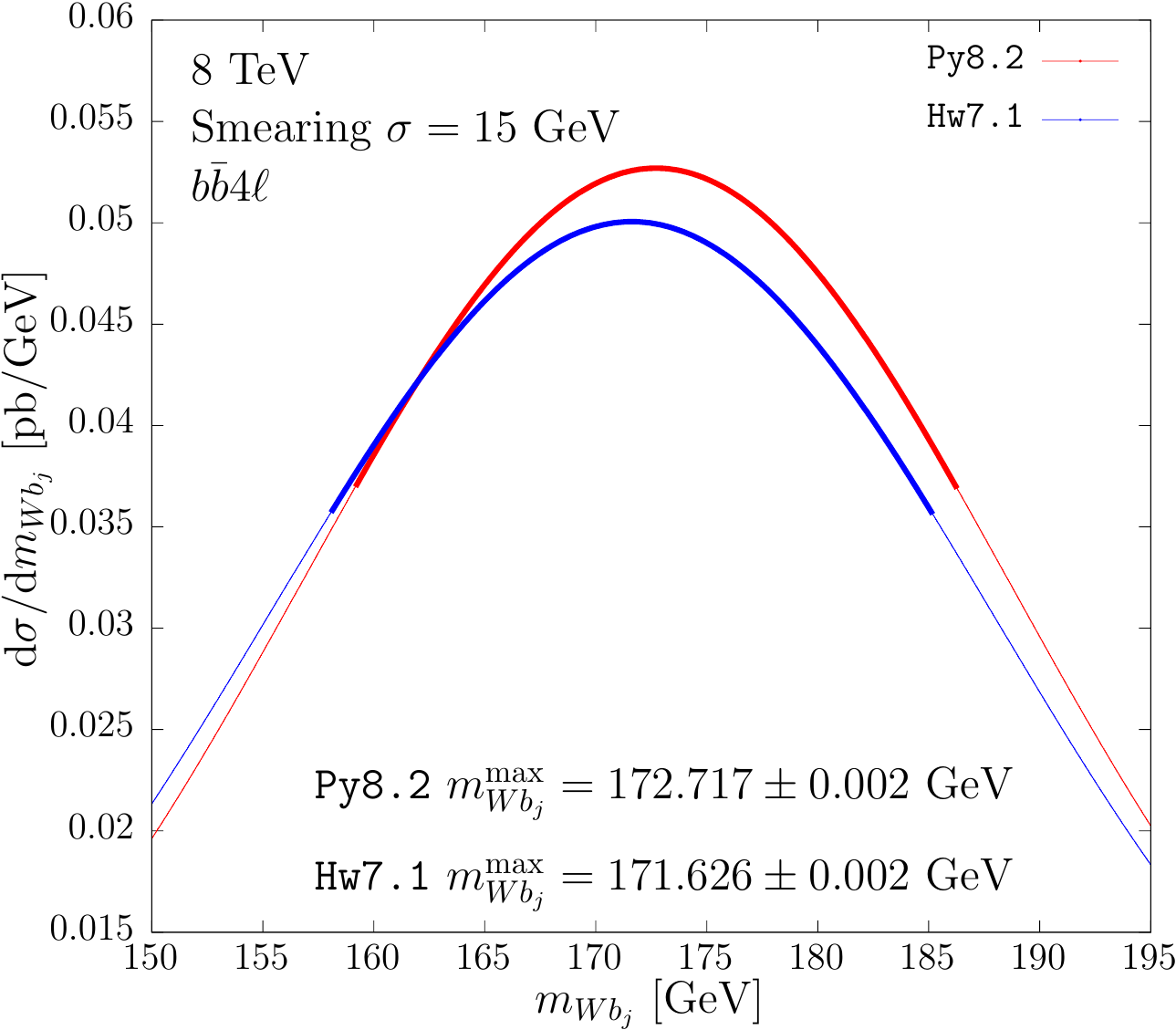}
\caption{Smeared ${d\sigma}/{d \mwbj}$ distribution obtained by
  matching the \bbfourl{} generator with \PythiaEightPtwo{} and
  \HerwigSevenPone{}.}
\label{fig:mwbjshapespyh7smeared}
\end{figure}

The origin of these large differences is better understood by looking
at the differential cross sections plotted in
Figs.~\ref{fig:mwbjshapespyh7} and~\ref{fig:mwbjshapespyh7smeared}.
 In Fig.~\ref{fig:mwbjshapespyh7} we plot the results for the
 non-smeared case, at the NLO+PS level~(left) and at the full hadron
 level~(right): while the peak position is nearly the same for both
 \PythiaEightPtwo{} and \HerwigSevenPone{}, the shape of the curves is
 very different around the peak, leading to a different mass peak
 position when smearing is applied, as displayed in
 Fig.~\ref{fig:mwbjshapespyh7smeared}.  We notice that in this last
 case we see a difference in shape also after smearing.  This suggests
 that at least one of the two generators may not describe the data
 fairly.

\begin{figure}[tb]
\centering
\includegraphics[width=\wfigsing]{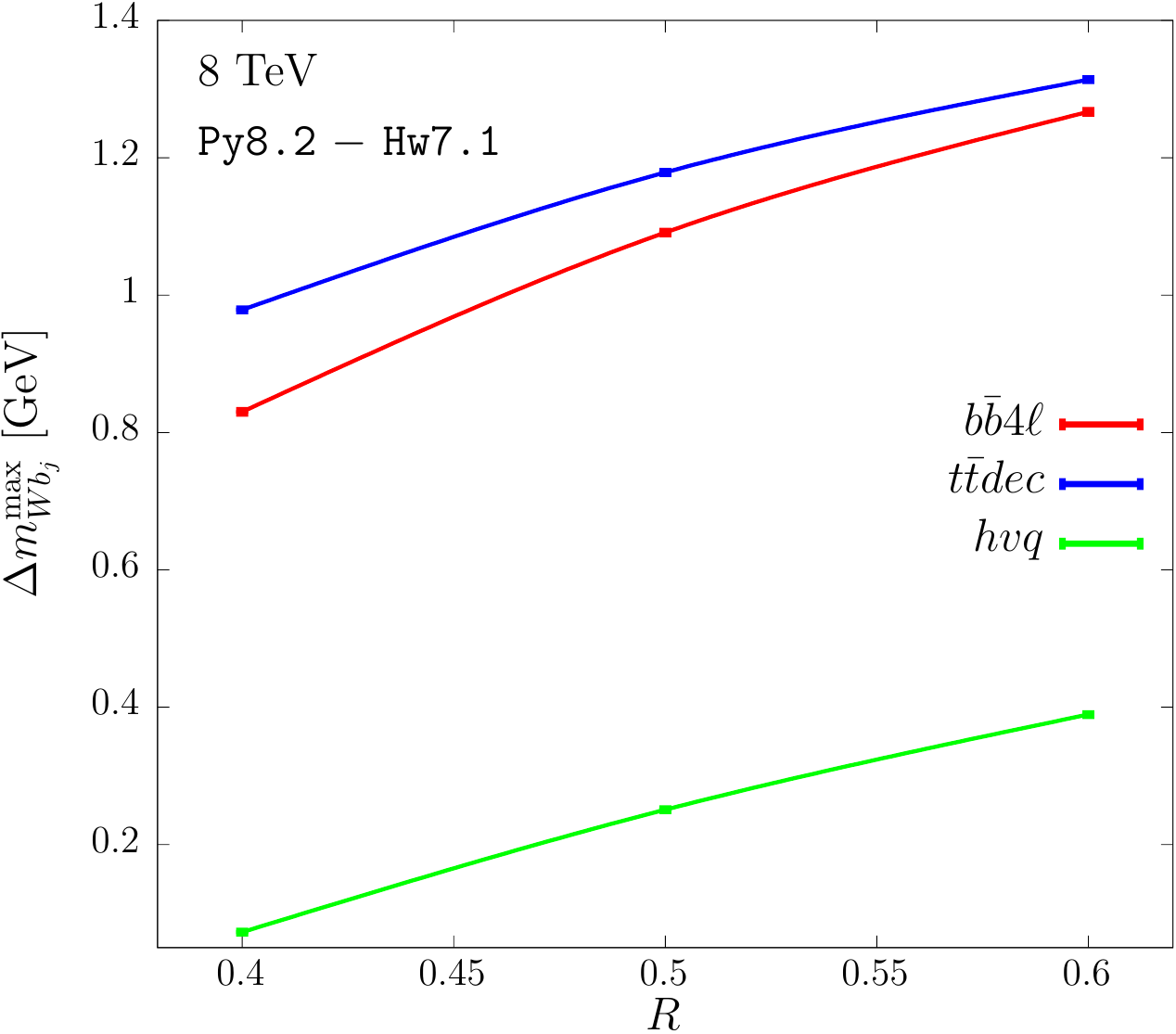}
 \caption{Differences of $\mwbjmax$ between the \PythiaEightPtwo{} and
   the \HerwigSevenPone{} showers, for the three generators, as a
   function of the jet radius.}
 \label{fig:R_mwbj_py8-hw7}
\end{figure}

Since we observe such large differences in the value of $\mwbjmax$ in
\HerwigSevenPone{} and \PythiaEightPtwo{}, we have also studied
whether sizeable differences are also present in the $\mwbjmax$
dependence upon the jet radius $R$.  The results are shown in
Tab.~\ref{tab:mass_extraction-shower-radius}, and displayed in
Fig.~\ref{fig:R_mwbj_py8-hw7}.
\begin{table}[tb]
\centering \resizebox{\textwidth}{!}
           { \begin{tabular}{l|c|c|c|c|c|c|}
 \cline{2-7}
 & \multicolumn{6}{ |c|}{ \phantom{\Big|}  \PythiaEightPtwo{} ${}-$ \HerwigSevenPone{}} \\
 \cline{2-7}
 &  \multicolumn{2}{ |c|}{ \phantom{\Big|} $R=0.4$}
 &  \multicolumn{2}{ |c|}{ \phantom{\Big|} $R=0.5$}
 &  \multicolumn{2}{ |c|}{ \phantom{\Big|} $R=0.6$} \\
 \cline{2-7}
 & \phantom{\Big|} No smearing & 15~GeV smearing
 & \phantom{\Big|} No smearing & 15~GeV smearing
 & \phantom{\Big|} No smearing & 15~GeV smearing \\
 \cline{1-7}
 \multicolumn{1}{ |c|  }{ \phantom{\Big|}\bbfourl{}}
 & $     -98
\pm      7$~MeV
 & $ +   830
\pm      2$~MeV
 & $ +    66
\pm      7$~MeV
 & $ +  1091
\pm      2$~MeV
 & $ +   253
\pm      8$~MeV
 & $ +  1267
\pm      2$~MeV
\\ \cline{1-7}
 \multicolumn{1}{ |c|  }{ \phantom{\Big|}\ttbnlodec{}}
 & $    -100
\pm      5$~MeV
 & $ +   979
\pm      2$~MeV
 & $ +    39
\pm      5$~MeV
 & $ +  1179
\pm      2$~MeV
 & $ +   210
\pm      6$~MeV
 & $ +  1314
\pm      2$~MeV
\\ \cline{1-7}
 \multicolumn{1}{ |c|  }{ \phantom{\Big|}\hvq{}}
 & $    -370
\pm      5$~MeV
 & $ +    73
\pm      2$~MeV
 & $    -235
\pm      5$~MeV
 & $ +   251
\pm      2$~MeV
 & $     -31
\pm      6$~MeV
 & $ +   389
\pm      2$~MeV
\\ \cline{1-7}
\end{tabular}
}
\caption{ Differences in the \mwbj{} peak position obtained matching
  the three generators with \PythiaEightPtwo{} and \HerwigSevenPone{},
  for three choices of the jet radius.}
\label{tab:mass_extraction-shower-radius}
\end{table}
In the case of the \bbfourl{} generator, the difference between
\PythiaEightPtwo{} and \HerwigSevenPone{} goes from \pyminushwRfour{}
to \pyminushwRsix~MeV. Thus, assuming for instance that
\PythiaEightPtwo{} fits the data perfectly, i.e.~that it extracts the
same value of the mass by fitting the $\mwbjmax$ values obtained with
the three different values of $R$, \HerwigSevenPone{} would extract at
$R=0.6$ a mass value that is larger by \pyminushwdeltaRfoursix{}~MeV
from the one extracted at $R=0.4$.  We stress that the differences in
the $R$ behaviour of $\mwbjmax$ may have the same origin as the
difference in the reconstructed mass value, since both effects may be
related to the amount of energy that enters the jet cone, and it is
not unlikely that, by tuning one of the two generators in such a way
that they both have the same $R$ dependence, their difference in
$\mwbjmax$ would also be reduced.\footnote{Similarly, one could fit
  appropriate calibration observables associated to the $b$-jet
  structure, along the lines of Ref.~\cite{Corcella:2017rpt}.} It is
unlikely, however that this would lead to a much improved agreement,
since the difference in slope is much less pronounced than the
difference in absolute value.

\subsection{Alternative settings in \HerwigSevenPone{}}
We have examined several variations in the \HerwigSevenPone{}
settings, in order to understand whether the \HerwigSevenPone{}
results are reasonably stable, or depend upon our particular settings.

\subsubsection{MEC and POWHEG options in \HerwigSevenPone{}}
\HerwigSevenPone{} applies matrix-element corrections by default, but
it also offers the possibility to replace the MEC with its internal
POWHEG method, when available, to achieve NLO accuracy in top decays.
These options are activated by the instructions\fignewline
\begin{verbatim}
set ShowerHandler:HardEmission None
\end{verbatim}
or
\begin{verbatim}
set ShowerHandler:HardEmission POWHEG
\end{verbatim}
respectively.  We have verified that, as expected, switching off the
matrix-element corrections does not significantly affect the
\bbfourl{} and \ttbnlodec{} results. In the case of the \hvq{}
generator, we can compare the default case, where MEC is on, with the
cases where POWHEG replaces MEC, and with the case where neither MEC
nor POWHEG is implemented.  The results are shown in
Tab.~\ref{tab:mass_extraction-hvq-mec-powheg-herwig}.
\begin{table}[tb]
\centering
{ \begin{tabular}{|c|c|c|}
 \cline{1-3}
 \hvq & \phantom{\Big|} No smearing  & 15~GeV smearing \\
 \cline{1-3}
 \multicolumn{1}{ |c|  }{ \phantom{\Big|} MEC ${}-$ no MEC}
  & $         307 \pm            6 $~MeV  & $        1371 \pm            2 $~MeV 
 \\ \cline{1-3}
 \multicolumn{1}{ |c|  }{ \phantom{\Big|} MEC ${}-$ {\tt POWHEG}}
  & $         244 \pm            6 $~MeV  & $         356 \pm            2 $~MeV 
 \\ \cline{1-3}
\end{tabular}
}
\caption{Differences in the $\mwbj{}$ peak position for the \hvq{}
  generator showered with \HerwigSevenPone{}, with MEC switched off
  (no MEC) or using the \HerwigSevenPone{} POWHEG option, with respect
  to our default setting, that has MEC switched on.}
\label{tab:mass_extraction-hvq-mec-powheg-herwig}
\end{table}
We notice that the inclusion of MEC enhances by more than 1.3~GeV the
peak position of the smeared distribution. A similar result was found
in \PythiaEightPtwo{} (see Tab.~\ref{tab:mwbj_MEC}), where the
difference was slightly less than 1~GeV.  The difference between the
POWHEG and MEC results is much below the 1~GeV level but not
negligible. This fact is hard to understand, since the POWHEG and MEC
procedures should only differ by a normalization factor.

We have seen previously that the three NLO+PS generators interfaced to
\PythiaEightPtwo{} yield fairly consistent results for the
reconstructed top mass peak. The same consistency is not found when
they are interfaced to \HerwigSevenPone{}. However, the best agreement
is found when the internal POWHEG option for top decay is activated in
\HerwigSevenPone{}, as can be seen in Tab.~\ref{tab:cmp_herwig_table}.
\begin{table}[tb]
\centering \resizebox{\textwidth}{!}
           { \begin{tabular}{l|c|c|c|c|}
 \cline{2-5}
 &  \multicolumn{2}{ |c|}{PS only}
 &  \multicolumn{2}{ |c|}{ \phantom{\Big|} full}\\
 \cline{2-5}
 & \phantom{\Big|} No smearing & 15~GeV smearing
 & \phantom{\Big|} No smearing & 15~GeV smearing \\
 \cline{1-5}
 \multicolumn{1}{ |c|  }{ \phantom{\Big|}  \bbfourl{}}
 & $172.512\pm  0.002$~GeV
 & $170.419\pm  0.002$~GeV
 & $172.727\pm  0.005$~GeV
 & $171.626\pm  0.002$~GeV
 \\ \cline{1-5}
 \multicolumn{1}{ |c|  }{ \phantom{\Big|}\ttbnlodec{} ${}-$ \bbfourl{}}
  &  $         -13 \pm            2 $~MeV
  &  $+          92 \pm            2 $~MeV
  &  $+          48 \pm            7 $~MeV
  &  $+          52 \pm            2 $~MeV
 \\ \cline{1-5}
 \multicolumn{1}{ |c|  }{ \phantom{\Big|}\hvq{} ${}-$ \bbfourl{}}
  &  $         -14 \pm            2 $~MeV
  &  $+         782 \pm            2 $~MeV
  &  $+         311 \pm            7 $~MeV
  &  $+         693 \pm            2 $~MeV
 \\ \cline{1-5}
 \multicolumn{1}{ |c|  }{ \phantom{\Big|}\hvq{}+PWG ${}-$ \bbfourl{}}
  &  $         -16 \pm            2 $~MeV
  &  $+         479 \pm            2 $~MeV
  &  $+          67 \pm            7 $~MeV
  &  $+         337 \pm            2 $~MeV
 \\ \cline{1-5}
\end{tabular}
}
\caption{Differences of \hvq{} and \ttbnlodec{} with respect to
  \bbfourl{}, all showered with \HerwigSevenPone{}. The result
  obtained using the \HerwigSevenPone{} internal POWHEG implementation
  of top decay, rather than MEC, labelled as \hvq+PWG, is also shown.}
\label{tab:cmp_herwig_table}
\end{table}
The difference between the POWHEG and MEC or POWHEG \HerwigSevenPone{}
results is puzzling, since they have the same formal accuracy. We will
comment about this issue later on.

\subsubsection{Veto procedures in \HerwigSevenPone{}}
\label{sec:HW7_different_showers_results}
\begin{table}[tb]
\centering \resizebox{\textwidth}{!}
           { \begin{tabular}{l|c|c|c|c|}
 \cline{2-5}
 &  \multicolumn{2}{ |c|}{ \phantom{\Big|} No smearing}
 &  \multicolumn{2}{ |c|}{15~GeV smearing} \\
 \cline{2-5}
 & \phantom{\Big|} {\tt FSV} & {\tt FSV} ${}-$ {\tt SV}
 & \phantom{\Big|} {\tt FSV} & {\tt FSV} ${}-$ {\tt SV}\\
 \cline{1-5}
 \multicolumn{1}{ |c|  }{ \phantom{\Big|}\bbfourl{}}
 & $172.776\pm   0.005$~GeV 
  &  $+          49 \pm            7 $~MeV
 & $171.829\pm   0.002$~GeV 
  &  $+         203 \pm            2 $~MeV
 \\ \cline{1-5}
 \multicolumn{1}{ |c|  }{ \phantom{\Big|}\ttbnlodec{}}
 & $172.810\pm   0.004$~GeV 
  &  $+          35 \pm            6 $~MeV
 & $171.906\pm   0.001$~GeV 
  &  $+         228 \pm            2 $~MeV
 \\ \cline{1-5}
\end{tabular}
}
\caption{$\mwbj{}$ peak position for $\mt$=172.5~GeV for \bbfourl{}
  and \ttbnlodec{} showered with \HerwigSevenPone{} using the {\tt
    FullShowerVeto}~({\tt FSV}) procedure. The differences with {\tt
    ShowerVeto}~({\tt SV}), that represents our default, are also
  shown. }
\label{tab:mass_extraction-matching-hw7}
\end{table}
As discussed in \writeApp.~\ref{sec:herwig}, \HerwigSevenPone{} offers
two different classes that implement the veto procedure: the {\tt
  ShowerVeto}, our default one, where the veto is performed at the
emission level, and the {\tt FullShowerVeto} class, where the veto is
performed at the end of the whole showering phase.  The corresponding
results are summarized in Tab.~\ref{tab:mass_extraction-matching-hw7}.
For both the \bbfourl{} and the \ttbnlodec{} the two procedures lead
to a 200~MeV difference in the peak position for the smeared
distributions.  The origin of such difference is not fully clear to
us. In part it may be ascribed to the fact that when using the {\tt
  ShowerVeto} class we mix two different definitions of transverse
momentum (the \HerwigSevenPone{} and the POWHEG one), and in part may
be due to the fact that in the {\tt FullShowerVeto} class the vetoing
is done on the basis of the shower structure after reshuffling has
been applied.

\subsubsection{Truncated showers}
In Ref.~\cite{Nason:2004rx} it was shown that, when interfacing a
POWHEG generator to an angular-ordered shower, in order to compensate
for the mismatch between the angular-ordered scale and the POWHEG
hardness, that is taken equal to the relative transverse momentum in
radiation, one should supply appropriate truncated showers. None of
our vetoing algorithms take them into account, but it turns out that
\HerwigSevenPone{} provides facilities to change the settings of the
initial showering scale according to the method introduced in
Ref.~\cite{Schofield:2011zi}, that, in our case, are equivalent to the
inclusion of truncated showers (see \writeApp\ref{sec:herwig}).  The
effects this inclusion, performed with the settings of
eq.~(\ref{eq:TSsettings}), for the \bbfourl{} and \ttbnlodec{}
generators are shown in
Tab.~\ref{tab:mass-extraction-truncated-shower-hw7}.
\begin{table}[tb]
\centering \resizebox{\textwidth}{!}
           { \begin{tabular}{l|c|c|c|c|}
 \cline{2-5}
 &  \multicolumn{2}{ |c|}{ \phantom{\Big|} No smearing}
 &  \multicolumn{2}{ |c|}{15~GeV smearing} \\
 \cline{2-5}
 & \phantom{\Big|} TS & TS ${}-$ default
 & \phantom{\Big|} TS  & TS ${}-$ default\\
 \cline{1-5}
 \multicolumn{1}{ |c|  }{ \phantom{\Big|}\bbfourl{}}
 & $172.730\pm   0.005$~GeV
  &  $+           3 \pm            8 $~MeV
 & $171.496\pm   0.002$~GeV
  &  $        -130 \pm            2 $~MeV
 \\ \cline{1-5}
 \multicolumn{1}{ |c|  }{ \phantom{\Big|}\ttbnlodec{}}
 & $172.786\pm   0.004$~GeV
  &  $+          12 \pm            6 $~MeV
 & $171.546\pm   0.001$~GeV
  &  $        -132 \pm            2 $~MeV
 \\ \cline{1-5}
\end{tabular}
}
\caption{$\mwbj{}$ peak position for $\mt$=172.5~GeV obtained with the
  \bbfourl{} and \ttbnlodec{} generators showered with
  \HerwigSevenPone{}, with the settings of eq.~(\ref{eq:TSsettings})
  (labelled as TS). The differences with the default results are also
  shown.}
\label{tab:mass-extraction-truncated-shower-hw7}
\end{table}
As we can see, this does not introduce dramatic changes in the peak
position: in fact the differences are negligible in the distributions
without smearing, and are roughly \hwTSsmearbbfourlttdec~MeV when
smearing is applied.  It should be noticed that these settings
slightly increase the difference with respect to the results obtained
with \PythiaEightPtwo{}.
 
\chapter{The energy of the $\boldsymbol{b}$ jet}
\label{sec:Ebjet}
In Ref.~\cite{Agashe:2016bok} it was proposed to extract $\mt$ using
the peak of the energy spectrum of the $b$ jet.  This method has been
investigated by the CMS collaboration in
Ref.~\cite{CMS-PAS-TOP-15-002}, where it was found
\begin{equation}
\mt = 172.29 \, \pm \,1.17{\rm \, (stat) \,} \pm 2.66 {\rm \,
  (syst)}~{\rm GeV}
\end{equation}

At leading order, the $b$ jet consists of the $b$ quark alone, and its
energy in the top rest frame, neglecting top-width effects, is fixed
and given by
\begin{equation}
\Ebjmax=\frac{\mt^2-m_W^2+m_b^2}{2\,\mt}\,,
\label{eq:ebjlo}
\end{equation}
i.e.~the spectrum is a delta function in the energy and its value is
independent from the top-production mechanism.  In the laboratory
frame, because of the variable boost that affects the top, the delta
function is smeared into a wider distribution, but it can be shown
that its peak position remains at $\Ebjmax$. On the basis of this
observation we are led to assume that also after the inclusion of
off-shell effects, radiative and non-perturbative corrections, the
relation between $\Ebjmax$ and the top pole-mass $\mt$ should be
largely insensitive to production dynamics.

We performed a study of the $\Ebjmax$ observable along the same lines
adopted for \mwbj{} in the previous section.  If the range of
variations of the top mass around a given central value $\mtc$ is
small enough, a linear relation between $\Ebjmax$ and the top mass
must hold, so that we can write
\begin{equation}
  \label{eq:B-for-ebj}
\Ebjmax(\mt)= \Ebjmax(\mtc) +B\,(\mt-\mtc)+\mathcal{O}(\mt-\mtc)^2.
\end{equation}

It was suggested in Ref.~\cite{CMS-PAS-TOP-15-002} that the $\Ebj$
distribution $\mathd \sigma/\mathd \Ebj$ is better fitted in terms of
$\log \Ebj$. Thus, in order to extract the peak position, we fitted
the energy distribution with a fourth order polynomial
\begin{equation}
y=a+b(x-x^{\rm max})^2+c(x-x^{\rm max})^3+d(x-x^{\rm max})^4\,,
\end{equation}
where $x=\log\Ebj$.  The fitting procedure is the same employed to
extract \mwbjmax{}, that is described in
Sec.~\ref{sec:Mwbjmax_extraction}

The parameter $B$ of eq.~\eqref{eq:B-for-ebj}, extracted from a linear
fit of the three \Ebjmax{} values corresponding to the three different
values of $\mt$ that we have considered (see Tab.~\ref{tab:samples})
using the \hvq{} generator showered by \PythiaEightPtwo{}, was found
to be
\begin{equation}
  \label{eq:B_Ebj}
  B= \BfromEbjhvq \pm \BerrfromEbjhvq \, ,
\end{equation}
compatible with the expected value of 0.5 from
eq.~\eqref{eq:ebjlo}.\footnote{When using the \bbfourl{} generator we
  obtain $B= \BfromEbjbbfourl \pm \BerrfromEbjbbfourl$, while with the
  \ttbnlodec{} one, we get $B= \BfromEbjttdec \pm \BerrfromEbjttdec$.
  When using \HerwigSevenPone{} instead of \PythiaEightPtwo{}, we find
  values compatible with the given ones within 10\%{}.}

\section{Comparison among different NLO+PS generators}
\begin{figure}[tb]
  \centering
  \includegraphics[width=\wfigsing]{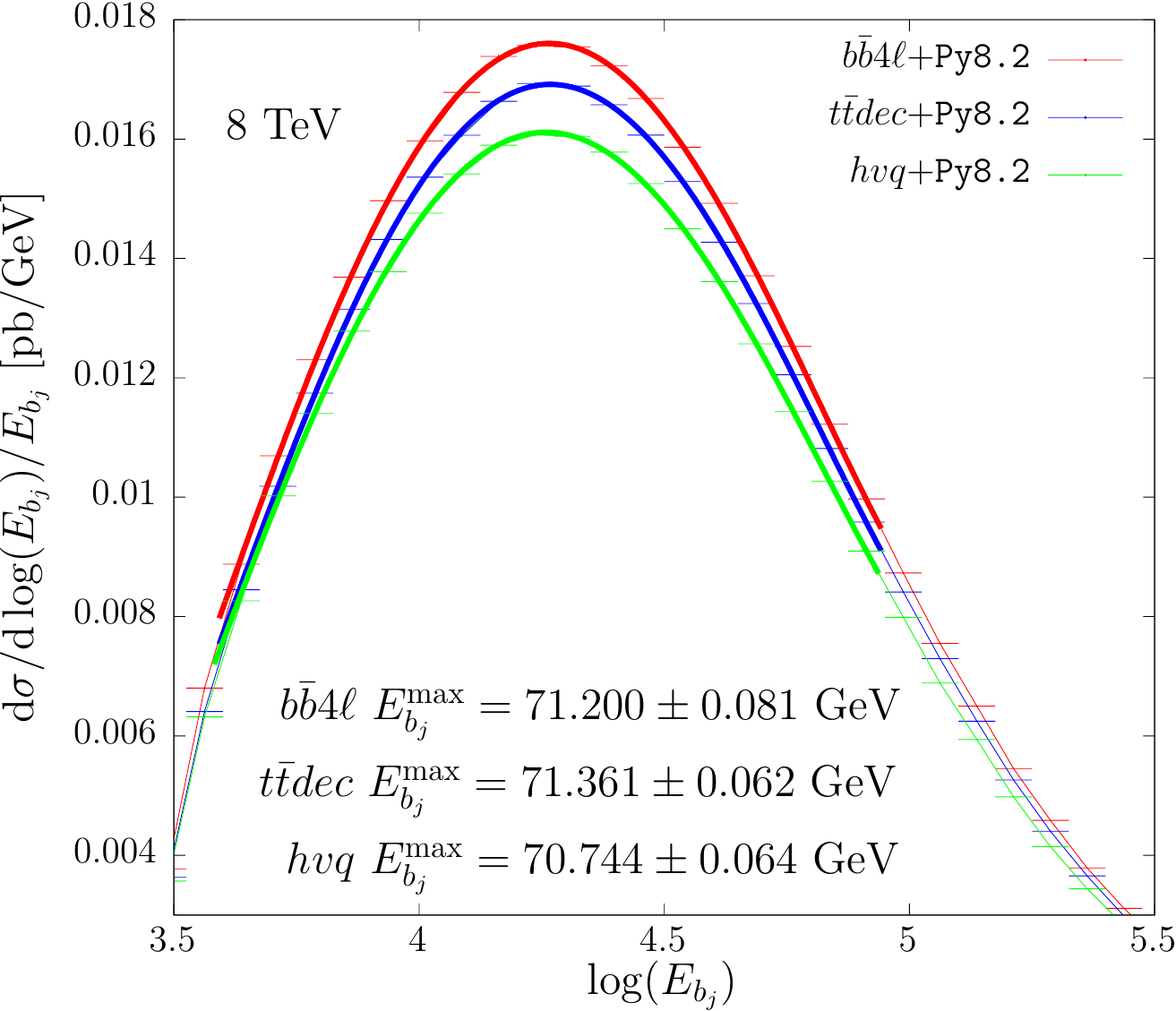}
  \caption{Logarithmic energy distribution obtained with the three
    generators interfaced to \PythiaEightPtwo, together with their
    polynomial fit, in the range displayed in the figure. The value of
    $\Ebjmax$ for each generator is also reported.}
\label{fig:Ebj_bb4l_ttdec_hvq}
\end{figure}

In Fig.~\ref{fig:Ebj_bb4l_ttdec_hvq} we plot the logarithmic energy
distribution for the three generators interfaced to \PythiaEight,
together with their polynomial fit.  The extracted \Ebj{} peaks from
the \bbfourl{} and the \ttbnlodec{} generators are compatible within
the statistical errors. On the other hand, the \hvq{} generator yields
a prediction which is roughly \Ebjbbfourlmhvq{} $\pm$
\Ebjbbfourlmhvqerr~MeV smaller than the \bbfourl{} one.  We thus
observe that the jet modelling implemented by \PythiaEightPtwo{} with
MEC seems to yield slightly less energetic jets. An effect going in
the same direction was also observed for the \mwbj{} observable (see
Tab.~\ref{tab:mass_extraction-errors}, the first column of the results
with smearing), although to a smaller extent.

\begin{table}[tb]
\centering
{ \begin{tabular}{l|c|c|}
 \cline{2-3}
 & $\phantom{\Big|}$ MEC  & MEC ${}-$ no MEC  \\ 
 \cline{1-3}
\multicolumn{1}{ |c|}{ $\phantom{\Big|}$ \bbfourl{}}
 & $ 71.200\pm  0.081 $~GeV
 & $+   170\pm    115 $~MeV
\\ \cline{1-3}
\multicolumn{1}{ |c|}{ $\phantom{\Big|}$ \ttbnlodec{}}
 & $ 71.361\pm  0.062 $~GeV
 & $    -69\pm     87 $~MeV
\\ \cline{1-3}
\multicolumn{1}{ |c|}{ $\phantom{\Big|}$ \hvq{}}
 & $ 70.744\pm  0.064 $~GeV
 & $+  1937\pm     92 $~MeV
\\ \cline{1-3}
\end{tabular}
}
\caption{\Ebj{} peak position obtained with the three generators
  showered with \PythiaEightPtwo{}. The differences between the peak
  positions extracted by switching on and off the matrix-element
  corrections are also shown.}
\label{tab:Ebj_MEC}
\end{table}
In Tab.~\ref{tab:Ebj_MEC} we have collected the values of $\Ebjmax$
computed with MEC, and the differences between the results with and
without MEC.  We notice that the MEC setting has little impact in the
\bbfourl{} and \ttbnlodec{} cases. On the other hand, in the \hvq{}
case the absence of MEC would have lead to an $\Ebjmax$ value about
2~GeV smaller than with MEC. We take this as another indication that
the implementation of radiation in top decay using MEC leads to
results that are much closer to the NLO+PS ones.

In Tab.~\ref{tab:Ebj_extraction-errors}
\begin{table*}[tb]
\centering
{ \begin{tabular}{l|c|c|c|c|c|}
 \cline{2-6}
 &$\phantom{\Big|}$  $\%$ ${}-$ \bbfourl{} & $(\muR, \muF)$ & PDF & $\as$ & stat \\
\cline{1-6}
\multicolumn{1}{ |c|}{ $\phantom{\Big|}$ \bbfourl{} } & $+     0 $~MeV & ${}_{-     15}^{+     22}$~MeV & -                                                                                                 & $\pm     35$~MeV  & $\pm     81$~MeV \\
\cline{1-6}
\multicolumn{1}{ |c|}{ $\phantom{\Big|}$ \ttbnlodec{} } & $+   161 $~MeV & ${}_{-     24}^{+     22}$~MeV & -                                                                                                 & $\pm     17$~MeV  & $\pm     62$~MeV \\
\cline{1-6}
\multicolumn{1}{ |c|}{ $\phantom{\Big|}$ \hvq{} } & $   -456 $~MeV & ${}_{-     47}^{+     32}$~MeV & $\pm     30$~MeV                                                                                  & $\pm     25$~MeV  & $\pm     64$~MeV \\
\cline{1-6}
\end{tabular}
}
\caption{Theoretical uncertainties for the $\Ebj${} peak position
  obtained with the three generators showered with \PythiaEightPtwo{}.
  The last column reports the statistical uncertainty of our results.}
\label{tab:Ebj_extraction-errors}
\end{table*}
we summarize our results together with the scale, PDF and $\as$
uncertainties, that are extracted with a procedure analogous to one
described for the \mwbj{} observable.  We also report the
corresponding statistical errors of our results.  We see that scale
and PDF variations have negligible impact on our observable, given the
small sensitivity on the production dynamics, the only important
change being associated with the choice of the NLO+PS generator.

We notice that our errors on scale and PDF variations are much smaller
than our statistical errors. On the other hand, these variations are
performed by reweighting techniques, that, because of correlations,
lead to errors in the differences that are much smaller than the error
on the individual term. In view of the small size of these variations,
we do not attempt to perform a better estimate of their error. On the
other hand, the variation of $\as$ do not benefit from this
cancellation, and are all below the statistical uncertainties.

As previously done for \mwbj{}, we have also investigated the
dependence of the \bjet{} peak positions on the jet radius. The
results are summarized in Tab.~\ref{tab:Ebj_cmp_bb4l_allradii}.
\begin{table*}[h!tb]
\centering
{ \begin{tabular}{l|c|c|c|}
 \cline{2-4}
 & $\phantom{\Big|}$ $R=0.4$ & $R=0.5$ & $R=0.6$ \\ 
 \cline{1-4}
\multicolumn{1}{ |c|}{ $\phantom{\Big|}$ \bbfourl{}}
 & $ 67.792\pm  0.089 $~GeV
 & $ 71.200\pm  0.081 $~GeV
 & $ 74.454\pm  0.076 $~GeV
\\ \cline{1-4}
\multicolumn{1}{ |c|}{ $\phantom{\Big|}$ \ttbnlodec{} ${}-$ \bbfourl{}}
 & $+   365\pm    110 $~MeV
 & $+   161\pm    102 $~MeV
 & $+    75\pm     97 $~MeV
\\ \cline{1-4}
\multicolumn{1}{ |c|}{ $\phantom{\Big|}$ \hvq{} ${}-$ \bbfourl{}}
 & $   -563\pm    110 $~MeV
 & $   -456\pm    103 $~MeV
 & $   -323\pm     97 $~MeV
\\ \cline{1-4}
\end{tabular}
}
\caption{\Ebj{} peak position obtained with the \bbfourl{} generator
  showered with \PythiaEightPtwo{}, for three choices of the jet
  radius. The differences with the \ttbnlodec{} and the \hvq{}
  generators are also shown.}
\label{tab:Ebj_cmp_bb4l_allradii}
\end{table*}
While we observe a marked change in \Ebjmax{}, that grows by
$\diffEbjcml$ and $\diffEbjumc$~GeV when going from $R=0.4$ to $0.5$
and from $0.5$ to $0.6$ respectively, \ttbnlodec{} and \hvq{} differ
by \bbfourl{} by much smaller amounts.  It is not clear whether such
small differences could be discriminated experimentally.

According to eqs.~(\ref{eq:delta_mt}) and~(\ref{eq:B_Ebj}), the
uncertainties that affect the value of the extracted top mass are
nearly twice the uncertainties on the \bjet{} energy. Considering the
difference for $R=0.5$ between \hvq{} and \bbfourl{} in
Tab.~\ref{tab:Ebj_cmp_bb4l_allradii}, we see that, by using \hvq{}
instead of \bbfourl{}, the extracted top mass would be roughly 900~MeV
larger. This should be compared with the corresponding difference of
about~150~MeV, that is shown in Tab.~\ref{tab:mass_extraction-radius},
for the smeared \mwbj{} case.  In any case, these difference are much
more smaller than the statical error of 2.66~GeV quoted by CMS in
Ref.~\cite{CMS-PAS-TOP-15-002}.

As before, we have checked the sensitivity of our result to variations
in the matching procedure in \PythiaEightPtwo{}, by studying the
difference between {\tt ScaleResonance} and {\tt FSREmission}
options. The differences turn out to be of the order of the
statistical error.

\section{Comparison with \HerwigSevenPone{}}
\begin{figure}[tb]
  \centering
  \includegraphics[width=\wfigsing]{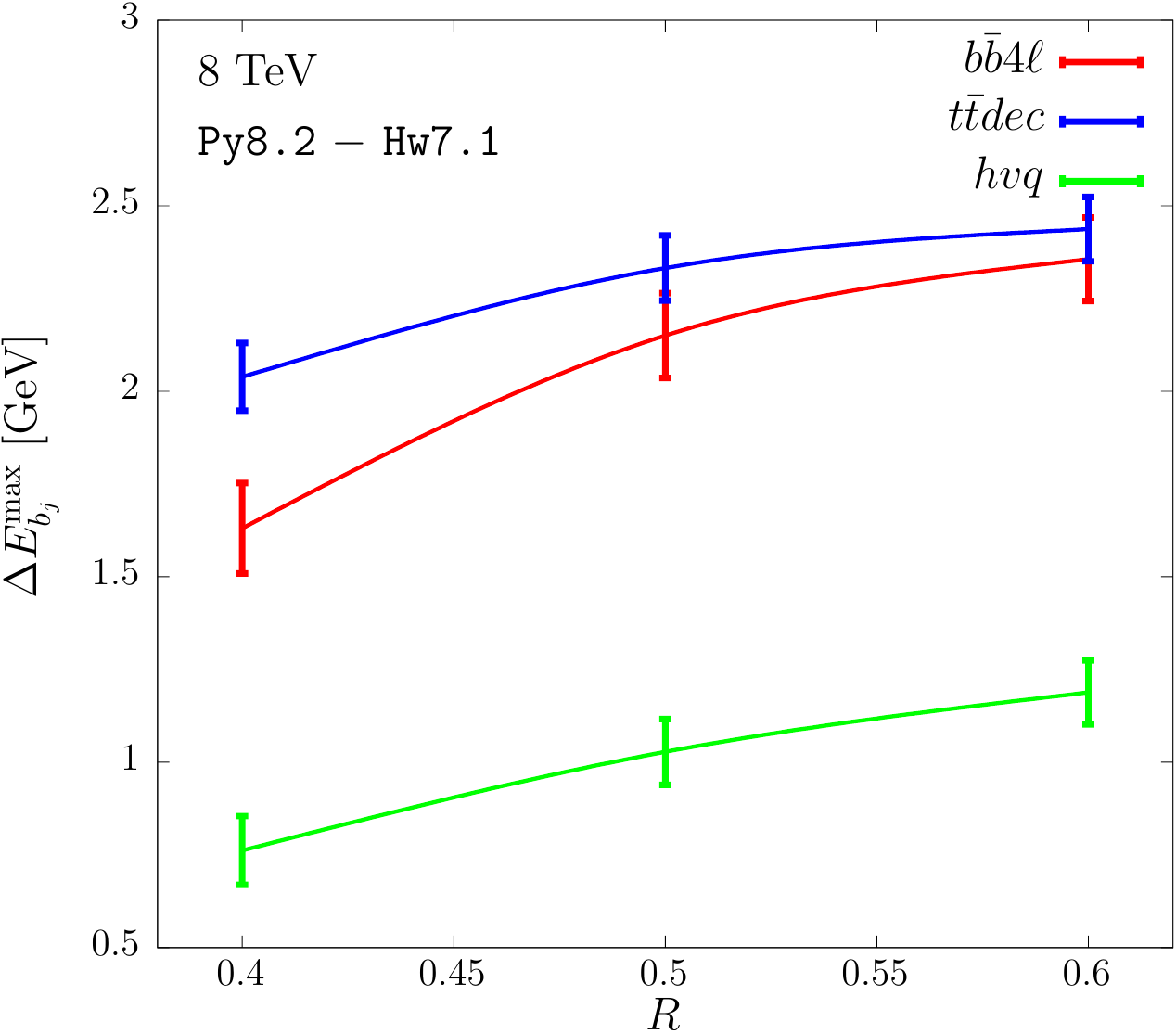}
 \caption{Differences of $\Ebjmax$ between the \PythiaEightPtwo{} and
   the \HerwigSevenPone{} showers, for the three generators, as a
   function of the jet radius.}
 \label{fig:R_Ebj_py8-hw7}
\end{figure}

In this section, we study the dependence of our results on the shower
MC program, comparing \HerwigSevenPone{} and \PythiaEightPtwo{}
predictions.  We extract the differences in the \Ebjmax{} position for
three values of the jet radius: $R=0.4$, 0.5 and 0.6. The results are
summarized in Tab.~\ref{tab:Ebj_py8-hw7}, where we also show the
results at the PS-only level,
\begin{table*}[tb]
  \centering \resizebox{\textwidth}{!}
             { \begin{tabular}{l|c|c|c|c|c|c|}
 \cline{2-7}
 & \multicolumn{6}{|c|}{$\phantom{\Big|}$ \PythiaEightPtwo{} ${}-$ \HerwigSevenPone{} [MeV]} \\
 \cline{2-7}
 & \multicolumn{2}{|c|}{$\phantom{\Big|}$ $R=0.4$}  & \multicolumn{2}{|c|}{$R=0.5$}  & \multicolumn{2}{|c|}{$R=0.6$} \\
 \cline{2-7}
 & $\phantom{\Big|}$ PS only & full & $\phantom{\Big|}$ PS only & full & $\phantom{\Big|}$ PS only & full \\ 
 \cline{1-7}
\multicolumn{1}{ |c|}{ $\phantom{\Big|}$ \bbfourl{}}
 & $+  1297\pm    120 $
 & $+  1631\pm    122 $
 & $+  1666\pm    117 $
 & $+  2150\pm    114 $
 & $+  1802\pm    114 $
 & $+  2356\pm    113 $
\\ \cline{1-7}
\multicolumn{1}{ |c|}{ $\phantom{\Big|}$ \ttbnlodec{}}
 & $+  1786\pm     91 $
 & $+  2039\pm     91 $
 & $+  2179\pm     88 $
 & $+  2332\pm     88 $
 & $+  2121\pm     89 $
 & $+  2437\pm     87 $
\\ \cline{1-7}
\multicolumn{1}{ |c|}{ $\phantom{\Big|}$ \hvq{}}
 & $+   515\pm     94 $
 & $+   762\pm     93 $
 & $+   707\pm     90 $
 & $+  1028\pm     89 $
 & $+   779\pm     87 $
 & $+  1188\pm     86 $
\\ \cline{1-7}
\end{tabular}
}
  \caption{Differences in the \Ebj{} peak position between the
    \PythiaEightPtwo{} and the \HerwigSevenPone{} showers applied to
    the three generators for three choices of the jet radius. The
    results at the NLO+PS level (PS only) are also shown.}
  \label{tab:Ebj_py8-hw7}
\end{table*}
and in Fig.~\ref{fig:R_Ebj_py8-hw7}.
From Tab.~\ref{tab:Ebj_py8-hw7} we clearly see that the \bbfourl{} and
the \ttbnlodec{} generators display larger discrepancies. For example,
for the central value $R=0.5$, we would get $\Delta \Ebjmax{}\approx
2$~GeV, that roughly corresponds to $\Delta \mt=-4$~GeV, much larger
than the 2.66~GeV statistical error quoted in
Ref.~\cite{CMS-PAS-TOP-15-002}.  In the case of the \hvq{} generator
the difference is near 1~GeV, implying that the extracted mass using
\hvq{}+\HerwigSevenPone{} would be 2~GeV bigger than the one obtained
with \hvq{}+\PythiaEightPtwo{}.

We find that the differences between \HerwigSevenPone{} and
\PythiaEightPtwo{} increases for larger jet radii.  Furthermore, by
looking at Fig.~\ref{fig:R_Ebj_py8-hw7}, we notice that the \bbfourl{}
generator displays a different $R$ dependence, as we have already
observed from Tab.~\ref{tab:Ebj_cmp_bb4l_allradii}.
Figure~\ref{fig:R_Ebj_py8-hw7} indicates that \bbfourl{} and
\ttbnlodec{} are in better agreement for larger values of the jet
radius. This was also observed for the peak of the \mwbj{} smeared
distribution~(Tab.~\ref{tab:mass_extraction-radius}).

We notice that, as in the case of the reconstructed mass peak, the
predominant contribution to the difference arise at the PS
level.

As for the previous cases, we have examined the variations due to a
different choice of the matching scheme in \HerwigSevenPone{}, that we
found to be below the 200~MeV level, and thus negligible in the
present context.

\chapter{Leptonic observables}
\label{sec:LepObs}
In this section, we investigate the extraction of the top mass from
the leptonic observables introduced in Ref.~\cite{Frixione:2014ala}.
This method has been recently studied by the ATLAS collaboration in
Ref.~\cite{Aaboud:2017ujq} that finds
\begin{equation}
m_t= 173.2 \,\pm \, 0.9 {\rm \, (stat) \,} \pm \, 0.8 {\rm \,(syst)\,}
\pm \, 1.2 {\rm \,(theo)}~{\rm GeV}.
\end{equation}

Following Ref.~\cite{Frixione:2014ala}, we consider the subsequent
five observables
\begin{equation}
  \begin{array}{l l l }
O_1 = \pT(\ell^+), & O_2= \pT(\ell^+\ell^-), & O_3= m(\ell^+\ell^-),
\\[2mm] O_4= E(\ell^+\ell^-),\quad & O_5= \pT(\ell^+)+\pT(\ell^-), &
  \end{array}
  \nonumber
\end{equation}
i.e.~the transverse momentum of the positive charged lepton, and the
transverse momentum, the invariant mass, the energy and the scalar sum
of the transverse momenta of the lepton pair.  We compute the average
value of the first three Mellin moments for each of the above
mentioned observables, $\langle (O_i)^j\rangle$, with $i=1,\dots,5$
and $j=1,2,3$.  We assume that, if we do not vary too much the range
of the top mass, we can write the linear relation
\begin{equation}
\langle (O_i)^j \rangle =O_{\rm c}^{(ij)} + B^{(ij)} \lq \(\mt\)^j-
\(\mtc\)^j \rq.
\label{eq:leptonicObs}
\end{equation}
For ease of notation, we will refer to $O_{\rm c}^{(ij)}$ and
$B^{(ij)}$ as $O_{\rm c}$ and $B$ in the following.  Their
determination will be discussed later.

We choose as reference sample the one generated with \bbfourl{}
matched with \PythiaEightPtwo{}, using $\mtc=172.5$~GeV as input mass
and the central choices for the PDF and scales. We indicate the values
of the observables computed with this generator as
${O}^{b\bar{b}4\ell}$, and with $O_{\rm c}'$ the values of the
observable computed either with an alternative generator or with
different generator settings, but using as input parameter the same
reference mass. The mass value that we would extract from the events
of the reference sample using the new generator is then given by
\begin{equation}
\label{eq:mtprime}  
\mt' = \left[\(\mtc\)^j -\frac{O'_{\rm c}-{O}^{b\bar{b}4\ell }_{\rm
      c}}{B} \right]^{1/j}\,.
\end{equation}

Since all the leptonic observables are statistical-correlated among
each other, we then performed a weighted average of all the results
considering as covariance matrix
\begin{equation}
V_{\alpha\beta} = \min \left[ \sigma^2\left(m_t^{(\alpha)}\right)\,,\;
  \sigma^2\left(m_t^{(\beta)}\right)\,, \;
  C_{\alpha\beta}\,\sigma(m_t^{(\alpha)})\,\sigma(m_t^{(\alpha)})
  \right]
\end{equation}
being $\sigma(m_t^{(\alpha)})$ the error on the extracted top-mass
relative to the observable $O_\alpha$ and
\begin{equation}
C_{\alpha\beta} = \frac{\langle O_\alpha \, O_\beta \rangle - \langle
  O_\alpha \rangle \langle O_\beta \rangle} {\sqrt{\langle O^2_\alpha
    -\langle O_\alpha \rangle^2 \rangle \, \langle O^2_\beta -\langle
    O_\beta \rangle^2 \rangle}}
\end{equation}
is the statistical correlation between $O_\alpha$ and $O_\beta$.  This
procedure has been taken from \writeApp{}B of
Ref.~\cite{Aaboud:2017ujq}.


\begin{table}[tb]
  \centering \resizebox{\textwidth}{!}
             { \begin{tabular}{lcc|c|c|c|c|}
\cline{1-7}
\multicolumn{1}{ |c|}{observable $\phantom{\Big|}$}  & \multicolumn{1}{
                                                       |c|}{gen} &
                                                                   \multicolumn{1}{ |c|}{ $\langle O_{\rm c}\rangle $ [GeV]}
 & $\%$ ${}-$ \bbfourl{} [MeV] & $(\muF, \muR )$ [MeV] & PDF [MeV] & $\as$ [MeV] \\
\cline{1-7}
 \\[-1.25em]
\cline{1-7}
\multicolumn{1}{ |c|}{}  & \multicolumn{1}{ |c|}{\bbfourl $\phantom{\Big|}$}
& $ 56.653\pm  0.050$
& -                                               
& ${}_{-     86}^{+     79}$
& -                                               
& $\pm     26\,\,( \pm     92) $   
\\ \cline{2-7}
\multicolumn{1}{ |c|}{$\langle \pT(\ell^+)\rangle $} & \multicolumn{1}{ |c|}{\ttbnlodec $\phantom{\Big|}$}
& $ 56.804\pm  0.033$
& $+   151\pm     60$                         
& ${}_{-     86}^{+     84}$
& -                                               
& $\pm     41\,\,( \pm     23) $   
\\ \cline{2-7}
\multicolumn{1}{ |c|}{} & \multicolumn{1}{ |c|}{\hvq $\phantom{\Big|}$ }
& $ 56.738\pm  0.032$
& $+    85\pm     59$                         
& ${}_{-     86}^{+     82}$
& $\pm    130$                                
& $\pm     49\,\,( \pm     23) $   
\\ \cline{2-7}
\cline{1-7}
 \\[-1.25em]
\cline{1-7}
\multicolumn{1}{ |c|}{}  & \multicolumn{1}{ |c|}{\bbfourl $\phantom{\Big|}$}
& $ 69.759\pm  0.059$
& -                                               
& ${}_{-    444}^{+    710}$
& -                                               
& $\pm     85\,\,( \pm    110) $   
\\ \cline{2-7}
\multicolumn{1}{ |c|}{$\langle \pT(\ell^+\ell^-)\rangle $} & \multicolumn{1}{ |c|}{\ttbnlodec $\phantom{\Big|}$}
& $ 69.660\pm  0.040$
& $   -100\pm     71$                         
& ${}_{-    361}^{+    538}$
& -                                               
& $\pm     78\,\,( \pm     28) $   
\\ \cline{2-7}
\multicolumn{1}{ |c|}{} & \multicolumn{1}{ |c|}{\hvq $\phantom{\Big|}$ }
& $ 69.201\pm  0.038$
& $   -558\pm     71$                         
& ${}_{-    367}^{+    553}$
& $\pm     95$                                
& $\pm     95\,\,( \pm     27) $   
\\ \cline{2-7}
\cline{1-7}
 \\[-1.25em]
\cline{1-7}
\multicolumn{1}{ |c|}{}  & \multicolumn{1}{ |c|}{\bbfourl $\phantom{\Big|}$}
& $108.685\pm  0.099$
& -                                               
& ${}_{-    341}^{+    234}$
& -                                               
& $\pm     57\,\,( \pm    191) $   
\\ \cline{2-7}
\multicolumn{1}{ |c|}{$\langle m(\ell^+\ell^-)\rangle $} & \multicolumn{1}{ |c|}{\ttbnlodec $\phantom{\Big|}$}
& $108.812\pm  0.065$
& $+   127\pm    119$                         
& ${}_{-    259}^{+    244}$
& -                                               
& $\pm     33\,\,( \pm     46) $   
\\ \cline{2-7}
\multicolumn{1}{ |c|}{} & \multicolumn{1}{ |c|}{\hvq $\phantom{\Big|}$ }
& $109.200\pm  0.064$
& $+   515\pm    118$                         
& ${}_{-    265}^{+    247}$
& $\pm    395$                                
& $\pm     68\,\,( \pm     45) $   
\\ \cline{2-7}
\cline{1-7}
 \\[-1.25em]
\cline{1-7}
\multicolumn{1}{ |c|}{}  & \multicolumn{1}{ |c|}{\bbfourl $\phantom{\Big|}$}
& $186.803\pm  0.163$
& -                                               
& ${}_{-    385}^{+    342}$
& -                                               
& $\pm    540\,\,( \pm    305) $   
\\ \cline{2-7}
\multicolumn{1}{ |c|}{$\langle E(\ell^+\ell^-)\rangle $} & \multicolumn{1}{ |c|}{\ttbnlodec $\phantom{\Big|}$}
& $187.005\pm  0.107$
& $+   201\pm    195$                         
& ${}_{-    434}^{+    448}$
& -                                               
& $\pm    474\,\,( \pm     76) $   
\\ \cline{2-7}
\multicolumn{1}{ |c|}{} & \multicolumn{1}{ |c|}{\hvq $\phantom{\Big|}$ }
& $186.809\pm  0.105$
& $+     6\pm    194$                         
& ${}_{-    427}^{+    441}$
& $\pm   1068$                                
& $\pm    559\,\,( \pm     74) $   
\\ \cline{2-7}
\cline{1-7}
 \\[-1.25em]
\cline{1-7}
\multicolumn{1}{ |c|}{}  & \multicolumn{1}{ |c|}{\bbfourl $\phantom{\Big|}$}
& $113.322\pm  0.095$
& -                                               
& ${}_{-    184}^{+    165}$
& -                                               
& $\pm     93\,\,( \pm    178) $   
\\ \cline{2-7}
\multicolumn{1}{ |c|}{$\langle \pT(\ell^+)+\pT(\ell^-)\rangle $} & \multicolumn{1}{ |c|}{\ttbnlodec $\phantom{\Big|}$}
& $113.598\pm  0.063$
& $+   276\pm    114$                         
& ${}_{-    174}^{+    165}$
& -                                               
& $\pm     72\,\,( \pm     44) $   
\\ \cline{2-7}
\multicolumn{1}{ |c|}{} & \multicolumn{1}{ |c|}{\hvq $\phantom{\Big|}$ }
& $113.425\pm  0.062$
& $+   104\pm    113$                         
& ${}_{-    177}^{+    163}$
& $\pm    259$                                
& $\pm    101\,\,( \pm     43) $   
\\ \cline{2-7}
\cline{1-7}
\end{tabular}
}
  \caption{The average values of each leptonic observable computed
    with \bbfourl{}, \ttbnlodec{} and \hvq{}, showered with
    \PythiaEightPtwo{}, for $\mt$=172.5~GeV, and their variations with
    respect to \bbfourl{} are shown in the first two columns.  The
    differences with respect to their corresponding central values due
    to scale and PDF variations are also shown in columns three and
    four.  Their $\as$ uncertainties, computed as described in
    Chap.~\ref{sec:pheno} are displayed in column five.  The
    statistical errors are also reported, except for the scale and PDF
    variations, where they have been estimated to be below 13\%{} of
    the quoted values.}
\label{tab:Olep-summary-py}
\end{table}

\begin{table}[tb]
  \centering \resizebox{\textwidth}{!}
             { \begin{tabular}{lcc|c|c|c|c|}
\cline{1-7}
\multicolumn{1}{ |c|}{observable $\phantom{\Big|}$}  & \multicolumn{1}{
                                                       |c|}{gen} &
                                                                   \multicolumn{1}{ |c|}{ $\langle O_{\rm c}\rangle $ [GeV]}
 & $\%$ ${}-$ \bbfourl{} [MeV] & $(\muF, \muR )$ [MeV] & PDF [MeV] & $\as$ [MeV] \\
\cline{1-7}
 \\[-1.25em]
\cline{1-7}
\multicolumn{1}{ |c|}{}  & \multicolumn{1}{ |c|}{\bbfourl $\phantom{\Big|}$}
& $ 56.104\pm  0.049$
& -                                               
& ${}_{-    106}^{+     92}$
& -                                               
& $\pm     20\,\,( \pm     91) $   
\\ \cline{2-7}
\multicolumn{1}{ |c|}{$\langle \pT(\ell^+)\rangle $} & \multicolumn{1}{ |c|}{\ttbnlodec $\phantom{\Big|}$}
& $ 56.199\pm  0.047$
& $+    95\pm     68$                         
& ${}_{-    105}^{+     90}$
& -                                               
& $\pm     23\,\,( \pm     23) $   
\\ \cline{2-7}
\multicolumn{1}{ |c|}{} & \multicolumn{1}{ |c|}{\hvq $\phantom{\Big|}$ }
& $ 56.399\pm  0.032$
& $+   295\pm     59$                         
& ${}_{-    100}^{+     87}$
& $\pm    222$                                
& $\pm     45\,\,( \pm     23) $   
\\ \cline{2-7}
\cline{1-7}
 \\[-1.25em]
\cline{1-7}
\multicolumn{1}{ |c|}{}  & \multicolumn{1}{ |c|}{\bbfourl $\phantom{\Big|}$}
& $ 68.665\pm  0.059$
& -                                               
& ${}_{-    372}^{+    587}$
& -                                               
& $\pm     54\,\,( \pm    108) $   
\\ \cline{2-7}
\multicolumn{1}{ |c|}{$\langle \pT(\ell^+\ell^-)\rangle $} & \multicolumn{1}{ |c|}{\ttbnlodec $\phantom{\Big|}$}
& $ 68.632\pm  0.051$
& $    -33\pm     78$                         
& ${}_{-    307}^{+    452}$
& -                                               
& $\pm     56\,\,( \pm     28) $   
\\ \cline{2-7}
\multicolumn{1}{ |c|}{} & \multicolumn{1}{ |c|}{\hvq $\phantom{\Big|}$ }
& $ 68.566\pm  0.038$
& $    -99\pm     70$                         
& ${}_{-    312}^{+    466}$
& $\pm    161$                                
& $\pm     91\,\,( \pm     27) $   
\\ \cline{2-7}
\cline{1-7}
 \\[-1.25em]
\cline{1-7}
\multicolumn{1}{ |c|}{}  & \multicolumn{1}{ |c|}{\bbfourl $\phantom{\Big|}$}
& $108.497\pm  0.099$
& -                                               
& ${}_{-    265}^{+    201}$
& -                                               
& $\pm     24\,\,( \pm    190) $   
\\ \cline{2-7}
\multicolumn{1}{ |c|}{$\langle m(\ell^+\ell^-)\rangle $} & \multicolumn{1}{ |c|}{\ttbnlodec $\phantom{\Big|}$}
& $108.076\pm  0.072$
& $   -422\pm    122$                         
& ${}_{-    250}^{+    240}$
& -                                               
& $\pm      2\,\,( \pm     46) $   
\\ \cline{2-7}
\multicolumn{1}{ |c|}{} & \multicolumn{1}{ |c|}{\hvq $\phantom{\Big|}$ }
& $109.056\pm  0.063$
& $+   559\pm    117$                         
& ${}_{-    258}^{+    247}$
& $\pm    683$                                
& $\pm     52\,\,( \pm     45) $   
\\ \cline{2-7}
\cline{1-7}
 \\[-1.25em]
\cline{1-7}
\multicolumn{1}{ |c|}{}  & \multicolumn{1}{ |c|}{\bbfourl $\phantom{\Big|}$}
& $185.540\pm  0.162$
& -                                               
& ${}_{-    380}^{+    337}$
& -                                               
& $\pm    504\,\,( \pm    304) $   
\\ \cline{2-7}
\multicolumn{1}{ |c|}{$\langle E(\ell^+\ell^-)\rangle $} & \multicolumn{1}{ |c|}{\ttbnlodec $\phantom{\Big|}$}
& $185.315\pm  0.118$
& $   -225\pm    200$                         
& ${}_{-    416}^{+    428}$
& -                                               
& $\pm    426\,\,( \pm     76) $   
\\ \cline{2-7}
\multicolumn{1}{ |c|}{} & \multicolumn{1}{ |c|}{\hvq $\phantom{\Big|}$ }
& $186.125\pm  0.104$
& $+   585\pm    192$                         
& ${}_{-    410}^{+    420}$
& $\pm   1842$                                
& $\pm    520\,\,( \pm     73) $   
\\ \cline{2-7}
\cline{1-7}
 \\[-1.25em]
\cline{1-7}
\multicolumn{1}{ |c|}{}  & \multicolumn{1}{ |c|}{\bbfourl $\phantom{\Big|}$}
& $112.280\pm  0.095$
& -                                               
& ${}_{-    218}^{+    188}$
& -                                               
& $\pm     52\,\,( \pm    177) $   
\\ \cline{2-7}
\multicolumn{1}{ |c|}{$\langle \pT(\ell^+)+\pT(\ell^-)\rangle $} & \multicolumn{1}{ |c|}{\ttbnlodec $\phantom{\Big|}$}
& $112.455\pm  0.077$
& $+   174\pm    122$                         
& ${}_{-    205}^{+    177}$
& -                                               
& $\pm     36\,\,( \pm     45) $   
\\ \cline{2-7}
\multicolumn{1}{ |c|}{} & \multicolumn{1}{ |c|}{\hvq $\phantom{\Big|}$ }
& $112.796\pm  0.061$
& $+   516\pm    112$                         
& ${}_{-    204}^{+    176}$
& $\pm    444$                                
& $\pm     97\,\,( \pm     43) $   
\\ \cline{2-7}
\cline{1-7}
\end{tabular}
}
  \caption{As in Tab.~\ref{tab:Olep-summary-py} but for
    \HerwigSevenPone{}.}
\label{tab:Olep-summary-hw}
\end{table}

We begin by showing in Tabs.~\ref{tab:Olep-summary-py}
and~\ref{tab:Olep-summary-hw} the average values of the leptonic
observables computed with our three NLO+PS generators interfaced with
\PythiaEightPtwo{} and \HerwigSevenPone{}. We show the central values,
the differences with respect to \bbfourl{}, and the upper and lower
results induced by scale, PDF and $\as$ variations.

The scale and PDF variations are performed by reweighting.  As a
consequence of that, the associated error is much smaller than the
statistical error on the cross section. In order to estimate it, we
have divided our sample of events in ten sub-samples, computed the
observables for each sub-sample, and carried out a straightforward
statistical analysis on the ten sets of results. We found errors that
never exceed the quoted value by more than 13\%.

For the PDF variation, we have verified that differences due to
variations in our reference PDF sets (see Chap.~\ref{sec:pheno}) are
very similar among the different generators. On the other hand, a full
error study using the {\tt PDF4LHC15\_nlo\_30\_pdfas} set was only
performed with the \hvq{} generator, and the associated errors exceed
by far the variation band that we obtain with our reference
sets. Thus, also in this case we quote the PDF variations only for
\hvq{}, implying that a very similar variation should also be present
for the others.  It is clear from the tables that the PDF
uncertainties are dominant for several observables, and scale
variations are also sizeable.

The large variations in the $\as$ column are not always conclusive
because of the large statistical errors (in parentheses), due to the
fact that we cannot perform this variation by reweighting. However,
unlike for the $\mwbj$ case, here the PDF dependence is not small, and
thus we cannot conclude that the $\as$ variation probes mainly the
sensitivity to the intensity of radiation in decay, since when we vary
$\as$ we change also the PDF set.

It is instead useful to look at the effect of MEC on the leptonic
observables, displayed in Tab.~\ref{tab:leptobs_MEC}.
\begin{table}[tb]
  \centering
  { \begin{tabular}{c|c|c|c|}
 \cline{2-4}
 &  \multicolumn{3}{ |c|}{$\phantom{\Big|}$  MEC ${}-$ no MEC} \\
 \cline{2-4}
 & $\phantom{\Big|}$  \tt \bbfourl & \ttbnlodec & \hvq{}\\
 \cline{1-4}
\multicolumn{1}{ |c|}{$\phantom{\Big|}$ $\langle \pT(\ell^+)\rangle $}
 & $+   117\pm
     74$~MeV 
 & $+    30\pm
     47$~MeV 
 & $+   342\pm
     46$~MeV 
\\ \cline{1-4}
\multicolumn{1}{ |c|}{$\phantom{\Big|}$ $\langle \pT(\ell^+\ell^-)\rangle $}
 & $+   167\pm
     89$~MeV 
 & $+    41\pm
     57$~MeV 
 & $+   544\pm
     55$~MeV 
\\ \cline{1-4}
\multicolumn{1}{ |c|}{$\phantom{\Big|}$ $\langle m(\ell^+\ell^-)\rangle $}
 & $+   171\pm
    149$~MeV 
 & $+   102\pm
     94$~MeV 
 & $+   631\pm
     91$~MeV 
\\ \cline{1-4}
\multicolumn{1}{ |c|}{$\phantom{\Big|}$ $\langle E(\ell^+\ell^-)\rangle $}
 & $+   372\pm
    243$~MeV 
 & $+   159\pm
    153$~MeV 
 & $+  1245\pm
    150$~MeV 
\\ \cline{1-4}
\multicolumn{1}{ |c|}{$\phantom{\Big|}$ $\langle \pT(\ell^+)+\pT(\ell^-)\rangle $}
 & $+   232\pm
    142$~MeV 
 & $+    85\pm
     89$~MeV 
 & $+   699\pm
     88$~MeV 
\\ \cline{1-4}
\end{tabular}
}
  \caption{Impact of MEC in \PythiaEightPtwo{} on the leptonic
    observables for the different NLO+PS generators.}
 \label{tab:leptobs_MEC}
\end{table}
We observe that in the \bbfourl{} and \ttbnlodec{} case the effect of
MEC is compatible with the statistical uncertainty.  In the \hvq{}
case we find instead sizeable effects. This is expected, since
large-angle radiation from the $b$ quark, by subtracting energy to the
whole $Wb$ system, affects significantly also leptonic observables.

In Ref.~\cite{Frixione:2014ala} it was observed that the observables
$\pt(\ell^+\ell^-)$ and $m(\ell^+\ell^-)$ had larger errors due to a
stronger sensitivity to radiative corrections, and were more sensitive
to spin-correlation effects.  We see a confirmation of this
observations in their larger errors due to scale variation, and in the
fact that for \hvq{} their central value is shifted with respect to
the \bbfourl{} and \ttbnlodec{} generators, that treat spin
correlations in a better way.

\begin{table}[tb!]
\centering
{ \begin{tabular}{cc|c|}
\cline{1-3}
\multicolumn{1}{|c|}{observable $\phantom{\Big|}$} & \multicolumn{1}{|c|}{generator} &  $B$  \\
\cline{1-3}
 \\[-1.25em]
\cline{1-3}
\multicolumn{1}{ |c|}{} & \multicolumn{1}{ |c|}{\bbfourl $\phantom{\Big|}$}
& $   0.17\pm   0.04$
\\ \cline{2-3}
\multicolumn{1}{ |c|}{$\langle \pT(\ell^+)\rangle $} & \multicolumn{1}{ |c|}{\ttbnlodec $\phantom{\Big|}$}
& $   0.19\pm   0.02$
\\ \cline{2-3}
\multicolumn{1}{ |c|}{} & \multicolumn{1}{ |c|}{\hvq $\phantom{\Big|}$ }
& $   0.19\pm   0.02$
\\ \cline{2-3}
 \cline{1-3}
 \\[-1.25em]
\cline{1-3}
\multicolumn{1}{ |c|}{} & \multicolumn{1}{ |c|}{\bbfourl $\phantom{\Big|}$}
& $   0.30\pm   0.05$
\\ \cline{2-3}
\multicolumn{1}{ |c|}{$\langle \pT(\ell^+\ell^-)\rangle $} & \multicolumn{1}{ |c|}{\ttbnlodec $\phantom{\Big|}$}
& $   0.30\pm   0.02$
\\ \cline{2-3}
\multicolumn{1}{ |c|}{} & \multicolumn{1}{ |c|}{\hvq $\phantom{\Big|}$ }
& $   0.29\pm   0.02$
\\ \cline{2-3}
 \cline{1-3}
 \\[-1.25em]
\cline{1-3}
\multicolumn{1}{ |c|}{} & \multicolumn{1}{ |c|}{\bbfourl $\phantom{\Big|}$}
& $   0.31\pm   0.08$
\\ \cline{2-3}
\multicolumn{1}{ |c|}{$\langle m(\ell^+\ell^-)\rangle $} & \multicolumn{1}{ |c|}{\ttbnlodec $\phantom{\Big|}$}
& $   0.31\pm   0.03$
\\ \cline{2-3}
\multicolumn{1}{ |c|}{} & \multicolumn{1}{ |c|}{\hvq $\phantom{\Big|}$ }
& $   0.33\pm   0.03$
\\ \cline{2-3}
 \cline{1-3}
 \\[-1.25em]
\cline{1-3}
\multicolumn{1}{ |c|}{} & \multicolumn{1}{ |c|}{\bbfourl $\phantom{\Big|}$}
& $   0.55\pm   0.14$
\\ \cline{2-3}
\multicolumn{1}{ |c|}{$\langle E(\ell^+\ell^-)\rangle $} & \multicolumn{1}{ |c|}{\ttbnlodec $\phantom{\Big|}$}
& $   0.56\pm   0.05$
\\ \cline{2-3}
\multicolumn{1}{ |c|}{} & \multicolumn{1}{ |c|}{\hvq $\phantom{\Big|}$ }
& $   0.56\pm   0.05$
\\ \cline{2-3}
 \cline{1-3}
 \\[-1.25em]
\cline{1-3}
\multicolumn{1}{ |c|}{} & \multicolumn{1}{ |c|}{\bbfourl $\phantom{\Big|}$}
& $   0.38\pm   0.08$
\\ \cline{2-3}
\multicolumn{1}{ |c|}{$\langle \pT(\ell^+)+\pT(\ell^-)\rangle $} & \multicolumn{1}{ |c|}{\ttbnlodec $\phantom{\Big|}$}
& $   0.39\pm   0.03$
\\ \cline{2-3}
\multicolumn{1}{ |c|}{} & \multicolumn{1}{ |c|}{\hvq $\phantom{\Big|}$ }
& $   0.39\pm   0.03$
\\ \cline{2-3}
 \cline{1-3}
\end{tabular}
}
\caption{Extracted $B$ coefficients for the three different generators
  showered with \PythiaEightPtwo{}.}
\label{tab:Bcoeffs-lept}
\end{table}

In Tab.~\ref{tab:Bcoeffs-lept} we show the extracted values of the $B$
coefficients for the first Mellin moment of each observable. The $B$
values corresponding to the different generators are compatible within
the statistical errors.  We thus choose the values computed with the
\hvq{} generator, that have the smallest error.  According to
eq.~(\ref{eq:mtprime}), we can translate a variation in an observable
into a variation of the extracted mass, that for the first Mellin
moment is simply obtained applying a $-1/B$ factor.  The results are
illustrated in Tab.~\ref{tab:mass_average_lept}.
\begin{table}[tb]
\begin{center}  
\resizebox{\textwidth}{!}
          { \begin{tabular}{c|c|c|c|c|c|c|}
 \cline{2-7}
\multicolumn{1}{c|}{} &\multicolumn{3}{|c|}{$\phantom{\Big|}$ $\mt$ extracted with \PythiaEightPtwo{}} &\multicolumn{3}{|c|}{ $\mt$ extracted with \HerwigSevenPone{}}
 \\ \cline{1-7}
\multicolumn{1}{|c|}{$\phantom{\Big|}$ observable}
 & \bbfourl{} & \ttbnlodec{} & \hvq{} & \bbfourl{} & \ttbnlodec{} & \hvq{}
 \\ \cline{1-7}
\multicolumn{1}{|c|}{$\phantom{\Big|}$$\langle \pT(\ell^+)\rangle $}
& $ 172.500_{-  0.825}^{+  0.845} $
& $ 171.719_{-  0.816}^{+  0.821} $
& $ 172.060_{-  0.811}^{+  0.822} $
& $ 175.340_{-  1.269}^{+  1.298} $
& $ 174.847_{-  1.263}^{+  1.293} $
& $ 173.817_{-  1.244}^{+  1.270} $
\\ \cline{1-7}
\multicolumn{1}{|c|}{$\phantom{\Big|}$$\langle \pT(\ell^+\ell^-)\rangle $}
& $ 172.500_{-  2.515}^{+  1.601} $
& $ 172.848_{-  1.915}^{+  1.315} $
& $ 174.451_{-  1.967}^{+  1.334} $
& $ 176.328_{-  2.141}^{+  1.433} $
& $ 176.442_{-  1.689}^{+  1.227} $
& $ 176.675_{-  1.728}^{+  1.235} $
\\ \cline{1-7}
\multicolumn{1}{|c|}{$\phantom{\Big|}$$\langle m(\ell^+\ell^-)\rangle $}
& $ 172.500_{-  1.419}^{+  1.605} $
& $ 172.116_{-  1.417}^{+  1.441} $
& $ 170.945_{-  1.420}^{+  1.450} $
& $ 173.068_{-  2.171}^{+  2.233} $
& $ 174.342_{-  2.198}^{+  2.208} $
& $ 171.379_{-  2.203}^{+  2.214} $
\\ \cline{1-7}
\multicolumn{1}{|c|}{$\phantom{\Big|}$$\langle E(\ell^+\ell^-)\rangle $}
& $ 172.500_{-  2.037}^{+  2.061} $
& $ 172.138_{-  2.091}^{+  2.081} $
& $ 172.490_{-  2.086}^{+  2.076} $
& $ 174.771_{-  3.378}^{+  3.393} $
& $ 175.176_{-  3.406}^{+  3.401} $
& $ 173.720_{-  3.401}^{+  3.397} $
\\ \cline{1-7}
\multicolumn{1}{|c|}{$\phantom{\Big|}$$\langle \pT(\ell^+)+\pT(\ell^-)\rangle $}
& $ 172.500_{-  0.827}^{+  0.852} $
& $ 171.791_{-  0.806}^{+  0.818} $
& $ 172.233_{-  0.802}^{+  0.821} $
& $ 175.178_{-  1.265}^{+  1.296} $
& $ 174.730_{-  1.246}^{+  1.275} $
& $ 173.851_{-  1.239}^{+  1.267} $
\\ \cline{1-7}
\multicolumn{1}{|c|}{$\phantom{\Big|}$$\langle \pT^2(\ell^+)\rangle $}
& $ 172.500_{-  0.960}^{+  0.977} $
& $ 171.657_{-  1.011}^{+  0.998} $
& $ 172.286_{-  1.007}^{+  0.991} $
& $ 175.816_{-  1.502}^{+  1.515} $
& $ 175.326_{-  1.524}^{+  1.541} $
& $ 174.424_{-  1.497}^{+  1.508} $
\\ \cline{1-7}
\multicolumn{1}{|c|}{$\phantom{\Big|}$$\langle \pT^2(\ell^+\ell^-)\rangle $}
& $ 172.500_{-  3.375}^{+  2.072} $
& $ 172.945_{-  2.585}^{+  1.716} $
& $ 174.738_{-  2.577}^{+  1.694} $
& $ 176.673_{-  2.725}^{+  1.770} $
& $ 176.864_{-  2.170}^{+  1.533} $
& $ 177.253_{-  2.199}^{+  1.532} $
\\ \cline{1-7}
\multicolumn{1}{|c|}{$\phantom{\Big|}$$\langle m^2(\ell^+\ell^-)\rangle $}
& $ 172.500_{-  1.643}^{+  1.787} $
& $ 172.119_{-  1.680}^{+  1.687} $
& $ 171.286_{-  1.695}^{+  1.702} $
& $ 173.511_{-  2.569}^{+  2.573} $
& $ 174.808_{-  2.595}^{+  2.571} $
& $ 172.082_{-  2.644}^{+  2.619} $
\\ \cline{1-7}
\multicolumn{1}{|c|}{$\phantom{\Big|}$$\langle E^2(\ell^+\ell^-)\rangle $}
& $ 172.500_{-  2.462}^{+  2.457} $
& $ 172.072_{-  2.534}^{+  2.490} $
& $ 172.611_{-  2.518}^{+  2.475} $
& $ 175.005_{-  4.067}^{+  3.992} $
& $ 175.339_{-  4.093}^{+  3.996} $
& $ 174.054_{-  4.117}^{+  4.019} $
\\ \cline{1-7}
\multicolumn{1}{|c|}{$\phantom{\Big|}$$\langle (\pT(\ell^+)+\pT(\ell^-))^2\rangle $}
& $ 172.500_{-  1.035}^{+  1.076} $
& $ 171.642_{-  1.004}^{+  1.036} $
& $ 172.198_{-  1.008}^{+  1.043} $
& $ 175.489_{-  1.552}^{+  1.608} $
& $ 174.982_{-  1.536}^{+  1.563} $
& $ 174.145_{-  1.539}^{+  1.566} $
\\ \cline{1-7}
\multicolumn{1}{|c|}{$\phantom{\Big|}$$\langle \pT^3(\ell^+)\rangle $}
& $ 172.500_{-  1.268}^{+  1.269} $
& $ 171.558_{-  1.302}^{+  1.273} $
& $ 172.626_{-  1.299}^{+  1.262} $
& $ 176.472_{-  1.817}^{+  1.801} $
& $ 175.877_{-  1.872}^{+  1.861} $
& $ 175.212_{-  1.823}^{+  1.798} $
\\ \cline{1-7}
\multicolumn{1}{|c|}{$\phantom{\Big|}$$\langle \pT^3(\ell^+\ell^-)\rangle $}
& $ 172.500_{-  4.970}^{+  2.912} $
& $ 173.092_{-  3.825}^{+  2.435} $
& $ 175.316_{-  3.692}^{+  2.333} $
& $ 177.424_{-  3.756}^{+  2.355} $
& $ 177.691_{-  3.038}^{+  2.075} $
& $ 178.410_{-  3.033}^{+  2.046} $
\\ \cline{1-7}
\multicolumn{1}{|c|}{$\phantom{\Big|}$$\langle m^3(\ell^+\ell^-)\rangle $}
& $ 172.500_{-  2.080}^{+  2.172} $
& $ 172.416_{-  2.099}^{+  2.089} $
& $ 171.834_{-  2.140}^{+  2.124} $
& $ 173.978_{-  3.243}^{+  3.170} $
& $ 175.662_{-  3.219}^{+  3.127} $
& $ 172.980_{-  3.339}^{+  3.237} $
\\ \cline{1-7}
\multicolumn{1}{|c|}{$\phantom{\Big|}$$\langle E^3(\ell^+\ell^-)\rangle $}
& $ 172.500_{-  3.022}^{+  2.958} $
& $ 172.003_{-  3.107}^{+  2.998} $
& $ 172.843_{-  3.070}^{+  2.963} $
& $ 175.349_{-  4.944}^{+  4.701} $
& $ 175.515_{-  4.972}^{+  4.704} $
& $ 174.576_{-  5.017}^{+  4.744} $
\\ \cline{1-7}
\multicolumn{1}{|c|}{$\phantom{\Big|}$$\langle (\pT(\ell^+)\!+\!\pT(\ell^-))^3\rangle $}
& $ 172.500_{-  1.428}^{+  1.511} $
& $ 171.431_{-  1.374}^{+  1.417} $
& $ 172.134_{-  1.373}^{+  1.422} $
& $ 175.963_{-  2.022}^{+  2.137} $
& $ 175.379_{-  1.995}^{+  2.011} $
& $ 174.558_{-  2.012}^{+  2.029} $
\\ \cline{1-7}
\multicolumn{1}{|c|}{$\phantom{\Big|}$ {\bf all observables}}
& $ \mathbf{172.500_{-  0.766}^{+  0.784} }$
& $ \mathbf{171.751_{-  0.751}^{+  0.751} }$
& $ \mathbf{172.238_{-  0.748}^{+  0.754} }$
& $ \mathbf{175.392_{-  1.138}^{+  1.045} }$
& $ \mathbf{175.452_{-  1.104}^{+  0.962} }$
& $ \mathbf{174.607_{-  1.097}^{+  0.961} }$
\\ \cline{1-7}
\multicolumn{1}{|c|}{$\phantom{\Big|}$ {\bf 1st moment}}
& $ \mathbf{172.500_{-  0.772}^{+  0.794} }$
& $ \mathbf{171.755_{-  0.756}^{+  0.764} }$
& $ \mathbf{172.247_{-  0.753}^{+  0.766} }$
& $ \mathbf{175.440_{-  1.184}^{+  1.102} }$
& $ \mathbf{175.445_{-  1.141}^{+  1.011} }$
& $ \mathbf{174.756_{-  1.135}^{+  1.010} }$
\\ \cline{1-7}
\end{tabular}
}
\caption{Extracted mass in GeV for all the generators, showered with
  \PythiaEightPtwo{} and \HerwigSevenPone{}, corresponding to the
  different leptonic observables, using as reference sample the
  \bbfourl{} one generated with $\mt=172.5$~GeV and showered with
  \PythiaEightPtwo{}.  The quoted errors are obtained by summing in
  quadrature the scale, PDF and the statistical errors.  The weighted
  average is also shown, for all the observables and considering only
  their first Mellin moment.}
\label{tab:mass_average_lept}
\end{center}
\end{table}
The errors shown have been obtained by summing in quadrature the
statistical error and the scale and PDF uncertainties. We have not
included the $\as$ variation in the error in order to avoid
over-counting, since, in the present case, is likely to be largely
dominated by the change in the associated PDF.

The overall errors on the last two lines of
Tab.~\ref{tab:mass_average_lept} are obtained with the same procedure
adopted in Ref.~\cite{Frixione:2014ala} to account for correlations
among the different observables.  We do not see excessive differences
among our three generators showered with the same MC
generator, while the differences between the \PythiaEightPtwo{} and
\HerwigSevenPone{} results are considerably large. This is also the
case for the \hvq{} generator, that has a much simpler interface to
both \PythiaEightPtwo{} and \HerwigSevenPone{}.

We observe in Tab.~\ref{tab:mass_average_lept} that the inclusion of
higher moments of the leptonic observables does not modify appreciably
the results from the first moments. This is a consequence of the large
error on the higher moments, and of the strong correlations among
different moments.

\begin{figure}[h!]
  \centering
  \includegraphics[width=0.65\textwidth]{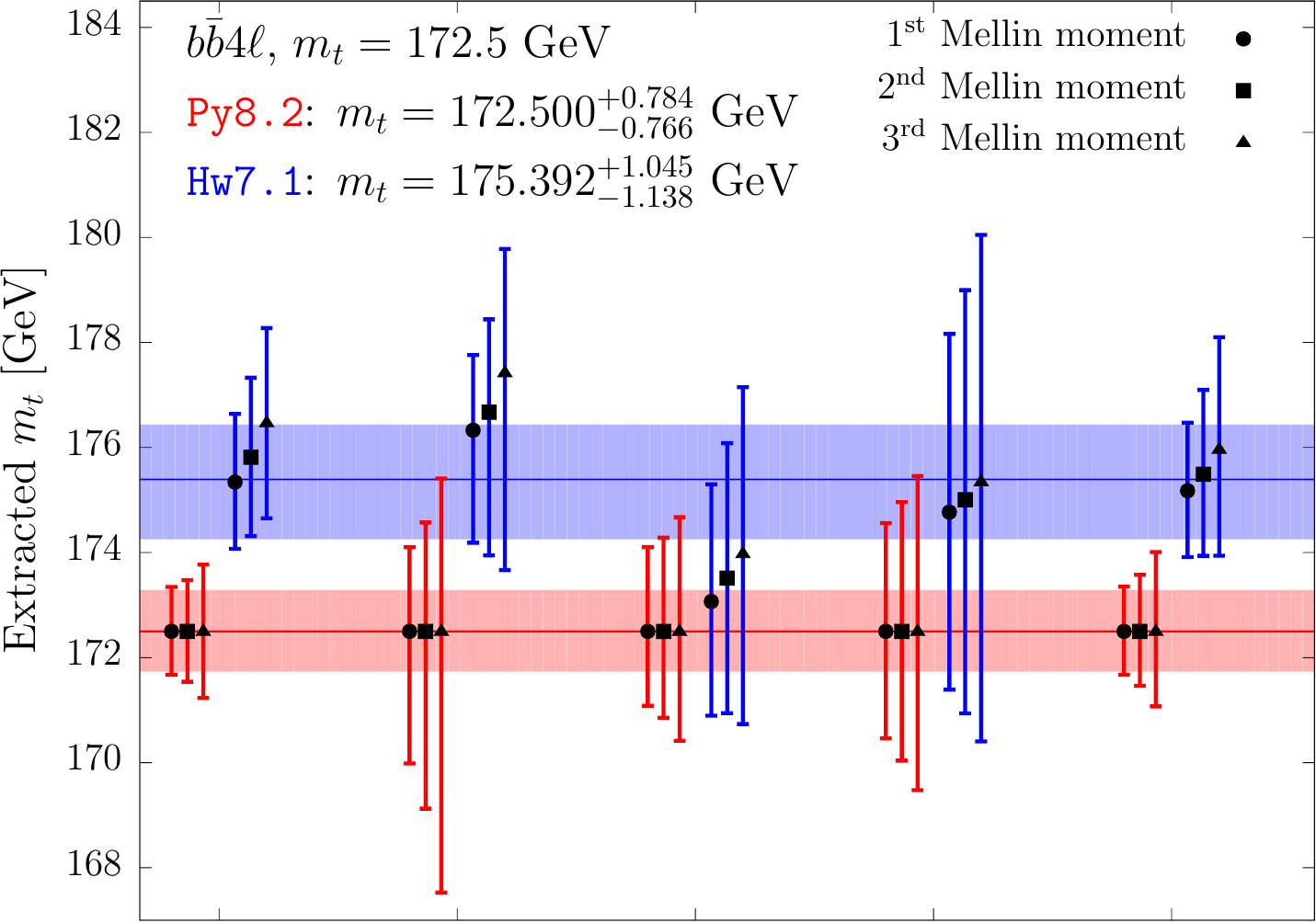}\\ \includegraphics[width=0.65\textwidth]{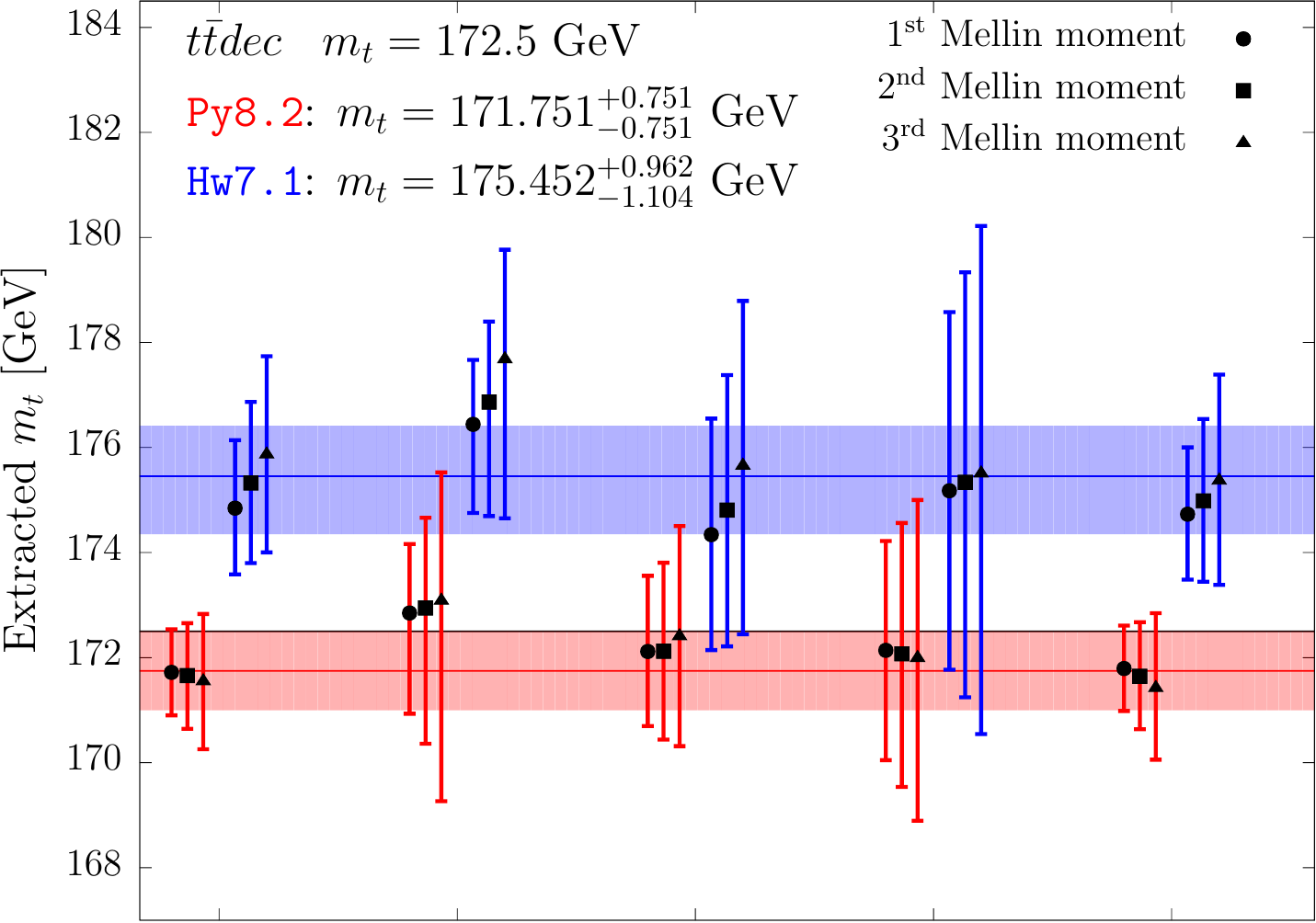}\\ \includegraphics[width=0.65\textwidth]{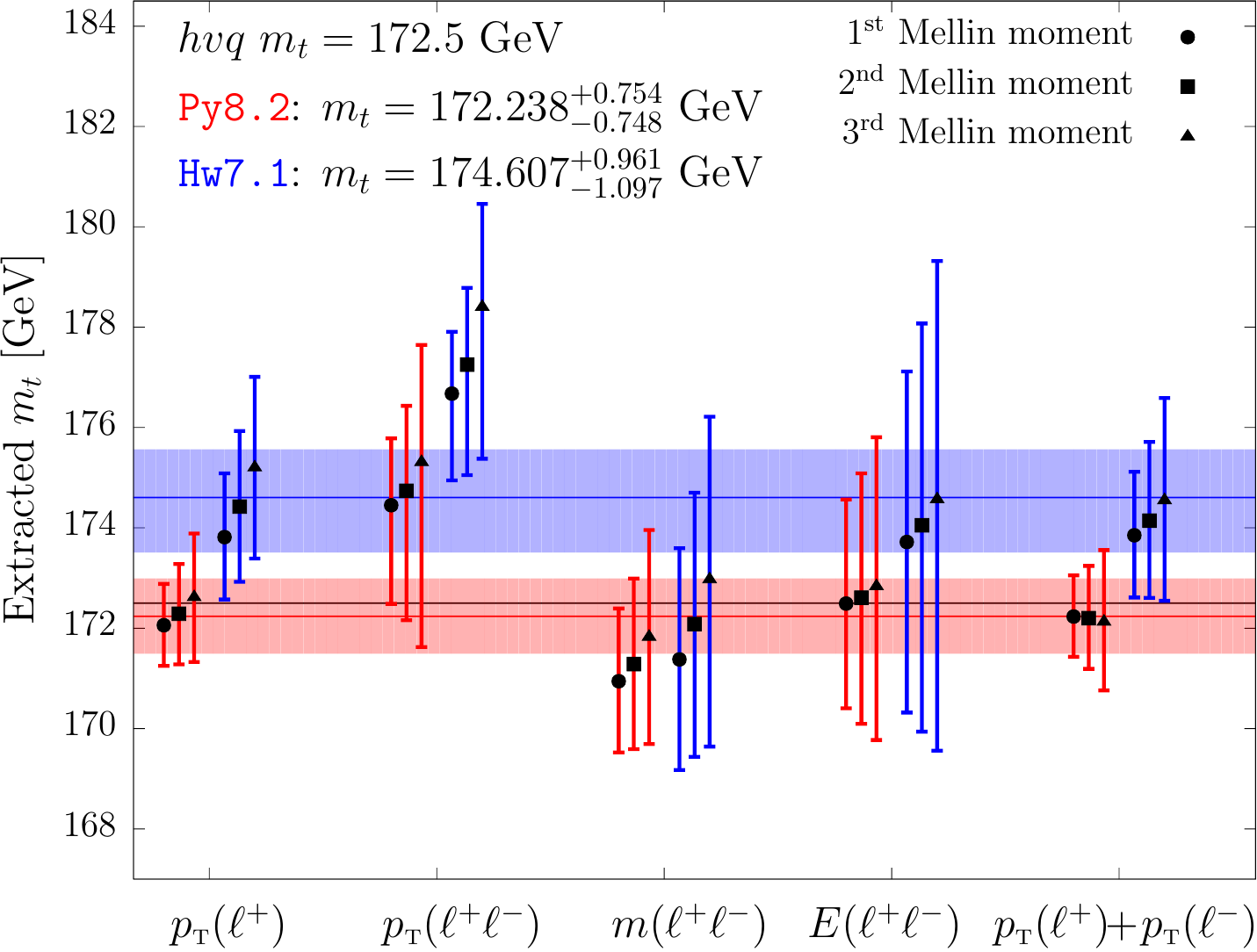}
\caption{ Extracted mass for the three generators matched with
  \PythiaEightPtwo{}~(red) and \HerwigSevenPone{}~(blue) using the
  first three Mellin moments of the five leptonic observables. The
  horizontal band represents the weighted average of the results, and
  the black horizontal line corresponds to $\mt=172.5$~GeV, which is
  the top mass value used in the \bbfourl{}+\PythiaEightPtwo{}
  reference sample.}
\label{fig:leptObs}
\end{figure}

The results in Tab.~\ref{tab:mass_average_lept} are also summarized in
Fig.~\ref{fig:leptObs}, where the discrepancy between
\PythiaEightPtwo{} and \HerwigSevenPone{} and the mutual consistency
of the different observables can be immediately appreciated.

As we did for \mwbjmax{} and \Ebjmax{}, also in the present case we
have computed the leptonic observables without including hadronization
effects, i.e.~at parton-shower only level, in order to determine
whether the differences between \PythiaEightPtwo{} and
\HerwigSevenPone{} are due to the shower or to the hadronization. Our
findings are summarized in Tab.~\ref{tab:lept-PSonly}.  Most of the
differences already arise at the shower level, this is not surprising
since the hadronization and the underlying event does not alter the
leptons momenta.

As for the previous observables, we have studied the effect of
changing the matching scheme in decay, by switching between our two
alternative matching schemes with \PythiaEightPtwo{} and
\HerwigSevenPone{}, and by considering the settings of
eq.~(\ref{eq:TSsettings}) in \HerwigSevenPone{}. In both cases we find
results that are consistent within statistical errors.

We also evaluate the impact of the {\tt PowhegHooks} veto for
radiation in production, the results are presented in
Tab.~\ref{tab:delta_pwgveto_leptObs}. Like for the \mwbjmax{} case,
the differences between the two alternatives are very small, in
particular, given the present statistical error, it is not negligible
only for the $\pT$ of the lepton pair.  Furthermore, the discrepancies
are roughly the same for each NLO generator.  Thus, our previous
conclusions will not be modified by the use of {\tt PowhegHooks} to
deal with radiation in the production process.

\begin{table}[h!]
  \centering 
      {\begin{tabular}{lc|c|c|}
\\ \cline{3-4}
 & & \multicolumn{2}{ |c|}{$\phantom{\Big|}$ \PythiaEightPlot{} ${}-$ \HerwigSevenPlot{} [MeV]}
\\ \cline{1-4}
 \multicolumn{1}{ |c|}{observable\!\!$\phantom{\Big|}$} & \multicolumn{1}{ |c|}{gen} & \multicolumn{1}{ |c|}{full} & \multicolumn{1}{ |c|}{PS only}
\\ \cline{1-4}
\\[-1.25em]
\cline{1-4}
\multicolumn{1}{ |c|}{} & \multicolumn{1}{ |c|}{\bbfourl $\phantom{\Big|}$}
 & $+   549 \pm
     70 $ 
 & $+   563 \pm
     71 $ 
\\ \cline{2-4}
\multicolumn{1}{ |c|}{$\langle \pT(\ell^+)\rangle $} & \multicolumn{1}{ |c|}{\ttbnlodec $\phantom{\Big|}$}
 & $+   605 \pm
     57 $ 
 & $+   609 \pm
     48 $ 
\\ \cline{2-4}
\multicolumn{1}{ |c|}{} & \multicolumn{1}{ |c|}{\hvq $\phantom{\Big|}$ }
 & $+   340 \pm
     45 $ 
 & $+   376 \pm
     46 $ 
\\ \cline{2-4}
\cline{1-4}
\\[-1.25em]
\cline{1-4}
\multicolumn{1}{ |c|}{} & \multicolumn{1}{ |c|}{\bbfourl $\phantom{\Big|}$}
 & $+  1094 \pm
     83 $ 
 & $+  1092 \pm
     84 $ 
\\ \cline{2-4}
\multicolumn{1}{ |c|}{$\langle \pT(\ell^+\ell^-)\rangle $} & \multicolumn{1}{ |c|}{\ttbnlodec $\phantom{\Big|}$}
 & $+  1027 \pm
     65 $ 
 & $+  1020 \pm
     59 $ 
\\ \cline{2-4}
\multicolumn{1}{ |c|}{} & \multicolumn{1}{ |c|}{\hvq $\phantom{\Big|}$ }
 & $+   636 \pm
     54 $ 
 & $+   662 \pm
     55 $ 
\\ \cline{2-4}
\cline{1-4}
\\[-1.25em]
\cline{1-4}
\multicolumn{1}{ |c|}{} & \multicolumn{1}{ |c|}{\bbfourl $\phantom{\Big|}$}
 & $+   188 \pm
    140 $ 
 & $+   286 \pm
    142 $ 
\\ \cline{2-4}
\multicolumn{1}{ |c|}{$\langle m(\ell^+\ell^-)\rangle $} & \multicolumn{1}{ |c|}{\ttbnlodec $\phantom{\Big|}$}
 & $+   736 \pm
     97 $ 
 & $+   814 \pm
     98 $ 
\\ \cline{2-4}
\multicolumn{1}{ |c|}{} & \multicolumn{1}{ |c|}{\hvq $\phantom{\Big|}$ }
 & $+   144 \pm
     90 $ 
 & $+   182 \pm
     91 $ 
\\ \cline{2-4}
\cline{1-4}
\\[-1.25em]
\cline{1-4}
\multicolumn{1}{ |c|}{} & \multicolumn{1}{ |c|}{\bbfourl $\phantom{\Big|}$}
 & $+  1263 \pm
    229 $ 
 & $+  1342 \pm
    232 $ 
\\ \cline{2-4}
\multicolumn{1}{ |c|}{$\langle E(\ell^+\ell^-)\rangle $} & \multicolumn{1}{ |c|}{\ttbnlodec $\phantom{\Big|}$}
 & $+  1690 \pm
    160 $ 
 & $+  1712 \pm
    159 $ 
\\ \cline{2-4}
\multicolumn{1}{ |c|}{} & \multicolumn{1}{ |c|}{\hvq $\phantom{\Big|}$ }
 & $+   684 \pm
    148 $ 
 & $+   719 \pm
    150 $ 
\\ \cline{2-4}
\cline{1-4}
\\[-1.25em]
\cline{1-4}
\multicolumn{1}{ |c|}{} & \multicolumn{1}{ |c|}{\bbfourl $\phantom{\Big|}$}
 & $+  1041 \pm
    134 $ 
 & $+  1091 \pm
    136 $ 
\\ \cline{2-4}
\multicolumn{1}{ |c|}{$\langle \pT(\ell^+)+\pT(\ell^-)\rangle $} & \multicolumn{1}{ |c|}{\ttbnlodec $\phantom{\Big|}$}
 & $+  1143 \pm
     99 $ 
 & $+  1173 \pm
     92 $ 
\\ \cline{2-4}
\multicolumn{1}{ |c|}{} & \multicolumn{1}{ |c|}{\hvq $\phantom{\Big|}$ }
 & $+   629 \pm
     86 $ 
 & $+   690 \pm
     88 $ 
\\ \cline{2-4}
\cline{1-4}
\\[-1.25em]
\cline{1-4}
\end{tabular}
}
\caption{Differences between the \PythiaEightPtwo{} and
  \HerwigSevenPone{} results for the leptonic observables, at full
  hadron level and at parton-level only.}
\label{tab:lept-PSonly}
\end{table} 
\begin{table}[h!]
\centering  \begin{tabular}{|c|c|c|c|}
\cline{1-4}
\multicolumn{4}{|c|}{ \phantom{\Big|} {\tt PowhegHooks} ${}{-}$ no {\tt PowhegHooks} [MeV]} \\
\cline{1-4}
\phantom{\Big|} observable & \bbfourl & \ttbnlodec & \hvq \\
\cline{1-4}
 $\phantom{\Big|} \langle \pT(\ell^+)\rangle $                         & $          57 \pm          70 $& $          74 \pm          47 $& $          50 \pm          46 $\\ \cline{1-4}
 $\phantom{\Big|} \langle \pT(\ell^+\ell^-)\rangle $                   & $         166 \pm          84 $& $         173 \pm          56 $& $         150 \pm          54 $\\ \cline{1-4}
 $\phantom{\Big|} \langle m(\ell^+\ell^-)\rangle $                     & $          25 \pm         140 $& $          16 \pm          91 $& $         -18 \pm          90 $\\ \cline{1-4}
 $\phantom{\Big|} \langle E(\ell^+\ell^-)\rangle $                     & $         145 \pm         230 $& $         143 \pm         152 $& $         123 \pm         149 $\\ \cline{1-4}
 $\phantom{\Big|} \langle \pT(\ell^+)+\pT(\ell^-)\rangle $             & $         123 \pm         135 $& $         144 \pm          89 $& $         107 \pm          87 $\\ \cline{1-4}
 \end{tabular}

\caption{Differences between the leptonic observables obtained using
  the {\tt POWHEG:veto = 1} and the {\tt POWHEG:veto = 0} settings for
  the three generators interfaced with \PythiaEightPtwo. }
\label{tab:delta_pwgveto_leptObs}
\end{table}

\chapter{Summary and outlook}
\label{sec:Summary}
In this second part of the thesis we have compared generators of
increasing accuracy for the production and decay of $t\bar{t}$ pairs
considering observables suitable for the measurement of the top mass.
The generators that we have considered are:
\begin{itemize}
\item The \hvq{} generator~\cite{Frixione:2007nw}, that implements NLO
  corrections in production for on-shell top quarks, and includes
  finite-width effects and spin correlations only in an approximate
  way, by smearing the on-shell kinematics with Breit-Wigner forms of
  appropriate width, and by generating the angular distribution of the
  decay products according to the associated tree-level matrix
  elements~\cite{Frixione:2007zp}.
\item The \ttbnlodec{} generator~\cite{Campbell:2014kua}, that
  implements NLO corrections in production and decay in the
  narrow-width approximation. Spin correlations are included at NLO
  accuracy. Finite width effects are implemented by reweighting the
  NLO results using the tree-level matrix elements for the associated
  Born-level process, including however all finite width non-resonant
  and interference effects at the Born level for the given final
  state.
\item The \bbfourl{} generator~\cite{Jezo:2016ujg}, that uses the full
  matrix elements for the production of the given final state,
  including all non-resonant diagrams and interference effects.  This
  includes interference of QCD radiation in production and decay.
\end{itemize}
The main focus has been the study of the mass distribution of a
particle-level reconstructed top, consisting of a lepton-neutrino pair
and a $b$-quark jet with the appropriate flavour. The peak position of
the mass of this system is our observable, that is loosely related to
the top mass. We considered its distributions both at the particle
level, and by assuming that experimental inaccuracies can be
summarized by a simple smearing with a resolution function, a Gaussian
with a width of 15~GeV, which is the typical resolution achieved on
the top mass by the LHC collaborations. This observable is an
oversimplified version of the mass observables that are used in direct
top-mass measurements, that are the methods that lead to the most
precise mass determinations.

We have found a very consistent picture in the comparison of our three
generators when they are interfaced to \PythiaEightPtwo{}, and thus we
begin by summarizing our results for this case.  We first recall
\emph{what we expect} from such comparison.  When comparing the \hvq{}
and the \ttNLOdec{} generators, we should remember that the latter has
certainly better accuracy in the description of spin correlations,
since it implements them correctly both at the leading and at the NLO
level. However, we do not expect spin correlations to play an
important role in the reconstructed top mass. As a further point, the
\ttNLOdec{} generator implements NLO corrections in decay. In the
\hvq{} generator, the decay is handled by the shower, where, by
default, \PythiaEightPtwo{} includes matrix-element corrections
(MEC). These differ formally from a full NLO correction only by a
normalization factor, that amounts to the NLO correction to the top
width. Thus, as long as the MEC are switched on, we do not expect
large differences between \hvq{} and \ttNLOdec{}.  As far as the
comparison between \ttNLOdec{} and \bbfourl{}, we expect the
difference to be given by NLO off-shell effects, and by interference
of radiation in production and decay, since these effects are not
implemented in \ttNLOdec{}. This comparison is particularly
interesting, since the interference between production and decay can
be considered as a ``perturbative precursor'' of colour reconnection
effects.

The results of these comparisons can be summarized as follows:
\begin{itemize}
\item The \ttNLOdec{} and the \bbfourl{} generators yield very similar
  results for most of the observables that we have considered, implying that
  NLO off-shell effects and interference between production and decay are
  modest.
\item
  As far as \mwbjmax{} (the peak of the reconstructed mass
  distribution) is concerned, the \ttNLOdec{} and the \hvq{}
  generators yield very similar results, confirming the fact that the
  MEC implementation in \PythiaEightPtwo{} has an effect very similar
  to the \POWHEG{} implementation of NLO corrections in decay in the
  \ttNLOdec{}. We have also observed that, if we switch off the MEC,
  the agreement between the two generators is spoiled.  More
  quantitatively, we find that the spread in the peak of the
  reconstructed mass at the particle level among the three NLO+PS
  generators is never above 30~MeV.  On the other hand, if resolution
  effects are accounted for with our smearing procedure, we find that
  the \hvq{} result is 147~MeV smaller, and the \ttbnlodec{} result
  140~MeV larger than the \bbfourl{} one. These values are safely
  below currently quoted errors for the top-mass measurements with
  direct methods.

  If we switch off the MEC in \PythiaEightPtwo{}, we find that the
  peak position at the particle level in the \hvq{} case is displaced
  by 61~MeV, while, if smearing effects are included, the shift is of
  $916$~MeV, a rather large value, that can however be disregarded as
  being due to the poor accuracy of the collinear approximation in $b$
  radiation when MEC corrections are off.
\item
  The jet-energy peak seems to be more sensitive to the modelling of
  radiation from the $b$ quark. In fact, while the \ttNLOdec{} and the
  \bbfourl{} results are quite consistent with each other, with the
  peak positions differing by less than 200~MeV, the \hvq{} result
  differs from them by more than 500~MeV. This would correspond to a
  difference in the extracted mass of the top quark roughly equal to
  twice that amount.  On the other hand, if the MEC in \hvq{} are
  switched off, the shift in the $b$-jet energy peak is more than
  1.9~GeV. This leads us to conclude that the impact of modelling of
  $b$ radiation on the $b$-jet peak is much stronger than in the
  reconstructed top mass peak.  We stress, however, that the
  difference between \hvq{} (with MEC on) and the other two generators
  is safely below the errors quoted in current
  measurements~\cite{CMS-PAS-TOP-15-002}.

\item
  For the leptonic observables, we generally see a reasonable
  agreement between the different generators. The largest differences
  are found in the \hvq{} case, for the $\pt(\ell^+\ell^-)$ and
  $m(\ell^+\ell^-)$, larger than 500~MeV with respect to the other
  two.  In Ref.~\cite{Frixione:2014ala} it was noticed that these
  observables had larger errors due to a stronger sensitivity to
  radiative corrections, and to spin-correlation effects, that are
  modelled incorrectly by \hvq{}.
\end{itemize}
Several sources of possible uncertainties have been explored in order
to check the reliability of these conclusions. First of all, two
different matching procedures for interfacing the \ttNLOdec{} and
\bbfourl{} generators to \PythiaEightPtwo{} have been implemented. For
example, for the reconstructed mass peak, we have checked that
switching between them leads to differences below 20~MeV for both
generators. The effect of scales, $\alpha_s$ and PDF uncertainties
have also been examined, and were found to yield very modest
variations in the reconstructed mass peak. It was found, in
particular, that scale variations lead to a negligible peak
displacement (below 7~MeV) in the \ttNLOdec{} and \hvq{} case, while
the effect is of ${}^{+86}_{-53}$~MeV for \bbfourl{}. The lack of
scale dependence in the \hvq{} and \ttNLOdec{} is easily understood as
being due to the fact that the peak shape is obtained by smearing an
on-shell distribution with a Breit-Wigner form, that does not depend
upon any scale, and it suggests that, in order to get realistic
scale-variation errors, the most accurate \bbfourl{} generator should
be used.  We have also computed results at the shower level, excluding
the effects of hadronization and multi-parton interactions, in order
to see if the consistent picture found at the hadron level is also
supported by the parton-level results, and we have found that this is
indeed the case.

We have thus seen that the overall picture of the comparison of our
three NLO+PS generators within the framework of the \PythiaEightPtwo{}
shower is quite simple and consistent.  For the most precise
observable, i.e.~the peak of the reconstructed mass distribution, it
leads to the conclusions that the use of the most accurate generator
may lead to a shift in the measured mass of at most 150~MeV, which is
well below the present uncertainties quoted by the experimental
collaborations.

Our study with \HerwigSevenPone{} instead reveals several problems. We
can summarize our findings as follows:
\begin{itemize}
\item The results obtained with \HerwigSevenPone{} differ
  substantially from those obtained with \PythiaEightPtwo{}. In
  particular, the peak of the reconstructed mass distribution at the
  particle level is shifted by -66 and -39~MeV in the \bbfourl{} and
  \ttNLOdec{} cases, and by +235 MeV in the \hvq{} case. When the
  experimental resolution is accounted for, using our smearing
  procedure, the shift raises to -1091 and -1179~MeV in the \bbfourl{}
  and \ttNLOdec{} cases, and to -251 MeV in the \hvq{} case.
\item The results obtained within the \HerwigSevenPone{} framework
  display large differences between the \hvq{} generator with respect
  to \bbfourl{} and \ttNLOdec{} ones. In particular, while the
  \ttNLOdec{} result exceeds the \bbfourl{} one only by about 50~MeV
  in both the particle level and smeared cases, \hvq{} exceeds
  \bbfourl{} by 311~MeV at particle level, and by 693~MeV after
  smearing.
\end{itemize}
These results are quite alarming. The shifts reach values that are
considerably larger than current experimental uncertainties.

In the \hvq{} case, which is the NLO+PS generator currently used for
top-mass studies by the experimental collaborations, the difference in
the mass-peak position between \HerwigSevenPone{} and
\PythiaEightPtwo{}, for the smeared distribution, is -251~MeV,
uncomfortably large but still below current errors.  One would then be
tempted to conclude that the large shifts may be linked to some
problems concerning the new generators. However, we also notice that
the same difference is +235~MeV when no smearing is applied, so it is
about as large in magnitude but with the opposite sign. This indicates
that the shape of the reconstructed mass distribution is considerably
different in the two shower models.  Lastly, if we use the internal
\POWHEG{} implementation of top decay (rather than the MEC) in
\HerwigSevenPone{}, the difference with respect to \PythiaEightPtwo{}
raises to 607~MeV. Thus, we conclude that in the \hvq{} case the
smaller difference between \HerwigSevenPone{} and \PythiaEightPtwo{}
is accidental, and is subject to considerable variations depending
upon the settings.

Also in this case we checked whether the MEC yield an improved
agreement between the \hvq{} and the other two generators, as was
observed for \PythiaEightPtwo{}.  We find that, by switching off MEC,
the \hvq{}+\HerwigSevenPone{} result decreases by 307~MeV at particle
level, and by 1371~MeV in the smeared case. These effects are
qualitatively similar to what was observed in
\PythiaEightPtwo{}. However, in the present case, when MEC are
switched off, the \hvq{} result exceeds the \bbfourl{} one by a
negligible amount at the particle level, and is lower than the
\bbfourl{} one by 678~MeV in the smeared case.

The discrepancy between \hvq{} and the other two generators is
mitigated if, instead of the MEC procedure, the internal \POWHEG{}
option of \HerwigSevenPone{} for top decay is used.  In this case, the
discrepancy between \hvq{} and \bbfourl{} is reduced to 244~MeV with
no smearing, and to 337~MeV with smearing. We thus see that the
consistency of the three NLO+PS generators interfaced to
\HerwigSevenPone{} is not optimal as in \PythiaEightPtwo{}. It is
however acceptable if the internal \POWHEG{} feature is used rather
than MEC in \HerwigSevenPone{}.

We have performed several studies to determine the origin of the
difference between \PythiaEightPtwo{} and \HerwigSevenPone{}, and to
check whether it could be attributed to some problem in our matching
procedure. They can be summarized as follows:
\begin{itemize}
\item
  We have shown that the difference is mostly due to the shower model,
  since it is already largely present at the parton level.
\item
  We have considered the $R$ dependence of the \HerwigSevenPone{}
  result. It differs from the one in \PythiaEightPtwo{}, leading to
  the hope that both generators may not represent the same set of data
  well, and tuning them may reduce their differences.  However, we
  have also noticed that the difference in slope is much smaller than
  the difference in size.
\item
  We have already mentioned that we have also compared results by
  making use of the internal \POWHEG{} implementation of top decay in
  \HerwigSevenPone{}, rather than using MEC. We have found
  non-negligible differences in this case.
\item
  We have implemented alternative veto procedure in the matching of
  \HerwigSevenPone{} with the NLO+PS generators. We found differences
  of the order of 200~MeV, not large enough to cover the discrepancy
  with \PythiaEightPtwo{}.
\item
  When interfacing \POWHEG{} generators to angular-ordered showers, in
  order to maintain the double-logarithmic accuracy of the shower, one
  should introduce the so called ``truncated
  showers''~\cite{Nason:2004rx}. One could then worry that the lack of
  truncated showers is at the origin of the discrepancies that we
  found.  Fortunately, \HerwigSevenPone{} offers some optional
  settings that are equivalent to the introduction of truncated
  showers. We found that these options lead to a shift of only 200~MeV
  in the peak position.
\end{itemize}
In summary, we found no indication that the discrepancy with
\PythiaEightPtwo{} is due to the specific matching procedure and
general settings that we have used in \HerwigSevenPone{}.

When comparing \HerwigSevenPone{} and \PythiaEightPtwo{} in the
computation of the $b$-jet energy peak, we have found even larger
differences: when using \bbfourl{} and \ttbnlodec{}, the shifts are of
the order of 2~GeV, while for \hvq{} the shift is around 1~GeV. They
correspond to differences in the extracted mass of around 4~GeV in the
first two cases, and 2~GeV in the last one.  This is not surprising,
in view of the stronger sensitivity of the $b$-jet peak to the shower
model.

Finally, when considering leptonic observables, we find again large
differences between \HerwigSevenPone{} and \PythiaEightPtwo{}. Most
differences already arise at the shower level. Notice that this is in
contrast with the naive view that leptonic observables should be less
dependent upon QCD radiation effects and jet modelling. The comparison
between \HerwigSevenPone{} and \PythiaEightPtwo{} for leptonic
observables can by appreciated by looking at Fig.~\ref{fig:leptObs},
representing the value of the extracted top mass from a sample
generated with \bbfourl{} interfaced to \PythiaEightPtwo{}.

\chapter{Conclusions}
\label{sec:Conc}
We focus our conclusions on the results obtained for the reconstructed
mass peak, since the issues that we have found there apply to the
direct top mass measurements, that are the most precise.  The
experimental collaborations extensively use the \hvq{} generator for
this kind of analyses, and since new generators of higher accuracy,
the \ttbnlodec{} and the \bbfourl{} ones, have become available, we
have addressed the question of whether the physics effects not
included in \hvq{} may lead to inaccuracies in the top-mass
determination. The answer to this question is quite simple and clear
when our generators are interfaced to \PythiaEightPtwo{}. The
differences that we find are large enough to justify the use of the
most accurate generators, but not large enough to drastically overturn
the conclusions of current measurements.  Notice that, since the
\hvq{} generator does not include NLO corrections in decays, we might
have expected a very different modelling of the $b$-jet in \hvq{} with
respect to the other two generators, leading to important shifts in
the extracted top mass value. It turns out, however, that the
\PythiaEightPtwo{} handling of top decay in \hvq{}, improved with the
matrix-element corrections, does in practice achieve NLO accuracy up
to an irrelevant normalization factor.

This nicely consistent picture does not hold anymore if we use
\HerwigSevenPone{} as shower generator. In particular, it seems that
the MEC implemented in \HerwigSevenPone{} do not have the same effect
as the handling of radiation in decay of our modern NLO+PS generators,
leading to values of the extracted top mass that can differ up to
about 700~MeV. Furthermore, interfacing our most accurate NLO+PS
generator (the \bbfourl{} one) to \HerwigSevenPone{} leads to an
extracted top mass of up to 1.2~GeV smaller with respect to the
corresponding result with \PythiaEightPtwo{}.

At this point we have two options:
\begin{itemize}
\item
  Dismiss the \HerwigSevenPone{} results, on the ground that its MEC
  handling of top decay does not match our modern generators.
\item
  Consider the \HerwigSevenPone{} result as a variation to be included
  as theoretical error.
\end{itemize}
We believe that the first option is not soundly motivated. In fact,
the implementation of MEC in \PythiaEightPtwo{} is also
\emph{technically} very close to what \POWHEG{} does. The hardest
radiation is essentially generated in the same way, and in both cases
the subsequent radiation is generated with a lower transverse
momentum. Thus the good agreement between the two is not
surprising. The case of \HerwigSevenPone{} is completely different,
since in angular-ordered showers the hardest radiation is not
necessarily the first~\cite{Seymour:1994df}. It is thus quite possible
that the differences we found when \HerwigSevenPone{} handles the
decay with MEC, with respect to the case when \POWHEG{} does, are due
to the fact that the two procedures, although \emph{formally}
equivalent (i.e.~both leading to NLO accuracy) are \emph{technically}
different. In this last case, their difference should be attributed to
uncontrolled higher-order effects, and should thus be considered as a
theoretical uncertainty.

A further question that this analysis raises is whether we should
consider the variation between the \PythiaEightPtwo{} and the
\HerwigSevenPone{} programs as an error that should be added to
current top-mass measurements. By doing so, current errors, that are
of the order of 500-600~MeV, would become larger than 1~GeV. We
believe that our crude modelling of the measurement process does not
allow us to draw this conclusion. The analysis procedures used in
direct measurements are much more complex, and involve adequate tuning
of the MC parameters and jet-energy calibration using hadronic $W$
decays in the same top events. It is not unlikely that these
procedures could lead to an increased consistency between the
\PythiaEightPtwo{} and \HerwigSevenPone{} results. However, in view of
what we have found in our study, it is difficult to trust the
theoretical errors currently given in the top quark mass determination
if alternative NLO+PS and shower generators combinations are not
considered.

\clearpage
\pagestyle{fancy}
\renewcommand{\chaptermark}[1]{%
\markboth{
{\appendixname}\ \thechapter.%
\ #1}{}}

\lhead{}
\chead{}
\rhead{\leftmark}
\lfoot{}
\cfoot{\thepage}
\rfoot{}
\renewcommand{\headrulewidth}{0.4pt}
\part{Appendix}
\appendix 
\chapter{Radiative corrections to quark lines}
\label{sec:self-energy}

In this appendix we describe how to treat radiative corrections to external
quark lines and how to compute the mass counterterm.

We denote by $\Sigma(\slashed{p}, \mb)$ the set of all the
one-particle-irreducible corrections to an heavy quark line whose
bare mass is $\mb$.
If we include all the possible self-energy~$\Sigma(\slashed{p}, \mb)$ insertions
into a free propagator we obtain the expression for the dressed propagator
\begin{eqnarray}
  G(\slashed{p}, \mb)
  & = & \frac{i}{\slashed{p} -\mb}+ \frac{i}{\slashed{p}
    -\mb}\Sigma(\slashed{p}, \mb)\frac{i}{\slashed{p} -\mb}  + \dots
  \nonumber  \\
  & = & \frac{i}{\slashed{p} -\mb} \sum_{i=0}^\infty \left[ i
    \Sigma(\slashed{p}) \frac{1}{\slashed{p}-\mb}\right]^n
  \, = \, \frac{i}{\slashed{p} -\mb-i\Sigma(\slashed{p}, \mb)}\,,
\end{eqnarray}
where we have used the fact that $\Sigma(\slashed{p}, \mb)$ commutes with
$\slashed{p}$. 
The location of the pole of the full propagator, i.e. the pole mass $\mpole$,
is equal to eigenvalue of the operator $\slashed{p}$ that satisfies
\begin{equation}
  \left[ \slashed{p} - \mb -i \Sigma(\slashed{p}, \mb)\right]_{\slashed{p}=m}
  \rightarrow \mpole = \mb + i \Sigma(\mpole, \mb).
\end{equation}
We can introduce a mass counterterm $\mpolec$
\begin{equation}
  \label{eq:mpolec}
  \mpolec\! \equiv - i \, \Sigma\!\left(\mpole, \mb \right),
\end{equation}
so that we can replace in our Lagrangian
\begin{equation}
  \mb \to \mpole + \mpolec.
\end{equation}
The Feynman rule for the mass counterterm is thus defined as the insertion,
in the fermion propagator, of the vertex $-i\mpolec$.

At order $\mathcal{O}(\as)$
\begin{equation}
 \Sigma\!\left(\mpole, \mb \right) \approx   \Sigma\!\left(\mb, \mb \right)
 \approx \Sigma\!\left(\mpole, \mpole\right),
\end{equation}
so that
\begin{equation}
  \label{eq:mpolec_NLO}
  \mpolec = - i \, \Sigma\!\left(\mpole, \mpole\right) + \mathcal{O}(\as^2).
\end{equation}
The dressed propagator expressed in terms of the pole mass is thus given by
\begin{equation}
  G^r(\slashed{p}, \mpole) =\frac{i}{\slashed{p} - \mpole
-i\Sigma(\slashed{p}, \mb)- \mpolec},
\end{equation}
that has a pole for $\slashed{p}=m$.
If we consider eigenvalues of $\slashed{p}$ slightly different from $\mpole$,
the propagator becomes
\begin{eqnarray}
  G^r(\slashed{p}, \mpole) &\approx&\frac{i}{\slashed{p} - \mpole
  -i\Sigma(m,\mb) - z\(\slashed{p} - \mpole\) - \mpolec} \nonumber \\
  &=&\frac{i}{\slashed{p}-\mpole}(1-z)^{-1}\equiv
  \frac{i}{\slashed{p}-\mpole}Z,
\end{eqnarray}
where we have defined
\begin{equation}
  \label{eq:zq}
  z\equiv i \frac{\partial \Sigma(\slashed{p}, \mb)}{\partial
\slashed{p}}\Big|_{\slashed{p}=\mpole}.
\end{equation}
If we introduce the rescaled field $\phi_r$
\begin{equation}
\phi = Z^{1/2} \phi_r,
\end{equation}
we have the effect of the fermion
self-energy correction to an external line is equivalent to multiply the
leading order amplitude for a factor $Z^{1/2}$.
At order $\mathcal{O}(\as)$
\begin{eqnarray}
\label{eq:Z_NLO}
Z = (1-z)^{-1} \approx 1+z, \qquad \sqrt{Z} \approx 1 + \frac{1}{2}z,
\end{eqnarray}
where, neglecting orders $\mathcal{O}(\as^2)$, we have
\begin{equation}
\label{eq:z_NLO}  
  z \approx i \frac{\partial \Sigma(\slashed{p}, \mpole)}{\partial
\slashed{p}}\Big|_{\slashed{p}=\mpole}.
\end{equation}
Thus, the contribution to the virtual amplitude containing
radiative corrections to an external quark line is given by
\begin{equation}
\frac{1}{2}z \mathcal{M}_b,
\end{equation}
being $\mathcal{M}_b$ the LO amplitude.

An alternative to the pole mass scheme is given by the \MSB{} scheme.
The mass counterterm is defined in order to absorb only the divergent part
of $\Sigma$ (performing the replacement $\mu^2 \to \mu^2 \frac{e^{ \gameul}}{4\pi}$)
\begin{eqnarray}
\label{eq:MSbar_mass_counterterm}
\mMSB^c(\mu) &\equiv& -i \,\Sigma^{\rm (d)}\!\left(\mMSB(\mu), \mb \right)
 =  -i \,\Sigma^{\rm (d)}\!\left(\mpole, \mpole \right) + \mathcal{O}(\as^2)\,,
\end{eqnarray}
where $\mMSB(\mu)$ is the \MSB{} mass evaluated at the scale $\mu$ and (d)
denotes the divergent part.
We have
\begin{eqnarray}
  && \mb =  \mpole + \mpolec = \mMSB(\mu) + \mMSB^c(\mu), \\
  && \mpole - \mMSB(\mu) = -\lq \mpolec -
  \mMSB^c(\mu)\rq =i \,\Sigma^{\rm (f)}\!\left(\mpole, \mpole \right) + \mathcal{O}(\as^2) ,
\end{eqnarray}
where (f) denotes the finite part.

\section{One-loop radiative corrections to heavy quark lines}

In the following we describe how to compute the one-loop on-shell top-quark
self energy at order $\as$.  We give a finite mass $ \lambda$ to the gluon, since
this result is necessary for the computation of the self energy at all orders
in $\as(\as n_f)^n$, as discussed in Sec.~\ref{sec:msbar2pole}. Since we adopt
the \MSB{} scheme, in our computation we perform the replacement
\begin{equation}
  \mu^2 \to \mu^2 \frac{e^{\ep\gameul}}{4\pi}
\end{equation}
so that, as we have seen in Sec.~\ref{sec:dressed_gluon}, the \MSB{}
counterterms have a very simple expression since they need to absorb only
the UV divergent part of the radiative corrections.

At order $\as{}$, the top-quark self-energy expression with a gluon of mass
$ \lambda$ in $d=4-2\ep$ dimensions is given by
\begin{align}
\!\!\!\!\!\!\!
\Sigma^{(1)}_\lambda(\slashed{p},  m)  = & \, -g^2 \CF \left(\frac{\mu^2}{4\pi}
      e^{\gameul} \right)^\ep \int \frac{\mathd^d k}{(2\pi)^d}
      \frac{\gamma^\alpha (\slashed{p}+\slashed{ k}+m)\gamma_\alpha
      }{\lq k^2- \lambda^2\rq\lq( k+p)^2-m^2\rq} \nonumber \\
      = & \, -g^2 \CF
      \left(\frac{\mu^2}{4\pi} e^{\gameul} \right)^\ep \int
      \frac{\mathd^d k}{(2\pi)^d} \frac{(2-d)\slashed{p} +d\,m +
        (2-d)\slashed{ k}}{\lq k^2- \lambda^2\rq \lq ( k+p)^2-m^2\rq} \label{eq:topSelfMassiveGluon_partial}
      \\  = & \, -g^2 \CF \left[(2-d)\slashed{p}
        \frac{(m^2- \lambda^2-p^2)B(p^2,  \lambda^2,
          m^2)+A( \lambda^2)-A(m^2)}{2p^2} \right. \nonumber \\
        &\,
        \phantom{-g^2 \CF \bigg[} +\left((2-d)\slashed{p}+d\,m\right)B(p^2,
           \lambda^2, m^2)\bigg]\,
  \label{eq:topSelfMassiveGluon} ,
\end{align}
where the suffix $\lambda$ signals the presence of a finite gluon-mass
$\lambda$ and $A$ and $B$ are the one-point and two-point Feynmam scalar
integrals
\begin{eqnarray}
  A(m^2) &=& \left(\frac{\mu^2}{4\pi}
      e^{\gameul} \right)^\ep \int \frac{\mathd^d k}{(2\pi)^d}
  \frac{1}{ k^2 -m^2}\,, \\
 B(p^2,  \lambda^2, m^2) &=& \left(\frac{\mu^2}{4\pi}
      e^{\gameul} \right)^\ep \int \frac{\mathd^d k}{(2\pi)^d}
  \frac{1}{\left[ k^2 -\lambda^2\right]\left[( k+p)^2-m^2\right]}\,.
\end{eqnarray}

\section{Mass counterterm}
From eq.~(\ref{eq:mpolec_NLO}), we see that the pole mass counterterm can be
obtained from the self energy evaluated for the eigenvalue of the operator
$\slashed{p}$ equal to $\mpole$.  We thus compute $\Sigma_\lambda^{(1)}(m,
m)$. As it is shown in Sec.~\ref{sec:msbar2pole}, we also need the exact
$d$-dimensional expression for the $ \lambda=0$ and $ \lambda \gg m$ cases,
while for a generic $\lambda$ value it is sufficient the $\mathcal{O}(\ep^0)$
expressions.

\subsection{Non-vanishing gluon mass}
We now calculate $\Sigma^{(1)}_\lambda(m, m)$ for a generic value of $ \lambda >0$.
Eq.~(\ref{eq:topSelfMassiveGluon}) becomes
\begin{eqnarray}
   \label{eq:topSelfMassiveGluon2}  
 \Sigma^{(1)}_\lambda(m,m)  &=& - g^2\, \CF\, m \,
\left[(2-d)\frac{- \lambda^2\, B(m^2,  \lambda^2,
                          m^2)+A( \lambda^2)-A(m^2)}{2 m ^2}\right. \nonumber \\
                    &&  \phantom{- g^2\, \CF\, m \, \Big[}    \left. +d\,B(m^2,  \lambda^2, m^2)\right]\qquad
\end{eqnarray}
and the scalar integrals assume the values
\begin{eqnarray}
  \label{eq:A_expanded}
A(m^2) & = & \frac{i}{(4\pi)^2} \, m^2 \, \left[\frac{1}{\ep}+
  \log\left(\frac{\mu^2}{m^2}\right) + 1 \right] +\mathcal{O}(\ep),\\
  \label{eq:B_expanded}
B(m^2,  \lambda^2, m^2) & = & \frac{i}{(4\pi)^2} \left[\frac{1}{\ep}+2+
  \log\left(\frac{\mu^2}{m^2}\right)+H\left(\frac{ \lambda^2}{m^2}\right) \right]+\mathcal{O}(\ep),
\end{eqnarray}
with
\begin{equation}
\label{eq:H}
H(x) =
\left\{
\begin{array}{ll}
 \displaystyle{  -\frac{x}{2} \log{x} - \sqrt{x(4-x)}
   \,\arctan\sqrt{\frac{4-x}{x}}} & x \leq 4\,,
 \\[2mm]
  \displaystyle{   -\frac{x}{2} \,\log{x} + \frac{1}{2}  \sqrt{x(x-4)} \log \frac{\sqrt{x} +
      \sqrt{x-4}}{\sqrt{x} - \sqrt{x-4}}} \quad \quad& x>4   \,.
\end{array}
\right.
\end{equation}
We can check that for small $x$
\begin{equation}
  \label{eq:H_smallx}
  H(x) = -\pi \sqrt{x} +\mathcal{O}(x)\,,
\end{equation}
while for large $x$
\begin{equation}
  \label{eq:H_largex}
  H(x) = -1 - \log(x) +\frac{1}{x}\left(\frac{1}{2}-\log(x)\right)+\mathcal{O}\(\frac{1}{x^2}\).
\end{equation}
Substituting the values of $A$ and $B$ in
eq.~(\ref{eq:topSelfMassiveGluon2}), we get
\begin{eqnarray}
 \label{eq:Sigma1_lambda} 
 \Sigma^{(1)}_\lambda(m,m)  &=& -i\,\frac{\as}{4\pi} \CF \,m\left[
    \frac{3}{\ep}+3\log\left(\frac{\mu^2}{m^2}\right)+4+
    \frac{ \lambda^2}{m^2}\left( 1 + \log \left( \frac{ \lambda^2}{m^2}\right)
  \right) \right. \qquad \nonumber \\
  && \phantom{-i\,\frac{\as}{4\pi} \CF \,m\Big[} \left. + \left(2+\frac{ \lambda^2}{m^2}\right)H\left( \frac{ \lambda^2}{m^2}\right)  \right]\,.\qquad 
\end{eqnarray}
For small $ \lambda$ we have
\begin{eqnarray}
  \label{eq:Sigma1_small_lambda}
\Sigma^{(1)}_\lambda(m,m) = - i \frac{\as}{4\pi} \CF
  \,m\left[ \frac{3}{\ep}+3\log\left(\frac{\mu^2}{m^2}\right)+4 \right] + \as
  \frac{\CF}{2}  \lambda + \mathcal{O}( \lambda^2)\,,\qquad
\end{eqnarray}
while for large $ \lambda$ 
\begin{eqnarray}
  \label{eq:Sigma1_large_lambda}
\Sigma^{(1)}_\lambda(m,m) = - i \frac{\as}{4\pi} \CF
  \,m\left[ \frac{3}{\ep}+3\log\left(\frac{\mu^2}{ \lambda^2}\right)+\frac{5}{2} \right] + \mathcal{O}\left(\frac{1}{ \lambda^2}\right)\,.\qquad
\end{eqnarray}
We thus have that the mass counterterm in the pole scheme, computed keeping a
fixed gluon mass $\lambda$, is given by
\begin{eqnarray}
 \mpolec_\lambda & =& - i\,\Sigma^{(1)}_\lambda(m,m) \nonumber \\
 & =& - \frac{\as}{4\pi} \CF \,m\left[
    \frac{3}{\ep}+3\log\left(\frac{\mu^2}{m^2}\right)+4+
    \frac{ \lambda^2}{m^2}\left( 1 + \log \left( \frac{ \lambda^2}{m^2}\right)
   \right) \right. \nonumber \\
  && \phantom{- \frac{\as}{4\pi} \CF \,m\Big[} \left.+ \left(2+\frac{ \lambda^2}{m^2}\right)H\left( \frac{ \lambda^2}{m^2}\right)
    \right]\,. 
\end{eqnarray}

\subsection{Massless gluon}
\label{sec:sigma1_0}
We are now interested in the particular case $ \lambda=0$.
By setting $ \lambda=0$ in eq.~(\ref{eq:topSelfMassiveGluon2}), we find
\begin{eqnarray} 
\Sigma^{(1)}_0(m, m)  &=& - g^2\, \CF\, m \,
\left[-(2-d)\frac{A(m^2)}{2 m ^2} +d\,B(m^2,  \lambda^2, m^2)\right]\,.
\end{eqnarray}
In this case, it is possible to use the exact $d$ dimensional expressions for
the scalar integrals
\begin{eqnarray}
  \label{eq:A_exact}
A(m^2) & = & \frac{i}{(4\pi)^2} \Gamma(1+\ep) \,{\rm
  e}^{\ep\gameul}\left(\frac{\mu^2}{m^2}\right)^\ep \frac{m^2}{\ep\,(1-\ep)}\,,
\\
  \label{eq:B0_exact}
B(m^2, 0, m^2) & = & \frac{i}{(4\pi)^2} \Gamma(1+\ep) \,{\rm
  e}^{\ep\gameul}\left(\frac{\mu^2}{m^2}\right)^\ep \frac{m^2}{\ep\,(1-2\ep)}\,,
\end{eqnarray}
so that we find
\begin{eqnarray}
  \label{eq:Sigma1_0} 
\Sigma^{(1)}_0(m,m) = -i\frac{\as}{4\pi} \CF \,m \left(
\frac{\mu^2}{m^2}\right)^\ep e^{\ep \gameul}\frac{\Gamma(1+\ep)}{\ep}\frac{3-2\ep}{1-2\ep}\,.
\end{eqnarray}
By expanding around $\ep=0$ we get
\begin{eqnarray}
 \label{eq:Sigma1_0_expanded}
\Sigma^{(1)}_0(m, m) & \approx & -i\,\frac{\as}{4\pi} \CF \,m\left[
  \frac{3}{\ep}+3\log\left(\frac{\mu^2}{m^2}\right)+4 \right]\,.
\end{eqnarray}
By comparing this expression with eq.~(\ref{eq:Sigma1_small_lambda}), we find
that the small $ \lambda$ result smoothly converges to the massless one.
Since $\Sigma^{(1)}_0(m,m)$ corresponds to the standard result peformed with massless gluons, we have
\begin{equation}
  \Sigma^{(1)}_0(m,m) \equiv \Sigma^{(1)}(m,m).
\end{equation}
The pole mass counterterm is thus given by
\begin{equation}
  \mpolec = -i \Sigma^{(1)}(m,m) = -\frac{\as}{4\pi} \CF \,m\left[
  \frac{3}{\ep}+3\log\left(\frac{\mu^2}{m^2}\right)+4 \right]+\mathcal{O}(\as^2).
\end{equation}
As shown in eq.~(\ref{eq:MSbar_mass_counterterm}), if we adopt the \MSB{}
scheme, the mass counterterm contains only the divergent part of
$\Sigma^{(1)}$:
\begin{equation}
  \mMSB^c(\mu) =  -\frac{\as}{4\pi} \CF \, \mMSB(\mu)\, \frac{3}{\ep}
  +\mathcal{O}(\as^2)\,,
\end{equation}
and
\begin{equation}
  \mpole - \mMSB(\mu) = - \left[\mpolec - \mMSB^c(\mu) \right ] = 
\frac{\as}{4\pi} \CF \,m\left[3\log\left(\frac{\mu^2}{m^2}\right)+4
  \right]+\mathcal{O}(\as^2).
\end{equation}

\subsection{Large gluon mass}
In order to evaluate the exact $d$ dimensional expression of
eq.~(\ref{eq:topSelfMassiveGluon2}) for large $ \lambda$ we just need to
compute $B(m^2, \lambda^2, m^2)$ neglecting terms of order
$\frac{m^4}{ \lambda^4}$
\begin{align}
 B(m^2, \lambda^2, m^2)  = & \,\left( \frac{\mu^2}{4\pi}\,{\rm
   e}^{\gameul}\right)^\ep \int \frac{\mathd^d  k}{(2 \pi)^d} \frac{1}{( k +
                        p)^2 - m^2 + i \eta} \frac{1}{ k^2 -  \lambda^2 + i \eta}\nonumber\\
                   = & \,\frac{i
   \,e^{\ep\gameul}\Gamma (\epsilon)}{(4 \pi)^{2}}\mu^{2\ep} \int_0^1 [- z
   (1 - z) \lambda^2 + (1 - z)  \lambda^2 + z m^2]^{- \epsilon} \mathd z
 \nonumber\\  = &  \,\frac{i
   \,e^{\ep\gameul}\Gamma (\epsilon)}{(4 \pi)^{2}}\mu^{2\ep}  \int_0^1 [z^2 m^2 + (1 - z)  \lambda^2]^{- \epsilon} \mathd z
 \nonumber\\  = & \, \frac{i}{(4\pi)^2} \left(
 \frac{\mu^2}{m^2}\right)^\ep\,e^{\ep \gameul} \, \Gamma(\ep) \int_0^1
 (\eta_+ - z)^{- \epsilon} (\eta_- - z)^{- \epsilon} \mathd z,
\end{align}
where we have defined
\begin{equation}
  \eta_+ = \frac{1 + \sqrt{1 - 4 \xi}}{2 \xi} \approx \frac{1}{\xi} - 1, \hspace{2em} \eta_- = \frac{1 - \sqrt{1 - 4 \xi}}{2 \xi} \approx 1 + \xi \hspace{1em},
\end{equation}
with $ \xi = m^2 /  \lambda^2$.
For small $\xi$ the integral yields
\begin{align}
  \!\!\!\!\!\!\!\!\!\!\!\!\!\!
  \int_0^1 (\eta_+ - z)^{- \epsilon} (\eta_- - z)^{- \epsilon} \mathd z  \approx &
  \, \xi^{\epsilon} \int_0^1 (1 - \xi (1 + z))^{- \epsilon} (1 - z + \xi)^{-
  \epsilon} \mathd z \nonumber\\
   \approx & \, \xi^{\epsilon} \int_0^1 (1 + \epsilon \, \xi (1 + z)) (1 - z + \xi)^{-
  \epsilon} \mathd z \nonumber\\
   \approx & \, \xi^{\epsilon} \left\{ \frac{(1 + \xi)^{1 - \epsilon} - \xi^{1 -
  \epsilon}}{1 - \epsilon} + \epsilon \, \xi \int_0^1 (1 + z) (1 - z)^{-
  \epsilon} \mathd z \right\} \nonumber\\
  = & \, \xi^{\epsilon} \left\{ \frac{1 + (1 - \epsilon) \xi - \xi^{1 -
  \epsilon}}{1 - \epsilon} + \epsilon \, \xi \frac{3 - \epsilon}{(1 - \epsilon)
  (2 - \epsilon)} \right\} \nonumber\\
  = & \, \xi^{\epsilon} \left\{ \frac{1}{1 - \epsilon} - \frac{\xi^{1 -
  \epsilon}}{1 - \epsilon} + \xi \left( 1 + \epsilon \frac{3 - \epsilon}{(1 -
  \epsilon) (2 - \epsilon)} \right) \right\} \nonumber\\
  = & \, \xi^{\epsilon} \left\{ \frac{1}{1 - \epsilon} - \frac{\xi^{1 -
  \epsilon}}{1 - \epsilon} + \xi \frac{2}{(1 - \epsilon) (2 - \epsilon)}
  \right\}, 
\end{align}
so
\begin{eqnarray}
\label{eq:B_large_lambda}
  B(m^2, \lambda^2, m^2) &=& \frac{i}{(4\pi)^2} \left(
                       \frac{\mu^2}{ \lambda^2}\right)^\ep\,e^{\ep \gameul} \, \frac{\Gamma(1+\ep)}{\ep}  \nonumber \\
                  && \times  \left[\frac{1}{1 - \epsilon} - \left( \frac{m^2}{ \lambda^2} \right)^{1 -
                      \epsilon}\frac{1}{1 - \epsilon}+ \frac{m^2}{ \lambda^2} \frac{2}{(1 -
                      \epsilon) (2 - \epsilon)} \right]\,. \qquad
\end{eqnarray}
By inserting eqs.~(\ref{eq:B_large_lambda}) and (\ref{eq:A_exact}) in 
eq.~(\ref{eq:topSelfMassiveGluon2}) and neglecting terms of the order
$ \lambda^{-2}$ we find
\begin{eqnarray}
\Sigma_{\lambda_\infty}^{(1)}(m, m) = -i\frac{\as}{4\pi} \CF \,m \left(
\frac{\mu^2}{ \lambda^2}\right)^\ep\frac{\Gamma(1+\ep)\,e^{\ep\gameul}}{\ep}\frac{2\,(3-2\ep)}{(1-\ep)(2-\ep)}\,,
\end{eqnarray}
where the subscript $\infty$ signals that $\Sigma$ has been computed for a large value of $ \lambda$ and subleading powers $1/ \lambda^2$ have been neglected.
By expanding $\Sigma_{\lambda_\infty}^{(1)}(m,  \lambda^2)$ around $\ep=0$ we get
\begin{eqnarray}
  \label{eq:Sigma1_large}
\Sigma_{\lambda_\infty}^{(1)}(m, m) \approx -i\,\frac{\as}{4\pi} \CF \,m\left[
    \frac{3}{\ep}+3\log\left(\frac{\mu^2}{ \lambda^2}\right)+\frac{5}{3} \right]\,,
\end{eqnarray}
as we found in eq.~(\ref{eq:Sigma1_large_lambda}).

\section{External field normalization}
From eqs.~(\ref{eq:Z_NLO}) and~(\ref{eq:z_NLO}) we have that the external field
normalization is given by 
\begin{equation}
Z=1+z\,,\qquad {\rm with} \quad z=i\frac{\partial
  \Sigma(\slashed{p}, \mpole)}{\partial \slashed{p}}\Big|_{\slashed{p}=\mpole}.
\end{equation}
We perform this computation at $\mathcal{O}(\as)$ keeping a finite gluon mass
$ \lambda$ and, separately, for $ \lambda=0$.  Conversely to the previous
case, the small $ \lambda$ limit is not guaranteed to approach smoothly the $
\lambda=0$ result, since to compute $z$ we deal both with IR and UV
singularities.

We rewrite eq.~(\ref{eq:topSelfMassiveGluon}) as
\begin{equation}
  \label{eq:Sigma_coefficients}
\Sigma^{(1)}_\lambda(\slashed{p}, m)=g^2 \CF \lg \alpha\, B(p^2,  \lambda^2, m^2)+\beta \lq A( \lambda^2)-A(m^2) \rq \rg
\end{equation}
with
\begin{eqnarray}
\alpha = \frac{d-2}{2p^2}(m^2- \lambda^2-p^2)\slashed{p}+(d-2)\slashed{p} -d \lambda, \qquad\qquad
\beta = \frac{d-2}{2p^2}\slashed{p}\,.
\end{eqnarray}
Since $\slashed{p}\slashed{p}=p^2$,  the derivative of with respect
$\slashed{p}$ evaluated at $\slashed{p}=m$ can be rewritten as
\begin{equation}
  \label{eq:pslash_derivative}
 \frac{\partial}{\partial \slashed{p}}
 \Bigg|_{\slashed{p}=m}=2m\frac{\partial}{\partial
   p^2}\Bigg|_{\slashed{p}=m}.
\end{equation}
Using eq.~(\ref{eq:pslash_derivative}) and
\begin{eqnarray}
\alpha(\slashed{p}=m) = -m\left[2+(1-\ep)\frac{ \lambda^2}{m^2}\right]\,,\quad \frac{\partial \alpha}{\partial \slashed{p}
}\Bigg|_{\slashed{p}=m} =(1-\ep)\frac{ \lambda^2}{m^2}\,, \quad
\frac{\partial \beta}{\partial \slashed{p} }\Bigg|_{\slashed{p}=m}=
\frac{\ep-1}{m^2}\,,\quad
\label{eq:coefficients}
\end{eqnarray}
eq.~(\ref{eq:Sigma_coefficients}) becomes
\begin{align}
\frac{\partial\Sigma^{(1)}_\lambda(\slashed{p},m) }{\partial
  \slashed{p}}\Bigg|_{\slashed{p}=m} = &
g^2\,\CF\left\{\frac{ \lambda^2}{m^2}(1-\ep)B(m^2,  \lambda^2,m^2) +
  \frac{\ep-1}{m^2}\left[ A( \lambda^2)-A(m^2)\right]\right.  \nonumber \\
& \qquad \left.-m\left[2+(1-\ep)\frac{ \lambda^2}{m^2}\right]\, 
2m \, \frac{\partial B(p^2,  \lambda^2, m^2)}{\partial p^2}\Big|_{p^2=m^2}
\right\}.  \qquad
\label{eq:Sigma_prime}
\end{align}
The calculation of
 \[\frac{\partial B(p^2,  \lambda^2, m^2)}{ \partial m^2}
\Big|_{p^2=m^2}\]  must be carried
out by distinguishing the two cases $ \lambda=0$ and $ \lambda >0$.

\subsection{Massless gluon}
We need to compute the derivative of $B(p^2,  \lambda^2, m^2)$ for $p^2=m^2$
in order to evaluate the derivative of $\Sigma^{(1)}$. If $ \lambda^2=0$, we
have the well-known result
\begin{eqnarray}
\frac{\partial B(p^2, 0, m^2)}{\partial p^2} \Bigg|_{p^2=m^2} 
&=&\frac{i}{(4\pi)^2}\left(\frac{\mu^{2}}{m^2}\right)^\ep \frac{1}{m^2}
\left[-\frac{1}{2\ep}-1\right] +\mathcal{O}(\ep) \nonumber \\ &=
&\frac{1}{(4\pi)^2}\frac{1}{m^2}
\left[-\frac{1}{2\ep}-1-\frac{1}{2}\log\left(\frac{\mu^{2}}{m^2}\right)
  \right]+\mathcal{O}(\ep).\qquad \label{eq:Bprime_zerom}
\end{eqnarray}
Substituting eqs.~(\ref{eq:Bprime_zerom}), (\ref{eq:A_expanded}) and the
$\mathcal{O}(\ep^0)$ expansion of eq.~(\ref{eq:B0_exact}) in
eq.~(\ref{eq:Sigma_prime}) and setting $ \lambda=0$ we find
\begin{eqnarray}
\frac{\partial \Sigma^{(1)}(\slashed{p}, m)}{\partial \slashed{p} }\Bigg|_{\slashed{p}=\mpole}=i\frac{\as}{4\pi}\left[\frac{3}{\ep}+4+3\log\left(\frac{\mu^2}{\mpole^2}\right)\right]\,.
\end{eqnarray}
So, from eq.~(\ref{eq:z_NLO}), we get
\begin{equation}
z = i \frac{\partial \Sigma^{(1)}(\slashed{p}, m)}{\partial \slashed{p}
}\Bigg|_{\slashed{p}=\mpole} =
-\frac{\as}{4\pi}\left[\frac{3}{\ep}+4+3\log\left(\frac{\mu^2}{\mpole^2}\right)\right]\,.
\label{eq:zt_0}
\end{equation}

\subsection{Massive gluon}
We now compute the derivative of the bubble integral for $ \lambda>0$.
\begin{align}
 & \frac{\partial B(p^2,  \lambda^2, m^2)}{\partial p^2} \Bigg|_{p^2=m^2}  =
\frac{\partial}{\partial p^2} \left[\left(\mu^2 \frac{e^{\ep \gameul}}{4\pi}
\right)^{\ep} \int \frac{\mathd^d  k}{(2\pi)^{d}}
\frac{1}{\left[ k^2-\lambda^2+i\eta\right]\left[( k+p)^2-\mpole^2+i\eta\right]}
  \right] 
\nonumber \\
& \qquad =
\frac{i e^{\ep
  \gameul}}{(4\pi)^2}\frac{\Gamma(1+\ep)}{\ep}\mu^{2\ep}\frac{\partial}{\partial
  p^2}\int_0^1 \mathd x \left(-x(1-x) p^2 +(1-x) \lambda^2+x\, m^2 +i\eta\right)^{-\ep}
  \Bigg|_{p^2=m^2}  \nonumber \\
& \qquad = \frac{i}{(4\pi)^2} \int_0^1 \frac{\mathd x \,x(1-x)}{\left[m^2\, x^2+(1-x)
  \lambda^2-i\eta\right]}+\mathcal{O}(\ep). 
\end{align}
The denominator can be rewritten as
\begin{eqnarray}
m^2\, x^2+(1-x) \lambda^2-i\eta \approx m^2\, x^2+(1-x)(\lambda^2-i\eta) =m^2(x-\alpha_+)(x-\alpha_-)\,,
\end{eqnarray}
with 
\begin{equation}
\alpha_\pm = \frac{(\lambda^2+i\eta)\pm \sqrt{(\lambda^2+i\eta)-4\,m^2(\lambda^2+i\eta)}}{2\,m^2}.
\end{equation}
We can suppose $\alpha_\pm$ negative thanks to the analytic continuation.
\begin{eqnarray}
\frac{\partial B(p^2, \lambda^2, m^2)}{\partial p^2} \Bigg|_{p^2=m^2} & = &\frac{i}{(4\pi)^2}\frac{1}{m^2}\int_0^1\frac{\mathd x \, x(1-x)}{(x-\alpha_+)(x-\alpha_-)} \nonumber \\
&= & \frac{i}{(4\pi)^2}\frac{1}{m^2}\frac{1}{\alpha_+-\alpha_-}\int_0^1 \mathd x\, x(1-x) \left[ \frac{1}{x-\alpha_+}-\frac{1}{x-\alpha_-}\right] \nonumber \\
&=& \frac{i}{(4\pi)^2}\frac{1}{m^2}\frac{1}{\alpha_+-\alpha_-}\left[I_1+I_2\right].
\end{eqnarray}
We have defined
\begin{eqnarray}
I_1&= &\int_0^1dx \frac{x(1-x)}{x-\alpha_+} = \int_0^1dx\ \frac{x(1-x)-\alpha_+(1-\alpha_+)+\alpha_+(1-\alpha_+)}{x-\alpha_+} \nonumber \\
&= & \int_0^1dx\ (1-\alpha_+-x) + \alpha_+(1-\alpha_+)\int_0^1dx\frac{1}{x-\alpha_+} \nonumber \\
&= & \frac{1}{2}-\alpha_+ +\alpha_+(1-\alpha_+)\log\left(\frac{\alpha_+-1}{\alpha_+}\right)\,,
\end{eqnarray}
and, similarly,
\begin{equation}
I_2= -\frac{1}{2}+\alpha_- -\alpha_-(1-\alpha_-)\log\left(\frac{\alpha_--1}{\alpha_-}\right).
\end{equation}
The final result is
\begin{eqnarray}
\frac{\partial B(p^2,  \lambda^2, m^2)}{\partial p^2} \Bigg|_{p^2=m^2}  &=&\frac{i}{(4\pi)^2}\left(\frac{\mu^{2}}{m^2}\right)^\ep \frac{1}{m^2} \left[-1+G\( \frac{\lambda^2}{m^2}\)\right]
\label{eq:Bprime_finitem}
\end{eqnarray}
with
\begin{equation}
  \label{eq:G}
\!\!\!
G(x) =
\left\{
\begin{array}{ll}
  \displaystyle{  -\frac{3-x}{4-x} \sqrt{x(4-x)}\left[
      \arctan\(\frac{2-x}{\sqrt{x(4-x)}}\)
      +\arctan\( \sqrt{\frac{x}{4-x}}\)\right]
  }  & x < 4\,, 
  \\[5mm]
  0 & x=4\,,
  \\[5mm]
  \displaystyle{ \frac{x-3}{2(x-4)} \sqrt{x(x-4)}\left[
      \log\(\frac{x-2-\sqrt{x(x-4)}}{x-2+\sqrt{x(x-4)}}\)
       +     \log\(\frac{x-\sqrt{x(x-4)}}{x+\sqrt{x(x-4)}}\)\right]
  } &x>4  \\
 \end{array}
\right.
\end{equation}
We can easily demonstrate that, for small $x$, 
\begin{equation}
G(x) =  -\frac{1}{2}\log\left(x\right) + \frac{3}{4}\pi
\sqrt{x}+\mathcal{O}\left(x^2\right).
\label{eq:G_smallx}
\end{equation}
Inserting eqs.~(\ref{eq:Bprime_finitem}), (\ref{eq:A_expanded}) and~(\ref{eq:B_expanded}) in eq.~(\ref{eq:Sigma_prime}) leads to
\begin{align}
\frac{\partial\Sigma^{(1)}_\lambda(\slashed{p}, m) }{\partial \slashed{p}}\Bigg|_{\slashed{p}=m}
 =&i
\frac{\as}{4\pi}\CF\, \lg
\frac{1}{\ep}+3\log\left(\frac{\mu^2}{m^2}\right)+4 -4\,G\(\frac{\lambda^2}{m^2}\)
\right.+\frac{ \lambda^2}{m^2}\lq \log\left(\frac{\mu^2}{m^2}\right) \right. \nonumber \\
&  \left.\left. +3 +H\left(\frac{ \lambda^2}{m^2}
\right)-2\,G\(\frac{\lambda^2}{m^2}\)-\log\left(\frac{m^2}{ \lambda^2}\right)\rq\rg. \qquad
\qquad
\label{eq:Sigma_derivative}
\end{align}
From eq.~(\ref{eq:z_NLO}) we have
\begin{eqnarray}
z_{\lambda} &=& - \frac{\as}{4\pi}\CF\, \lg
\frac{1}{\ep}+3\log\left(\frac{\mu^2}{m^2}\right)+4 -4\,G\(\frac{\lambda^2}{m^2}\)
\right.+\frac{ \lambda^2}{m^2}\lq \log\left(\frac{\mu^2}{m^2}\right) \right. \nonumber \\
&&  \left.\left. +3 +H\left(\frac{ \lambda^2}{m^2}
\right)-2\,G\(\frac{\lambda^2}{m^2}\)-\log\left(\frac{m^2}{ \lambda^2}\right)\rq\rg. \qquad
\qquad
\label{eq:zt_lambda}
\qquad
\end{eqnarray}
From eqs.~(\ref{eq:H_smallx}) and~(\ref{eq:G_smallx}) we can extract the
small $ \lambda$ behaviour
\begin{equation}
\!
z_{\lambda} =-
\frac{\as}{4\pi}\CF\,\lq  \frac{1}{\ep}+2\log\left(\frac{\mu^2}{
  \lambda^2}\right)+3\log\left(\frac{\mu^2}{m^2}\right)+4+\log\left(\frac{\mu^2}{m^2}\right)-3\pi
\frac{ \lambda}{m}\rq+\mathcal{O}\left(\lambda^2\right). \!\!\!\!\!
\end{equation}

\section{Radiative corrections to external massless quark lines}
We now compute the field normalization for a massless $b$ quark.  At order
$\as{}$, the self-energy for a massless quark computed with a gluon of mass
$ \lambda$ is given by setting $m=0$ in eq.~\ref{eq:topSelfMassiveGluon}.
\begin{eqnarray}
\Sigma^{(1)}_\lambda(\slashed{p}, 0) = -g^2 \CF(2-d) \slashed{p}\left[
          \frac{A( \lambda^2)+(p^2- \lambda^2)B(p^2, \lambda^2,0)}{2\,p^2}\right]\,.
        \label{eq:Sigma1_b}
\end{eqnarray}
The tadpole integral is given by eq.~(\ref{eq:A_expanded}), while the bubble one is given by
\begin{eqnarray}
  B(p^2, \lambda^2,0) & = &  \frac{i}{(4\pi)^2}\lq
  \frac{1}{\ep}+\log\(\frac{\mu^2}{ \lambda^2}\) + 2-
  \left(1-\frac{ \lambda^2}{p^2}\right)\log\left(1-\frac{p^2}{ \lambda^2}\right)\right]
\nonumber \\
& = & \frac{i}{(4\pi)^2}\lq  \frac{1}{\ep}+\log\(\frac{\mu^2}{ \lambda^2}\) +
1+\frac{1}{2}\frac{p^2}{m^2} \rq +\mathcal{O}\(\frac{p^4}{ \lambda^4} \)\,.
\end{eqnarray}
Thus, for small $p^2$
\begin{eqnarray}
  \Sigma^{(1)}_\lambda(\slashed{p}, 0)= i \frac{\as}{4\pi} \CF
  \slashed{p}\left[\frac{1}{\ep}+\log\(\frac{\mu^2}{ \lambda^2}\)-\frac{1}{2}+\mathcal{O}\(\frac{p^2}{ \lambda^2} \)
    \right]\,.
\end{eqnarray}
We can now compute the mass counterterm and the field normalization constant
for a massless $b$ quark
\begin{eqnarray}
  m_\lambda^c &=& i \Sigma^{(1)}_\lambda(0, 0) = 0\,,\\
  z_{\lambda}  &=& i \frac{\partial\Sigma^{(1)}_\lambda(\slashed{p}, 0)}{\partial
    \slashed{p}}\Big|_{\slashed{p}=0} =- \frac{\as}{4\pi} \CF\left[\frac{1}{\ep}+\log\(\frac{\mu^2}{ \lambda^2}\)-\frac{1}{2}
    \right]\,.
  \label{eq:zb_lambda}
\end{eqnarray}
The above computation of $z_\lambda$ cannot be used to infer the value for $
\lambda=0$. Indeed in this case eq.~(\ref{eq:Sigma1_b}) becomes
\begin{eqnarray}
  \Sigma^{(1)}(\slashed{p}, 0) & = & -g^2(d-2)\slashed{p} \frac{B(p^2,0,0)}{2}
\end{eqnarray}
and, since $B(0,0,0)=0$,
\begin{eqnarray}
  m^c = 0, \qquad
  z   = 0\,.
  \label{eq:zb_0}
\end{eqnarray}

\chapter{A useful dispersive relation}
We want to apply a dispersive relation to the function
\begin{equation}
 f( k^2)= \frac{1}{ k^2}  \frac{1}{1 + \Pi ( k^2, m_q^2, \mu^2) -
     \Pi_{\tmop{ct}}}\,,
 \end{equation}
 where $ k^2$ is a complex number far from the real positive axis.  The
 procedure is similar to the one proposed in Ref.~\cite{Beneke:1994qe}.  We
 notice that $f( k^2)$ has a pole at $  k^2=0$ and a branch cut for
 $ k^2> 4m_q^2$.
\begin{figure}[tb]
  \centering
  \includegraphics[width=0.65\textwidth]{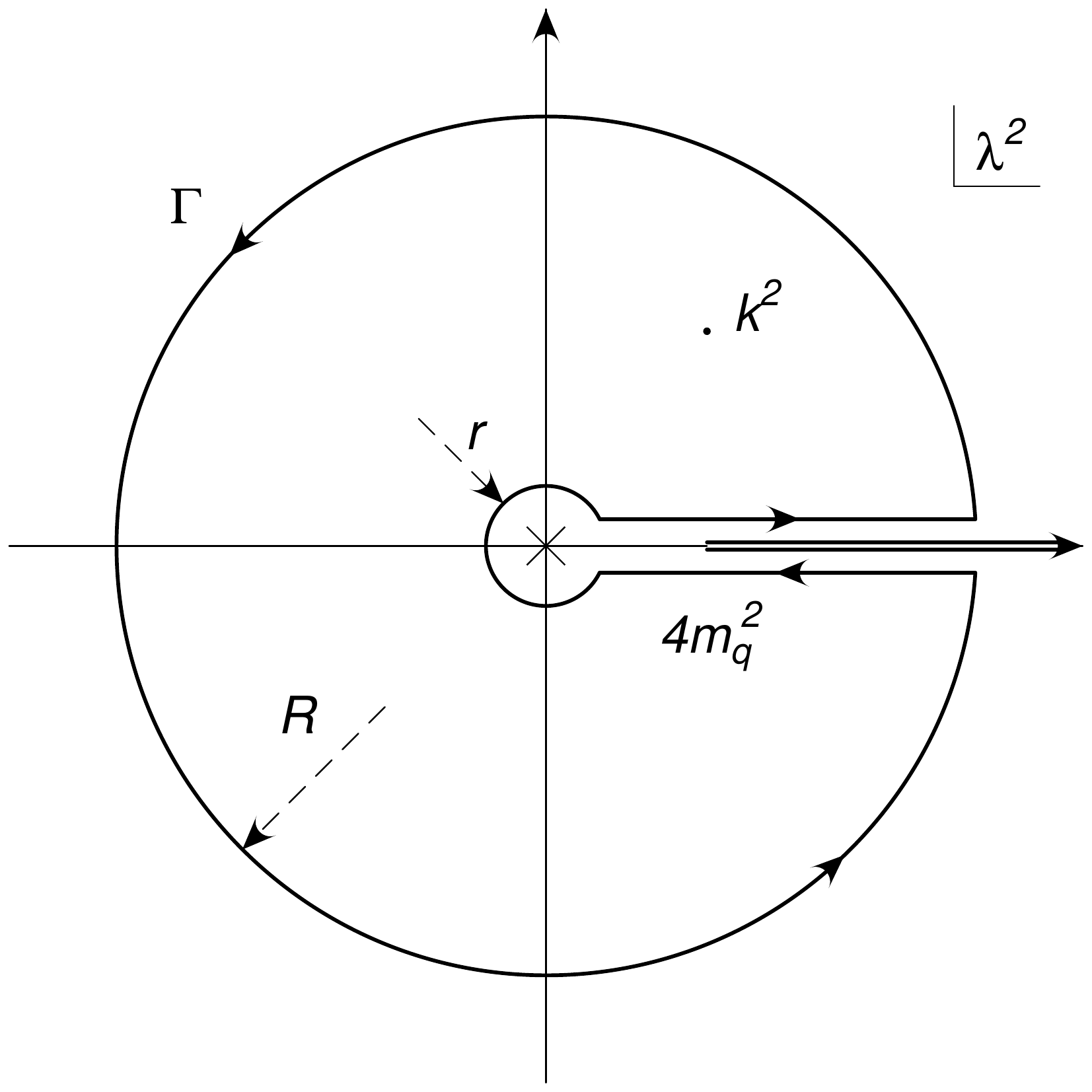}
  \caption{Integration path.}
  \label{fig:integr_path}
\end{figure}
We choose as integration path the closed contour $\Gamma$, depicted in
Fig.~\ref{fig:integr_path}.  Applying the residue theorem we have
\begin{equation}
  \label{eq:residue}
 f( k^2) = \frac{1}{2\pi i} \oint_\Gamma \mathd \lambda^2 \frac{1}{ \lambda^2-
   k^2}\, f(\lambda^2) = \frac{1}{2\,\pi\,i}\oint_\Gamma d \lambda^2
 \frac{1}{ \lambda^2- k^2}\, \frac{1}{\lambda^2}\,\frac{1}{1 + \Pi ( \lambda^2, m_q^2, \mu^2) -
   \Pi_{\tmop{ct}}} ,
\end{equation}
where the path $\Gamma$ is displayed in fig.~\ref{fig:integr_path}.  Notice
that we have ignored the presence of the Landau singularity at
\begin{equation}
  \label{eq:LandauPole}
   \lambda^2_L = -\mu^2\exp\left(-\frac{1}{b_0\as}+C\right)
 \end{equation}
 correponding to the vanishing of the denominator in the expression
 \begin{equation} \label{eq:vacpol33}
   \frac{1}{1+\Pi(k^2,m_q^2,\mu^2)-\Pi_{\rm ct}}=\sum_{i=0}^\infty
   \left(-\Pi(k^2,m_q^2,\mu^2)+\Pi_{\rm ct}\right)^i\,.
 \end{equation}
 This is because we are only interested in the formal power expansion in $\as$
 of our result, and no such singularity is present in the coefficients
 of the geometric expansion in eq.~(\ref{eq:vacpol33}).

The contribution along the circle of radius $R$ in eq.~(\ref{eq:residue})
goes to zero in the $R\to\infty$ limit.  The contribution given by the small
circle of radius $r$ is given by
\begin{equation}
  I_r =\frac{1}{ k^2}  \frac{1}{1 + \Pi (0, m_q^2, \mu^2)}.
\end{equation}
We are thus left with
\begin{eqnarray}
f( k^2) & = & \frac{1}{2\,\pi\,i} \left[ \int_{+\infty}^{0} \frac{\mathd
  \lambda^2}{\lambda^2- k^2} f(\lambda^2-i\eta) + \int_{0}^{+\infty} \frac{\mathd
    \lambda^2}{\lambda^2- k^2} f(\lambda^2+i\eta) \right] \nonumber \\
&& + \frac{1}{ k^2}  \frac{1}{1 + \Pi (0, m_q^2, \mu^2)}
\end{eqnarray}
where $i\eta$ is an infinitesimal imaginary part that enables us to avoid the
branch cut on the real axis. For positive values of $\lambda^2$ 
\begin{equation}
f(\lambda^2+i\eta)-f(\lambda^2-i\eta) = \frac{ 2\,i}{\lambda^2} \,{\rm Im}    \left[\frac{1}{1 + \Pi (\lambda^2, m_q^2, \mu^2) -
    \Pi_{\tmop{ct}}}\right]\,,
\end{equation}
and the imaginary part vanishes for $k^2 \le 4 m_q^2$. We can thus write
\begin{eqnarray}
  \label{eq:virt_trick}
f( k^2) &=& \frac{1}{ k^2}  \frac{1}{1 + \Pi
    ( k^2, m_q^2, \mu^2) - \Pi_{\tmop{ct}}}
\nonumber\\
& = &  -\frac{1}{\pi} \int_{4 m_q^2}^{+\infty}\!\!  \frac{\mathd \lambda^2}{\lambda^2}
 \, \frac{1}{  k^2 -  \lambda^2}\, {\rm Im} \lq 
 \frac{1}{1 + \Pi 
   ( \lambda^2, m_q^2, \mu^2)  - \Pi_{\tmop{ct}}} \rq \nonumber \\
&& +  \frac{1}{ k^2}  \frac{1}{1 + \Pi (0, m_q^2, \mu^2)}\,.
\end{eqnarray}
If $ k^2$ is a real-positive number, we can still employ
eq.~(\ref{eq:virt_trick}) since we need to include a small positive
imaginary part $i\eta$ coming from the Feynman prescription, so that
$k^2+i\eta$ is indeed a complex number. 

Since the imaginary part of $\Pi$ vanishes for $\lambda^2<4m_q^2$ and
\begin{align}
& {\rm Im} \lq \frac{1}{\lambda^2+i\eta}  \frac{1}{1 + \Pi 
  ( \lambda^2, m_q^2, \mu^2)  - \Pi_{\tmop{ct}}} \rq \nonumber \\
 & \qquad = -\pi \,\delta(\lambda^2) \,\frac{1}{ 1 + \Pi (0, m_q^2, \mu^2)} + \frac{1}{\lambda^2} \,{\rm Im} \lq 
 \frac{1}{1 + \Pi 
   ( \lambda^2, m_q^2, \mu^2)  - \Pi_{\tmop{ct}}} \rq,
 \label{eq:imag_den}
\end{align}
eq.~(\ref{eq:virt_trick}) can be rewritten as:
\begin{eqnarray}
  \label{eq:virt_trick2}
f( k^2) &=& \frac{1}{ k^2}  \frac{1}{1 + \Pi
    ( k^2, m_q^2, \mu^2) - \Pi_{\tmop{ct}}}
\nonumber\\
& = &  -\frac{1}{\pi} \int_{0^-}^{+\infty}\!\!  d \lambda^2
 \, \frac{1}{  k^2 -  \lambda^2}\,{\rm Im} \lq \frac{1}{\lambda^2+i\eta} \, 
 \frac{1}{1 + \Pi 
   ( \lambda^2, m_q^2, \mu^2)  - \Pi_{\tmop{ct}}} \rq, \phantom{aaaaa}
\end{eqnarray}
where the $0^-$ lower boundary underlines the fact that the delta function
arising from the imaginary part of $1/(\lambda^2+i\eta)$ must be included.

The dispersive relation we presented here is useful to evaluate virtual
contributions in which the gluon line has been dressed with the insertion of
infinite fermionic bubbles. When IR divergences are not present, the form
in eq.~(\ref{eq:virt_trick2}) can be safely employed, since the $\lambda \to
0$ limit is smooth and does not require extra care.  Furthermore, in this
case, it is also possible to set to zero the quark-mass regulator $m_q$.
When IR singularities are present, the $\lambda=0$ value must be handled
separately, since it requires for example dimensional regularization to deal
with soft and collinear divergences. In this case, eq.~(\ref{eq:virt_trick})
is more appropriate since it shows a clear separation between the $\lambda=0$
and the $\lambda > 2 m_q$ regions. 

\chapter{Single-top production in the narrow-width approximation}
\label{sec:NWA}
In this appendix we compute the NLO corrections to the total cross section
for the process $W^{*} \to t\,\bar{b} \to W\,b\, \bar{b}$ in the narrow-width
approximation~(NWA), i.e. for a vanishing top width, keeping a finite gluon
mass $ k$. The top mass is defined in the on-shell (real) pole scheme. We
denote it with $m_0$ since we reserve $m$ for the complex top mass.  We can
write
\begin{equation}
\sigma\!\(W^{*} \to W\, b\, \bar{b}\) = \sigma\!\(W^{*} \to t\,\bar{b}\)
\frac{\Gamma(t\to W\, b)}{\Gamma_t}\,,
\end{equation}
where
\begin{equation}
  \Gamma_t = 1.3279~{\rm GeV}.
\end{equation}
At one loop, in the NWA,
\begin{equation}
\sigma^{(1)}\!\(W^{*} \to W\, b\, \bar{b}\) = \sigma^{(1)}\!\(W^{*} \to
t\,\bar{b}\) \frac{\Gamma^{(0)}(t\to W\, b)}{\Gamma_t}+\sigma^{(0)}\!\(W^{*}
\to t\,\bar{b}\) \frac{\Gamma^{(1)}(t\to W\, b)}{\Gamma_t}\,.
\end{equation}
As we have shown in eq.~(\ref{eq:sigma_final}), to compute the all-orders
corrections to a physical tree-level quantity in the large-$n_f$ limit, we
need the expression of its NLO corrections calculated keeping a fixed gluon
mass $\lambda$.\footnote{The term $\Delta(\lambda)$ of
  eq.~(\ref{eq:Deltak2_final}) is identically 0 since we do not impose any
  selection cuts.}  We denote this quantity with $T(\lambda)$. The linear
$\lambda$ dependence of $T$ is responsible for the presence of linear
renormalons.

\section{Production cross section}
At LO, neglecting all the couplings, the production cross section is given by
\begin{equation}
  \sigma^{(0)}\(W^{*} \to t\,\bar{b}\) = \frac{2\(s-m_0^2\)\(2\,s+m_0^2\)}{s}
  \left(1-\frac{m_0^2}{s}\right) \frac{1}{16\pi \,s}\,.
\end{equation}
Its derivative with respect to $m_0$ is given by
\begin{equation}
 \frac{\partial \sigma^{(0)}\(W^{*} \to t\,\bar{b}\)}{\partial m_0} = -\frac{3}{4\pi}\frac{m_0\(s-m_0^2\)\(m_0^2+s\)}{s^3}\,.
\end{equation}
We computed separately the virtual and the real cross sections for several
positive values of the gluon mass $\lambda$ and we denote with $T(\lambda)$
their sum.  The external field normalization constant of the $\bar{b}$ and of
the $t$ quark, used in the expression of the UV normalized virtual
contribution, are given by eqs.~(\ref{eq:zb_lambda}) and~(\ref{eq:zt_lambda})
respectively.  We used the {\tt COLLIER}~\cite{Denner:2016kdg} library for
the evaluation of the one-loop scalar integrals.  The $\lambda=0$ computation
was performed separately within the \RES{} framework.  In this case the
external $b$ normalization constant is 0, while the top one is given in
eq.~(\ref{eq:zt_0}).

\begin{figure}[htb]
  \centering
  \includegraphics[width=0.65\textwidth]{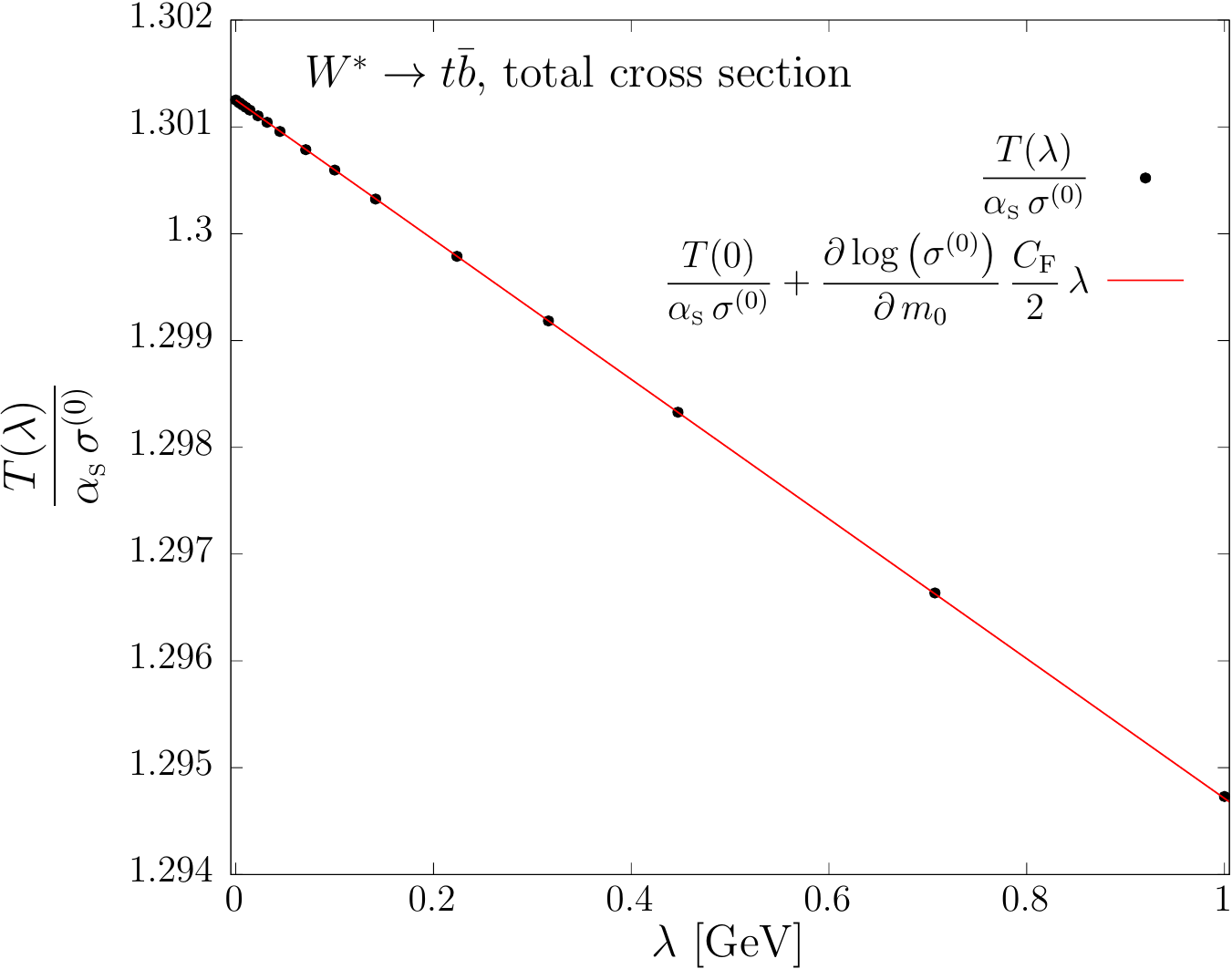}
    \caption{NLO corrections to the cross section for the process $W^* \to t
      \,\bar{b}$ computed with a finite gluon mass $\lambda$.}
  \label{fig:W_tb_xsec}
\end{figure}

We display our results in Fig.~\ref{fig:W_tb_xsec}, where we also plot the
straight line
\begin{equation}
 \frac{1}{\as\,\sigma^{(0)}\!\(W^{*} \to t\,\bar{b}\)} \left[T(0)+ \frac{\partial \sigma^{(0)}\!\(W^{*} \to t\,\bar{b}\)}{\partial\,
  m_0}\frac{\CF}{2}\,\as\, \lambda \right],
\end{equation}
that interpolates fairly well the points we have computed for small-$k$
values.  Since to move from the pole to the mass scheme, we can naively
replace $T(\lambda) \to T(\lambda) -\frac{\partial \sigma^{(0)}\(W^{*} \to
  t\,\bar{b}\)}{\partial\, m_0}\frac{\CF}{2} \lambda$, we notice that, if the
top mass is expressed in terms of the \MSB{} one, the inclusive cross section
is free from linear renormalons.  This behaviour is expected, since it is a
totally inclusive decay of a colour-neutral system.

\section{Top decay width}
\begin{figure}[htb]
  \centering
  \includegraphics[width=0.65\textwidth]{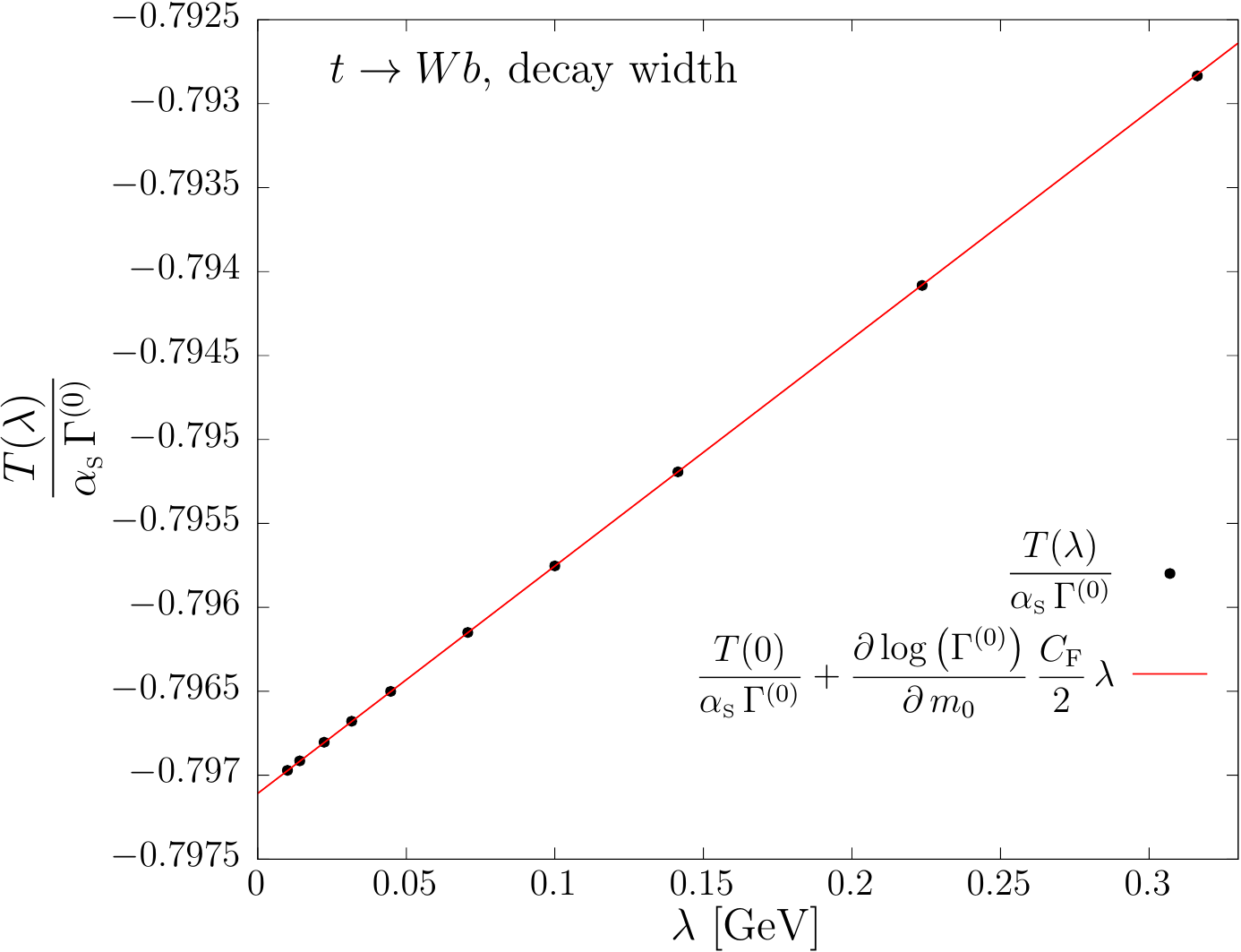}
  \caption{NLO corrections to the top decay width computed with a finite
    gluon mass $\lambda$.}
  \label{fig:Gammat}
\end{figure}

At LO, neglecting all the couplings, the decay width of the top quark is
given by
\begin{equation}
  \Gamma^{(0)}\(t \to W \, b\) =
  m_0^3\,\(1-\frac{m_{\sss W}^2}{m_0^2}\)^2\(1+\frac{2m_{\sss W}^2}{m_0^2} \)\frac{1}{8\pi}\,.
\end{equation}
Its derivative with respect to $m_0$ is given by
\begin{equation}
  \frac{\partial \Gamma^{(0)}\( t \to W\,b\)}{\partial m_0} = 3\(m_0^2-m_{\sss
    W}^2\)\(m_0^4+m_0^2\,m_{\sss W}^2+2m_{\sss W}^4\)\, \frac{1}{8\pi m_0^4}\,.
\end{equation}
As we did for the production cross section, we computed the virtual and the
real corrections separately keeping a fixed gluon mass $\lambda$ and we
denoted with $T(\lambda)$ their sum.  From Fig.~\ref{fig:Gammat} it is clear
that, if the top mass is expressed in the \MSB{}-mass scheme, the top decay
width is free from linear renormalons.  Although not obvious, this
cancellation is expected since the absence of linear renormalons in the
heavy-particles decay-width expressed in terms of the \MSB{} mass was already
shown in Refs.~\cite{Bigi:1994em, Beneke:1994qe, Beneke:1994bc}.

\section{Considerations}
We have just seen that both $\sigma(W^*\to t \bar{b})$ and $\Gamma\(t \to W
\, b\)$ are linear-renormalon free if expressed in terms of the \MSB{} mass,
so it is the cross section $\sigma(W^*\to t \bar{b} \to W \, b\, \bar{b})$ in
NWA. In Sec.~\ref{sec:Xtot_wocuts} we showed that this behaviour is achieved
also for finite top width, since the top quark can never be on-shell if a
complex mass scheme is used.  A formal demonstration of the cancellation of
the linear sensitivity in case of finite top width is given in
appendix~\ref{sec:Ew_linear_sensitivity}.

\chapter{Cancellation of the linear sensitivity in the total cross section
  and in leptonic observables}
\label{sec:Ew_linear_sensitivity}
In order to discuss the cancellation of the linear sensitivity in the total
cross section and in the energy of the $W$ boson, $E_{\sss W}$, when finite
top-width effects are included, we make use of the old-fashioned perturbation
theory.  To this purpose, we first briefly summarize the main features of
this approach in Appendix~\ref{sec:OFPT} and we recall the conditions for
Landau threshold singularities in Appendix~\ref{sec:landau}.  In
Appendix~\ref{sec:Ew_OFPT} we show that, as long as $\Gamma_t>0$, the Landau
singularities are not present and thus, if we adopt a short-distance mass, no
linear $m_g$-term can arise, being $m_g$ the mass of the gluon.

\section{Old-fashioned perturbation theory}
\label{sec:OFPT}

In the old-fashioned (time-ordered) perturbation theory the propagator
denominators in a Feynman diagram are split into an advanced and a retarded
part
\begin{equation}
 \frac{i}{k^2 - m^2 + i \eta} = \frac{i}{2 E_{k, m}}  \left[
   \frac{1}{k^0 - E_{k, m} + i \eta} + \frac{1}{- k^0 - E_{k, m} + i
   \eta} \right] ,
   \label{eq:propagator}
\end{equation}
where
\begin{equation}
E_{k, m} = \sqrt{\underline{k}^2+m^2},
\end{equation}
and $\underline{k}=|\vec{k}|$.
The time Fourier transform of the first term vanishes for negative time, while
for the second term it vanishes for positive time
\begin{eqnarray}
  \int \frac{\mathd k^0}{2 \pi}  \frac{i \exp (- i k^0 t)}{k^0 - E_{k, m} + i
  \eta} & = & \theta (t) \exp \!\(- i E_{k, m} \,t\)\,,\\
  \int \frac{\mathd k^0}{2 \pi}  \frac{i \exp (- i k^0 t)}{- k^0 - E_{k, m} +
  i \eta} & = & \theta (- t) \exp \!\(i E_{k, m} \,t\)\,.
\end{eqnarray}
These results can be easily demonstrated by promoting the integration path into
a semicircle integral and applying the residue theorem.  

We observe that the first term in eq.~(\ref{eq:propagator}), that
corresponds to an advanced propagator, propagates positive energies in the
future, while the second one, i.e. the retarded one, propagates negative
energies in the past.

In presence of an unstable particle, the denominator is given by
\begin{equation} \frac{i}{k^2 - m^2 + i m \Gamma} = \frac{i}{2 E_{k, m, \Gamma}}  \left[
   \frac{1}{k^0 - E_{k, m, \Gamma}} + \frac{1}{- k^0 - E_{k, m, \Gamma}}
   \right], \end{equation}
where
\begin{equation}
  E_{k, m, \Gamma} = \sqrt{\underline{k}^2 + m^2 - i m \Gamma}\,,
\end{equation}
so that $E_{k, m, \Gamma}$ has a negative imaginary part.
As a consequence, we also have
\begin{eqnarray*}
  \int \frac{\mathd k^0}{2 \pi}  \frac{i \exp (- i k^0 t)}{k^0 - E_{k, m,
  \Gamma} + i \eta} & = & \theta (t) \exp\!\(- i E_{k, m, \Gamma}^0 \,t\)\,,\\
  \int \frac{\mathd k^0}{2 \pi}  \frac{i \exp (- i k^0 t)}{- k^0 - E_{k, m,
  \Gamma} + i \eta} & = & \theta (- t) \exp \!\(i E_{k, m, \Gamma}^0 \,t\)\,,
\end{eqnarray*}
and both functions have exponential damping for large positive (negative)
time.
We notice that the $i \eta$ factor is not necessary in intermediate states
containing the unstable particle, because they can never be on-shell.

We can write any Feynman diagram assigning an independent momentum to each
propagator, and include a factor $(2 \pi)^4 \delta^4 \left( \sum k \right)$
to each vertex to ensure four-momentum conservation, where $\sum k$ is the
sum of all four-momenta entering the vertex. Furthermore, the time component
of the $\delta$ function can be rewritten as
\begin{equation}
(2 \pi)^4 \delta^4 \left( \sum k \right) = \int_{- \infty}^{+\infty} \mathd
   t \exp \left( - i t \sum k_0 \right) \,(2\pi)^3\delta^3 \left( \sum \vec{k} \right) .
\end{equation}
Thus each advanced or retarded propagator will carry the
factor
\begin{equation} 
\frac{1}{2 E_{k, m}}  \frac{\exp \!\lq- i (t_2 - t_1) k_0\rq}{\pm k^0 - E_{k, m}
   + i \eta}, 
\end{equation}
where $t_1$ and $t_2$ are the time variables associated with the beginning and
the end of the propagator, that are chosen accordingly to the direction of
$\vec{k}$. Performing the $k^0$ integration, we get
\begin{equation} 
\int_{- \infty}^{\infty}  \frac{\mathd k_0}{2 \pi} \frac{i}{2 E_{k, m}} 
   \frac{\exp \! \lq - i (t_2 - t_1) k_0 \rq}{\pm k^0 - E_{k, m} + i \eta} = \theta
   (\pm (t_2 - t_1)) \exp (\mp i E_{k, m} t)\, \frac{1}{2\,E_{k,m}}\,, 
   \label{eq:E_onshell}
   \end{equation}
with $k^0$ replaced by its on shell (positive or negative) value in all other
occurrences (i.e.~propagator numerators or vertex factors).

By splitting each propagator into an advanced and retarted part, a single
Feynman diagram will split into a sum of diagrams with all possible different
time orderings for the vertices. We consider a graph with a given ordering, and
label the time of each vertex as
\begin{equation} t_0 < t_1 < \ldots < t_n \,, \end{equation}
and we can collect all exponential in the form
\begin{equation} 
\exp (- i t_0 E_{V_0}) \ldots \exp (- i t_n E_{V_n})\,, 
\end{equation}
where by $E_{V_i}$ we mean the total energy entering the $i^{\tmop{th}}$
vertex.
Integrating in $t_n$ from $t_{n - 1}$ to infinity yields a factor
\begin{equation} 
\int_{t_{n - 1}}^{+\infty} \mathd t_n \, \exp(-it_nE_{V_n}) =- i \frac{\exp (- i t_{n - 1} E_{V_n})}{E_{V_n} - i \eta} . 
\end{equation}
Then, performing the $\mathd t_{n - 1}$ integration yields
\begin{equation} 
\displaystyle
\frac{-i}{E_{V_n} - i \eta}\int_{t_{n - 2}}^{+\infty} \mathd t_n \, \exp\left[-it_n(E_{V_n}+E_{V_n-1})\right]=(- i)^2 \frac{\exp \left[- i t_{n - 2}  (E_{V_{n - 1}} + E_{V_n})\right]}{(E_{V_n} - i \eta)(E_{V_{n - 1}} +
   E_{V_n} - i \eta)}, 
\end{equation}
and so on. The last integral in $\mathd t_0$ is unrestricted, yielding
\begin{equation} 
\frac{(-i)^n\,2 \pi\, \delta (E_{V_0} + E_{V_1} + \ldots E_{V_n})}{(E_{V_n} - i \eta)\ldots(E_{V_n} +E_{V_{n-1}} \ldots +E_{V_1}- i \eta)}, 
\end{equation}
i.e. the total energy entering in all the vertices must be zero. The
delta function is usually removed from the amplitude.

\begin{figure}[htb]
\centering
\includegraphics[width=0.6\textwidth]{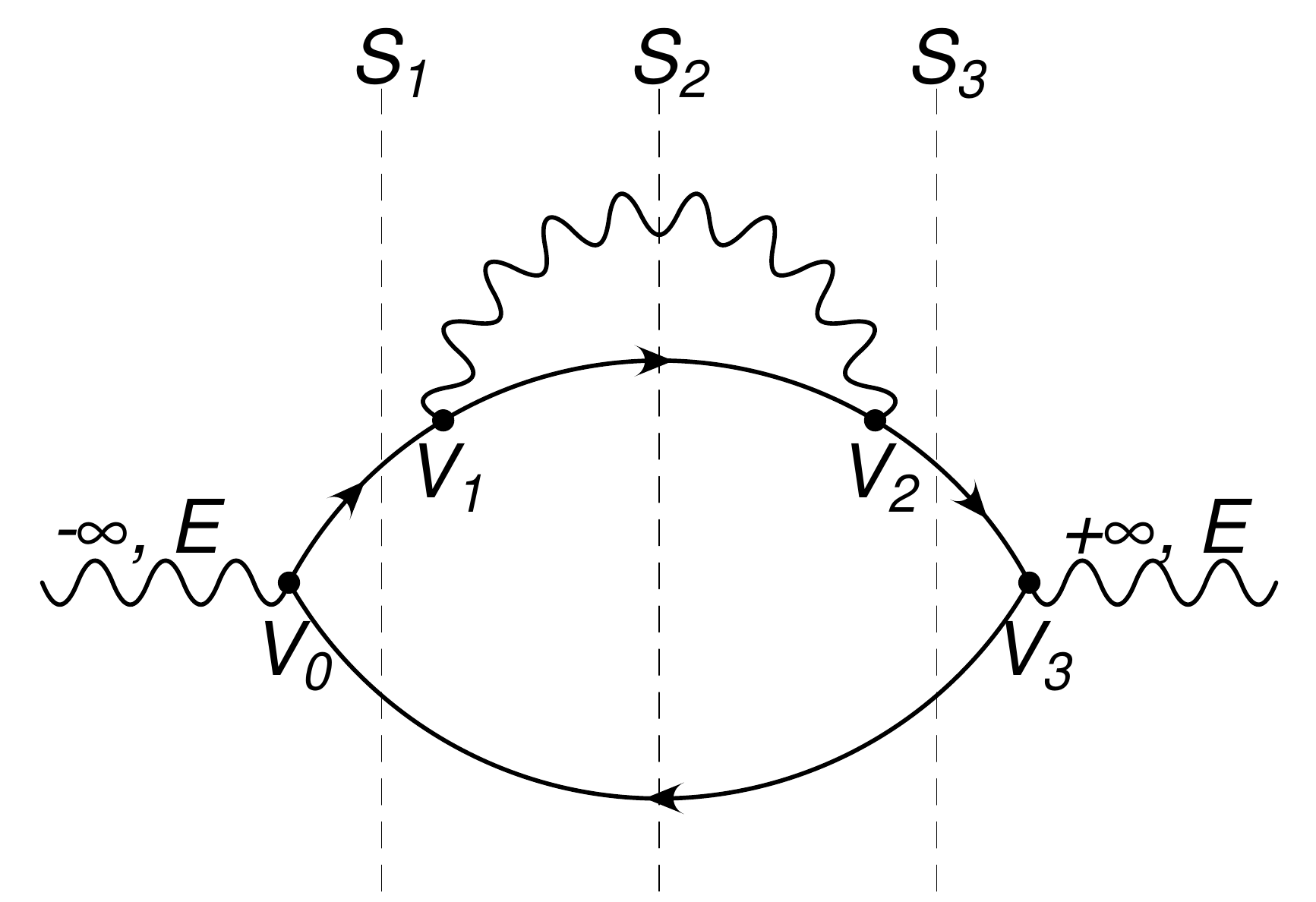}
\caption{Example of time ordered graph.}
\label{fig:example}
\end{figure}

In order to have a clearer picture of the resulting graph, we can attach a
line coming from $t = - \infty$ to all vertices with an entering external
momentum, and a line going to $t = + \infty$ from all vertices with outgoing
external momenta. Then we imagine we cut our Feynman graphs with lines at constant
time between any pair or time-ordered vertices $i - 1, i$. This line defines
the intermediate state $S_i$. We define the energy of the intermediate state
$S_i$ as the sum $E_{S_i}$ of the energy flowing in all cut lines (including
those from $- \infty$ or to $+ \infty$) from smaller to larger times. Then,
the denominator of the $i^{\tmop{th}}$ vertex is
\begin{equation} 
  \frac{i}{E - E_{S_i} + i \eta},
  \label{eq:denEsi}
\end{equation}
where $E = E_{S_{- 1}} = E_{S_{n + 1}}$ is the energy of the intermediate
state before all vertices, and the energy of the intermediate state after all
vertices, that are equal by momentum conservation.
In fact, in general the denominator arising from the $n^{\tmop{th}}$ vertex is
\begin{equation} 
\frac{- i}{E_{V_n} - i \eta}.
\end{equation}
Since $E_n$ is the total energy entering the last vertex, it must equal the
energy of the $S_n$ intermediate state, minus the energy of the $S_{n + 1}$
state
\begin{equation} 
\frac{- i}{E_{V_n} - i \eta} = \frac{- i}{E_{S_n} - E - i \eta} =
   \frac{i}{E - E_{S_n} + i \eta} . 
\end{equation}
The denominator arising from the $n - 1$ vertex is instead
\begin{equation} 
\frac{- i}{E_{V_n} + E_{V_{n - 1}} - i \eta} . 
\end{equation}
Arguing as before, $E_{V_{n - 1}} = E_{S_{n - 1}} - E_{S_n}$, so
\begin{equation} 
\frac{- i}{E_{V_n} + E_{V_{n - 1}} - i \eta} = \frac{- i}{E_{S_n} - E +
   E_{S_{n - 1}} - E_{S_n} + i \eta} = \frac{i}{E - E_{S_{n - 1}} + i
   \eta}, 
   \end{equation}
and so on.
Indeed, if we take for example  the Feynman diagram of Fig.~\ref{fig:example}
and we write the energy entering in each vertex, we have
\begin{itemize}
\item $E_{V_0}=(E_{-\infty})-(E_{q,1}+E_{\bar{q}}) = E_{S_0}-E_{S_1} = E-E_{S_1}$
\item $E_{V_1}= E_{q,1} - E_{W} - E_{q,2} = (E_{q,1} +E_{\bar{q}})- (E_{W}+E_{q,2}+E_{\bar{q}}) = E_{S_1}-E_{S_2}$;
\item $E_{V_2}= E_{W}+E_{q,2}-E_{q,3} = (E_{W}+E_{q,2}+E_{\bar{q}})-(E_{q,3}+E_{\bar{q}})=E_{S_2}-E_{S_3}$;
\item $E_{V_3}= (E_{q,3}+ E_{\bar{q}}) - (E_{+\infty}) =E_{S_3}-E_{S_4}=E_{S_3}-E$.
\end{itemize}
We can now formulate the rules for representing a given Feynman graph as the
sum of contributions coming from old-fashioned perturbation theory:
\begin{enumerate}
  \item For a given Feynman graph, consider all possible ordering of its
  vertices. Write the whole Feynman graph formula without the inclusion of the
  $i / (k^2 - m^2 + i \eta)$ factors.
  
  \item For each line joining two vertices provide the factor
  \begin{equation} \frac{1}{2 E_{k, m}} = \frac{1}{2 \sqrt{\underline{k}^2 + m^2}}, 
  \end{equation}
  where $\underline{k}$ is the modulus of the three-momentum flowing in the
  graph. Assume that the energy $E_{k, m}$ flowing in propagators from
  smaller to larger time vertices is positive.  This factor indeed multiplies
  the advanced/retarded propagator in eq.~(\ref{eq:propagator}).
  
  \item In all numerator factors of the Feynman graph, substitute the time
  component of the momenta flowing in the propagators with its on-shell value,
  with the sign determined according to the convention of the previous point.
  This is a consequence of eq.~(\ref{eq:E_onshell}).
  
  \item Assign 3-momenta as usual, with 3-momentum conservation at each
  vertex, and
  \begin{equation} 
  \frac{\mathd^3 \vec{k}}{(2 \pi)^3} 
  \end{equation}
  momentum integration.
  
  \item Add lines from $t = - \infty$ to each vertex with entering external
  momentum. 
  
  \item Add lines to $t = \infty$ to each vertex with exiting external
  momentum.
  
  \item Consider all intermediate states between any pair of nearby vertices.
  For each intermediate state provide a factor
  \begin{equation}
    \frac{i}{E - E_{S_i} + i \eta},
  \end{equation}
  where $E$ is the total energy coming from $- \infty$ (or going to $+
  \infty$) and $E_{S_i}$ is the energy of the state $S_i$.
\end{enumerate}
Notice that each energy denominator carries a factor of $i$, and so does each
vertex. Since the number of energy denominators is equal to the number of
vertices minus 1, one $i$ will survive in the product.

\section{Landau singularities}
\label{sec:landau}
Landau singularities~\cite{Eden:1966dnq,Landau:1959fi} are naturally
explained in the old-fashioned perturbation theory. In general, we
have a denominator of the form
\begin{equation} 
\frac{1}{E - \sum_{i = 1}^n  \sqrt{| \vec{k}_i |^2 + m_i^2} + i \eta}, \qquad E \sim \sum_{i = 1}^n m_i . 
   \end{equation}
The so-called Landau anomalous threshold can arise when a sequel of
singular denominators appear at the same time. Let us assume we have a
single loop where the 3-momentum $\vec{k}$ flows, and we have a set of
denominators
\begin{equation} 
\prod_{i = 1}^n D_i = \prod_{i = 1}^n \frac{1}{E_i - \sum_{j_i = 1}^{n_i} \sqrt{(\vec{q}_{j} +
   \vec{k})^2 + m_{j}^2} + i \eta} . 
\end{equation}
Assuming that the term $E_i$ can be different for each denominator amounts to
the assumption that we may have external momenta entering at any vertex. We
now assume that at some value of $\vec{k} = \vec{k}_0$ all denominators vanish
at the same time. 
A necessary and sufficient condition for this to be an
avoidable singularity is that upon integrating in any component of $\vec{k}$
all $i \eta$ pole are on the same side of the integration plane near
$\vec{k}_0$. 
We expand a energy denominator for $\vec{k} = \vec{k}_0 + \delta \vec{k}$.
\begin{eqnarray} 
D_i &=& \frac{1}{E_i - \sum_{j_i = 1}^{n_i} \sqrt{(\vec{q}_{j}
    + \vec{k}_0+\delta\vec{k})^2 + m_{j}^2} + i \eta} \nonumber \\ &\approx&
\frac{1}{E_i - \sum_{j_i = 1}^{n_i} \sqrt{(\vec{q}_{j} + \vec{k}_0)^2 +
    m_{j}^2} - \delta \vec{k} \cdot\sum_ {j = 1}^{n_i} \frac{\vec{q}_{j} +
    \vec{k}_0}{\sqrt{(\vec{q}_{j} + \vec{k}_0)^2 + m_{j}^2}}+ i
  \eta}\nonumber \\ &=& \frac{1}{- \delta \vec{k}\cdot \sum_ {j = 1}^{n_i}
  \frac{\vec{q}_{j} + \vec{k}_0}{\sqrt{(\vec{q}_{j} + \vec{k}_0)^2 +
      m_{j}^2}}+ i \eta} = \frac{-1}{(\vec{k}-\vec{k_0})\cdot\sum_ {j =
    1}^{n_i} \frac{\vec{q}_{j} \vec{k}_0}{\sqrt{(\vec{q}_{j} + \vec{k}_0)^2 +
      m_{j}^2}}- i \eta}.\qquad
\end{eqnarray}
In one dimension the sign of the imaginary part of the complex pole is given by
\begin{equation}
 -  {\rm sign}\left( \sum_ {j = 1}^{n_i} \frac{q_j+k_0}{\sqrt{({q}_{j} + {k}_0)^2 + m_{j}^2}}\right)\,,
\end{equation}
so, if this sum has the same sign for all the $D_i$, the
singularity is avoidable since it lies in the same part of the complex plane
for all the denominators.  In 3 dimensions, if there is a direction in the
3-dimensional integration space (corresponding to the 3-momenta flowing in
the loop) such that, integrating along it, it leaves the singularities of all
the $D_i$ on the same side of the complex plane, the
integration contour can be deformed away from the singularity, so that the
denominators cannot contribute to the singularities.  This is achieved if
there is at least one direction of $\delta \vec{k}$ such that
\begin{equation}
  \delta \vec{k} \cdot \sum_{j = 1}^{n_i} \frac{\vec{q}_{j} +
  \vec{k}_0}{\sqrt{(\vec{q}_{j} + \vec{k}_0)^2 + m_{j}^2}}
  \label{eq:samesign}
\end{equation}
have the same sign for all $i$. Thus, all vectors
given by the sum in (\ref{eq:samesign}) have components of the same sign in at
least one direction and so no null linear combination
\begin{equation}
  \sum_{i = 1}^n \lambda_i \sum_{j = 1}^{n_i} \frac{\vec{q}_{j} +
  \vec{k}_0}{\sqrt{(\vec{q}_{j} + \vec{k}_0)^2 + m_{j}^2}} = 0
  \label{eq:landaucond}
\end{equation}
must exist for any sequence of $\lambda_i > 0$. Conversely, if a null linear
combination exists, (\ref{eq:samesign}) cannot be satisfied for any choice of
$\delta \vec{k}$. This is the so-called Landau anomalous threshold singularity.

\begin{figure}[tb]
  \centering
    \includegraphics[width=0.6\textwidth]{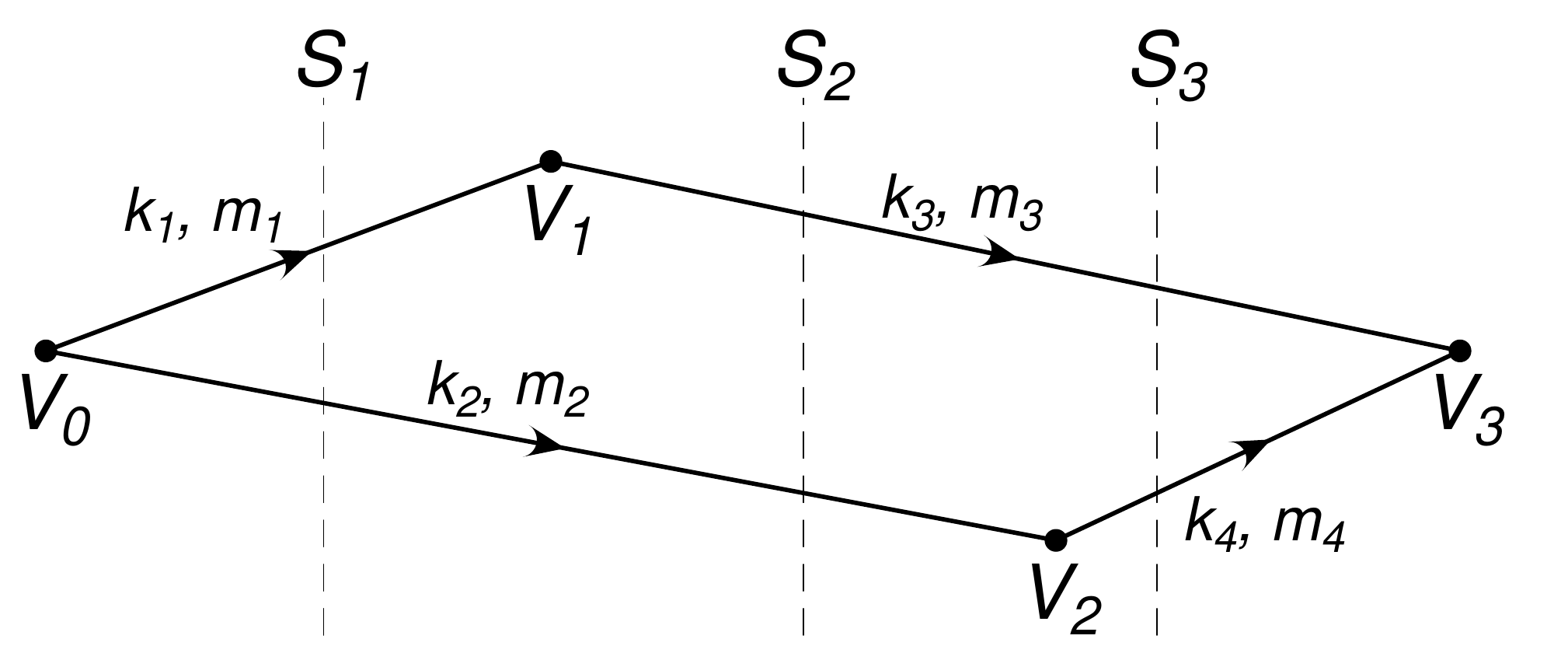}
  \caption{Illustration of eq.~(\ref{eq:landaucond}).}
  \label{fig:landau}
\end{figure}

The physical interpretation of eq.~(\ref{eq:landaucond}) is quite interesting.
We interpret the $\lambda_i$ as the time between the first vertex on the left
and on the right of the $i^{\tmop{th}}$ intermediate state. The ratios in the
sum in (\ref{eq:landaucond}) are just the velocities of the particles, so,
velocities times time equal the displacements. So, the sum of all
displacements of internal particles must be zero. Since all internal particles
are nearly on-shell (i.e. their displacement can be as large as one likes),
this means that their displacements must be compatible with kinematic
constraints. This is better illustrated with the example of
Fig.~\ref{fig:landau}.
The condition for the singularity is
\begin{equation} \(\vec{v}_1 + \vec{v}_2\) \lambda_1 + \(\vec{v}_3 + \vec{v}_2\) \lambda_2 +
   \(\vec{v}_3 + \vec{v}_4\) \lambda_3 = 0,\qquad \mbox{with }\;\vec{v}_i = \frac{\vec{k}_i}{\sqrt{|
   \vec{k}_i |^2 + m_i^2}}\,, 
   \end{equation}
for positive $\lambda_{1 \ldots 3}$. This is the same as
\begin{equation} 
\vec{v}_1 \lambda_1 + \vec{v}_3 (\lambda_3 + \lambda_2) = - \vec{v}_2
   (\lambda_1 + \lambda_2) - \vec{v}_4 \lambda_3, 
   \end{equation}
i.e. the displacement of particle 1 between its initial and final vertex, plus
the displacement of particle 3 between its initial and final vertex is equal
and opposite to the subsequent displacement of particles 2 and 4. In other
words, the momenta should be such that the particles meet again after having
come apart.

\section{The specific case}
\label{sec:Ew_OFPT}
\begin{figure}[h]
\centering
  \includegraphics[width=0.6\textwidth]{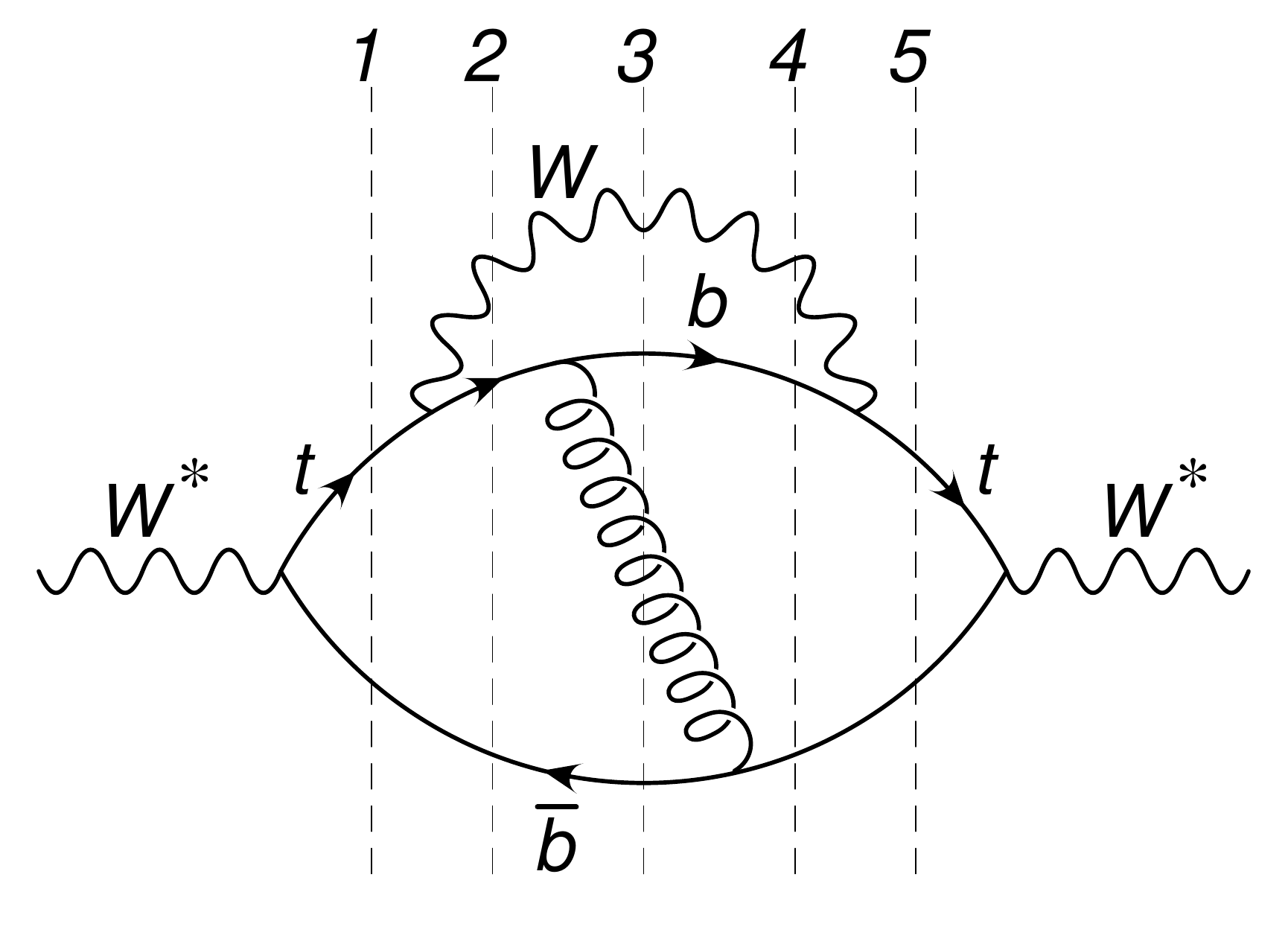}
  \caption{Example of NLO contribution to the process $W^{\ast} \rightarrow t
    \bar{b} \rightarrow W b \bar{b}$.}
  \label{fig:wstart}
\end{figure}

We now consider our $W^{\ast} \rightarrow \bar{b} + (t \rightarrow W + b)$
process. An example of NLO diagram is shown in Fig.~\ref{fig:wstart}.
The corresponding contribution to the cross section is obtained by
setting either one of the 2, 3, 4 intermediate states on the energy shell, and
changing the sign of the $i \eta$ in the denominators to the right of the
cut. We then define
{ \allowdisplaybreaks
\begin{eqnarray}
  D_1 & = & \frac{1}{E - E_{t, 1} - E_{\bar{b}, 1}}\,,
  \\
  D_2 & = & \frac{1}{E - E_{\sss W} - E_{b, 2} - E_{\bar{b}, 1} + i \eta}\,,
  \\
  D_3 & = & \frac{1}{E - E_{\sss W} - E_{b, 3} - E_{\bar{b}, 1} - E_{g, 3} + i \eta}\,,
  \\
  D_4 & = &  \frac{1}{E - E_{\sss W} - E_{b, 3} - E_{\bar{b}, 4} + i \eta}\,,
  \\
  D_5 & = & \frac{1}{E - E_{t, 5} - E_{\bar{b}, 5}}\,,
\end{eqnarray}
}
where
\begin{eqnarray}
  E_{t, i} & = & \sqrt{\vec{k}_{t, i}^2 + m^2 - i m \Gamma_t}\,,
  \\
  E_{l, i} & = & \sqrt{\vec{k}_{l, i}^2}\,, \qquad \qquad l = b, \bar{b}, g,
  \,
  \\
  E_{\sss W} & = & \sqrt{\vec{k}_{\sss W}^2 + m^2}\, .
\end{eqnarray}
Notice that the top energy has an imaginary part, so that no $i \eta$ is
needed in the denominators containing it.  We never include the corresponding
cuts since the top width prevents this particle from being on-shell. Thus,
the only intermediate states contributing to cuts will be the ones that do
not include the top.  According to the optical theorem, the cross section is
given by the imaginary part of a sum of contributions like the one displayed
in Fig.~\ref{fig:wstart}.  Since we do not include contributions in which the
top is on-shell, in the integrand for the cross section we have
\begin{equation}
D_1\, \tmop{Im} [D_2 D_3 D_4] D_5^{*}= D_1 \big[ {\rm Im}(D_2) D_3^{*}\,
  D_4^{*} + D_2\, {\rm Im}(D_3) D_4^{*} + D_2\, D_3\, {\rm Im}(D_4) \big]
D_5^{*}\, .
\end{equation}
When performing the 3-momentum integral for the loops not including the $W$
line, one can come close to the singularity in the denominators. However, if
there is a direction in the 9-dimensional integration space (corresponding to
the three 3-momenta flowing in the loops) such that, integrating along it, it
leaves the singularities of $D_2$, $D_3$ and $D_4$ on the same side of the
complex plane, the integration contour can be deformed away from the
singularities, so that the denominators cannot contribute to mass
singularities. The singularity for small $\lambda$ will thus be determined only by the
remaining factor
\begin{equation}
  \frac{\mathd^3 k_g}{\sqrt{\vec{k}_g^2 + \lambda^2}}\,,
\end{equation}
that gives a quadratic sensitivity to the gluon mass $\lambda$. 

As we have already discussed, the only cases when an
appropriate deformation of the contour does not exist correspond to Landau
singularities. 

\begin{figure}[h]
  \centering
  \includegraphics[width=0.6\textwidth]{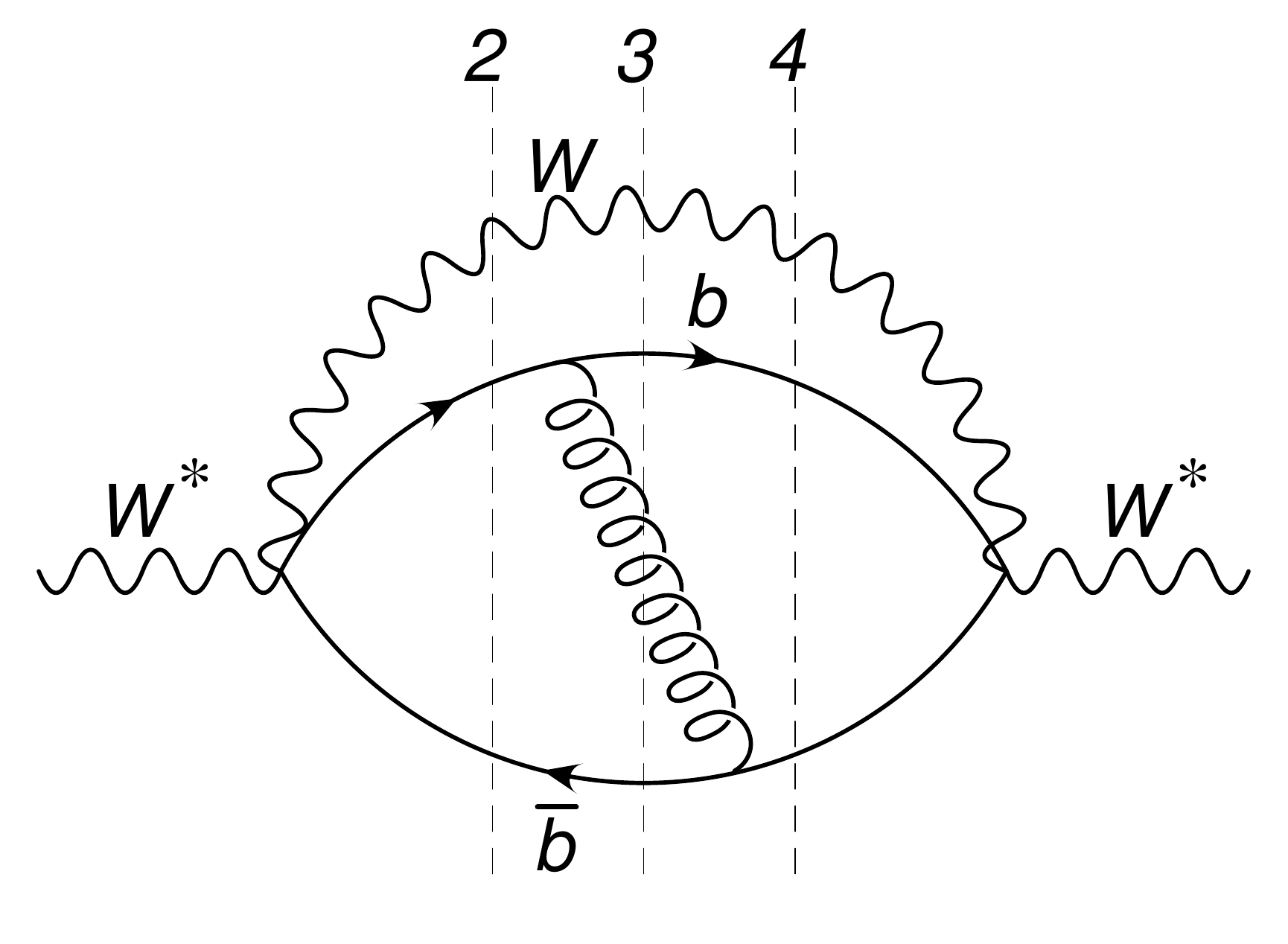}
  \caption{\label{fig:wstart-landau1} The reduced graph to look for Landau singularities
    in the graph of Fig.~\ref{fig:wstart}.}
\end{figure}

In order to explore the possible Landau configuration, one can start with the
graph of Fig.~\ref{fig:wstart} with the top lines shrunk to a point. In fact,
the top is always off-shell, and it cannot propagate over large
distances. The remaining configuration is shown in Fig.~\ref{fig:wstart-landau1}.
In order for the 2, 3 and 4 intermediate states to be on-shell at the same
time, either both the $b$ and the $\bar{b}$ are collinear to the gluon, or
the gluon is soft. In the first case, also the two $b$ quarks and the gluon are
collinear, and are all travelling in the opposite direction with respect to
the $W$. Thus, they cannot meet at the same point on the last vertex to the
right.  On the other hand, if the gluon is soft, the $b$, the $\bar{b}$ and
the $W$ produced at the primary vertex have momenta that sum to zero, so,
again their velocities will make them diverge.  In both cases
eq.~(\ref{eq:landaucond}) cannot be satisfied and these configurations lead
to an avoidable singularity.  One can try to shrink other propagators to a
point to see if two intermediate states can be on-shell.
Shrinking a $b$ and a $\bar{b}$ line to a point leads to configuration with
two massless system (either $b$ quarks of collinear $b g$ systems) and a $W$,
that again cannot meet at the same point.  Shrinking the $W$, the two $b$ or
the two $\bar{b}$ lines to a point shrinks the whole graph to a point,
leading to nothing.  Thus, no Landau configuration can exist, so one infers
that the $m_g$ sensitivity of the total cross section is at least quadratic.

We can repeat the same reasoning including a factor $E_{\sss W}$ in our Feynman
graph. The argument runs as before, and so even for the average energy of the
$W$ one expects that the sensitivity to the gluon mass is at least
quadratic. Notice that in order for this to work, one needs that the $E_{\sss W}$
factor is the same for all cuts, which is in fact the case.

This argument fails if $\Gamma_t$ is sent to zero. Indeed in this case $D_1$
and $D_5^*$ can be simultaneously singular but their $i \eta$ have opposite
signs.  Under these condition, the pinch is clearly unavoidable.

As a last point, we recall that the total cross section is free of linear
$m_g$ sensitivity also in the limit of zero width. This happens because, in
the zero-width limit, the cross section factorizes into a production cross
section times a decay width, and both of them are free of linear sensitivity
to $m_g$ if the mass is in a short-distance scheme. The same, however, does
not hold for the average $E_{\sss W}$. In fact, the cancellation of mass
singularities in $\Gamma_t$ cannot be proven in the same fashioned adopted here,
since logarithmic divergences are also present in the wave-function
renormalization, and cannot be treated in a straightforward way in the
old-fashioned perturbation theory.

\chapter{The \POWHEGBOX{} framework}
\label{sec:powheg}

The fixed order computations needed to evaluate the factors $T(\lambda)$ and
$\widetilde{T}(\lambda)$ of eqs.~(\ref{eq:sigma_final})
and~(\ref{eq:final_M_wcuts}) and the NLO generators described in
chap.~\ref{sec:generators} are implemented in the \POWHEGBOX{}.  The
\POWHEGBOX{} is a framework that can be used to generate NLO events that
can be interfaced to modern parton shower~(PS) according to the POWHEG
method~\cite{Nason:2004rx,Frixione:2007vw,Alioli:2010xd}. The latest release,
i.e.~the \RES{}~\cite{Jezo:2015aia}, has been developed to handle processes
containing coloured resonances that can emit.

\section{NLO computations in the \POWHEGBOX{}~{\tt(RES)}}
\label{sec:NLOpowheg}
The \POWHEGBOX{} can be used as an integrator for fixed order NLO
computations.
The user has to provide, according to
Ref.~\cite{Alioli:2010xd}:
\begin{itemize}
\item the Born squared amplitude;
\item the finite part of the virtual squared amplitude;
\item the real squared amplitude;
\item the Born phase space~(in the \RES{} it can be generated
  automatically);
\item the flavour structure of the Born and of the real processes;
\item the Born colour- and spin-correlated squared
  amplitudes.
\end{itemize}

We describe the main features for the computations of a $2\to n$
process at NLO for a generic hadron-hadron collision.  We introduce
$\bold{\Phi}_n$ to be the set of variables
\begin{equation}
\bold{\Phi}_n = \left\{ x_1, x_2, p_1, \dots p_n\right\},
\end{equation}
being $x_{1,2}$ the fractions of the hadrons energies carried by the
incoming partons and $p_1, \dots, p_n$ the four-momenta of the
outgoing particles.  We define the luminosity $\mathcal{L} =
f_{1}^{i}(x_1)f_{2}^{j}(x_2)$, where $f_{1}^{i}(x_1)$ is the parton
distribution function relative to the first hadron and it describes
the probability of finding a parton $i$ carrying a fraction $x_1$ of
the hadron energy.  If one of the incoming particle is a lepton, we
replace $f_{1}^{i}(x_1)$ with $\delta(x_1-1)$.  We also denote with
$\mathcal{B}$ the partonic Born cross section, with $\mathcal{V}_{\rm
  b}$ the partonic UV-renormalized virtual cross section and with
$\mathcal{R}$ the partonic real cross section.  The total NLO cross
section is given by
\begin{eqnarray}
\sigma^{\rm NLO} & = & \int \mathd
\bold{\Phi}_n\;\mathcal{L}\left[\mathcal{B}(\bold{\Phi}_n) +
  \mathcal{V}_{\rm b}(\bold{\Phi}_n) \right] +\int \mathd
\bold{\Phi}_{n+1}\;\mathcal{L} \; \mathcal{R}(\bold{\Phi}_{ n+1})
\\ &&+\sum_{i=1}^2 \int\mathd \bold{\Phi}_n \int_0^1
\mathd{z}\;\mathcal{L} \;\mathcal{G}_i(z,\bold{\Phi}_n)\,, \nonumber
\label{eq:sigNLO}
\end{eqnarray}
where the $\mathcal{G}_i$ terms are introduced by the renormalization
of $\mathcal{L}$ and they ensure the cancellation of the initial-state
collinear singularities contained in $\mathcal{R}$. If the incoming
particle $i$ is a lepton, $\mathcal{G}_i=0$.  For ease of notation, we
define
\begin{equation}
\mathcal{V}(\bold{\Phi}_n) = \mathcal{V}_{\rm b}(\bold{\Phi}_n) +
\sum_{i=1}^2 \int_0^1 \mathd{z} \;\mathcal{G}_i(z,\bold{\Phi}_n).
\end{equation}
The Kinoshita-Lee-Nauenberg~(KLN)~\cite{Kinoshita:1962ur,Lee:1964is} theorem
guarantees us that the infrared divergences contained in $\mathcal{R}$
and in $\mathcal{V}$ cancel, yielding to a finite prediction for
$\sigma^{\rm NLO}$.

In order to handle numerically eq.~\eqref{eq:sigNLO}, a counterterm
$\mathcal{C}(\bold{\Phi}_{\rm n+1})$ is introduced, that approaches the real
cross section in the soft and collinear limit, such as
$\mathcal{R}(\bold{\Phi}_{\rm n+1}) -\mathcal{C}(\bold{\Phi}_{\rm n+1})$ is
finite. The phase space $\bold{\Phi}_{\rm n+1}$ can be rewritten as
\begin{equation}
\bold{\Phi}_{\rm n+1} = \left\{\bold{\Phi}_{n}, \Phi_{\rm rad}\right\}\,,
\end{equation}
where $\bold{\Phi}_{n}$ is the phase space of the \emph{underlying}
Born configuration and $\Phi_{\rm rad}$ is a set of 3 kinematic
variables that describe the phase space of the extra parton emitted in
the $n+1$-body configuration.  The total cross section now takes the
form
\begin{eqnarray}
\sigma^{\rm NLO} & = & \int \mathd
\bold{\Phi}_n\;\mathcal{L}\;\mathcal{B}(\bold{\Phi}_n) + \int \mathd
\bold{\Phi}_n \;\mathcal{L}\left[\mathcal{V}(\bold{\Phi}_n) + \int
  \mathd\Phi_{\rm rad} \;\mathcal{C}(\bold{\Phi}_{n+1})\right]
\\ &&+\int \mathd \bold{\Phi}_{n+1}\;\mathcal{L}
\left[\mathcal{R}(\bold{\Phi}_{n+1}) -
  \mathcal{C}(\bold{\Phi}_{n+1})\right]\,, \nonumber
\label{eq:sigNLO2}
\end{eqnarray}
where the expression in the squared brackets is finite in $d=4$
dimensions and thus can be evaluated numerically.

With this subtraction formalism, we can compute predictions for any
generic infrared-safe observable $O$, i.e.~such that
$O_{n+1}(\bold{\Phi}_{n+1}) \to O_n(\bold{\Phi}_n)$ when one parton
becomes soft or collinear to another one.  More explicitly, we have
\begin{eqnarray}
O & = & \int \mathd
\bold{\Phi}_n\;\mathcal{L}\;\mathcal{B}(\bold{\Phi}_n)
\;O_n(\bold{\Phi}_n) \\ &&+ \int \mathd \bold{\Phi}_n
\;\mathcal{L}\left[\mathcal{V}(\bold{\Phi}_n) + \int \mathd\Phi_{\rm
    rad}\; \mathcal{C}(\bold{\Phi}_{n+1})\right] O_n(\bold{\Phi}_n)
\nonumber \\ &&+\int \mathd \bold{\Phi}_{n+1}\;\mathcal{L}
\left[\mathcal{R}(\bold{\Phi}_{n+1}) \;O_{n+1}(\bold{\Phi}_{n+1}) -
  \mathcal{C}(\bold{\Phi}_{n+1})\;O_n(\bold{\Phi}_n)\right]\,. \nonumber
\label{eq:obsNLO}
\end{eqnarray}

In the \POWHEGBOX{} framework, the counterterms $\mathcal{C}$ introduced
eq.~\eqref{eq:sigNLO2} is evaluated using the Frixione-Kunszt-Signer~(FKS)
subtraction~\cite{Frixione:1995ms} method.  The real cross section is written
as a sum of terms, each of them containing at most one collinear and one soft
singularity associated with one parton (the FKS parton).
We label with $i$ the region of the phase space where the
parton $i$ becomes soft or collinear to an incoming parton, $ij$ the
region where $i$ and $j$ are collinear.  We introduce non-negative
functions $d_i$ and $d_{ij}$, that in the \POWHEGBOX{} are defined as
\begin{equation}
 d_i = E^2_i (1-\cos^2\theta_i)\,, \qquad d_{ij} =
 \frac{2\,E_i\,E_j}{(E_i+E_j)^2}\, p_i \cdot p_j\,,
\end{equation}
being $E_{i}$ and $ p_{i}$ the energies and the four-momenta of the parton $i$,
$\theta_i$ the angle between the parton $i$ and the beams axis, everything
computed in the partonic rest frame.  This enables us to introduce the
$\mathcal{S}$ functions defined as
\begin{equation}
\displaystyle \mathcal{S}_i = \frac{1}{d_i\;\mathcal{D}}\,, \qquad
\mathcal{S}_{ij} = \frac{1}{d_{ij}\;\mathcal{D}}.
\end{equation}
where
\begin{equation}
\mathcal{D} = \sum_i \frac{1}{d_i}+ \sum_{ij} \frac{1}{d_{ij}}\,.
\end{equation}
The real cross section $\mathcal{R}$ is then split into a sum of
contributions associated to the several singular regions:
\begin{equation}
\mathcal{R} =  \sum_{i} \mathcal{R}_i+\sum_{ij}\mathcal{R}_{ij}\,,
\end{equation}
with
\begin{equation}
\mathcal{R}_i = \mathcal{S}_i \mathcal{R}\,,\qquad \mathcal{R}_{ij} =
\mathcal{S}_{ij} \mathcal{R}.
\end{equation}
This decomposition offers us the possibility to choose a different
parametrization for the $n+1$-body phase space for each contribution
$\alpha= i, ij$.
\begin{eqnarray}
\int \mathd \bold{\Phi}_{n+1}\;\mathcal{L}
\left[\mathcal{R}(\bold{\Phi}_{n+1}) -
  \mathcal{C}(\bold{\Phi}_{n+1})\right] = \sum_{\alpha} \int \mathd
\bold{\Phi}^{\alpha}_{n+1}\;\mathcal{L}
\left[\mathcal{R}^{\alpha}(\bold{\Phi}^{\alpha}_{n+1}) -
  \mathcal{C}^{\alpha}(\bold{\Phi}^{\alpha}_{n+1})\right].
\end{eqnarray}
A fully detailed description of this method can be found in
Ref.~\cite{Frixione:2007vw}.  What we want to underline is that for each
region $\alpha$, we have a subtraction term $\mathcal{C}^{\alpha}$, that
approaches $\mathcal{R}^{\alpha}$ when $\alpha$ is singular.  The subtraction
terms $\mathcal{C}^{\alpha}$ are automatically built by \POWHEGBOX{} from the
Born squared amplitude\footnote{More precisely, also the spin- and
  colour-correlated squared amplitudes are necessary. For further details,
  see~\cite{Alioli:2010xd}.}
\begin{equation}
\mathcal{C}^{\alpha}(\bold{\Phi}^{\alpha}_{n+1}) =
\tilde{\mathcal{C}}(\bold{\Phi}^{\alpha}_{n}, \Phi_{\rm
  rad})\;\;\mathcal{B}^{\alpha}(\bold{\Phi}^{\alpha}_n)\,
\label{eq:Cborn}
\end{equation}
and $\tilde{\mathcal{C}}(\bold{\Phi}^{\alpha}_{n}, \Phi^{\rm \alpha}_{\rm
  rad})$ is a function that can be easily integrated analytically in
$d=4-2\epsilon$ dimensions over the radiation phase space.  The
\emph{underlying} Born phase space $\bold{\Phi}^{\alpha}_n$ associated with
$\bold{\Phi}^{\alpha}_{n+1}$ is reconstructed using a mapping $M_\alpha
(\bold{\Phi}^{\alpha}_{n+1})$. When the parton $i$ is soft or collinear to
the parton $j$, $M_\alpha (\bold{\Phi}^{\alpha}_{n+1})$ should fulfill the
following property:
\begin{equation}
\bold{\Phi}^{\alpha}_{n+1} =\left\{x_1, x_2, \dots, p_i, p_j, \dots\right\}
\to \bold{\Phi}^{\alpha}_{n} =\left\{x^\prime_1, x^\prime_2, \dots, p_i+ p_j,
\dots\right\},
\end{equation}
where the incoming partons energy-fraction differs in case of an
initial-state emission.
The inverse mapping $M^{-1}_\alpha (\bold{\Phi}^{\alpha}_{n+1})$ is used to
generate the real emission phase space from $\bold{\Phi}^{\alpha}_{n}$ and
the three radiation variables.


From eq.~\eqref{eq:Cborn} we see that the subtraction term is proportional to
the Born and thus it will contains the denominators like
$(p^2-m^2)^2+\Gamma^2m^2$, being $m$ and $\Gamma$ the pole mass and the decay
width of resonance and $p^2$ its virtuality obtained from the momenta of its
decay products as computed in the $n$-body kinematic, while
$\mathcal{R}^\alpha$ contains the same denominators but evaluated using the
full $n+1$-body kinematic.  In the collinear limit, the momenta of the
\emph{underlying} Born configurations are assigned by merging the two
collinear partons and then applying some reshuffling to ensure that all the
external particles are on their mass shell. The procedure described in Sec.~5
of Ref.~\cite{Frixione:2007vw} does not preserve $p^2$ and the resonance
momentum will differ of an amount of order $m^2/E$, being $E$ the energy of
the resonance. As it is shown in Ref.~\cite{Jezo:2015aia}, the cancellation
of the divergences contained in $\mathcal{R}^\alpha$ and in
$\mathcal{C}^\alpha$ takes place only if $m^2 < \Gamma\;E$. We clearly see
that for small decay width the convergence is problematic, and it is
completely spoiled in the zero-width limit.

To this purpose, a new version of the \POWHEGBOX, namely the
\RES~\cite{Jezo:2015aia}, has been built. Given a flavour structure all the
possible intermediate resonances histories are assigned. The mapping between
the real phase space and the \emph{underlying} Born configuration preserves
the virtuality of the assigned intermediate resonances.

A similar procedure, applied to the dipole subtraction formalism, is
described in Ref.~\cite{Hoche:2018ouj}.

\section{NLO events with the POWHEG method}
\label{app:powhegMethod}
The POWHEG method, proposed for the first time in Ref.~\cite{Nason:2004rx},
enables to consistently match NLO computations with PS algorithms.

 In order to preserve the NLO accuracy of inclusive observables, the events
 generated according to the POWHEG method have weight
\begin{equation}
\overline{B}(\bold{\Phi}_n) = B(\bold{\Phi}_n )+V(\bold{\Phi}_n ) +
\int \mathd \Phi_{\rm rad} \,R(\bold{\Phi}_{n+1}) \,,
\end{equation}
where we have introduced
\begin{equation}
  B = \mathcal{L\,B}\,, \qquad V = \mathcal{L\,V}\,, \qquad R =
  \mathcal{L\,R}\,.
  \end{equation}
The Sudakov form factor, that describes the non-emission probability,
is defined by
\begin{equation}
  \label{eq:PowhegSudakov}
  \Delta\big(p_{\perp},\bold{\Phi}_{n} \big) = 
  \prod_\alpha\Delta_\alpha\big(p_{\perp},\bold{\Phi}_{n} \big)
 \end{equation}
with
\begin{equation}
 \Delta_\alpha\big(p_{\perp},\bold{\Phi}_{n} \big) =
  \exp\left\{
  \frac{-\left[\int \mathd \Phi^\alpha_{\rm rad}\,
      R^\alpha(\bold{\Phi}^\alpha_{n+1}
      )\,\theta\big(p_{\perp}-k_{\perp}(\bold{\Phi}^\alpha_{n+1})\big)\right]_{\bold{\Phi}^\alpha_n=\bold{\Phi}_n}}{B(\bold{\Phi}_n)}\right\}, \qquad
\end{equation}
with $R^\alpha =\mathcal{L\,R}^\alpha$ and the hardness of the emission
$p_\perp$ is defined to be the transverse momentum of the emitted parton
with respect to the emitter. Thus, the differential
cross section for the first emission becomes
\begin{equation}
\mathd \sigma =\overline{B}(\bold{\Phi}_n)\;\mathd \bold{\Phi}_n \Big[
  \Delta(p_{\perp,\rm min},
  \bold{\Phi}_n)+\frac{R(\bold{\Phi}_{n+1})}{B(\bold{\Phi}_n)}
  \Delta\big(p_{\perp}(\bold{\Phi}_{n+1}),\bold{\Phi}_{n} \big)\;\mathd
  \Phi_{\rm rad}\;\Big]\,.
\label{eq:sigNLOPS}
\end{equation}
The event generated by \POWHEGBOX{} using eq.~\eqref{eq:sigNLOPS} is then
completed by a standard shower MC program, like {\tt Pythia} or {\tt Herwig}.
The core of a shower MC is represented by the PS that generates the
subsequent emissions in the collinear limit. In order not to spoil the NLO
accuracy of the result, a veto algorithm is implemented in such a way that
all the emissions with a transverse momentum larger that the POWHEG one are
discarded. Further details are given in appendix~\ref{sec:interfacePS}.  The SMC
provides also a model for the underlying event, that is generally described
in terms of multiple-parton interactions~(MPI), and for the hadronization.
The \POWHEGBOX{} framework offers the possibility to separate the real
contribution $R$ in two pieces, one containing all the infrared
singularities and one finite for $p_{\perp} \to 0$
\begin{equation}
R = R_s + R_{f}
\end{equation}
with
\begin{equation}
R_s = \frac{h^2}{\pt^2+h^2}R\,, \quad R_{f} =
\frac{\pt^2}{\pt^2+h^2}R\,,
\label{eq:hdamp}
\end{equation}
where the parameter $h$ is called {\tt hdamp} in the POWHEG jargon.
Only $R_s$ is used in the computation of eq.~\eqref{eq:sigNLOPS},
while the $R_f$ contribution, that corresponds to high transverse
momentum radiation, is evaluated separately and it is employed to
generate Born-like events called \emph{remnants}.  It is
also possible to have a process that admits real corrections
that are not associated with any singular region. Those contributions
are handled separately in an analogous way to the \emph{remnants}.
The POWHEG formula thus becomes
\begin{eqnarray}
\mathd \sigma &=&\overline{B}(\bold{\Phi}_n)\;\mathd \bold{\Phi}_n
\Big[ \Delta(p_{\perp,\rm min},
  \bold{\Phi}_n)+\frac{R_s(\bold{\Phi}_{n+1})}{B(\bold{\Phi}_n)}
  \Delta\big(p_{\perp}(\bold{\Phi}_{n+1}),\bold{\Phi}_n\big)\;\mathd
  \Phi_{\rm rad}\;\Big] \nonumber\\ &&+ R_{f}(\bold{\Phi}_{n+1})\,
\mathd \bold{\Phi}_{n+1}\,,
\end{eqnarray}
where $\overline{B}$ and $\Delta$ are evaluated only using $R_s$ and
$R_f$ collects the either the \emph{remnants} or the finite real
processes.

In the standard POWHEG formalism, radiation is generated using the Sudakov
form factor defined in eq.~\ref{eq:PowhegSudakov} and it contains the ratio
$R(\bold{\Phi}_{n+1})/B(\bold{\Phi}_n)$. If the resonance virtualities are
not the same when building the \emph{underlying} Born phase space, large
ratios that badly violate the collinear approximation can arise.  The
improved version of the FKS method implemented in \RES{} overcomes this
problem.

Furthermore, the PS that will complete the event should be instructed
to preserve the mass of the resonances. In the \RES{} framework this
is achieved by providing a resonance assignment to radiation.

To each Born configuration a resonance structure $f_b$ is assigned. The only
contribution $\alpha$ we consider are those where the collinear partons both
arises from the same resonance (the production process is considered as a
resonance). We label with $i$ all the resonances contained in $f_b$ and we
denote by $\alpha_i$ a singular contribution where an emission is originated
from the decay products of $i$. The improved resonance-aware POWHEG formalism
allow us to write
\begin{eqnarray}
  \mathd \sigma& = &\sum_{f_b}
  \overline{B}_{f_b}(\bold{\Phi}_n)\; \mathd \bold{\Phi}_n \prod_{i
    \in f_b}\Bigg[ \Delta_{i}(p_{\perp, \rm min}) \nonumber \\ && +
    \Delta_{i}\left(p_{\perp}(\bold{\Phi}_{n+1}),
    \bold{\Phi}_{n}\right)\sum_{\alpha_i}
    \frac{\left[R_{\alpha_i}(\bold{\Phi}^{\alpha_i}_{n+1})\;\mathd
        \Phi^{\alpha_i}_{\rm
          rad}\right]_{\bold{\Phi}^{\alpha_i}_{n}=\bold{\Phi}_n}}{B_{f_b}(\bold{\Phi}_n)}
    \Bigg]\,.
 \label{eq:NLOPSres}
\end{eqnarray}
Radiation is now generated according to the POWHEG Sudakov form factor both
for the production and for all resonance decays that involve coloured
partons.  This feature also offers the opportunity to modify the standard
POWHEG single-radiation approach: instead of keeping only hardest radiation
from one of all possible origins, the \RES{} can generate simultaneously the
hardest radiation in production and in each resonance decay.  The Les
Houches~(LH) events can thus contain more radiated partons, one for
production and one for each resonance.

Multiple-radiation events have to be completed by a shower MC program, that
has to generate radiation from each origin without exceeding the hardness of
the corresponding POWHEG one, thus requiring an interface that goes beyond
the simple Les Houches standard~\cite{Boos:2001cv}.  The two most widely used
PS programs, {\tt Pythia} and {\tt Herwig}, already implement a veto
algorithm for radiation in production in order to guarantee that the POWHEG
emission is the hardest, while radiation in decay is left, by default,
unrestricted.

\section{NLO+PS matching}
\label{sec:interfacePS}

We briefly describe now how the NLO events generated with \POWHEGBOX{} {(\tt
  RES)} can be completed by the shower MC programs \PythiaEightPtwo{} or
\HerwigSevenPone{}.

\subsection{Interface to shower generators}
According to the POWHEG method, the event needs to be completed by a standard
shower MC program.  The subsequent emissions included by the PS must be
softer than the POWHEG generated one.  In the standard Les Houches Interface
for User Processes~(LHIUP)~\cite{Boos:2001cv}, each generated event has a
hardness parameter associated with it, called {\tt scalup}. The \POWHEGBOX{}
sets this parameter is set equal to the relative transverse momentum of the
generated radiation $\pt$.  By default, radiation in decay is left
unrestricted and the starting scale is set to be of the order of the
resonance virtuality.  The PS preserves the virtuality of the resonance, like
the \RES{} does when generating the real emission, if its decay products have
the resonance as mother particle in the LH event record.

A generic method for interfacing POWHEG processes that include
radiation in decaying resonances with PS generators was introduced in
Refs.~\cite{Campbell:2014kua} and~\cite{Jezo:2016ujg}: radiation from
the top decay products is left unrestricted and when the event is
completed. A veto is applied \emph{a posteriori}: if any radiation
in the decaying resonance shower is found to be harder than the POWHEG
generated one, the event is showered again.

The practical implementation of the veto procedure depends on whether
we are using a dipole, as in \PythiaEightPtwo{}, or an angular-ordered
shower, as in \HerwigSevenPone{}. In the following we describe the
implementation of our veto procedures. Conversely to the method
introduced in Ref.~\cite{Campbell:2014kua}, we do not need the whole
event to be completed before applying the veto, thus saving a lot of
computational time.

\subsection{\PythiaEightPtwo}
\label{app:pythia}
The \PythiaEightPtwo{}~\cite{Sjostrand:2014zea} code implements a
$p_\perp$ ordered shower.  Thus, the matching with \POWHEGBOX{} is
natural because it is enough to require that the starting scale of the
shower evolution is {\tt scalup}, i.e.~the transverse momentum of the
POWHEG emission coming from the production process.

When multiple emissions are concerned, the user can define a starting
scale for each resonance, implementing the virtual functions {\tt
  canSetResonanceScale} and {\tt scaleResonance}. In this way also
emissions from a resonance can have an upper limit defined by the
corresponding POWHEG emission, like it is done by default for
radiation from the production process. This solution was adopted for
example in Ref.~\cite{Jezo:2016ujg}.

However, the $p_{\perp}$ definitions employed by \Pythia{} and by the
\POWHEGBOX{} are slightly different. To overcome this issue, the
\PythiaEightPtwo{} program offers the possibility to use the {\tt
  PowhegHooks} class.  Each time an emission takes place, its transverse
momentum is re-computed with the \POWHEGBOX{} definition: if it is higher
than {\tt scalup}, the emission is discarded.

We have implemented a new class, {\tt PowhegHooksBB4L}, that can be
used for the case of $t\bar{t}$ production and decay.  We rewrote the
{\tt FSREmission} method to guarantee that when an emission is
originated from the production process, the {\tt PowhegHooks}
machinery can be employed\footnote{We have checked that the {\tt
    PowhegHooks} veto in production does not alter significantly the
  results presented in this work, thus we have not adopted it.},
otherwise, when it comes from the $t$ (or the $\bar{t}$) the momenta
of the emitter and of the emitted particles $i$ and $j$ are boosted in
the top frame and the transverse momentum of the radiation is computed
using
\begin{equation}
p^2_\perp = 2 p_b \cdot p_g \frac{E_g}{E_b}\,,
\label{eq:bEmitter}
\end{equation} 
if the emitter is a $b$ (or $\bar{b}$) heavy quark, using
\begin{equation}
p^2_\perp = 2 p_i \cdot p_j \frac{E_i E_j}{(E_i+E_j)^2}\,,
\label{eq:masslessEmitter}
\end{equation} 
otherwise.  If it is harder than the corresponding POWHEG one, this
emission is discarded.  The {\tt FSREmission} veto algorithm
represents our default choice for handling radiation in decay.

\subsection{\HerwigSevenPone}
\label{sec:herwig}
\begin{figure}[t]
\centering
\includegraphics[width=0.65\textwidth]{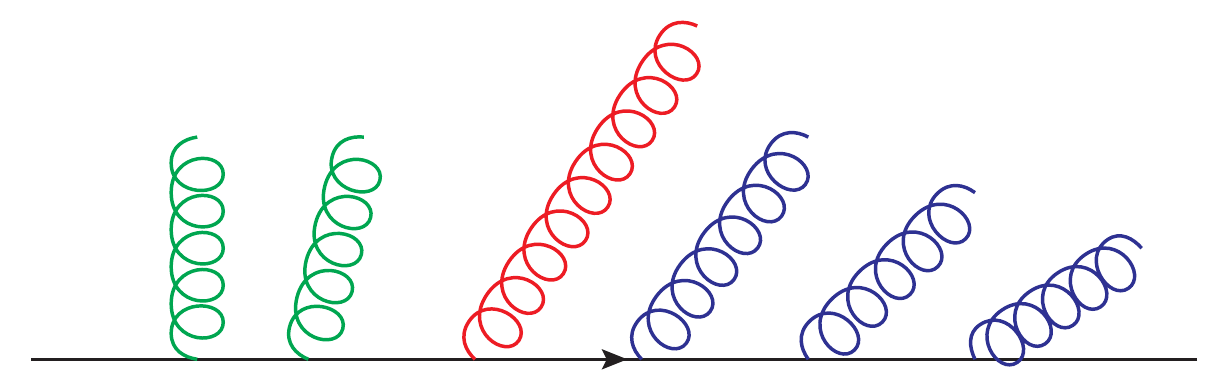}
\caption{Angular ordered PS: in green the truncated-vetoed shower, in
  red the hardest emission and in blue the remaining vetoed shower.}
\label{fig:angularOrderedPS}
\end{figure}

The \HerwigSevenPone{} package~\cite{Bahr:2008pv,Bellm:2015jjp} implements an
angular ordered PS. In the collinear limit, the variable $q$ that
parametrizes the hardness of the emission is given by
\begin{equation}
q \approx E \theta\,,
\end{equation}
with $E$ the energy of the emitter before the emission and $\theta$
the angle between the two emitted particles.  A brief sketch of the
evolution of an angular ordered PS is represented graphically in
Fig.~\ref{fig:angularOrderedPS}.  The initial stages are characterized
by large-angle soft radiation~(green), while the hardest emission,
i.e.~the one with largest $p_{\perp}$~(red), may appear later.

The procedure on how POWHEG-style NLO emissions can be matched with an
angular ordered shower is discussed in detail in
Ref.~\cite{Nason:2004rx}. Here we give only a brief summary.  The hardest
emission originated from a parton $i$ satisfies the following properties:
\begin{itemize}
\item it can be found by following the hardest line each time a
  branching takes place;
\item if $i$ is a gluon, the probability that the hardest emission
  takes place after a $g\to q\bar{q}$ can be neglected, since this
  splitting is not soft-singular;
\item if $i$ is a quark, following the hardest line coincides with following
  the quark line since the probability that a gluon carries the main part of
  the energy after a $q \to q g$ splitting is power suppressed.
\end{itemize}
When matching an event generated with \POWHEGBOX{} with an angular ordered
PS, the scale $q$ of the POWHEG emission should be computed.  The two partons
corresponding to the POWHEG emission should be merged and an angular ordered
PS associated with the paired parton, starting from $q_0$, the maximum
allowed scale, down to $q$, should be implemented. All the emissions with a
transverse momentum larger than {\tt scalup}, i.e. the $p_{\perp}$ of the
POWHEG emission, must be vetoed. This shower is called truncated-vetoed
shower and corresponds to the green contribution portrayed in
Fig~\ref{fig:angularOrderedPS}.  When the scale $q$ is reached, all the
subsequent vetoed showers (blue emissions in Fig.~\ref{fig:angularOrderedPS})
are implemented.  Since truncated-vetoed showers are known to give rise only
to a little contribution\footnote{In the POWHEG simulations where truncated
  showers have been implemented and studied~\cite{LatundeDada:2006gx,Hamilton:2008pd,Hamilton:2009za,Hamilton:2009ne,Hamilton:2010mb}
  their effect was found to be basically negligible.}, \HerwigSevenPone{}
neglects them by default, simply requiring that all the emissions have a
transverse momentum smaller than {\tt scalup}.

However, by doing so, all the large angle emissions with a scale
larger than $q$ would be missing. The veto technique introduced in
Ref.~\cite{Schofield:2011zi} and activated in \HerwigSevenPone{} with
the settings
\begin{equation}
  \label{eq:TSsettings}
  \begin{split}
& \mbox{\tt set PartnerFinder:PartnerMethod Maximum} \\ & \mbox{\tt
      set PartnerFinder:ScaleChoice Different}
  \end{split}
\end{equation}
performs a task equivalent to the implementation of truncated showers.  For
example if a $q\to qg$ splitting is the POWHEG hardest emission, the initial
angle for radiation from the gluon is taken as the maximum angle between the
gluon and its two colour partners, i.e.~the incoming and the outgoing quark
$q$. This leads to unrestricted radiation from the gluon.  However, the
colour factor $C_A$ associated with this radiation is reduced by a factor of
two if the scale of the emission $t^\prime$ is larger than the POWHEG one
$t$, while it is restored to $C_A$ for smaller angles. Since $C_A/2 \approx
C_F$ in the large $N_c$ limit, we see that this is equivalent to the
inclusion of a truncated-vetoed shower from the incoming $q$ quark down to a
scale $t$.  If not specified, in this work the settings of
eq.~(\ref{eq:TSsettings}) have not been used.

The \HerwigSevenPone{} code offers two alternative solutions for the
implementation of a veto algorithm that can act on the emissions coming from
resonances. The two methods are equivalent at the leading-logarithmic
accuracy level, i.e. at all orders in powers of $\as \log^2(p_{\perp})$.

It is possible to implement the virtual function {\tt vetoTimeLike} belonging
to the {\tt ShowerVeto} class. We have built the {\tt BB4LShowerVeto} class
that inherits the {\tt vetoTimeLike} method from {\tt ShowerVeto}. Each time
an emission that has a $t$ or a $\bar{t}$ quark as shower progenitor takes
place, it is rejected if its transverse momentum is larger than the one of
the corresponding POWHEG emission.  This implementation of the veto is the
analogous of the default behaviour employed to treat radiation from the
production process.  The drawback of this method is that we are unable to
compute the $p_{\perp}$ of the emission we are inspecting using
eqs.~(\ref{eq:bEmitter}) and~(\ref{eq:masslessEmitter}), since the momenta of
the emitted particles have not been generated yet, so we must rely on the
\HerwigSevenPone{} definition of transverse momentum.  However, the two
definitions are the same at the leading-logarithmic accuracy level.

An alternative is represented by the virtual function {\tt vetoShower}, which
belongs to the {\tt FullShowerVeto} class.  As the name of the class
suggests, this veto is applied at the end of the full showering phase, before
the hadronization takes place.  In our implementation of the {\tt
  BB4LFullShowerVeto} class, the hardest emissions of the showers initiated
by the coloured (anti-)top decay products, i.e. the (anti-)bottom and
eventually a gluon, is searched. If we encounter an emission whose transverse
momentum is larger than the corresponding POWHEG one, the event is showered
again.  This method enables us to use the POWHEG $p_{\perp}$ definition,
since we have access to the momenta of the particles. However, since some
momentum reshuffling must be applied to keep the external particles on their
mass-shell after a $1\to2$ branching, the transverse momentum of the
intermediate emissions turns out to be computed with off-shell partons. This
has an impact only beyond the leading-logarithmic accuracy level.

If not specified, we adopted {\tt BB4LShowerVeto} to perform the veto.

\chapter{The treatment of remnants for multiple-emission processes}
\label{app:remnants}
As we have seen in \writeApp{}\ref{app:powhegMethod}, in POWHEG it is
possible to separate the real cross section, in a given singular
region $\alpha$, into two contributions $R^\alpha_s$ and $R^\alpha_f$,
where $R^{\alpha}_f$ does not contain any singularities, while
$R^{\alpha}_s $ is singular.  Only $R^{\alpha}_s$ is exponentiated in
the Sudakov form factor and used for the computation of $\tilde{B}$,
while the leftover $R^{\alpha}_f$, dubbed the remnant contribution, is
finite upon phase space integration~\cite{Nason:2004rx}.

In all our three NLO generators it is possible to achieve this
separation for initial-state radiation~(ISR) emissions by setting the
parameter {\tt hdamp}\footnote{We used an {\tt hdamp} value equal to
  the input top-quark mass, i.e.~the {\tt qmass} parameter for the
  \hvq{} generator, {\tt tmass} for \bbfourl{} and \ttbnlodec{}.  } in
the {\tt powheg.input} file.  Denoting with $\aprod$ the production
region, eq.~(\ref{eq:hdamp}) tells us that $R^{\aprod}_s$ and
$R^{\aprod}_f$ are defined as
\begin{align}
R^{\aprod}_s = & \frac{({\tt hdamp})^2}{({\tt hdamp})^2+(\pt^{\aprod})^2}R^{\aprod}\,,
\\ \;\;\; R^{\aprod}_f = &
\frac{(\pt^{\aprod})^2}{({\tt hdamp})^2+(\pt^{\aprod})^2}R^{\aprod}\,,
\end{align}
where $\pt^{\aprod}$ is the transverse momentum of the emitted parton
relative to the beam axis.  The {\tt scalup} variable contained in the
Les Houches event, that is used by the parton shower program to veto
emissions harder than the \POWHEG{} one, is set equal to
$\pt^{\aprod}$.

Since remnant events are non-singular, the associated radiation has
transverse momenta of the order of the partonic center-of-mass energy.
We can thus define {\tt scalup} as
\begin{equation}
{\tt scalup} = \frac{\hat{s}}{2}\,.
\end{equation}
We have checked that, by using as {\tt scalup} the default \POWHEG{}
scale (i.e.~the transverse momentum of the radiated parton) the
$\mwbjmax{}$ and the $\Ebjmax$ values are very close to the ones we have
presented in this paper. This is consistent with the expectation that
these observables should be relatively insensitive to radiation in
production, that in our case is always treated as ISR.  The same holds
for the leptonic observable $m(\ell^+\ell^-)$.  For the remaining
ones, a higher sensitivity to ISR effects is not excluded, and in fact
the differences of the first Mellin moments reported in
Tab.~\ref{tab:Olep-summary-py} with the corresponding ones obtained
with the default {\tt scalup} value, for the \hvq{} generator showered
with \PythiaEightPtwo{}, are given by
\begin{equation}
  \begin{split}
  \Delta \langle\pt(\ell^+) \rangle & = 125\pm 46~{\rm MeV}\,,
  \\ \Delta \langle\pt(\ell^++\ell^-) \rangle & = 298\pm 54~{\rm
    MeV}\,, \\ \Delta \langle E(\ell^+ \ell^-) \rangle & = 214\pm
  149~{\rm MeV}\,, \\ \Delta \langle\pt(\ell^+)+\pt(\ell^-) \rangle &
  = 219\pm 87~{\rm MeV}\,.
  \end{split}
\end{equation}
In comparison with Tab.~\ref{tab:Olep-summary-py}, we see that these
variations are of the same order or smaller than those arising from
scale and PDF uncertainties.

In the \bbfourl{} code, when ISR remnants are generated, no radiation
in decay is produced.\footnote{This behaviour may be changed in the
  future.} Thus, in this case, radiation off the resonances is fully
handled by the parton shower, without the use of a veto algorithm to
limit the $\pt$ of the radiated partons.

The \ttbnlodec{} generator does instead implement radiation in decay
also for remnants, and thus in this case vetoing is performed as for
the standard events.

The absence of emissions from the $t$ and $\bar{t}$ resonances in
remnant events for the \bbfourl{} generator, in contrast with the
\ttbnlodec{} one, is probably the reason why the former generator
displays a slightly larger sensitivity to matrix-element
corrections~(see Tabs.~\ref{tab:mwbj_MEC}, \ref{tab:Ebj_MEC}
and~\ref{tab:leptobs_MEC}).

To summarize:
\begin{itemize}
\item \hvq{}: Emissions in decay are never vetoed. For remnant events
  the {\tt scalup} value used to limit radiation in production is set
  to $\sqrt{\hat{s}}/2$.
\item \ttbnlodec{}: Emissions in decay are always vetoed. For remnant
  events the {\tt scalup} value is set to $\sqrt{\hat{s}}/2$.
\item \bbfourl{}: Emissions in decay are always vetoed except if the
  event is a remnant, in which case they are never vetoed. For remnant
  events the {\tt scalup} value is set to $\sqrt{\hat{s}}/2$.
\end{itemize}

\section{Sensitivity of the results on {\tt hdamp}}
\label{app:hdamp}
For our top-mass studied we have choosen
\begin{equation}
  {\tt hdamp} = m_t,
\label{eq:hdamp1}
\end{equation}
being the top mass $m_t$ a scale of our hard process. However, since we focus on
top-pair production, also
\begin{equation}
  {\tt hdamp} = 2m_t
\label{eq:hdamp2}
\end{equation}
seems an equally natural choice. We thus investigate the sensitivity of our
result on the value of {\tt hdamp}. Sice this parameter affects only the
treatment of ISR, we restricted our comparison to the \hvq{} generator, on
the ground that all the generators under analysis describe ISR with the same
accuracy.  We thus generated a sample of $10^6$ events with the setting in
eq.~(\ref{eq:hdamp2}).  The impact of the two choices in
eqs.~(\ref{eq:hdamp1}, \ref{eq:hdamp2}) is shown in
Tab.~\ref{tab:delta_hdamp}.

We notice that all the observables suggest a $m_t$ value larger for the ${\tt
  hdamp}= m_t$ case ($\Delta m_t = - \Delta O / B$), apart from the energy
of the \bjet{} and the mass of the lepton pair where, however, the standard
deviation has the same size of the difference.  The discrepancy between the
two alternatives is in general very small, in particular if we consider the
reconstructed mass, where the two predictions differ by 3-4~MeV.  The energy
of the lepton pair corresponds to a mass shift of $0.45$~GeV, but with an
error of $0.33$~GeV.  Given our statistical error, the only observable that
shows a non-negligible difference is the transverse momentum of the lepton
pair, where we find a $m_t$ shift of $0.6\pm 0.2$~GeV.
In any case, the variations produced by the different choices of {\tt hdamp}
are smaller than those induced by scale and PDF variations (see Tab.~\ref{tab:Olep-summary-py}).

\begin{table}[hb]
\centering  \begin{tabular}{|c|c|c|c|}
\cline{1-4}
\multicolumn{4}{|c|}{ \hvq{}, pwgveto=0, NLO+PS, GeV} \\
\cline{1-4}
\phantom{\Big|} observable & ${\tt hdamp_1}=m_t$ & ${\tt hdamp_2}=2m_t$ &
${\tt hdamp_1}-{\tt hdamp_2} $ \\
\cline{1-4}
$\phantom{\Big|} \mwbjmax$ no smearing                                & $172.498\pm  0.001$& $172.501\pm  0.002$& $ -0.003\pm  0.002$\\ \cline{1-4}
$\phantom{\Big|} \mwbjmax$ smearing                                   & $171.315\pm  0.001$& $171.319\pm  0.002$& $ -0.004\pm  0.002$\\ \cline{1-4}
$\phantom{\Big|} \Ebjmax$                                             & $  69.36\pm   0.06$& $  69.24\pm   0.09$& $   +0.12\pm   0.11$\\ \cline{1-4}
$\phantom{\Big|} \langle \pT(\ell^+)\rangle $                         & $  56.68\pm   0.03$& $  56.71\pm   0.04$& $  -0.03\pm   0.06$\\ \cline{1-4}
$\phantom{\Big|} \langle \pT(\ell^+\ell^-)\rangle $                   & $  69.16\pm   0.04$& $  69.35\pm   0.05$& $  -0.19\pm   0.07$\\ \cline{1-4}
$\phantom{\Big|} \langle m(\ell^+\ell^-)\rangle $                     & $ 109.07\pm   0.06$& $ 109.00\pm   0.09$& $   +0.07\pm   0.11$\\ \cline{1-4}
$\phantom{\Big|} \langle E(\ell^+\ell^-)\rangle $                     & $ 186.70\pm   0.11$& $ 186.95\pm   0.14$& $  -0.25\pm   0.18$\\ \cline{1-4}
$\phantom{\Big|} \langle \pT(\ell^+)+\pT(\ell^-)\rangle $             & $ 113.30\pm   0.06$& $ 113.38\pm   0.08$& $  -0.08\pm   0.10$\\ \cline{1-4}
 \end{tabular}

\caption{Differences between the leptonic observables obtained using
  the {\tt hdamp}$=m_t$ and the {\tt hdamp}$=2 m_t$  settings for
  the \hvq{} generator interfaced with \PythiaEightPtwo. The results shown
  are at the NLO+PS level, i.e. without the inclusion of the hadronization
  and of the underlying event.}
\label{tab:delta_hdamp}
\end{table}

\clearpage
\let\capmark\chaptermark
\renewcommand{\capmark}[1]{\markboth{#1}{}}
\addcontentsline{toc}{chapter}{Bibliography}
\capmark{Bibliography}

\providecommand{\href}[2]{#2}\begingroup\raggedright\endgroup

\end{document}